\documentclass[a4paper,18pt]{extarticle} 
\usepackage{geometry}                		
\usepackage{graphicx}					
\usepackage{amssymb}
\usepackage{indentfirst}
\usepackage{amsmath}
\usepackage{amsthm}
\usepackage{siunitx}
\usepackage{lscape}
\usepackage{subcaption}

\usepackage[linesnumbered,ruled]{algorithm2e}

\newcommand{\E}{\mathrm{E}}
\newcommand{\Var}{\mathrm{Var}}

\newcommand{\iid}{\textrm{i.i.d.}}
\providecommand{\keywords}[1]{\textbf{\textit{Keywords: }} #1}

\title{Adaptive Sequential MCMC for Combined State and Parameter Estimation} %

\author{Zhanglong Cao, David Bryant, Matthew Parry}

\begin{document}
\maketitle

\begin{abstract}
In the case of a linear state space model, we implement an MCMC sampler with two phases. In the learning phase, a self-tuning sampler is used to learn the parameter mean and covariance structure. In the estimation phase, the parameter mean and covariance structure informs the proposal mechanism and is also used in a delayed-acceptance algorithm. Information on the resulting state of the system is given by a Gaussian mixture. In on-line mode, the algorithm is adaptive and uses a sliding window approach  to accelerate sampling speed and to maintain appropriate acceptance rates. We apply the algorithm to joined state and parameter estimation in the case of irregularly sampled GPS time series data. 

\end{abstract}


\section{Introduction}

Data assimilation is a sequential process, by which the observations are incorporated into a numerical model describing the evolution of this system throughout the whole process. It is applied in many fields, particularly in weather forecasting and hydrology. The quality of the numerical model determines the accuracy of this system, which requires sequential combined state and parameters inferences. An enormous literature has been done on discussing pure state estimation, however, less research is talking about estimating combined state and parameters, particularly in a sequential updating way. 

Sequential Monte Carlo method is well studied in the scientific literature and quite prevalent in academic research in the last decades. It allows us to specify complex, non-linear time series patterns and enables performing real-time Bayesian estimations when it is coupled with Dynamic Generalized Linear Models \cite{vieira2016online}. However, model's parameters are unknown in real-world application and it is a limit for standard SMC. Extensions to this algorithm have been done by researchers. Kitagawa \cite{kitagawa1998self} proposed a self-organizing filter and augmenting the state vector with unknown parameters. The state and parameters are estimated simultaneously by either a non-Gaussian filter or a particle filter. Liu and West \cite{liu2001combined} proposed an improved particle filter to kill degeneracy, which is a normal issue in static parameters estimation. They are using a kernel smoothing approximation, with a correction factor to account for over-dispersion. Alternatively, Storvik \cite{storvik2002particle} proposed a new filter algorithm by assuming the posterior depends on a set of sufficient statistics, which can be updated recursively. However, this approach only applies to parameters with conjugate priors \cite{stroud2016bayesian}. Particle learning was first introduced in \cite{carvalho2010particle}. Unlike Storvik filter, it is using sufficient statistics solely to estimate parameters and promises to reduce particle impoverishment. These particle-like methods are all using more or less sampling and resampling algorithms to update particles recursively. 

Jonathan proposed in \cite{stroud2016bayesian} an SMC algorithm by using ensemble Kalman filter framework for high dimensional space models with observations. Their approach combines information about the parameters from data at different time points in a formal way using Bayesian updating. In \cite{polson2008practical}, the authors rely on a fixed-lag length of data approximation to filtering and sequential parameter learning in a general dynamic state-space model. This approach allows for sequential parameter learning where importance sampling has difficulties and avoids degeneracies in particle filtering. A new adaptive Markov Chain Monte Carlo method yields a quick and flexible way for estimating posterior distribution in parameter estimation \cite{haario1999adaptive}. This new Adaptive Proposal method depends on historical data, is introduced to avoid the difficulties of tunning the proposal distribution in Metropolis-Hastings methods.

In this chapter, I'm proposing an adaptive Delayed-Acceptance Metropolis-Hastings algorithm to estimate the posterior distribution for combined state and parameters with two phases. In the learning phase, a self-tuning random walk Metropolis-Hastings sampler is used to learn the parameter mean and covariance structure. In the estimation phase, the parameter mean and covariance structure informs the proposed mechanism and is also used in a delayed-acceptance algorithm, which greatly improves sampling efficiency. Information on the resulting state of the system is given by a Gaussian mixture. To keep the algorithm a higher computing efficiency for on-line estimation, it is suggested to cut off historical data and to use a fixed length of data up to the current state, like a window sliding along time. At the end of this chapter, an application of this algorithm on irregularly sampled GPS time series data is presented. 

%

\section{Bayesian Inference on Combined State and Parameters}

In a general state-space model of the following form, either the forward map $F$ in hidden states or the observation transition matrix $G$ is linear or non-linear. We are considering the model 
\begin{align}\label{obserY}
\mbox{Observation:}\hspace*{0.3cm}   & y_t=G(x_t,\theta), \\
\mbox{Hidden State:}\hspace*{0.3cm} & x_t=F(x_{t-1},\theta),\label{hiddX}
\end{align}
where $G$ and $F$ are linear processes with Gaussian white noises $\epsilon\sim N(0,R(\theta))$ and $\epsilon'\sim N(0,Q(\theta))$. This model has an initial state $p(x_0\mid \theta)$ and a prior distribution of the parameter $p(\theta)$ is known or can be estimated. Therefore, for a general Bayesian filtering problem with known static parameter $\theta$, it requires computing the posterior distribution of current state $p(x_t \mid y_{1:t})$ at each time $t=1,\dots, T$ by marginalizing the previous state
\begin{equation*}
p(x_t\mid y_{1:t}) = \int p(x_t\mid x_{t-1},y_{1:t})p(x_{t-1}\mid y_{1:t}) dx_{t-1}, 
\end{equation*}
where $y_{1:t} = \{y_1,\dots,y_t\}$ is the observation information up to time $t$. However, if $\theta$ is unknown, one has to marginalize the posterior distribution for parameter by 
\begin{align}\label{objecfun}
p(x_t \mid y_{1:t}) = \int p(x_t \mid y_{1:t},\theta)p(\theta\mid y_{1:t})d\theta.
\end{align}
The approach in equation (\ref{objecfun}) relies on the two terms : (i) a conditional posterior distribution for the states given parameters and observations; (ii) a marginal posterior distribution for parameter $\theta$. Several methods can be used in finding the second term, such as cross validation, Expectation Maximization algorithm, Gibbs sampling, Metropolis-Hastings algorithm and so on. A Monte Carlo method is popular in research area solving this problem. Monte Carlo method is an algorithm that relies on repeated random sampling to obtain numerical results. To compute an integration of $\int f(x)dx$, one has to sampling as many independent $x_i \mbox{ } (i = 1,\dots, N)$ as possible and numerically to find $\frac{1}{N}\sum_i f(x_i)$ to approximate the target function. In the target function, we draw samples of $\theta$ and use a numerical way to calculate its posterior $p(\theta\mid y_{1:t})$.

Additionally, the marginal posterior distribution for the parameter can be written in two different ways: 
\begin{align}\label{M1}
p(\theta \mid y_{1:t}) &\propto p(y_{1:t}\mid\theta)p(\theta),\\
p(\theta \mid y_{1:t}) &\propto p(y_t\mid y_{1:t-1}, \theta)p(\theta\mid y_{1:t-1}). \label{M2}
\end{align}
The above formula (\ref{M1}) is a standard Bayesian inference requiring a prior distribution $p(\theta)$. It can be used in off-line methods, in which $\hat{\theta}$ is inferred by iterating over a fixed observation record $y_{1:t}$. In contrast, formula (\ref{M2}) is defined in a recursive way over time depending on the previous posterior at time $t-1$, which is known as on-line method. $\hat{\theta}$ is estimated sequentially as a new observation $y_{t+1}$ becomes available.

Therefore, the question becomes finding an efficient way to sampling $\theta$, such as Importance sampling \cite{hammersley1964percolation} \cite{geweke1989bayesian}, Rejection sampling \cite{casella2004generalized} \cite{martino2010generalized}, Gibbs sampling \cite{geman1984stochastic}, Metropolis-Hastings method \cite{metropolis1953equation} \cite{hastings1970monte} and so on. 

\subsection{Log-likelihood Function of Parameter Posterior}\label{sectionlogParameter}

To sample $\theta$, firstly we should find its distribution function by starting from the joint covariance matrix of $x_{0:t}$ and $y_{1:t}$. With a given $\theta$, suppose the joint covariance matrix is in the form of 
\begin{equation}\label{generaljointmatrix}
\begin{bmatrix} \begin{matrix} x_{1:t}\\ y_{1:t}  \end{matrix} \biggr\rvert \theta \end{bmatrix}
\sim N\left(0, \Sigma_t \right),
\end{equation}
where $x_{1:t}$ represents the hidden states $\{x_0,x_1,\dots,x_t\}$, $y_{1:t}$ represents observed $\{y_1,\dots,y_t\}$ and $\theta$ is the set of all known and unknown parameters. The inverse of the covariance matrix $\Sigma_t^{-1}$ is the procedure matrix. In fact, it is a block matrix in the form of 
\begin{align*} \Sigma_t^{-1}=
\begin{bmatrix}
A_t& -B_t \\ -B_t^\top & B_t
\end{bmatrix}, 
\end{align*}
where $A_t$ is a $t \times t$ matrix of forward map hidden states, $B_t$ is a $t\times t$ matrix of observation errors up to time $t$. The structure of the matrices, such as bandwidth, sparse density, depending on the structure of the model. Temporally, we are using $A$ and $B$ to represent the matrices  $A_t$ and $B_t$ here. Then we may find the covariance matrix easily by calculating the inverse of the procedure matrix 
\begin{align*}
\Sigma &= \begin{bmatrix}
(A-B^\top B^{-1}B) ^{-1} & -(A-B^\top B^{-1}B)^{-1}B^\top B^{-1}\\
- B^{-1}B(A-B^\top B^{-1}B)^{-1} & (B-B^\top A^{-1}B) ^{-1}
\end{bmatrix} \\
&= \begin{bmatrix}
(A-B) ^{-1} & (A-B)^{-1}\\
(A-B)^{-1} & (I- A^{-1}B) ^{-1}B^{-1}
\end{bmatrix} \\
&\triangleq \begin{bmatrix}
\Sigma_{XX} & \Sigma_{XY} \\
\Sigma_{YX}  &\Sigma_{YY} 
\end{bmatrix}.
\end{align*}
Because of the covariance  $\Sigma_{YY} =  (I-A^{-1}B)^{-1}B^{-1}$, therefore the inverse is 
\begin{align*}
\Sigma_{YY}^{-1} &= B(I-A^{-1}B)= BA^{-1}\Sigma_{XX}^{-1}.
\end{align*}
Given the Choleski decomposition $LL^\top = A$, we have
\begin{align*}
\Sigma_{YY}^{-1} &=BL^{-\top}L^{-1}\Sigma_{XX}^{-1}\\
&=(L^{-1}B)^\top(L^{-1}\Sigma_{XX}^{-1}) 
\end{align*}
More usefully, by given another Choleski decomposition $RR^\top=A-B=\Sigma_{XX}^{-1}$,
\begin{align}\label{sigmayy01}
\begin{split}
y_{1:t}^\top \Sigma_{YY}^{-1} y_{1:t} &= (L^{-1}By_{1:t})^\top(L^{-1}\Sigma_{XX}^{-1}y_{1:t})\\
&\triangleq W^\top (L^{-1}\Sigma_{XX}^{-1}y_{1:t})\\
\end{split}
\end{align}
\begin{align}\label{sigmayy02}
\begin{split}
\det\Sigma_{YY}^{-1} &= \det B \det L^{-\top}\det L^{-1}\det R\det R^\top\\
&= \det B(\det L^{-1})^2(\det R)^2.
\end{split}
\end{align}
From the objective function (\ref{M1}), the posterior distribution of $\theta$ is 
\begin{align*}
p(\theta \mid y_{1:t}) &\propto p(y_{1:t}\mid\theta)p(\theta) \propto e^{-\frac{1}{2} y_{1:t} \Sigma_{YY}^{-1} y_{1:t} } \sqrt{\det \Sigma_{YY}^{-1}} p(\theta).
\end{align*}
Then by taking natural logarithm on the posterior of $\theta$ and using the useful solutions in equations (\ref{sigmayy01}) and (\ref{sigmayy02}), we will have
\begin{align}
\ln L(\theta) &= -\frac{1}{2}y_{1:t}^\top\Sigma_{YY}^{-1}y_{1:t}+\frac{1}{2}\sum\ln\mbox{tr}(B)-\sum\ln\mbox{tr}(L)+\sum\ln\mbox{tr}(R) + \ln p(\theta).
\end{align}

\subsection{The Forecast Distribution}\label{sectionforecast}

From equation (\ref{M2}), a sequential way for estimating the forecast distribution is needed. Suppose it is 
\begin{equation}
p(y_{t}\mid y_{1:t-1},\theta) \sim N\left( \bar{\mu}_{t},\bar{\sigma}_{t} \right). 
\end{equation}
Look back to the covariance matrices of observations that we found in the previous section 
\begin{align*}
p(y_{1:t-1},\theta) &\sim N\left( 0,\Sigma_{YY}^{(t-1)} \right),\\
p(y_{t},y_{1:t-1},\theta) &\sim N\left( 0,\Sigma_{YY}^{(t)} \right),
\end{align*}
where the covariance matrix of the joint distribution is $\Sigma_{YY}^{(t)} = (I_{t}-A_{t}^{-1}B_{t})^{-1}B_{t}^{-1}$, $I_t$ is a $t\times t$ identity matrix. Then, by taking its inverse, we will get 
\begin{align*}
\Sigma_{YY}^{(t) (-1)} &= B_{t}(I_{t}-A_{t}^{-1}B_{t}) \\
&= B_{t}(B_{t}^{-1}-A_{t}^{-1})B_{t} \\
&\triangleq \begin{bmatrix} 
B_t & 0 \\ 0 & B_1 \end{bmatrix}
\begin{bmatrix} 
Z_{t} & b_{t} \\
b_{t}^\top & K_{t}
\end{bmatrix} \begin{bmatrix} 
B_t & 0 \\ 0 & B_1\end{bmatrix}
\end{align*}
where $Z_{t}$ is a $t \times t$ matrix, $ b_{t} $ is a $t \times 1$ matrix and $K_{t}$ is a $1 \times 1$ matrix. Thus by taking its inverse again, we will get 
\begin{align*} \Sigma_{YY}^{(t)}= \left[ \begin{matrix}
B_t^{-1} (Z_{t}-b_{t}K_{t}^{-1}b_{t}^\top)^{-1}B_t^{-1}  & - B_t^{-1}  Z_{t}^{-1}b_{t}(K_{t}-b_{t}^\top Z_{t}^{-1}b_{t})^{-1}B_1^{-1} \\
-B_1^{-1}  K_{t}^{-1}b_{t}^\top (Z_{t}-b_{t}K_{t}^{-1}b_{t}^\top)^{-1}B_t^{-1}  & B_1^{-1}  (K_{t}-b_{t}^\top Z_{t}^{-1}b_{t})^{-1}B_1^{-1} 
\end{matrix}\right].
\end{align*}
So, from the above covariance matrix, we can find the mean and variance of $p(y_{t}\mid y_{1:t-1},\theta)$ are 
\begin{align}
\bar{\mu}_{t} & =  B_1^{-1}K_{t}^{-1}b_{t}^\top B_{t-1}^{-1}y_{1:t-1} ,\\
\bar{\sigma}_{t} & =  B_1^{-1}K_{t}B_1^{-1}  .
\end{align}

\subsection{The Estimation Distribution}\label{generalEstDistr}

From the joint distribution (\ref{generaljointmatrix}), one can find the best estimation with a given $\theta$ by
\begin{align*}
\hat{x}_{1:t} \mid y_{1:t},\theta &\sim N \left( A_{t}^{-1}B_{t}y_{1:t}, A_{t}^{-1} \right) \\
&\sim N(L^{-\top}L^{-1}B_{t}y_{1:t-1},L^{-\top}L^{-1})\\
&\sim N(L^{-\top}W,L^{-\top}L^{-1}).
\end{align*}
Consequently 
\begin{align*}
\hat{x}_{1:t} = L^{-\top}(W+Z),
\end{align*}
where $Z \sim N(0, I(\epsilon))$ is independent and identically distributed and drawn from a zero-mean normal distribution with variance $ I(\epsilon)$. 

For sole $x_{t}$, its joint distribution with $y_{1:t}$ is 
\begin{align*}
x_{t}, y_{1:t}\mid \theta \sim N\left( 0, \begin{bmatrix}
C_{t}^\top(A_{t}-B_{t}) ^{-1}C_{t} & C_{t}^\top (A_{t}-B_{t})^{-1}\\
(A_{t}-B_{t})^{-1}C_{t} & (I- A_{t}^{-1}B_{t}) ^{-1}B_{t}^{-1}
\end{bmatrix} \right),
\end{align*}
where $C_t^\top = \begin{bmatrix}0 & \cdots & 0 & 1\end{bmatrix}$ helps to  achieve the last element in the matrix. Thus the filtering distribution of the state is 
\begin{align*}
p(x_{t}\mid y_{1:t},\theta) \sim N\left( \mu_{t}^{(x)},\Var(x_{t}) \right),
\end{align*}
where, after simplifying, the mean and variance are  
\begin{align}\label{generalmux}
\mu_{t}^{(x)} & = C_{t}^\top A_{t}^{-1}B_{t}y_{1:t} ,\\
\Var(x_{t})& =C_{t}^\top A_{t}^{-1}C_{t}. \label{generalSigx}
\end{align}

Generally, researchers would like to find the combined estimation for $x_t$ and $\theta$ at time $t$ by
\begin{equation*}
p(x_t, \theta \mid y_{1:t}) = p(x_t\mid y_{1:t},\theta)p(\theta\mid y_{1:t}).
\end{equation*}
Differently, from the target equation (\ref{objecfun}), the state inference containing $N$ samples is a mixture Gaussian distribution in the following form 
\begin{equation}\label{mixtureGaussian}
p(x_t \mid y_{1:t}) = \int p(x_t\mid y_{1:t},\theta) p(\theta\mid y_{1:t})d\theta \dot{=} \frac{1}{N}\sum_{i=1}^{N}p(x_{t}\mid\theta^{(i)},y_{1:t}). 
\end{equation}
Suppose $x_t\mid y_{1:t},\theta_i \sim N\left( \mu_{ti}^{(x)},\Var(x_t)_i \right)$ is found from equation (\ref{generalmux}) and (\ref{generalSigx}) for each $\theta_i$, then its mean is 
\begin{equation}\label{mixturemean}
\mu_t^{(x)} = \frac{1}{N} \sum_i \mu_{ti}^{(x)} 
\end{equation}
and  the unconditional variance of $x_t$, by law of total variance, is 
\begin{equation}\label{mixturevariance}
\begin{split}
\Var(x_t) &= \E(\Var(x_t\mid y_{1:t},\theta)) + \Var(\E(x_t\mid y_{1:t},\theta))   \\
&= \frac{1}{N} \sum_i \left( \mu_{ti}^{(x)}  \mu_{ti}^{(x)\top} +\Var(x_t)_i\right) -\frac{1}{N^2} \left(  \sum_i  \mu_{ti}^{(x)} \right) \left( \sum_i \mu_{ti}^{(x)} \right) ^\top.
\end{split}
\end{equation}

\section{Random Walk Metropolis-Hastings algorithm}

Metropolis-Hastings algorithm is an important class of Markov Chain Monte Carlo (MCMC) algorithms \cite{smith1993bayesian} \cite{tierney1994markov} \cite{gilks1995markov}.  This algorithm has been used extensively in physics but was little known to others until M\"{u}ller \cite{muller1991generic} and Tierney \cite{tierney1994markov} expounded the value of this algorithm to statisticians. The algorithm is extremely powerful and versatile and has been included in a list of "The Top 10 Algorithms"  with the greatest influence on the development and practice of science and engineering in the 20th century \cite{dongarra2000guest} \cite{medova2008bayesian}. 

Given essentially a probability distribution $\pi$ (the "target distribution"), MH algorithm provides a way to generate a Markov Chain $x_1, x_2,\ldots, x_t$, who has the target distribution as a stationary distribution, for the uncertain parameters $x$ requiring only that this density can be calculated at $x$. Suppose that we can evaluate $\pi(x)$ for any $x$. The transition probabilities should satisfy the detailed balance condition
\begin{equation*}
\pi(x^{(t)})q(x', x^{(t)}) = \pi(x')q(x^{(t)}, x'),
\end{equation*}
which means that the transition from the current state $\pi(x^{(t)})$ to the new state $\pi(x')$ has the same probability as that 
from $\pi(x')$ to $\pi(x^{(t)})$. In sampling method, drawing $x_i$ first and then drawing $x_j$ should have the same probability as drawing $x_j$ and then drawing $x_i$. However, in most situations, the details balance condition is not satisfied. Therefore, we introduce a function $\alpha(x,y)$ satisfying 
\begin{equation*}
\pi(x')q(x', x^{(t)})\alpha(x',x^{(t)}) = \pi(x^{(t)})q(x^{(t)}, x')\alpha(x^{(t)},x').
\end{equation*}
In this way, a tentative new state $x'$ is generated from the proposal density $q(x';x^{(t)})$ and it is then accepted or rejected according to acceptance probability 
\begin{equation}\label{alphabalance}
\alpha=\frac{\pi(x')}{\pi(x^{(t)})}\frac{q(x^{(t)}, x')}{q(x', x^{(t)})}.
\end{equation}
If $\alpha \geq 1$, then the new state is accepted. Otherwise, the new state is accepted with probability $\alpha$.

Here comes an issues of how to choose $q(\cdot\mid x^{(t)})$. The most widely used subclass of MCMC algorithms is based around the Random Walk Metropolis (RWM). The RWM updating scheme was first applied by Metropolis \cite{metropolis1953equation} and proceeds as follows. Given a current value of the $d$-dimensional Markov chain $x^{(t)}$, a new value $x'$ is obtained by proposing a jump $\epsilon = \mid  x' - x^{(t)}\mid  $ from the pre-specified Lebesgue density 
\begin{equation}\label{stepsizeep}
\tilde{\gamma}(\epsilon^\star;\lambda) = \frac{1}{\lambda^d}\gamma \left( \frac{\epsilon^\star}{\lambda} \right),
\end{equation}
with $\gamma(\epsilon) = \gamma(-\epsilon)$ for all $\epsilon$. Here $\lambda>0$ governs the overall size of the proposed jump and plays a crucial role in determining the efficiency of any algorithm. In a random walk, the proposal density function $q(\cdot)$ can be chosen for some suitable normal distribution, and hence $q(x'\mid x^{(t)})=N(x'\mid x^{(t)},\epsilon^2)$ and $q(x^{(t)}\mid x')=N(x^{(t)}\mid x',\epsilon^2)$ cancel in the above equation (\ref{alphabalance}) \cite{sherlock2016adaptive}. Therefore, to decide whether to accept the new state, we compute the quantity
\begin{equation}
\alpha=\min \left\lbrace 1,\frac{\pi(x') q( x^{(t)}\mid x') }{\pi(x^{(t)})  q( x'\mid x^{(t)} ) }  \right\rbrace= \min \left\lbrace 1,\frac{\pi(x')  }{\pi(x^{(t)}) }  \right\rbrace.
\end{equation}
If the proposed value is accepted it becomes the next current value $x^{(t+1)}= x'$; otherwise the current value is left unchanged $x^{(t+1)} = x^{(t)}$ \cite{sherlock2010random}.

\subsection{Self-tuning Metropolis-Hastings Algorithm}

\begin{figure}[h]
\centering
 \begin{subfigure}[b]{0.32\textwidth}
     \includegraphics[width=\textwidth]{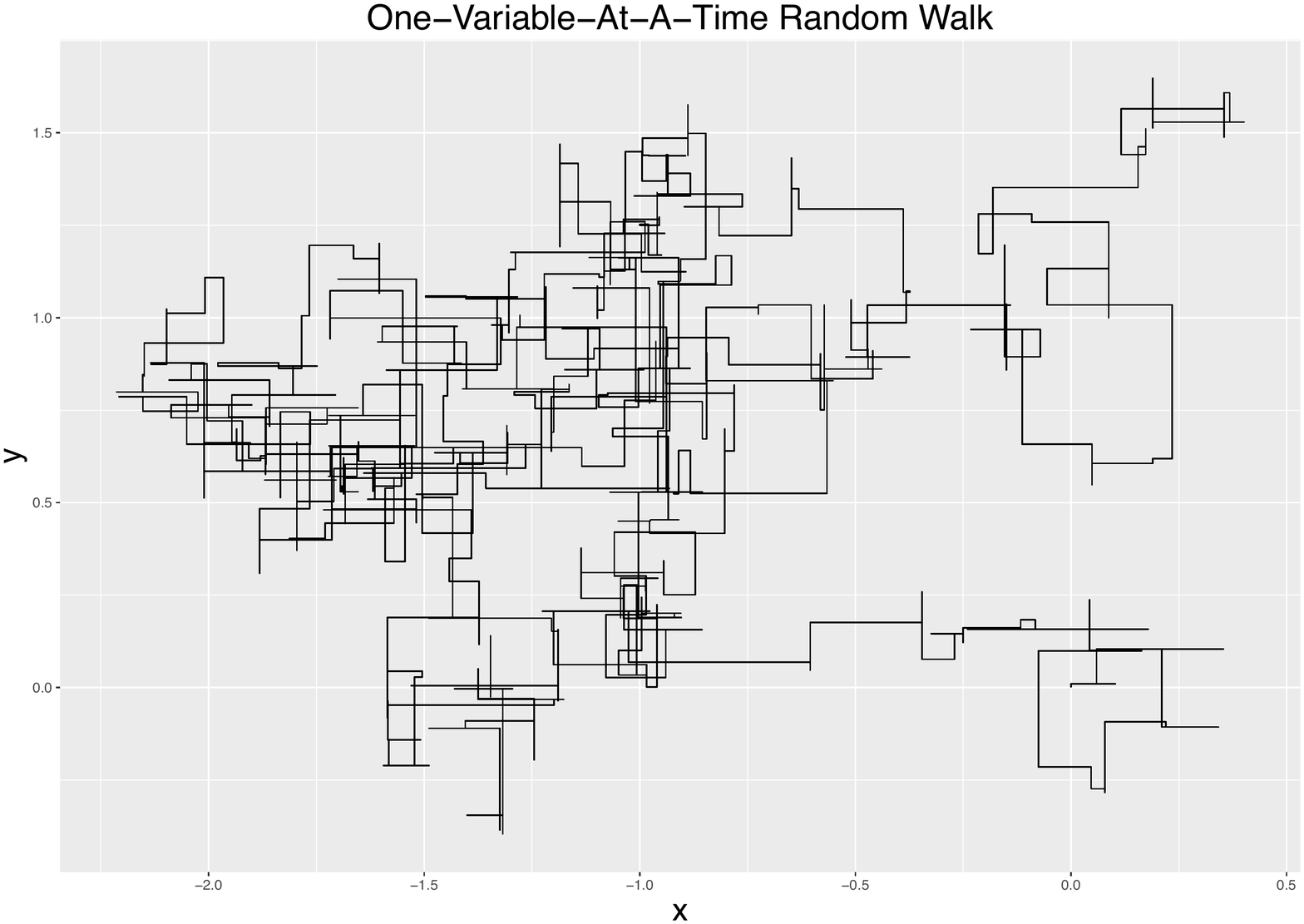}
     \caption{\footnotesize One-variable-at-a-time Random Walk.}\label{MCMConevariableRW}
\end{subfigure}
\begin{subfigure}[b]{0.32\textwidth}
     \includegraphics[width=\textwidth]{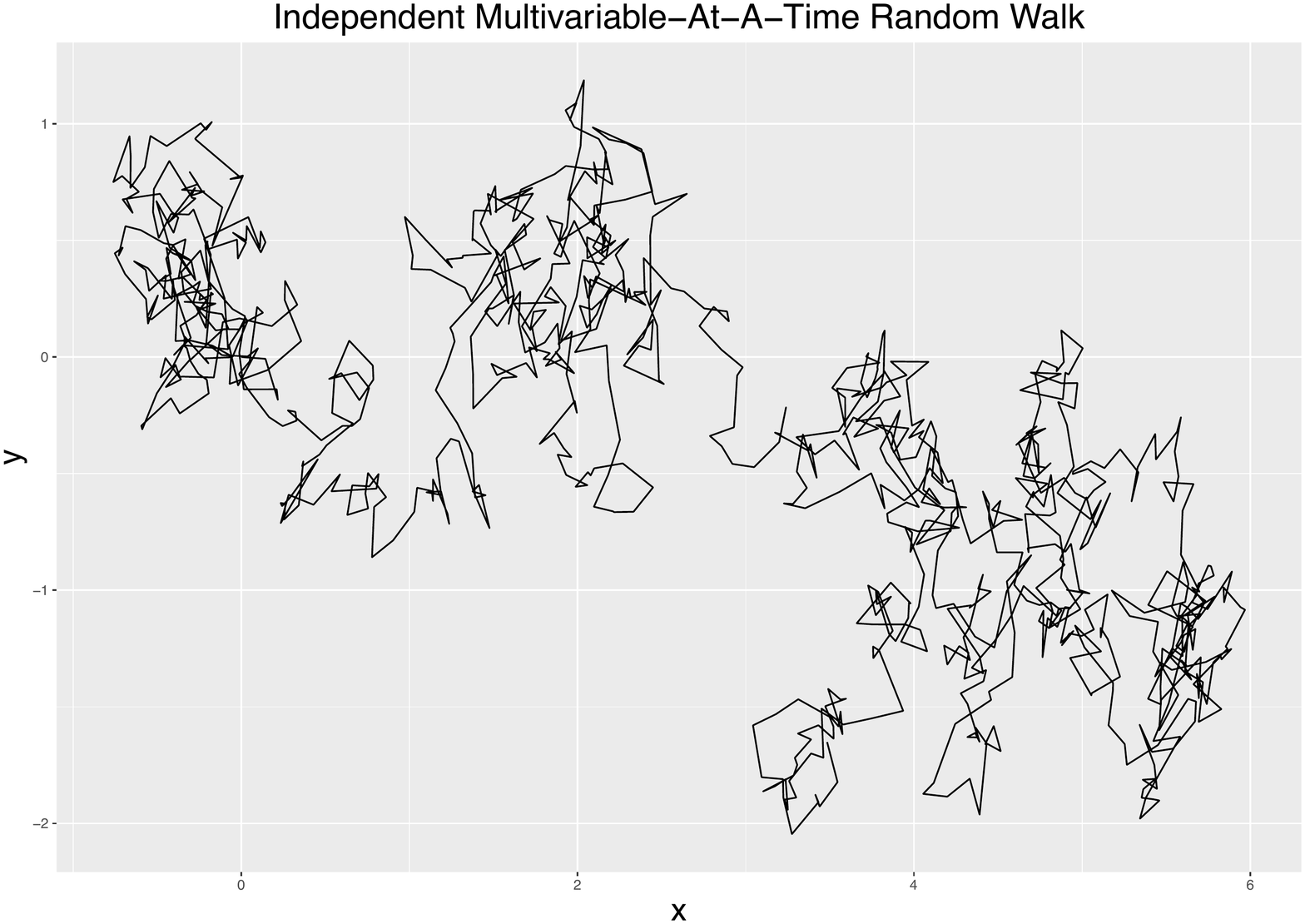}
    \caption{\footnotesize Independent Multi-variable-at-a-time Random Walk.}\label{MCMCMultivariableRW}
\end{subfigure}
\begin{subfigure}[b]{0.32\textwidth}
     \includegraphics[width=\textwidth]{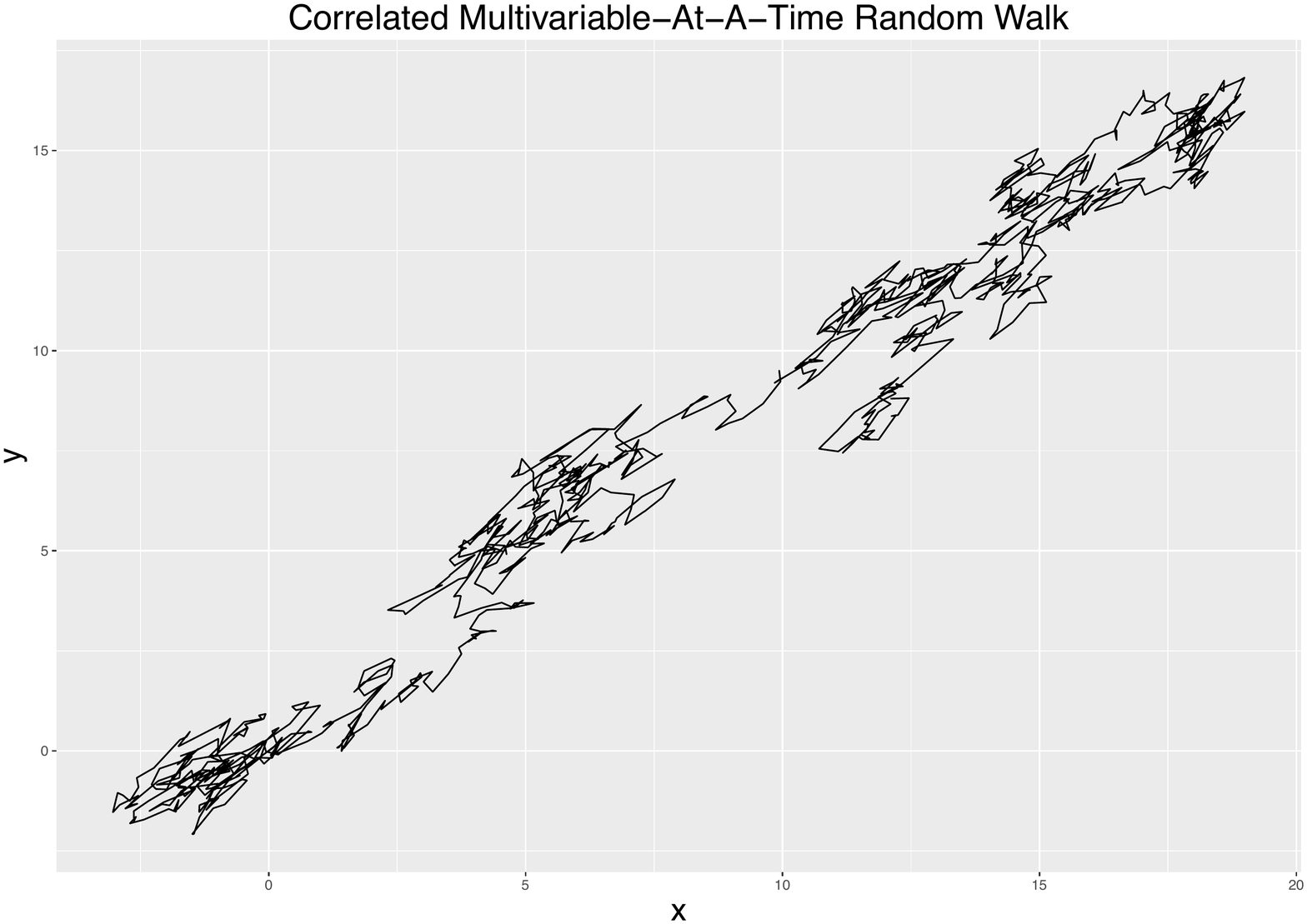}   
    \caption{\footnotesize Correlated Multi-variable-at-a-time Random Walk.}\label{MCMCCorrelatedRW}
\end{subfigure}
\caption{Examples of 2-Dimension Random Walk Metropolis-Hastings algorithm. Figure \ref{MCMConevariableRW} is using one-variable-at-a-time proposal Random Walk. At each time, only one variable is changed and the other one stay constant. Figure \ref{MCMCMultivariableRW} and \ref{MCMCCorrelatedRW} are using multi-variable-at-a-time Random Walk. The difference is in figure \ref{MCMCMultivariableRW}, every forward step are proposed independently, but in \ref{MCMCCorrelatedRW} are proposed according to the covariance matrix. }
\label{randomwalk}
\end{figure}

In this section, I am proposing a Self-tuning MH algorithm with one-variable-at-a-time Random Walk, which can tune step sizes on its own to gain the target acceptance rates, to estimate the structure of parameters in a $d$-dimensional space. Supposing all the parameters are independent, the idea of this algorithm is that in each iteration, only one parameter is proposed and the others are kept unchanged. After sampling, take $n$ samples out of the total amount of $N$ as new sequences. In figure \ref{randomwalk}, examples of different proposing methods are compared. 
To gain the target acceptance rates $\alpha_i (i = 1, \dots, d)$, the step sizes $s_i$ for each parameter can be tuned automatically. The concept of the algorithm is if the proposal is accepted, then we have more confidence on the direction and step size that were made. In this scenario, the next movement should be further, that means the step size $s_{t+1}$ in the next step is bigger than $s_t$; otherwise, a conservative proposal is made with a shorter distance, which is $s_{t+1}\leq s_t$. 

Supposing $a$ and $b$ are non-negative numbers indicating the distances of a forward movement, the new step size $s_{t+1}$ from current $s_t$ is 
\begin{align}\ln s_{t+1} = 
\begin{cases}
\ln s_t + a & \mbox{with probability } \alpha \\
\ln s_t - b & \mbox{with probability } 1 - \alpha 
\end{cases},
\end{align}
where the logarithm guarantees the step size is positive. 
By taking its expectation  
\begin{align*}
\E(\ln s_{t+1}\mid \ln s_t) = \alpha(\ln s_t+a) + (1-\alpha)(\ln s_t-b), 
\end{align*}
and simplifying to 
\begin{align*}
\mu= \alpha(\mu+a) + (1-\alpha)(\mu-b), 
\end{align*}
we can find that 
\begin{equation}\label{autostepab}
a = \frac{1-\alpha}{\alpha}  b. 
\end{equation}
Thus, if the proposal is accepted, the step size $s_t$ is tuned to $s_{t+1}=s_te^a$, otherwise $s_{t+1}=s_t/e^b$. 

The complete one-variable-at-a-time MH is illustrated in the following table: 

\begin{algorithm}[H]\label{algoonevarible}
Initialization: Given an arbitrary positive step size $s_i^{(1)}$ for each parameter. Set up a value for $b$ and find $a$ by using formula (\ref{autostepab}). 
Set up a target acceptance rate $\alpha_i$ for each parameter, where $i = 1,\dots, d$. \\
Run sampling algorithm: \For{$k$ from 1 to $N$}{
Randomly select a parameter $\theta_i^{(k)}$, propose a new one by $\theta_i'\sim N(\theta_i^{(k)}, \epsilon s_i^{(k)})$ and leave the rest unchanged.\label{stRWMHselect}\\
Accept $\theta_i'$ with probability $\alpha=\min\left\lbrace  1,\frac{\pi(\theta')q(\theta^{(k)},\theta')}{\pi(\theta^{(k)})q(\theta', \theta^{(k)})}  \right\rbrace$. \\
If it is accepted, tune step size to $s_i^{(k+1)}=s_i^{(k)}e^a$, otherwise $s_i^{(k+1)}=s_i^{(k)}/e^b$. \\
Set $k=k+1$ and move to step \ref{stRWMHselect} until $N$.\\
}
Take $n$ samples out from $N$ with equal spaced index for each parameter being a new sequence. 
\caption{Self-tuning Random Walk Metropolis-Hastings Algorithm.}
\end{algorithm}

The advantage of the algorithm (\ref{algoonevarible}) is that it returns a more accurate estimation for $\theta$ and it is more reliable to learn the structure of parameter space. However, if $\pi(\cdot)$ is in an irregular structure, the algorithm is really time-consuming and that cause a lower efficiency. To accelerate the computation, we are introducing the Delayed Acceptance Metropolis-Hastings Algorithm.

\subsection{Adaptive Delayed Acceptance Metropolis-Hastings Algorithm}

The DA-MH algorithm proposed in \cite{christen2005markov} is a two-stage Metropolis-Hastings algorithm in which, typically, proposed parameter values are accepted or rejected at the first stage based on a computationally cheap surrogate $\hat{\pi}(x)$ for the likelihood $\pi(x)$. In stage one, the quantity $\alpha_1$ is found by a standard MH acceptance formula 
\begin{equation*}
\alpha_1=\min\left\lbrace  1,\frac{\hat{\pi}(x')q(x^{(t)}, x')}{\hat{\pi}(x^{(t)})q(x', x^{(t)})}  \right\rbrace ,
\end{equation*}
where $\hat{\pi}(\cdot)$ is a cheap estimation for $x$ and a simple form is $\hat{\pi}(\cdot)=N(\cdot\mid \hat{x},\epsilon)$. Once $\alpha_1$ is accepted, the process goes into stage two and the acceptance probability $\alpha_2$ is
\begin{equation}\label{dahalpha2}
\alpha_2=\min \left\lbrace  1,\frac{\pi(x')\hat{\pi}(x^{(t)}) }{\pi(x^{(t)})\hat{\pi}(x')} \right\rbrace,
\end{equation}
where the overall acceptance probability $\alpha_1\alpha_2$ ensures that detailed balance is satisfied with respect to $\pi(\cdot)$; however if a rejection occurs at stage one then the expensive evaluation of $\pi(x)$ at stage two is unnecessary.

For a symmetric proposal density kernel $q(x', x^{(t)})$ such as is used in the random walk MH algorithm, the acceptance probability in stage one is simplified to
\begin{equation} \label{dahalpha1}
\alpha_1= \min \left\lbrace 1,\frac{\pi(x')  }{\pi(x^{(t)}) }  \right\rbrace.
\end{equation}
If the true posterior is available then the delayed-acceptance Metropolis-Hastings algorithm is obtained by substituting this for the unbiased stochastic approximation in (\ref{dahalpha2}) \cite{sherlock2015efficiency}.

To accelerate the MH algorithm, Delayed-Acceptance MH requires a cheap approximate estimation $\hat{\pi}(\cdot)$ in formula (\ref{dahalpha1}). Intuitively, the approximation should be efficient with respect to time and accuracy to the true posterior $\pi(\cdot)$. A sensible option is assuming the parameter distribution at each time $t$ is following a normal distribution with mean $m_t$ and covariance $C_t$. So the posterior density is given by 
\begin{equation*}
\hat{\pi}(\theta\mid y_{1:t}) \propto \exp\left( -\frac{1}{2}(\theta-m_t)^\top C_t^{-1}(\theta-m_t)\right). 
\end{equation*}
A lazy $C_t$ is using identity matrix, in which way all the parameters are independent. In terms of $m_t$, in most of circumstances, 0 is not an idea choice. To find an optimal or suboptimal $m_t$ and $C_t$, several algorithms have been discussed. In \cite{stroud2016bayesian}, the author is using a second-order expansion of $l(\theta)$ at the mode and the mean and covariance become $m_t=\arg \max l(\theta)$ and $C_t = - \left[ \frac{\partial l(\theta)}{\partial \theta_i \partial \theta_j} \right]_{\theta=m_t}^{-1}$ respectively. The drawback of this estimation is a global optimum is not guaranteed. In \cite{mathew2012bayesian}, the author proposed a fast adaptive MCMC sampling algorithm, which is a consist of two phases. In the learning phase, they use hybrid Gibbs sampler to learn the covariance structure of the variance components. In phase two, the covariance structure is used to formulate an effective proposal distribution for a MH algorithm.

Likewise, we are suggesting that use a batch of data with length $L<t$ to learn the parameter space by using self-tuning random walk MH algorithm in the learning phase first. This algorithm tunes each parameter at its own optimal step size and explores the surface in different directions. When the process is done, we have a sense of Hyper-surface of $\theta\approx\hat{\theta}$ and its mean $\hat{\mu}\approx m_L$ and covariance $\hat{\Sigma}\approx C_L$ can be estimated. Then we can move to the second phase: Delayed-Acceptance MH algorithm. The new $\theta'$ is proposed from  $N(\theta^{(t)}\mid m_L,C_L)$, which is in the following form 
\begin{equation}
\theta' = \theta^{(t)} + R\epsilon z,
\end{equation}
where $R^\top R = C_L$ is the Cholesky decomposition, $\epsilon$ is the tuned step size and $z\sim N(0,1)$ is Gaussian white noise. This proposing method reduces the impact of drawing $\theta'$ from a correlation space.


\subsection{Efficiency of Metropolis-Hastings Algorithm}\label{effMHA}

\begin{figure}[h]
\centering
 \begin{subfigure}[b]{0.32\textwidth}
     \includegraphics[width=\textwidth]{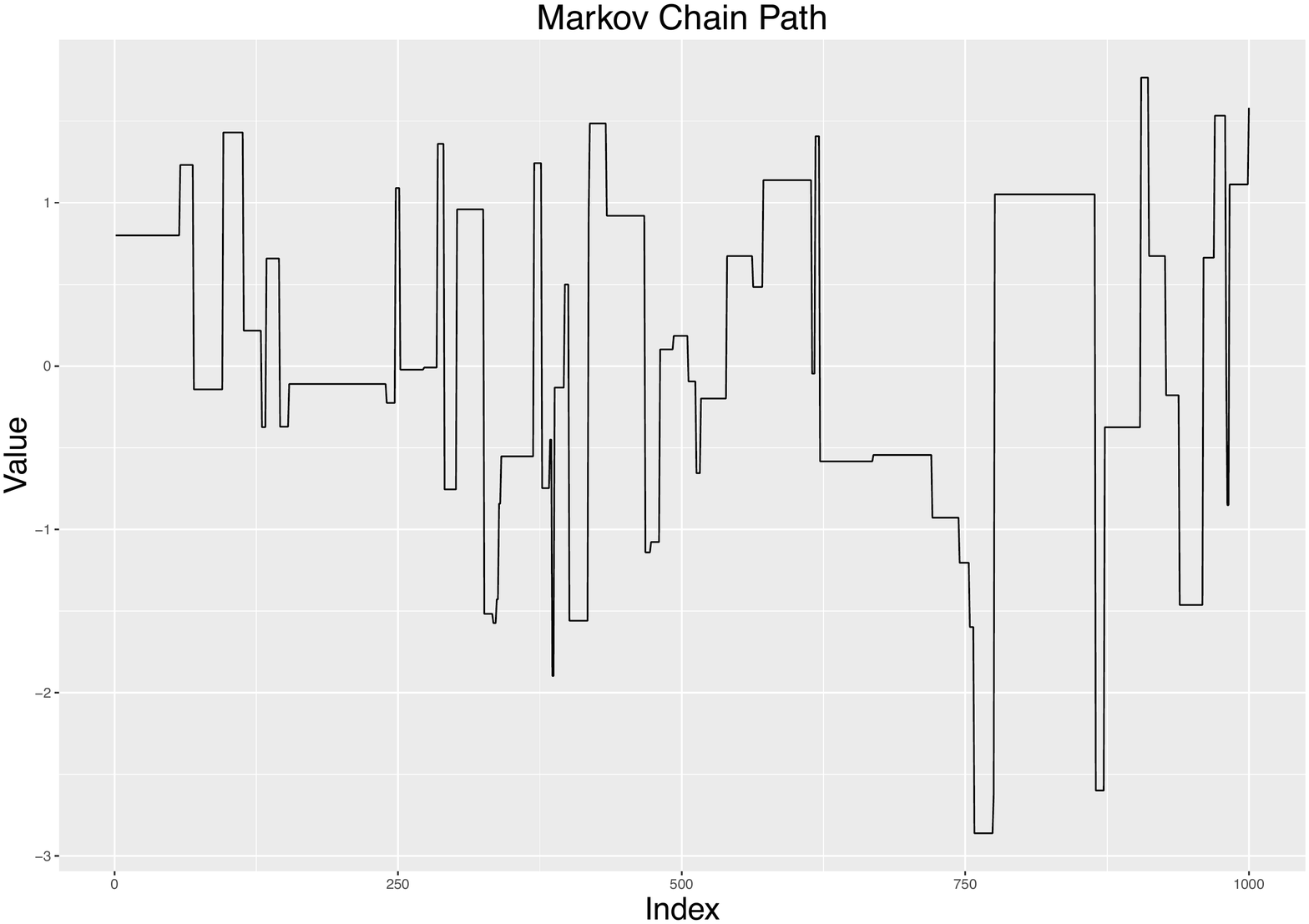}
     \includegraphics[width=\textwidth]{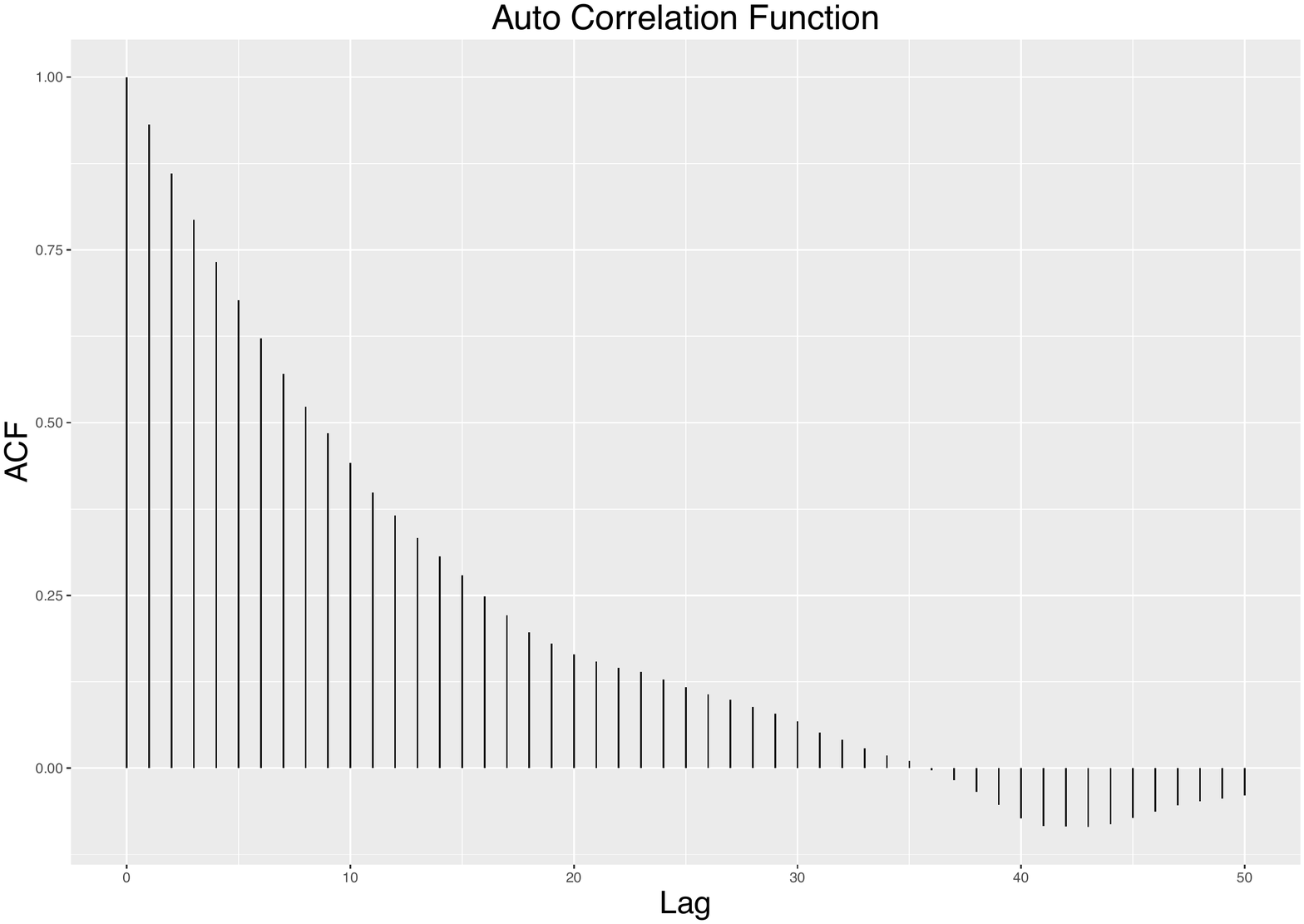}
     \caption{With a large step size}\label{MCMClargestep}
\end{subfigure}
\begin{subfigure}[b]{0.32\textwidth}
    \includegraphics[width=\textwidth]{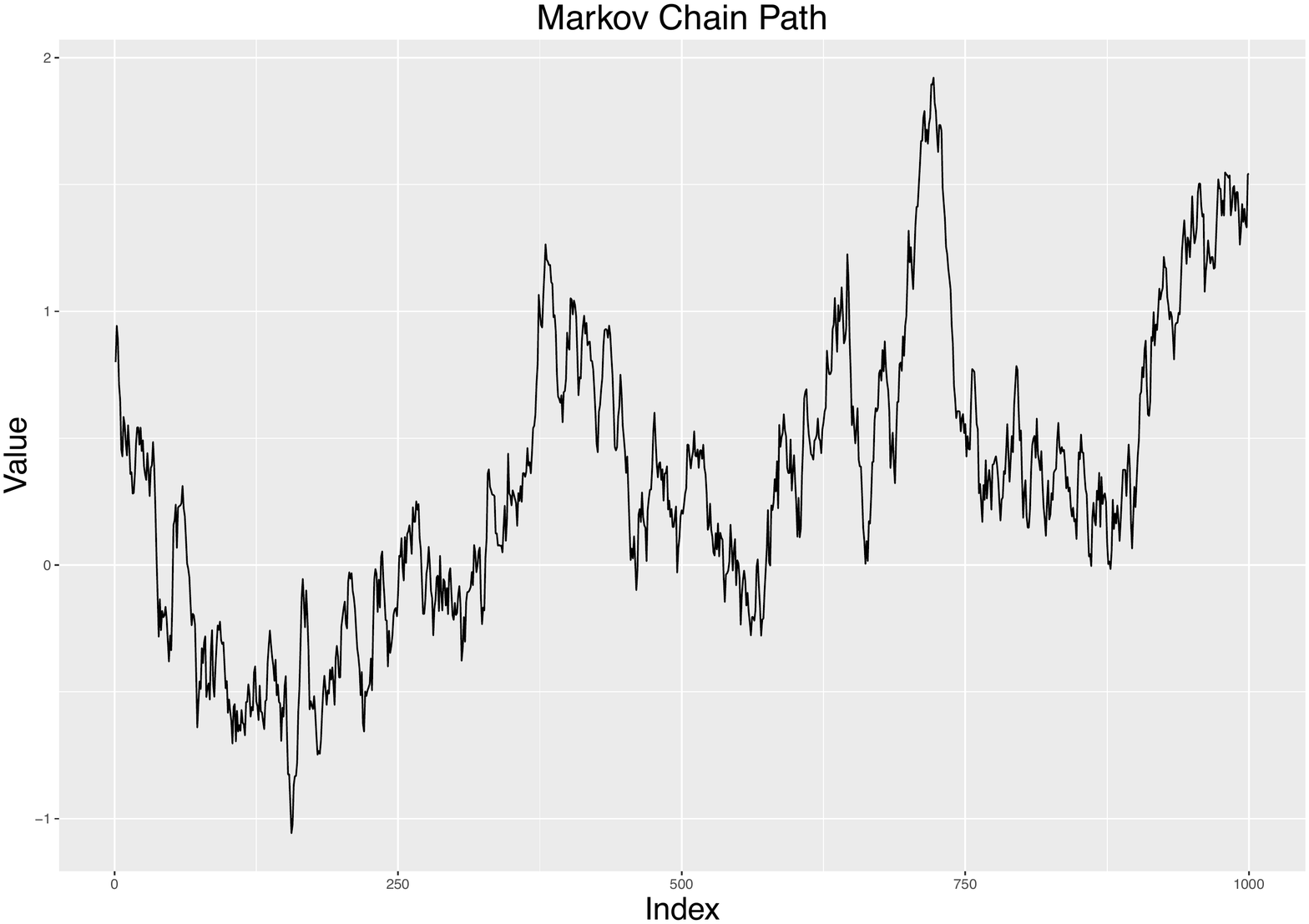}
    \includegraphics[width=\textwidth]{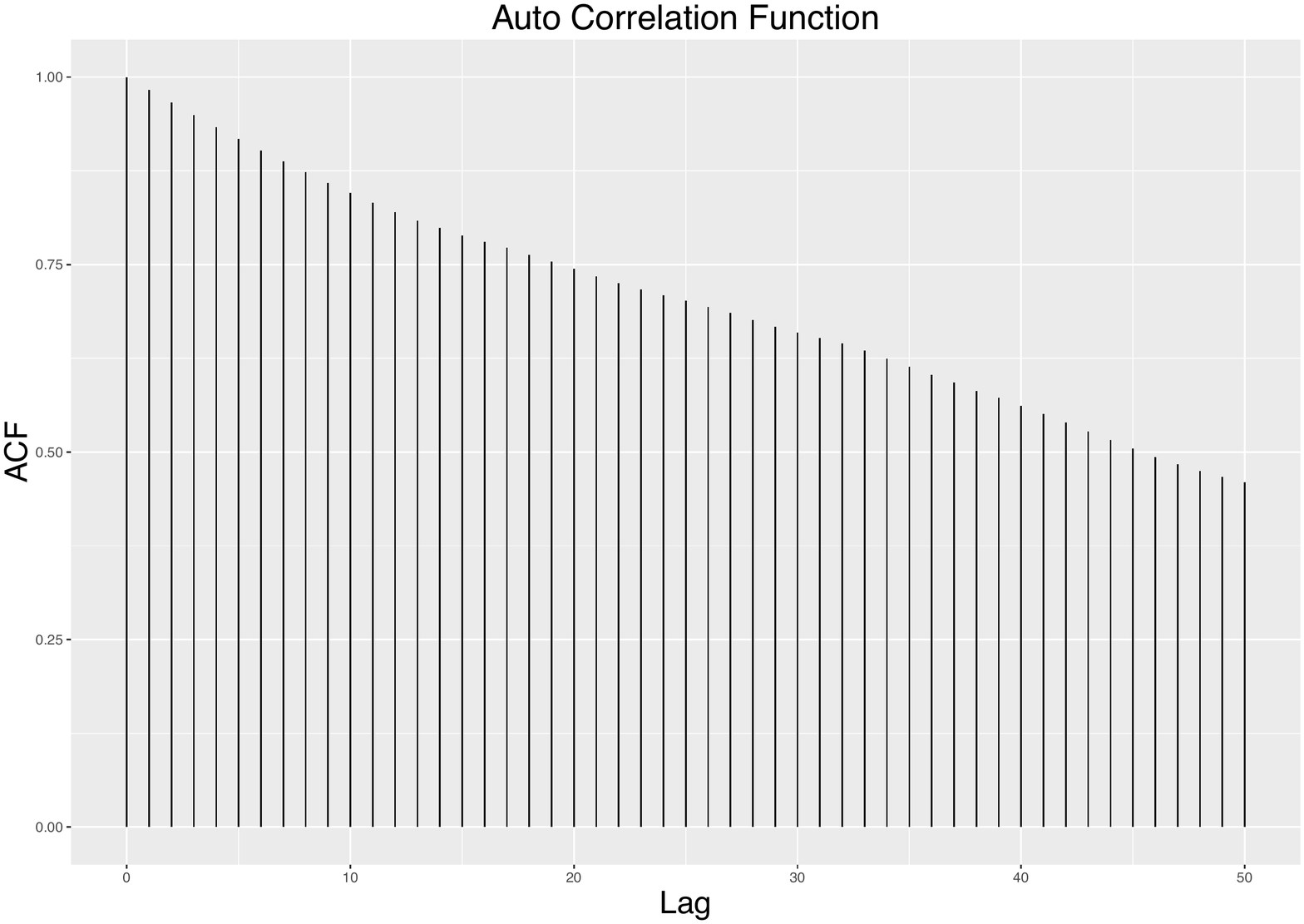}
    \caption{With a small step size}\label{MCMCsmallstep}
\end{subfigure}
\begin{subfigure}[b]{0.32\textwidth}
    \includegraphics[width=\textwidth]{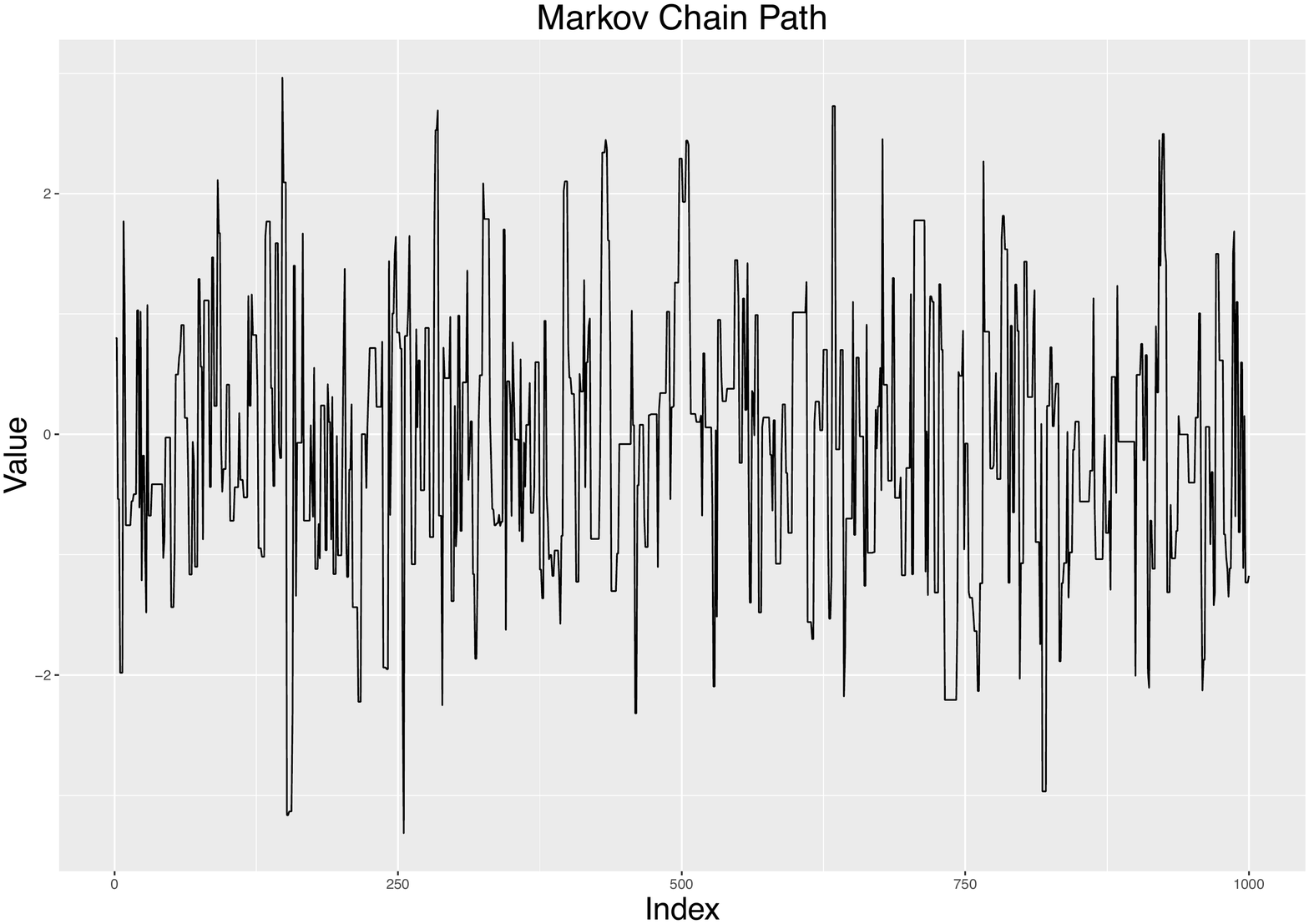}
    \includegraphics[width=\textwidth]{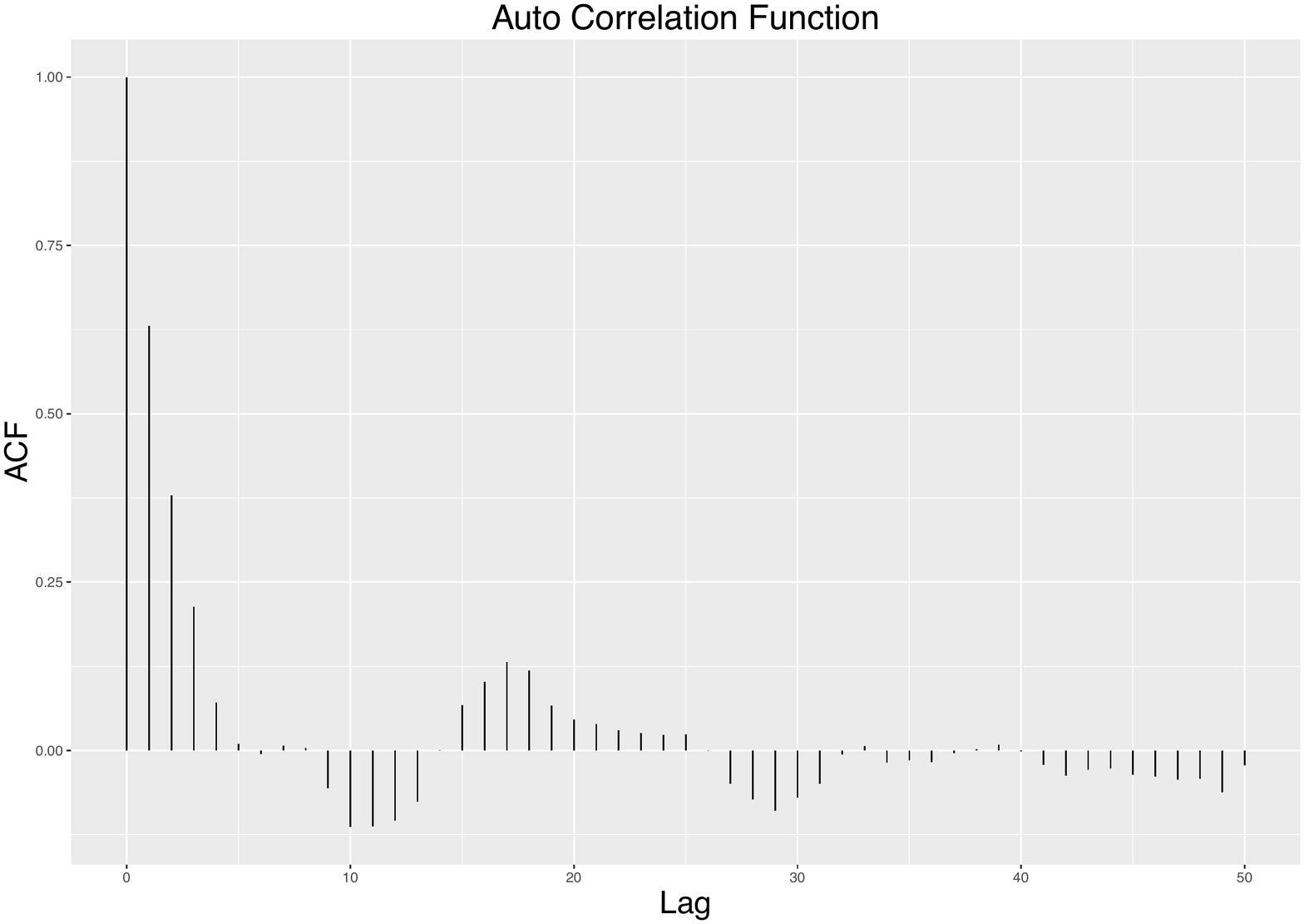}
    \caption{With a proper step size}\label{MCMCproperstep}
\end{subfigure}
\caption{Metropolis algorithm sampling for a single parameter with: \ref{MCMClargestep} a large step size, \ref{MCMCsmallstep} a small step size, \ref{MCMCproperstep} an appropriate step size. The upper plots show the sample chain and lower plots indicate the autocorrelation for each case.}
\label{largesmallstepsize}
\end{figure}

In equation (\ref{stepsizeep}), the jump size $\epsilon$ determines the efficiency of RWM algorithm. For a general RWM, it is intuitively clear that we can make the algorithm arbitrarily poor by making $\epsilon$ either very large or very small \cite{sherlock2010random}. Assuming $\epsilon$ is extremely large, the proposal $x'\sim N(x^{(t)},\epsilon)$, for example, is taken a further distance from current value $x^{(t)}$. Therefore, the algorithm will reject most of its proposed moves and stay where it was for a few iterations. On the other hand, if $\epsilon$ is extremely small, the algorithm will keep accepting the proposed $x'$ since $\alpha$ is always approximately be 1 because of the continuity of $\pi(x)$ and $q(\cdot)$ \cite{roberts2001optimal}. Thus, RWM takes a long time to explore the posterior space and converge to its stationary distribution. So, the balance between these two extreme situations must exist. This appropriate step size $\hat{\epsilon}$ is optimal, sometimes is suboptimal, the solution to gain a Markov chain. Figure (\ref{largesmallstepsize}) illustrates the performances of RWM with different step size $\epsilon$. From these plots we may see that either too large or too small $\epsilon$ causes high correlation chains, indicating bad samples in sampling algorithm. An appropriate $\epsilon$ decorrelate samples and returns a stationary chain, which is said to be high efficiency.


Plenty of work has been done to determine the efficiency of Metropolis-Hastings algorithm in recent years. Gelman, Roberts, and Gilks \cite{gelman1996efficient} work with algorithms consisting of a single Metropolis move (not multi-variable-at-a-time), and obtain many interesting results for the $d$-dimensional spherical multivariate normal problem with symmetric proposal distributions, including that the optimal scale is approximately $2.4/\sqrt{d}$ times the scale of target distribution, which implies optimal acceptance rates of $0.44$ for $d = 1$ and $0.23$ for $d\rightarrow \infty$ \cite{gilks1995markov}. Roberts and Rosenthal (2001) \cite{roberts2001optimal} evaluate scalings that are optimal (in the sense of integrated autocorrelation times) asymptotically in the number of components. They find that an acceptance rate of 0.234 is optimal in many random walk Metropolis situations, but their studies are also restricted to algorithms that consist of only a single step in each iteration, and in any case, they conclude that acceptance rates between 0.15 and 0.5 do not cost much efficiency. Other researchers \cite{roberts1997weak} \cite{bedard2007weak}, \cite{beskos2009optimal}, \cite{sherlock2009optimal}, \cite{sherlock2013optimal} have been tackled for various shapes of target on choosing the optimal scale of the RWM proposal and led to the similar rule: choose the scale so that the acceptance rate is approximately 0.234. Although nearly all of the theoretical results are based upon limiting arguments in high dimension, the rule of thumb appears to be applicable even in relatively low dimensions \cite{sherlock2010random}.

In terms of the step size $\epsilon$, it is pointed out that for a stochastic approximation procedure, its step size sequence $\{\epsilon_i\}$ should satisfy $\sum_{i=1}^\infty \epsilon_i=\infty $ and $\sum_{i=1}^\infty \epsilon_i^{1+\lambda}<\infty $ for some $\lambda>0$. The former condition somehow ensures that any point of $X$ can eventually be reached, while the second condition ensures that the noise is contained and does not prevent convergence \cite{andrieu2008tutorial}. Sherlock, Fearnhead, and Roberts \cite{sherlock2010random} tune various algorithms to attain target acceptance rates, and their Algorithm 2 tunes step sizes of univariate updates to attain the optimal efficiency of Markov chain at the acceptance rates between 0.4 and 0.45. Additionally, Graves in \cite{graves2011automatic} mentioned that it is certain that one could use the actual arctangent relationship to try to choose a good $\epsilon$: in the univariate example, if $\alpha$ is the desired acceptance rate, then $\epsilon = 2\sigma / \tan (\pi/2\alpha)$, where $\sigma$ is the posterior standard deviation, will be obtained. In fact, some explorations infer a linear relationship between acceptance rate and step size, which is $\mbox{logit}(\alpha) \approx 0.76-1.12\log \epsilon/\sigma$, and the slope of the relationship is nearly equal to the constant -1.12 independently. However, in multi-variable-at-a-time RWM, one expects that the proper interpretation of $\sigma$ is not the posterior standard deviation but the average conditional standard deviation, which is presumably more difficult to estimate from a Metropolis algorithm. In a higher $d$-dimensional space, or propose multi-variable-at-a-time, suppose $\Sigma$ is known or could be estimated, then $X'$ can be proposed from $q\sim N(X,\epsilon^2\Sigma)$. Thus the optimal step size $\epsilon$ is required. A concessive way of RWM in high dimension is proposing one-variable-at-a-time and treating them as one dimension space individually. In any case, however, the behavior of RWM on a multivariate normal distribution is governed by its covariance matrix $\Sigma$, and it is better than using a fixed $N(X,\epsilon^2I_d)$ distribution\cite{roberts2001optimal}.

To explore the efficiency of a MCMC process, we introduce some notions first. For an arbitrary square integrable function $g$, Gareth, Roberts and Jeffrey \cite{roberts2001optimal} define its integrated autocorrelation time by 
\begin{equation*}
\tau_g = 1+ 2\sum_{i=1}^{\infty} \mathrm{Corr}\left( g(X_0),g(X_i) \right),
\end{equation*}
where $X_0$ is assumed to be distributed according to $\pi$. Because central limit theorem, the variance of the estimator $\bar{g} = \sum_{i=1}^{n}g(X_i)/n$ for estimating $\mathrm{E}\left(g(X)\right)$ is approximately $\Var_\pi\left(g(X)\right)\times \tau_g/n$. The variance tells us the accuracy of the estimator $\bar{g}$. The smaller it is, the faster the chain converge. Therefore, they suggest that the efficiency of Markov chains can be found by comparing the reciprocal of their integrated autocorrelation time, which is 
\begin{equation*}
e_g(\sigma)\propto \left( \Var_\pi (g(X))\tau_g \right)^{-1}. 
\end{equation*}
However, the disadvantage of their method is that the measurement of efficiency is highly dependent on the function $g$. Instead, an alternative approach is using Effective Sample Size (ESS) \cite{kass1998markov} \cite{robert2004monte}. Given a Markov chain having $n$ iterations, the ESS measures the size of \iid . samples with the same standard error, which is defined in \cite{gong2016practical} in the following form of  
\begin{equation*}
\mbox{ESS} =  \frac{n}{1+2\sum_{k=1}^{\infty}\rho_k(X)} \approxeq \frac{n}{1+2\sum_{k=1}^{k_\text{\tiny cut}}\rho_k(X)}= \frac{n}{\tau}, 
\end{equation*}
where $n$ is the number of samples, $k_\mathtt{cut}$ is lag of the first $\rho_k<0.01 \mbox{ or } 0.05$ , and $\tau$ is the integrated autocorrelation time. Moreover, a wide support among both statisticians \cite{geyer1992practical} and physicists \cite{sokal1997monte} are using the following cost of an independent sample to evaluate the performance of MCMC, that is 
\begin{equation*}
\frac{n}{\mbox{ESS}}\times \mbox{cost per step} = \tau \times  \mbox{cost per step}.
\end{equation*} 

Being inspired by their research, we now define the Efficiency in Unit Time (EffUT)  and ESS in Unit Time (ESSUT) as follows: 
\begin{align}
\mbox{EffUT}     &= \frac{e_g}{T},\\
\mbox{ESSUT} &= \frac{\mbox{ESS}}{T},
\end{align} 
where $T$ represents the computation time, which is also known as running time. The computation time is the length of time, in minutes or hours, etc, required to perform a computational process. The best Markov chain with an appropriate step size $\epsilon$ should not only have a lower correlation, as illustrated in Figure (\ref{largesmallstepsize}), but also have less time-consuming. The standard efficiency $e_g$ and ESS do not depend on the computation time, but EffUT and ESSUT do. The best-tuned step size gains the balance between the size of effective proposed samples and cost of time.

\section{Simulation Studies}

In this section, we consider the model in regular and irregular spaced time difference separately. For an one dimensional state-space model, we consider the hidden state process $\{x_t, t\geq 1\}$ is a stationary and ergodic Markov process and transited by $F(x'\mid x)$. In this paper, we assume that the state of a system has an interpretation as the summary of the past one-step behavior of the system. The states are not observed directly but by another process $\{y_t, t\geq 1\}$, which is assumed depending on $\{x_t\}$ by the process $G(y\mid x)$ only and independent with each other. When observed on discrete time $T_1,\dots,T_k$, the model is summarized on the directed acyclic following graph  
\begin{align*}
\begin{matrix}
\mbox{State}  & x_0     &  \rightarrow& x_1   & \rightarrow \cdots  & x_k  & \rightarrow \cdots & x_t & \rightarrow \cdots\\
          & &       & \downarrow &         &\downarrow &        &\downarrow &   \\
\mbox{Observation}& && y_1               &          & y_k               &        & y_t               &   \\
          & &      & \downarrow &          &\downarrow  &        &\downarrow &   \\
\mbox{Time } & &       & T_1               &          & T_k               &        & T_t               &   \\
\end{matrix}
\end{align*}
We define $\Delta_k = T_k-T_{k-1}$. If $\Delta_t$ is a constant, we retrieve a standard  $\textit{AR(1)}$ model process with regular spaced time steps; if $\Delta_t$ is not constant, then the model becomes more complicated with irregular spaced time steps. 


\subsection{Simulation on Regular Time Series Data}

If the time steps are even spaced, the model can be written as a simple linear model in the following 
\begin{align*}
y_t\mid x_t      &\sim N(x_t,\sigma^2) \\
x_t\mid x_{t-1} &\sim N(\phi x_{t-1},\tau^2),
\end{align*}
where $\sigma$ and $\tau$ are \iid  errors occurring in processes and $\phi$ is a static process parameter in forward map. An initial value $x_0\sim N(0,L)$ is known.

To get the joint distribution for $x_{0:t}$ and $y_{1:t}$
\begin{equation*}
\left[ \begin{matrix} x\\y  \end{matrix}\bigg\rvert \theta \right]
\sim N\left(0, \Sigma  \right),
\end{equation*}
where $\theta = \{\phi,\sigma,\tau\}$, we should start from the procedure matrix $\Sigma^{-1}$, which looks like 
\begin{equation*}
\begin{bmatrix}
\frac{1}{L^2}+\frac{\phi^2}{\tau^2} & \frac{-\phi}{\tau^2} & \cdots & 0 & 0 & 0& \cdots & 0\\
\frac{-\phi}{\tau^2}   & \frac{1+\phi^2}{\tau^2}+\frac{1}{\sigma^2}& \cdots & 0 & -\frac{1}{\sigma^2} &0 & \cdots & 0 \\
0 & \frac{-\phi}{\tau^2}   &  \cdots & 0 & 0& -\frac{1}{\sigma^2} & \cdots & 0\\
\vdots & \vdots & \ddots & \vdots & \vdots & \vdots & \ddots & \vdots \\
0 & 0   &  \cdots & \frac{1}{\tau^2}+\frac{1}{\sigma^2} & 0 & 0 & \cdots &-\frac{1}{\sigma^2}\\
0 & -\frac{1}{\sigma^2}  & \cdots & 0 & \frac{1}{\sigma^2} & 0 & \cdots & 0 \\
0& 0 & \cdots & 0 & 0 &  \frac{1}{\sigma^2} & \cdots & 0\\
\vdots & \vdots & \ddots & \vdots & \vdots & \vdots & \ddots & \vdots\\
0 & 0& \cdots &-\frac{1}{\sigma^2} & 0 & 0 & \cdots &  \frac{1}{\sigma^2}
\end{bmatrix},
\end{equation*}
and denoted as $\Sigma^{-1}=\begin{bmatrix} A & -B \\ -B & B \end{bmatrix}$. Its inverse is the covariance matrix 
\begin{equation}
\Sigma=\begin{bmatrix} (A-B)^{-1} &  (A-B)^{-1} \\ (A-B)^{-1} & (I-A^{-1}B)^{-1}B^{-1} \end{bmatrix} \triangleq \begin{bmatrix}
\Sigma_{XX} & \Sigma_{XY}  \\ \Sigma_{YX} & \Sigma_{YY} 
\end{bmatrix},
\end{equation}
where $B$ is a $t\times t$ diagonal matrix with elements $\frac{1}{\sigma^2}$. The covariance matrices $\Sigma_{XX} =  (A-B)^{-1}$ and $\Sigma_{YY} =  (I-A^{-1}B)^{-1}B^{-1}$ are easily found.

\subsubsection*{Parameters Estimation}

In formula (\ref{M1}), the parameter posterior is estimated with observation data $y_{1:t}$. By using the algorithm \ref{algoonevarible}, although it may take a longer time, we will achieve a precise estimation. Similarly with section \ref{sectionlogParameter}, from the objective function, the posterior distribution of $\theta$ is 
\begin{equation*}
p(\theta \mid Y) \propto p(Y\mid\theta)p(\theta) \propto e^{-\frac{1}{2} Y \Sigma_{YY}^{-1} Y } \sqrt{\det \Sigma_{YY}^{-1}} p(\theta).
\end{equation*}
Then by taking natural logarithm on the posterior of $\theta$ and using the useful solutions in equations (\ref{sigmayy01}) and (\ref{sigmayy02}), we will have
\begin{equation}\label{linearlogL}
\ln L(\theta) = -\frac{1}{2}Y^\top\Sigma_{YY}^{-1}Y+\frac{1}{2}\sum\ln\mbox{tr}(B)-\sum\ln\mbox{tr}(L)+\sum\ln\mbox{tr}(R) + \ln p(\theta).
\end{equation}

In a simple linear case, we are choosing the parameter $\theta = \{\phi=0.9,\tau^2=0.5,\sigma^2=1\}$ as the author did in \cite{lopes2011particle} and using $n=500$ dataset, setting initial $L=0$. Instead of inferring $\tau$ and $\sigma$, we are estimating $\nu_1 = \ln \tau^2$ and $\nu_2 = \ln \sigma^2$ in the RW-MH to avoid singular proposals. After the process, the parameters can be transformed back to original scale. Therefore, the new parameter  $\theta^* =  \{\phi,\nu_1,\nu_2\} = \{\phi,\ln\tau^2,\ln\sigma^2\}$. 

\begin{figure}[h]
\centering
 \begin{subfigure}[b]{0.32\textwidth}
     \includegraphics[width=\textwidth]{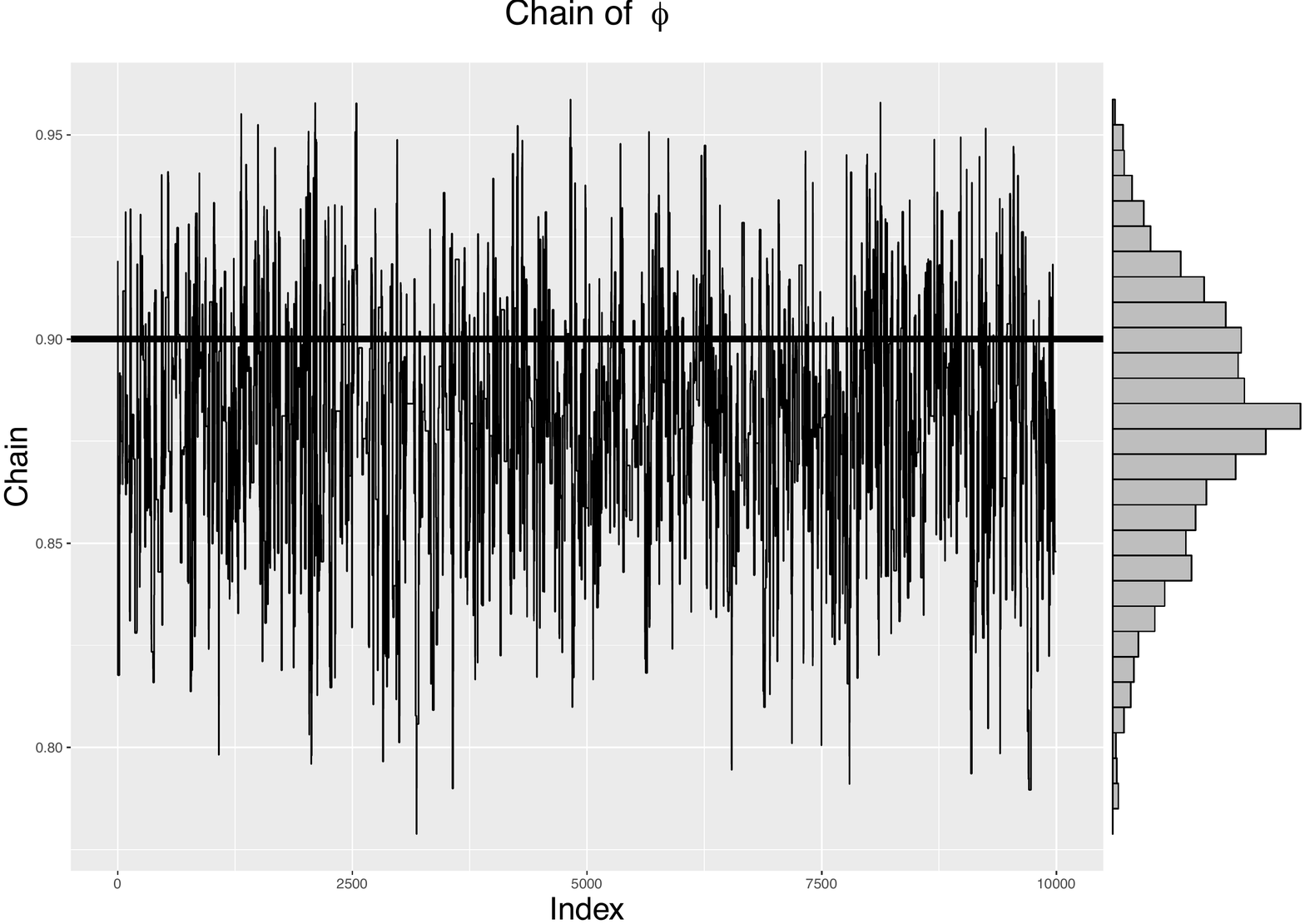}
     \caption{Trace plot of $\phi$.}
\end{subfigure}
\begin{subfigure}[b]{0.32\textwidth}
    \includegraphics[width=\textwidth]{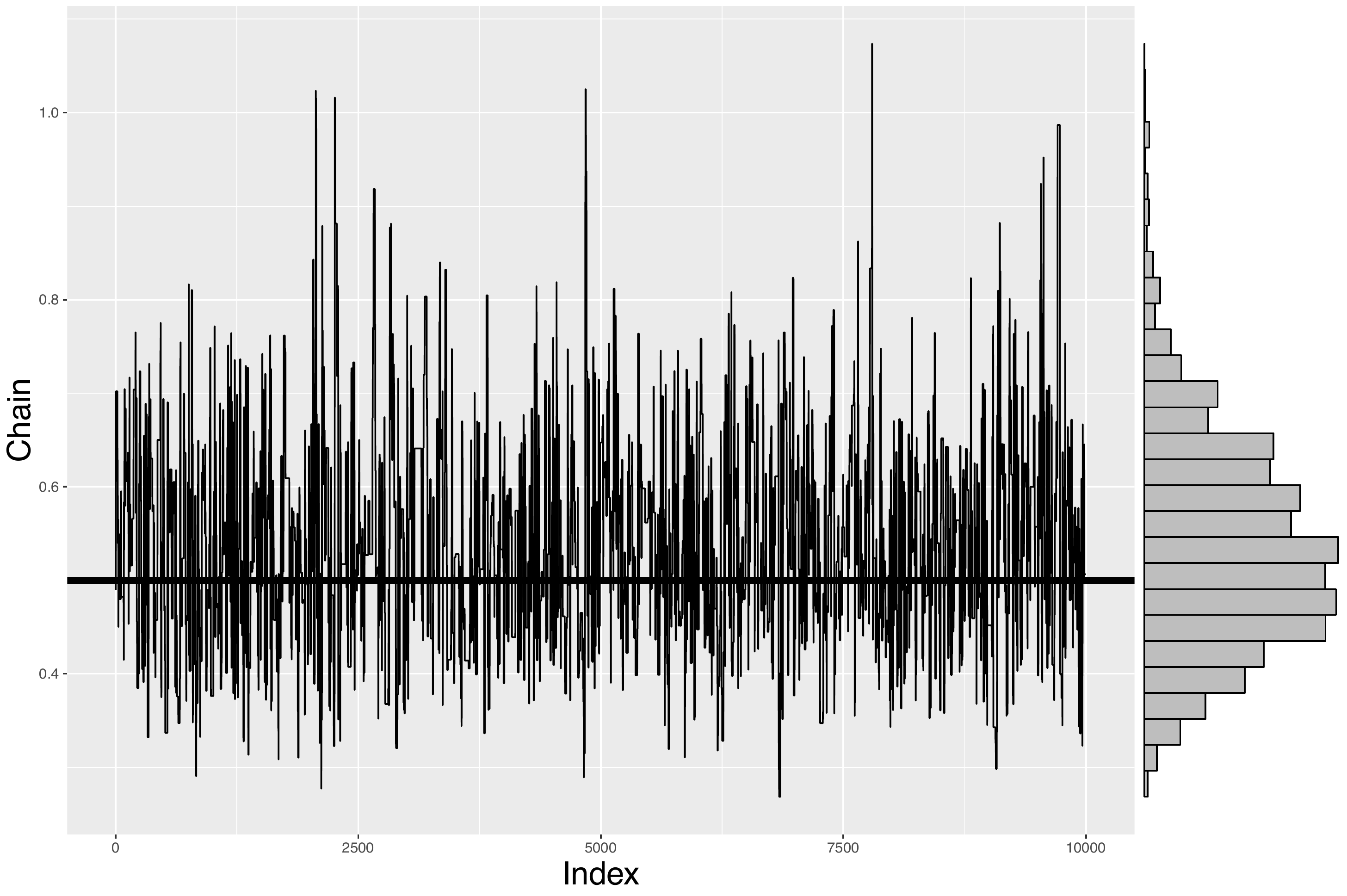}
     \caption{Trace plot of $\tau^2$.}
\end{subfigure}
\begin{subfigure}[b]{0.32\textwidth} \
    \includegraphics[width=\textwidth]{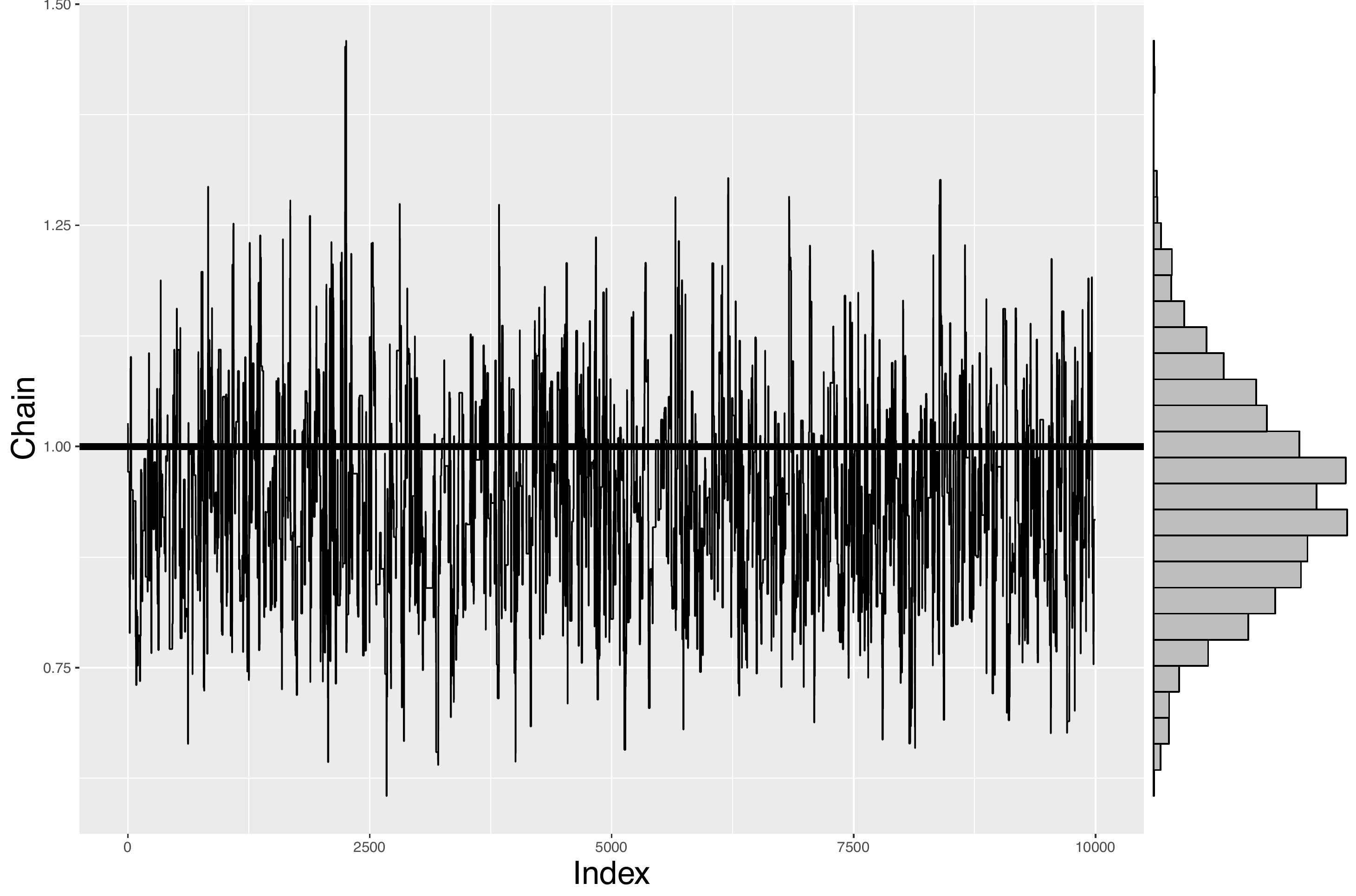}
     \caption{Trace plot of $\sigma^2$.}
\end{subfigure}
\caption{Linear simulation with true parameter $\theta = \{\phi=0.9,\tau^2=0.5,\sigma^2=1\}$. By transforming to original scale, the estimation is $\hat{\theta}=\{ \phi = 0.8810, \tau^2 = 0.5247,\sigma^2= 0.9416\}$. }
\label{linearmarginplots}
\end{figure}

Buy using algorithm (\ref{algoonevarible}) and aiming the optimal acceptance rate at 0.44, after 10\,000 iterations we get the acceptance rates for each parameters are $\alpha_\phi = 0.4409, \alpha_{\nu_1}= 0.4289$ and $\alpha_{\nu_2}= 0.4505$, and the estimations are $\phi =0.8794, \nu_1= -0.6471$ and $\nu_2= -0.0639$ respectively. Thus, we have the cheap surrogate $\hat{\pi}(\cdot)$. Keep going to the DA-MH with another 10\,000 iterations, the algorithm returns the best estimation with $\alpha_1=0.1896$ and $\alpha_2 = 0.8782$. In figure \ref{linearmarginplots}, the trace plots illustrates that the Markov Chain of $\hat{\theta}$ is stably fluctuating around the true $\theta$. 

\subsubsection*{Recursive Forecast Distribution}\label{sectionlinearRecursive}

Calculating the log-posterior of parameters requires finding out the forecast distribution of $p(y_{1:t}\mid y_{1:t-1},\theta)$. A general way is using the joint distribution of $y_{t}$ and $y_{1:t-1}$, which is $p(y_{1:t}\mid \theta)\sim N(0,\Sigma_{YY})$, and following the procedure in section \ref{sectionforecast} to work out the inverse matrix of a multivariate normal distribution. For example, one may find the inverse of the covariance matrix 
\begin{align*}
\Sigma_{YY}^{-1} = B_t(I-A_t^{-1}B_t) =\frac{1}{\sigma^4}(\sigma^2 I_t-A_t^{-1}) \triangleq \frac{1}{\sigma^4} \left[\begin{matrix} 
Z_{t} & b_{t} \\
b_{t}^\top & K_{t}
\end{matrix} \right].
\end{align*}
Therefore, the original form of this covariance is 
\begin{align*} \Sigma_{YY} =\sigma^4 \left[ \begin{matrix}
(Z_t-b_tK_t^{-1}b_t^\top)^{-1} & -Z_t^{-1}b_t(K_t-b_t^\top Z_t^{-1}b_t)^{-1}\\
-K_t^{-1}b_t^\top (Z_t-b_tK_t^{-1}b_t^\top)^{-1} & (K_t-b_t^\top Z_t^{-1}b_t)^{-1}
\end{matrix}\right].
\end{align*}
By denoting $C_{t}^\top = \begin{bmatrix} 0 & \cdots & 0 & 1\end{bmatrix}$ and post-multiplying $\Sigma_{YY}^{-1}$, we will have  
\begin{equation}\label{beforeSMformula}
\Sigma_{YY}^{-1} C_{t}= \frac{1}{\sigma^4}(\sigma^2 I-A^{-1}) C_{t}= \frac{1}{\sigma^4} \left[\begin{matrix} b_{t} \\ K_{t} \end{matrix} \right].
\end{equation}

A recursive way of calculating $b_t$ and $K_t$ is to use the Sherman-Morrison-Woodbury formula. In the late 1940s and the 1950s, Sherman and Morrison\cite{sherman1950adjustment}, Woodbury \cite{woodbury1950inverting}, Bartlett \cite{bartlett1951inverse} and Bodewig \cite{bodewig1959matrix} discovered the following result. The original Sherman-Morrison-Woodbury (for short SMW) formula has been used to consider the inverse of matrices \cite{deng2011generalization}. In this paper, we will consider the more generalized case. 

Theorem 1.1 (Sherman-Morrison-Woodbury). Let $A \in B(H)$ and $G \in B(K)$ both be invertible, and $Y, Z \in B(K, H)$. Then $A + YGZ^*$ is invertible if and only if $G^{-1} + Z^∗A^{-1}Y$ is invertible. In which case,
\begin{equation}\label{SMWformula}
(A+YGZ^*)^{-1}= A^{-1}-A^{-1}Y(G^{-1}+Z^∗A^{-1}Y)^{-1}Z^∗A^{-1}.
\end{equation}
A simple form of SMW formula is Sherman-Morrison formula represented in the following statement \cite{bartlett1951inverse}:
Suppose $A\in R^{n\times n}$ is an invertible square matrix and $u,v\in R^n$ are column vectors. Then $A+uv\top$ is invertible $\iff 1+u^\top A^{-1}v\neq 0$. If $A+uv\top$ is invertible, then its inverse is given by
\begin{equation}\label{SMformula}
(A+uv^{T})^{-1}=A^{-1}-{A^{-1}uv^{T}A^{-1} \over 1+v^{T}A^{-1}u}.
\end{equation}

By using the formula, one can find a recursive way to update $K_{t}$ and $b_{t-1}$, which is 
\begin{align}
K_{t}  &=\frac{\sigma^4}{\tau^2+\sigma^2+\phi^2(\sigma^2-K_{t-1})},\\
b_{t} &= \begin{bmatrix}
\frac{b_{t-1}\phi K_{t}}{\sigma^2} \\ \frac{K_{t}(\sigma^2+\tau^2)-\sigma^4 }{\phi\sigma^2}
\end{bmatrix}. 
\end{align}
With the above formula, the recursive way of updating the mean and covariance is in the following formula: 
\begin{align}
\bar{\mu}_{t}       & = \frac{\phi}{\sigma^2}K_{t-1}\bar{\mu}_{t-1} + \phi (1 - \frac{K_{t-1}}{\sigma^2})y_{t-1}, \\
\bar{\Sigma}_{t}  &= \sigma^4K_{t}^{-1},
\end{align}
where $K_1=\frac{\sigma^4}{\sigma^2+\tau^2+L^2\phi^2}$. For calculation details, we refer readers to appendices (\ref{linearcalculation}).

\subsubsection*{The Estimation Distribution}

As introduced in section \ref{generalEstDistr}, from the joint distribution of $x_{1:t}$ and $y_{1:t}$, one can find the best estimation with a given $\theta$ by
\begin{align*}
\hat{x}_{1:t} \mid y_{1:t},\theta \sim N(L^{-\top}W,L^{-\top}L^{-1}),
\end{align*}
where $W = L^{-1}B_{t}y_{1:t-1}$. 
Consequently 
\begin{align*}
\hat{x}_{1:t} = L^{-\top}(W+Z),
\end{align*}
where $Z \sim N(0, I(\epsilon))$ is independent and identically distributed and drawn from a zero-mean normal distribution with variance $ I(\epsilon)$. Moreover, the mixture Gaussian distribution  $p(x_t \mid y_{1:t})$ can be found by 
\begin{align}
\mu_t^{(x)} &= \frac{1}{N} \sum_i \mu_{ti}^{(x)} \label{linearmu}  \\
\Var(x_t) &= \frac{1}{N} \sum_i \left( \mu_{ti}^{(x)}  \mu_{ti}^{(x)\top} +\Var(x_t)_i\right) -\frac{1}{N^2} \left(  \sum_i  \mu_{ti}^{(x)} \right) \left( \sum_i \mu_{ti}^{(x)} \right) ^\top.\label{linearsigma} 
\end{align}

To find $\mu_{ti}^{(x)}$ and $\Var(x_t)_i$, we will use the joint distribution of $x_{t}$ and $y_{1:t}$, which is $p(x_{t}, y_{1:t}  \mid  \theta)\sim N(0,\Gamma)$ and 
\begin{equation*}
\Gamma=\begin{bmatrix} C_{t}^\top(A_t-B_t)^{-1}C_{t} & C_{t}^\top(A_t-B_t)^{-1}\\(A_t-B_t)^{-1}C_{t} & (I_t-A_t^{-1}B_t)^{-1}B_t^{-1} \end{bmatrix}.
\end{equation*}
Because of 
\begin{align*}
C_{t}^\top A_{t}^{-1} = \left[\begin{matrix} - b_{t}^\top & \sigma^2- K_{t} \end{matrix} \right],
\end{align*}
thus, for any given $\theta$, we have $\hat{x}_{t}\mid y_{1:t},\theta \sim N\left( \mu_{t}^{(x)},\Var(x_t) \right)$, where
\begin{align}
\mu_{t}^{(x)} &  =  \frac{K_{t}\bar{\mu}_{t}}{\sigma^2}+(1-\frac{K_{t}}{\sigma^2})y_{t} \\
\Var(x_t)&= \sigma^2-K_{t}.
\end{align}
By substituting them into the equation (\ref{linearmu}) and (\ref{linearsigma}), the estimated $\hat{x}_t$ is easily got. For calculation details, we refer readers to appendices (\ref{linearcalculation}).

\begin{figure}[h]
\centering
\begin{subfigure}[b]{0.45\textwidth}
    \includegraphics[width=\textwidth]{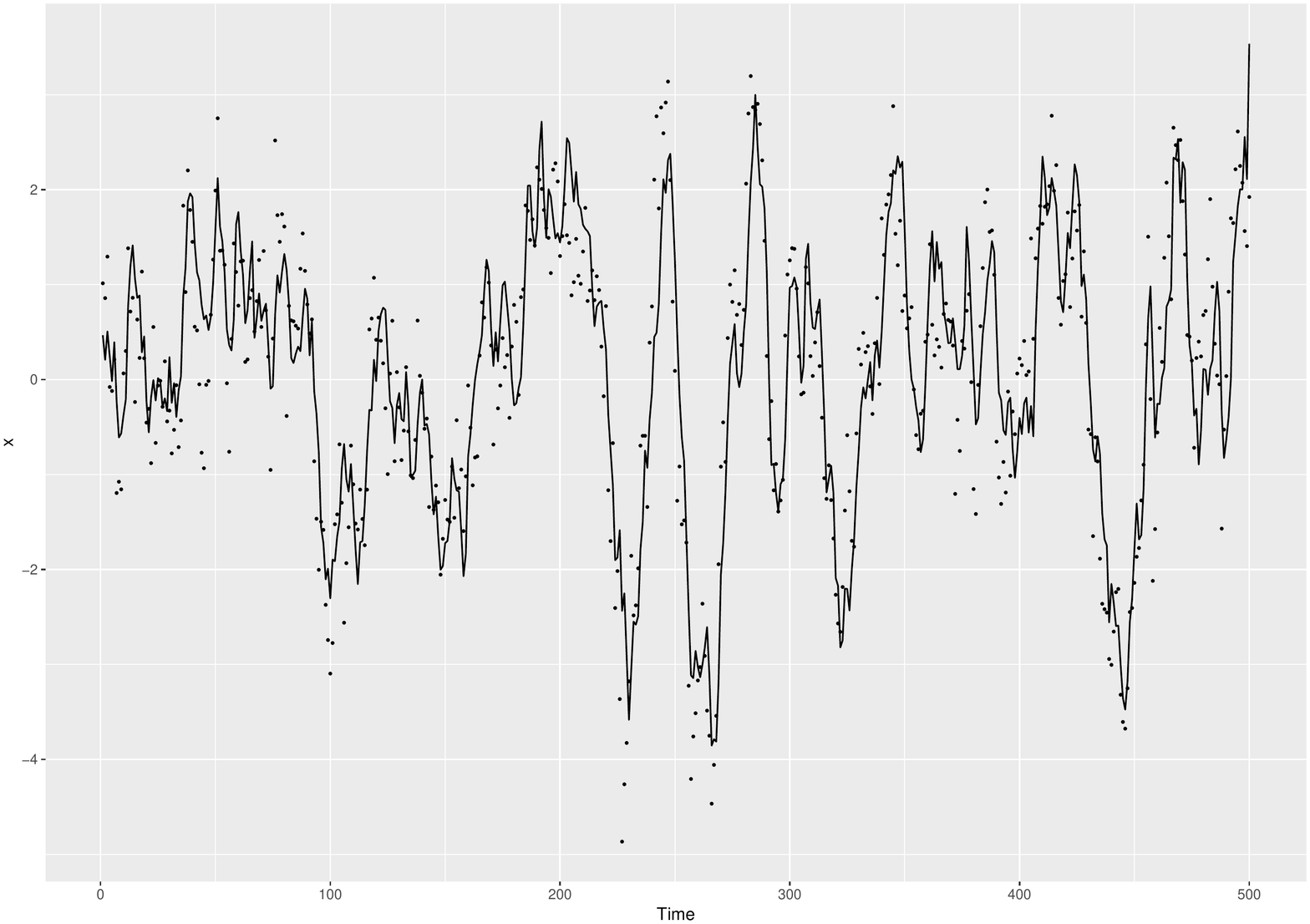}
     \caption{Estimation the whole $x_{1:t}$}\label{MCMClinearsimuXall}
\end{subfigure}
\begin{subfigure}[b]{0.45\textwidth}
	\includegraphics[width=\textwidth]{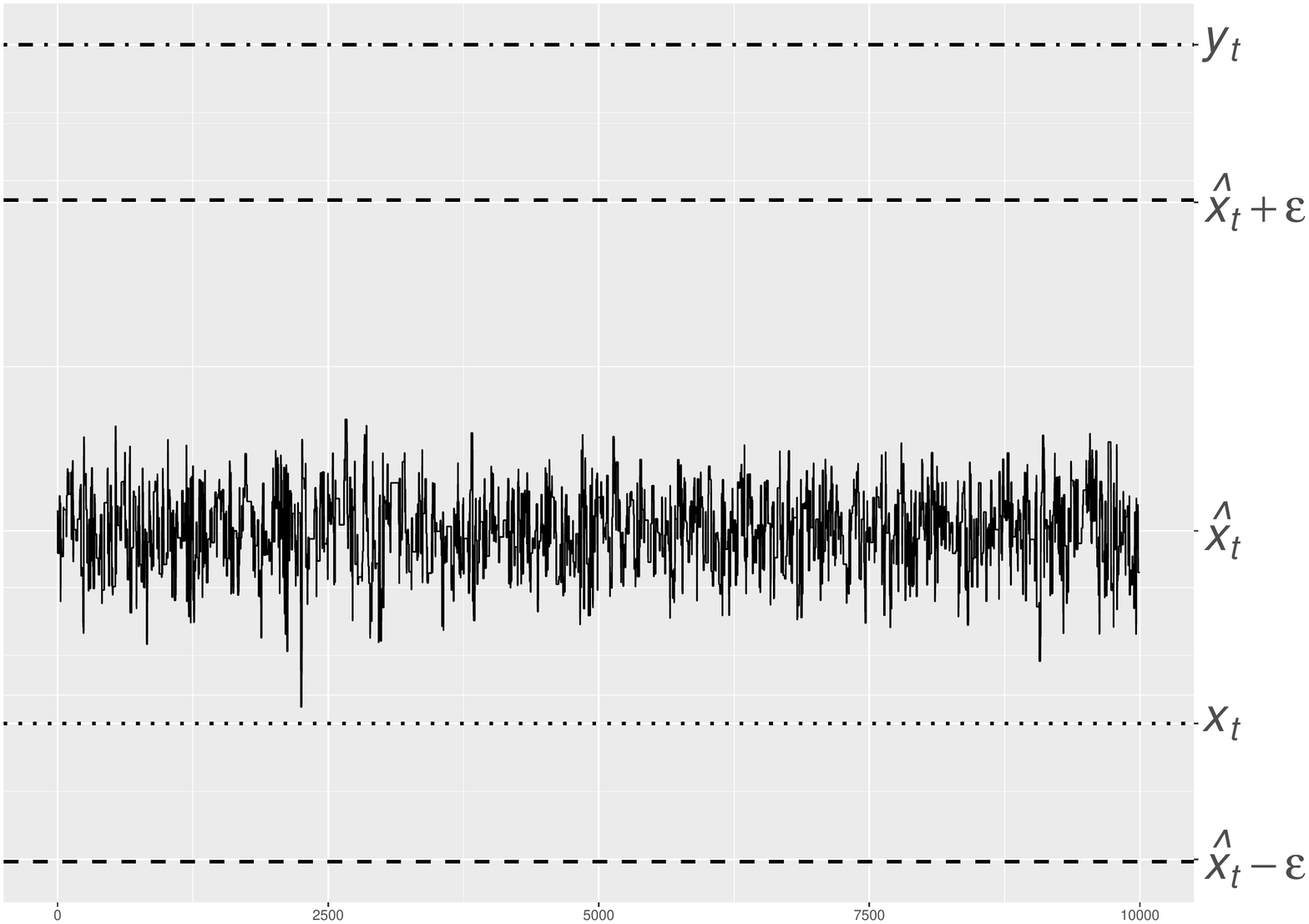}
     \caption{Estimation of a single $x_t$}\label{MCMClinearsimuXt2}
\end{subfigure}
\caption{Linear simulation of $x_{1:t}$ and single $x_t$.In figure \ref{MCMClinearsimuXall}, the dots is the true $x_{1:t}$ and the solid line is the estimation $\hat{x}_{1:t}$. In figure \ref{MCMClinearsimuXt2}, the estimation $\hat{x}_t$ is very close to the true $x$. In fact, the true $x$ falls in the interval $\lbrack \hat{x}-\varepsilon,\hat{x}+\varepsilon\rbrack$.}
\label{linearmarginXt}
\end{figure}

\subsection{Simulation on Irregular Time Series Data}

Irregularly sampled time series data is painful for scientists and researchers. In spatial data analysis, several satellites and buoy networks provide continuous observations of wind speed, sea surface temperature, ocean currents, etc. However, data was recorded with irregular time-step, with generally several data each day but also sometimes gaps of several days without any data. In \cite{tandeo2011linear}, the author adopts a continuous-time state-space model to analyze this kind of irregular time-step data, in which the state is supposed to be an Ornstein-Uhlenbeck process. 

The OU process is an adaptation of Brownian Motion, which models the movement of a free particle through a liquid and was first developed by Albert Einstein \cite{einstein1956investigations}. 
By considering the velocity $u_t$ of a Brownian motion at time $t$, over a small time interval, two factors affect the change in velocity: the frictional resistance of the surrounding medium whose effect is proportional to $u_t$ and the random impact of neighboring particles whose effect can be represented by a standard Wiener process. Thus, because mass times velocity equals force, the process in a differential equation form is 
\begin{equation*}
mdu_t = -\omega u_tdt+dW_t,
\end{equation*}
where $\omega>0$ is called the friction coefficient and $m>0$ is the mass. If we define $\gamma = \omega /m$ and $\lambda = 1/m$, we obtain the OU process \cite{vaughan2015goodness}, which was first introduced with the following differential equation:
\begin{equation*}
du_t= -\gamma u_tdt+\lambda dW_t.
\end{equation*}

The OU process is used to describe the velocity of a particle in a fluid and is encountered in statistical mechanics. It is the model of choice for random movement toward a concentration point. It is sometimes called a continuous-time Gauss Markov process, where a Gauss Markov process is a stochastic process that satisfies the requirements for both a Gaussian process and a Markov process. Because a Wiener process is both a Gaussian process and a Markov process, in addition to being a stationary independent increment process, it can be considered a Gauss-Markov process with independent increments \cite{kijima1997markov}. 

To apply OU process on irregular sampling data, we assume that the latent process $\{x_{1:t}\}$ is a simple OU process, that is a stationary solution of the following stochastic differential equation : 
\begin{equation}\label{linearOUequation}
dx_t= -\gamma x_tdt+\lambda dW_t, 
\end{equation}
where $W_t$ is a standard Brownian motion, $\gamma>0$ represents the slowly evolving transfer between two neighbor data and $\lambda$ is the forward transition variability. It is not hard to find the solution of equation (\ref{linearOUequation}) is 
\begin{equation*}
x_t = x_{t-1}e^{-\gamma t} +\int_{0}^{t} \lambda e^{-\gamma (t-s)}dW_s. 
\end{equation*}
For any arbitrary time step $t$, the general form of the process satisfies 
\begin{equation}
x_t = x_{t-1}e^{-\gamma \Delta_t} + \tau,
\end{equation}
where $\Delta_t = T_t-T_{t-1}$ is the time difference between two consecutive data points, $\tau$ is a Gaussian white noise with mean zero and variances $\frac{\lambda^2}{2\gamma}\left(1-e^{-2\gamma\Delta_t}\right)$. 

The observed $y_{1:t}$ is measured by 
\begin{equation}
y_t = Hx_t + \varepsilon,
\end{equation}
where $\varepsilon\sim N(0,\sigma)$ is a Gaussian white noise. 

To run simulations, we firstly generate irregular time lag sequence $\{\Delta_t\}$ from an \textit{Inverse Gamma} distribution with parameters $\alpha=2, \beta=0.1$. Then the following parameters were chosen for the numerical simulation: $\gamma = 0.5$, $\lambda^2 = 0.1$, $\sigma^2=1$.

\begin{figure}[h]
\centering
\includegraphics[width=0.45\textwidth,height=5cm]{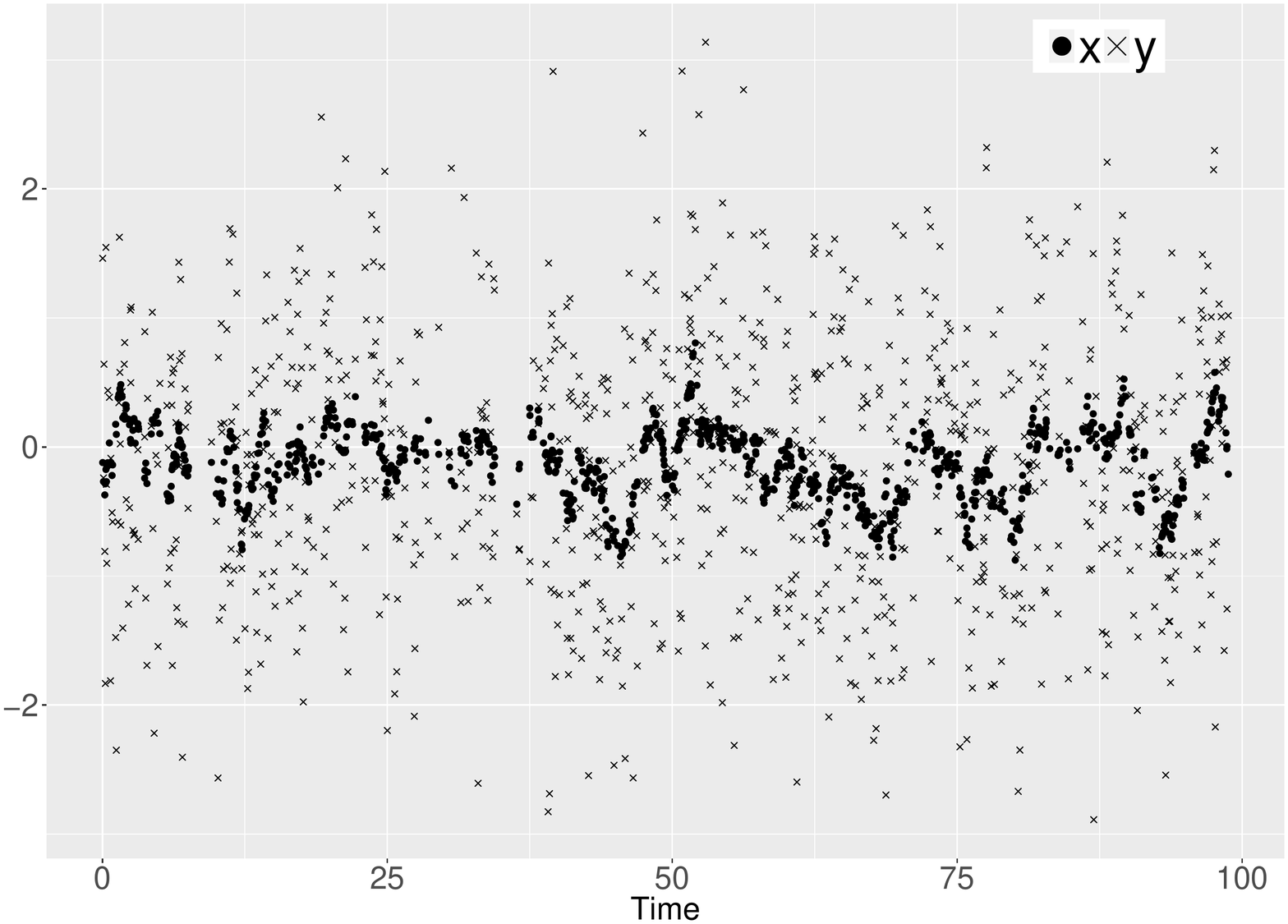}
\includegraphics[width=0.45\textwidth,,height=5cm]{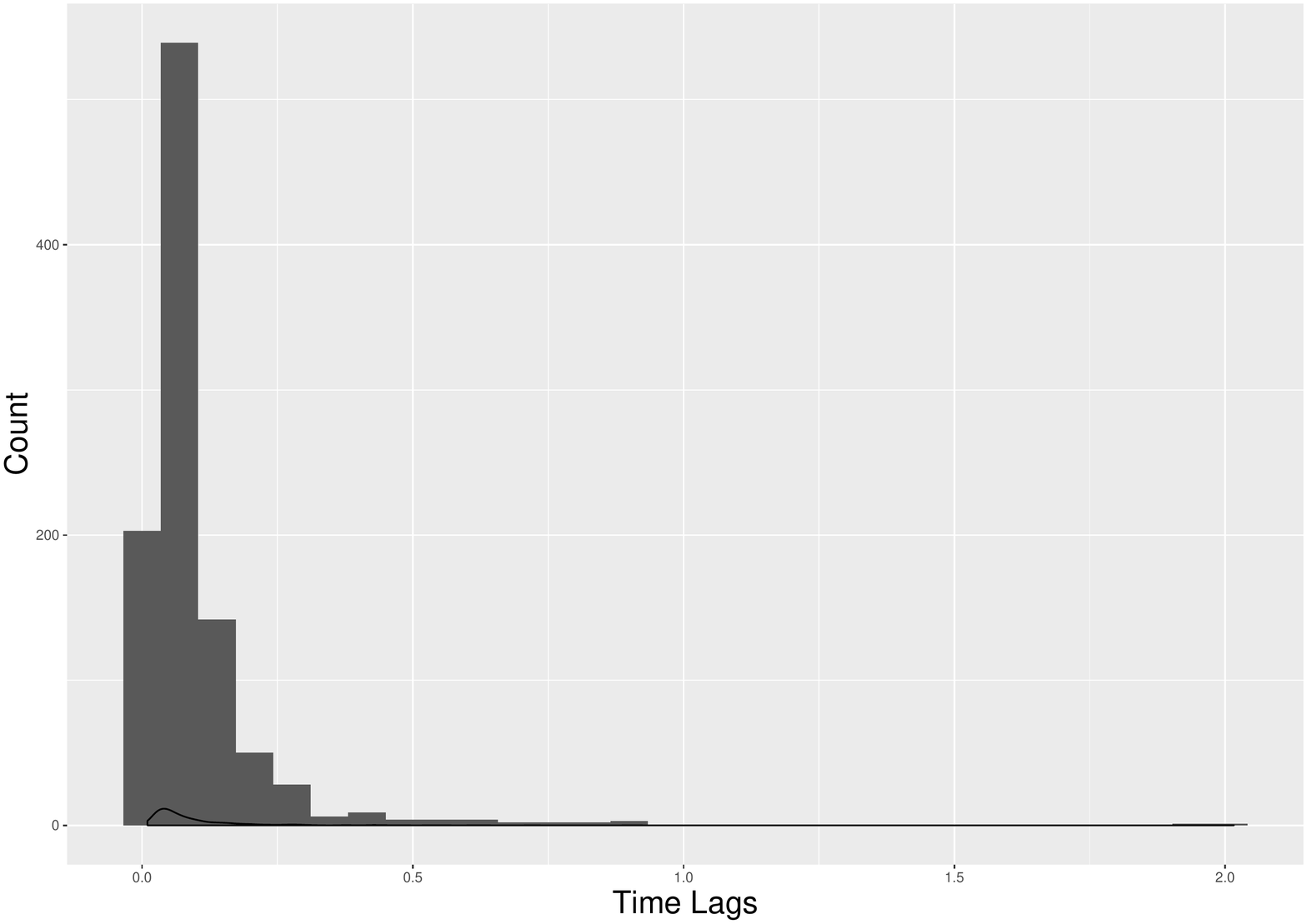}
\caption{Simulated data. The solid dots indicate the true state $x$ and cross dots indicate observation $y$. Irregular time lag $\Delta_t$ are generated from \textit{Inverse Gamma}(2,0.1) distribution.}
\label{simuOUreview}
\end{figure}

Similarly, we can get the joint distribution for $x_{0:t}$ and $y_{1:t}$ 
\begin{equation*}
\left[ \begin{matrix} x\\y  \end{matrix}\bigg\rvert \theta \right]
\sim N\left(0, \Sigma  \right),
\end{equation*}
from the procedure matrix 
\begin{equation*}
\begin{bmatrix}
\frac{1}{L^2}+\frac{\phi_1^2}{\tau_1^2} & \frac{-\phi_1}{\tau_1^2} & \cdots & 0 & 0 & 0& \cdots & 0\\ 
\frac{-\phi_1}{\tau_1^2}   &\frac{1}{\tau_1^2}+\frac{\phi_2^2}{\tau_2^2}+\frac{1}{\sigma^2}& \cdots & 0 & -\frac{1}{\sigma^2} &0 & \cdots & 0 \\
0 & \frac{-\phi_2}{\tau_2^2}   &  \cdots & 0 & 0& -\frac{1}{\sigma^2} & \cdots & 0\\
\vdots & \vdots & \ddots & \vdots & \vdots & \vdots & \ddots & \vdots \\
0 & 0   &  \cdots & \frac{1}{\tau_t^2}+\frac{1}{\sigma^2} & 0 & 0 & \cdots &-\frac{1}{\sigma^2}\\
0 & -\frac{1}{\sigma^2}  & \cdots & 0 & \frac{1}{\sigma^2} & 0 & \cdots & 0 \\
0& 0 & \cdots & 0 & 0 &  \frac{1}{\sigma^2} & \cdots & 0\\
\vdots & \vdots & \ddots & \vdots & \vdots & \vdots & \ddots & \vdots\\
0 & 0& \cdots &-\frac{1}{\sigma^2} & 0 & 0 & \cdots &  \frac{1}{\sigma^2}
\end{bmatrix},
\end{equation*}
where $\phi_t = e^{-\gamma\Delta_t}, \tau^2_t = \frac{\lambda^2}{2\gamma}\left(1-e^{-2\gamma\Delta_t}\right)$, $\theta$ represents unknown parameters. Denoted by $\Sigma^{-1}=\begin{bmatrix} A_t & -B_t \\ -B_t & B_t\end{bmatrix}$, covariance matrix is 
\begin{equation}
\Sigma=\begin{bmatrix} (A_t-B_t)^{-1} &  (A_t-B_t)^{-1} \\ (A_t-B_t)^{-1} & (I-A_t^{-1}B_t)^{-1}B_t^{-1} \end{bmatrix} \triangleq \begin{bmatrix}
\Sigma_{XX} & \Sigma_{XY}  \\ \Sigma_{YX} & \Sigma_{YY} 
\end{bmatrix},
\end{equation}
where $B_t$ is a $t\times t$ diagonal matrix with elements $\frac{1}{\sigma^2}$. The covariance matrices $\Sigma_{XX} =  (A_t-B_t)^{-1}$ and $\Sigma_{YY} =  (I-A_t^{-1}B_t)^{-1}B_t^{-1}$.

\subsubsection*{Parameters Estimation}

To use the algorithm \ref{algoonevarible}, similarly with section \ref{sectionlogParameter}, we firstly need to find the posterior distribution of $\theta$ with observations $y_{1:t}$, which in fact is 
\begin{equation*}
p(\theta \mid Y) \propto p(Y\mid\theta)p(\theta) \propto e^{-\frac{1}{2} Y \Sigma_{YY}^{-1} Y } \sqrt{\det \Sigma_{YY}^{-1}} p(\theta).
\end{equation*}
By taking natural logarithm on the posterior of $\theta$ and using the useful solutions in equations (\ref{sigmayy01}) and (\ref{sigmayy02}), we have 
\begin{equation}\label{simuOUlogL}
\ln L(\theta) = -\frac{1}{2}Y^\top\Sigma_{YY}^{-1}Y+\frac{1}{2}\sum\ln\mbox{tr}(B)-\sum\ln\mbox{tr}(L)+\sum\ln\mbox{tr}(R) + \ln p(\theta).
\end{equation}
Because of all parameters are positive, we are estimating $\nu_1=\ln\lambda$, $\nu_2=\ln\gamma^2$ and $\nu_3=\ln\sigma^2$ instead. When the estimation process is done, we can transform them back to the original scale by taking exponential. 

\begin{figure}[h]
\centering
 \begin{subfigure}[b]{0.3\textwidth}
     \includegraphics[width=\textwidth]{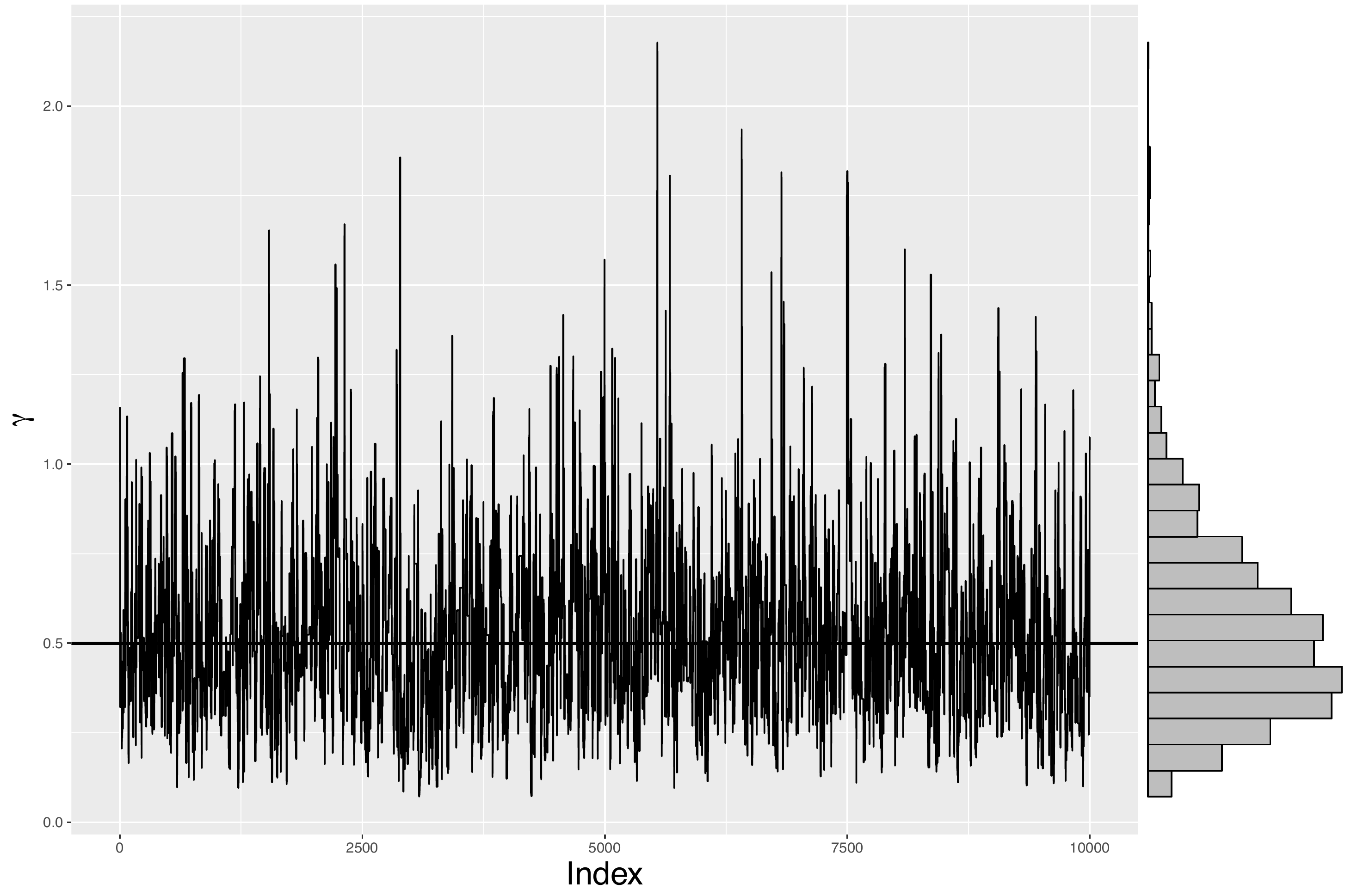}
     \caption{Trace plot of $\gamma$.}
\end{subfigure}
\begin{subfigure}[b]{0.3\textwidth}
    \includegraphics[width=\textwidth]{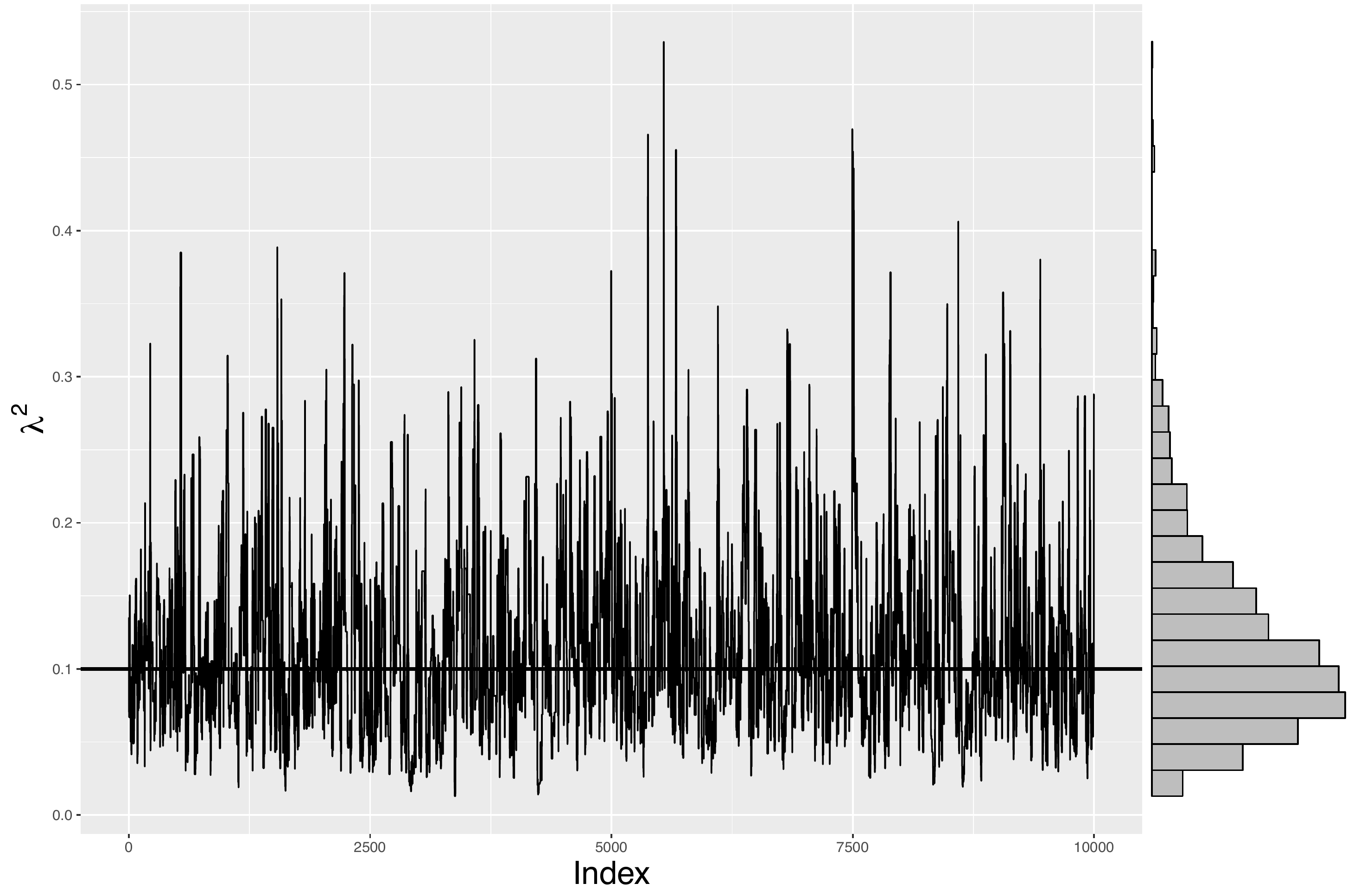}
     \caption{Trace plot of $\lambda^2$.}
\end{subfigure}
\begin{subfigure}[b]{0.3\textwidth}
    \includegraphics[width=\textwidth]{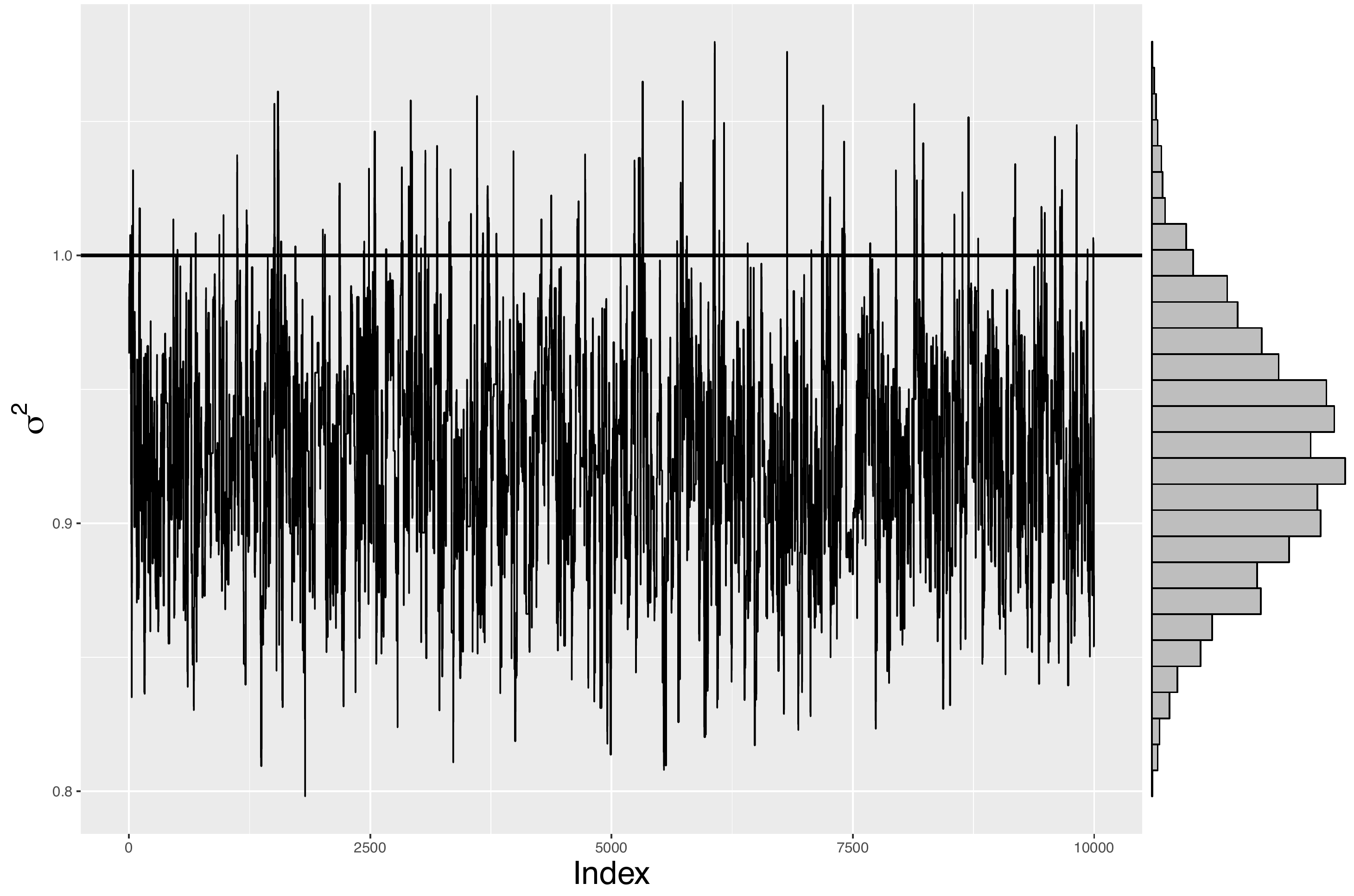}
     \caption{Trace plot of $\sigma^2$.}
\end{subfigure}
\caption{Irregular time step OU process simulation. The estimation of $\hat{\theta}$ is $\{\gamma=0.4841, \lambda^2=0.1032, \sigma^2=0.9276\}$. In the plots, the horizontal dark lines are the true $\theta$. }
\label{simuOUmarginplots}
\end{figure}

After running the whole process, it gives us the best estimation $\hat{\theta} = \{\gamma=0.4841, \lambda^2=0.1032, \sigma^2=0.9276\}$. In figure \ref{simuOUmarginplots}, we can see that the $\theta$ chains are skew to the true value with tails.

\subsubsection*{Recursive Calculation And State Estimation}

Follow the procedure in section \ref{sectionforecast} and do similar calculation with section \ref{sectionlinearRecursive}, one can find a recursive way to update $K_{t}$ and $b_{t-1}$, which are 
\begin{align} \label{linearOUK}
K_{t}  &=\frac{\sigma^4}{\tau_t^2+\sigma^2+\phi_t^2(\sigma^2-K_{t-1})},\\
b_{t} &= \begin{bmatrix}
\frac{b_{t-1}\phi_t K_{t}}{\sigma^2} \\ \frac{K_{t}(\sigma^2+\tau_t^2)-\sigma^4 }{\phi_t\sigma^2}
\end{bmatrix}. 
\end{align}
With the above formula, the recursive way of updating the mean and covariance are 
\begin{align} \label{linearOUmu}
\bar{\mu}_{t}       & = \frac{\phi_t}{\sigma^2}K_{t-1}\bar{\mu}_{t-1} + \phi_t (1 - \frac{K_{t-1}}{\sigma^2})y_{t-1}, \\
\bar{\Sigma}_{t}  &= \sigma^4K_{t}^{-1}, \label{linearOUsigma}
\end{align}
where $K_1=\frac{\sigma^4}{\sigma^2+\tau_1^2+L^2\phi_1^2}$. 

Additionally, as introduced in section \ref{generalEstDistr}, the best estimation of $x_{1:t}$ with a given $\theta$ is 
\begin{align*}
\hat{x}_{1:t} \mid y_{1:t},\theta \sim N(L^{-\top}W,L^{-\top}L^{-1}),
\end{align*}
where $W = L^{-1}B_{t}y_{1:t-1}$, and the mixture Gaussian distribution for $p(x_t \mid y_{1:t})$ is 
\begin{align}
\mu_t^{(x)} &= \frac{1}{N} \sum_i \mu_{ti}^{(x)}  \\
\Var(x_t) &= \frac{1}{N} \sum_i \left( \mu_{ti}^{(x)}  \mu_{ti}^{(x)\top} +\Var(x_t)_i\right) -\frac{1}{N^2} \left(  \sum_i  \mu_{ti}^{(x)} \right) \left( \sum_i \mu_{ti}^{(x)} \right) ^\top,
\end{align}
The same as we did in section \ref{sectionlinearRecursive}, for any given $\theta$, we have $\hat{x}_{t}\mid y_{1:t},\theta \sim N\left( \mu_{t}^{(x)},\Var(x_t) \right)$, where
\begin{align*}
\mu_{t}^{(x)} &  =  \frac{K_{t}\bar{\mu}_{t}}{\sigma^2}+(1-\frac{K_{t}}{\sigma^2})y_{t} \\
\Var(x_t)&= \sigma^2-K_{t}.
\end{align*}
By substituting them into the equation (\ref{linearmu}) and (\ref{linearsigma}), the estimated $\hat{x}_t$ is easily got. The difference at this time is the $\mu_{t}^{(x)}$ and $\Var(x_t)$ are dependent on time lag $\Delta_t$, that can be seen from formula (\ref{linearOUK}) and (\ref{linearOUmu}).

\begin{figure}[h]
\centering
\begin{subfigure}[h]{0.45\textwidth}
    \includegraphics[width=\textwidth]{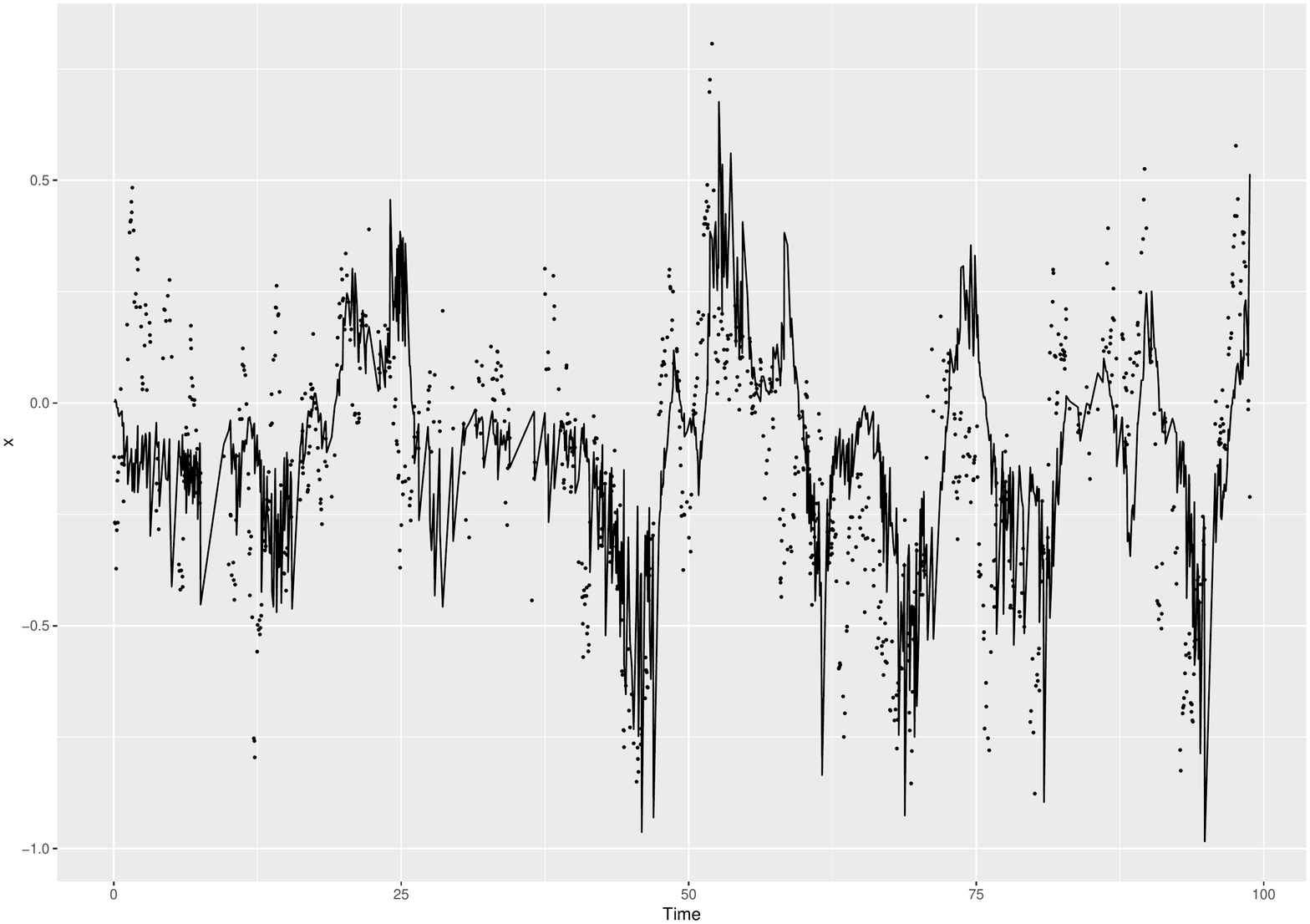}
     \caption{Batch method of estimation $x_{1:t}$}\label{MCMCOUallX}
\end{subfigure}
\begin{subfigure}[h]{0.45\textwidth}
    \includegraphics[width=\textwidth]{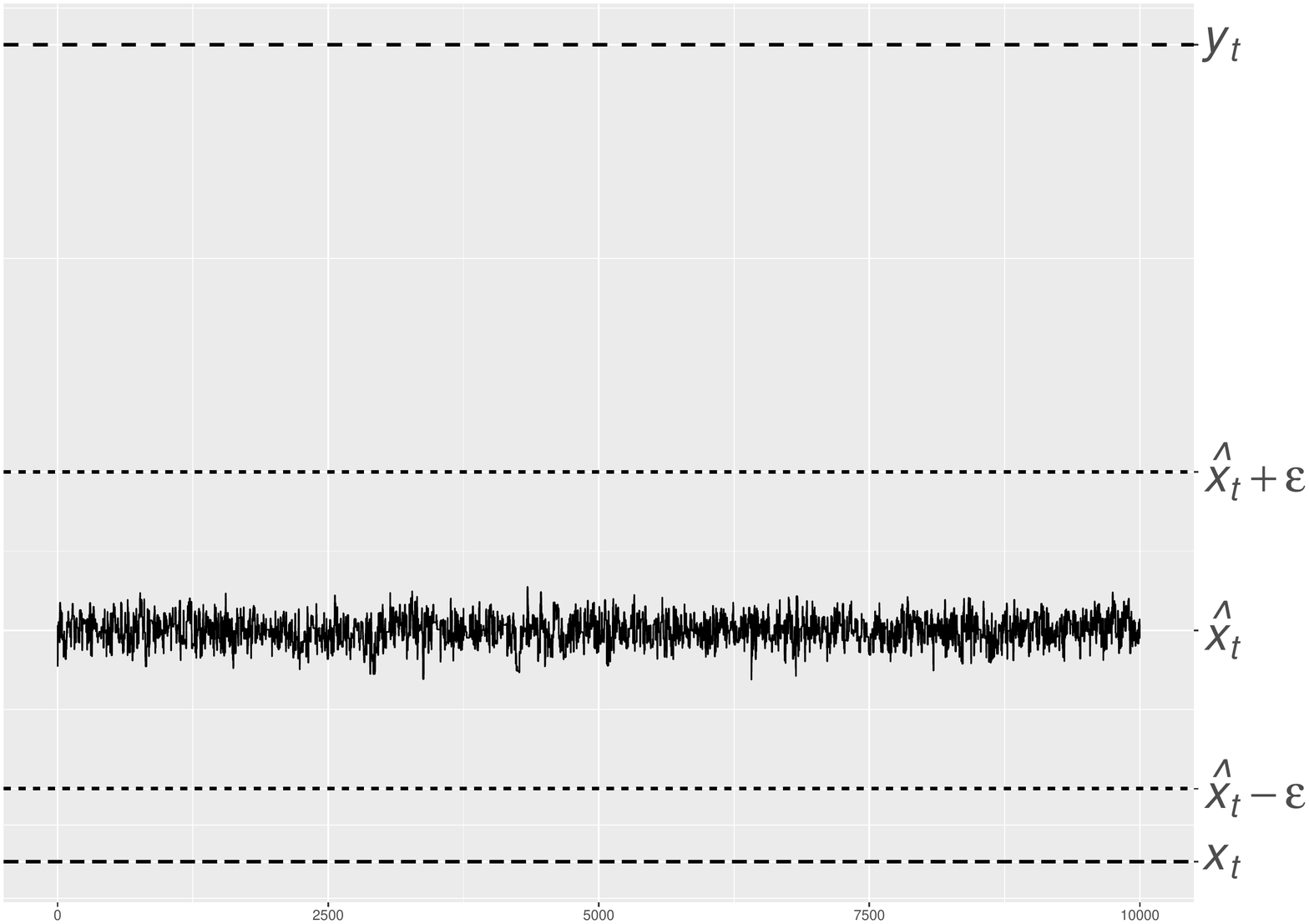}
     \caption{Sequential method of estimation $x_t$}\label{MCMCOUallXt2}
\end{subfigure}
\caption{Irregular time step OU process simulation of $x_{1:t}$ and sole $x_t$. In figure \ref{MCMCOUallX}, the dots is the true $x_{1:t}$ and the solid line is the estimation $\hat{x}_{1:t}$. In figure \ref{MCMCOUallXt2}, the chain in solid line is the estimation $\hat{x}_t$; dotted line is the true value of $x$; dot-dash line on top is the observed value of $y$; dashed lines are the estimated error. }
\label{simuOUxt}
\end{figure}

\section{High Dimensional OU-Process Application}\label{SectionHighDimensionalOU}

Tractors moving on an orchard are mounted with GPS units, which are recording data and transfer to the remote server. This data infers longitude, latitude, bearing, etc, with unevenly spaced time mark. However, one dimensional OU process containing either only position or velocity is not enough to infer a complex movement. 

\begin{figure}[h]
\centering
\includegraphics[width=0.45\textwidth]{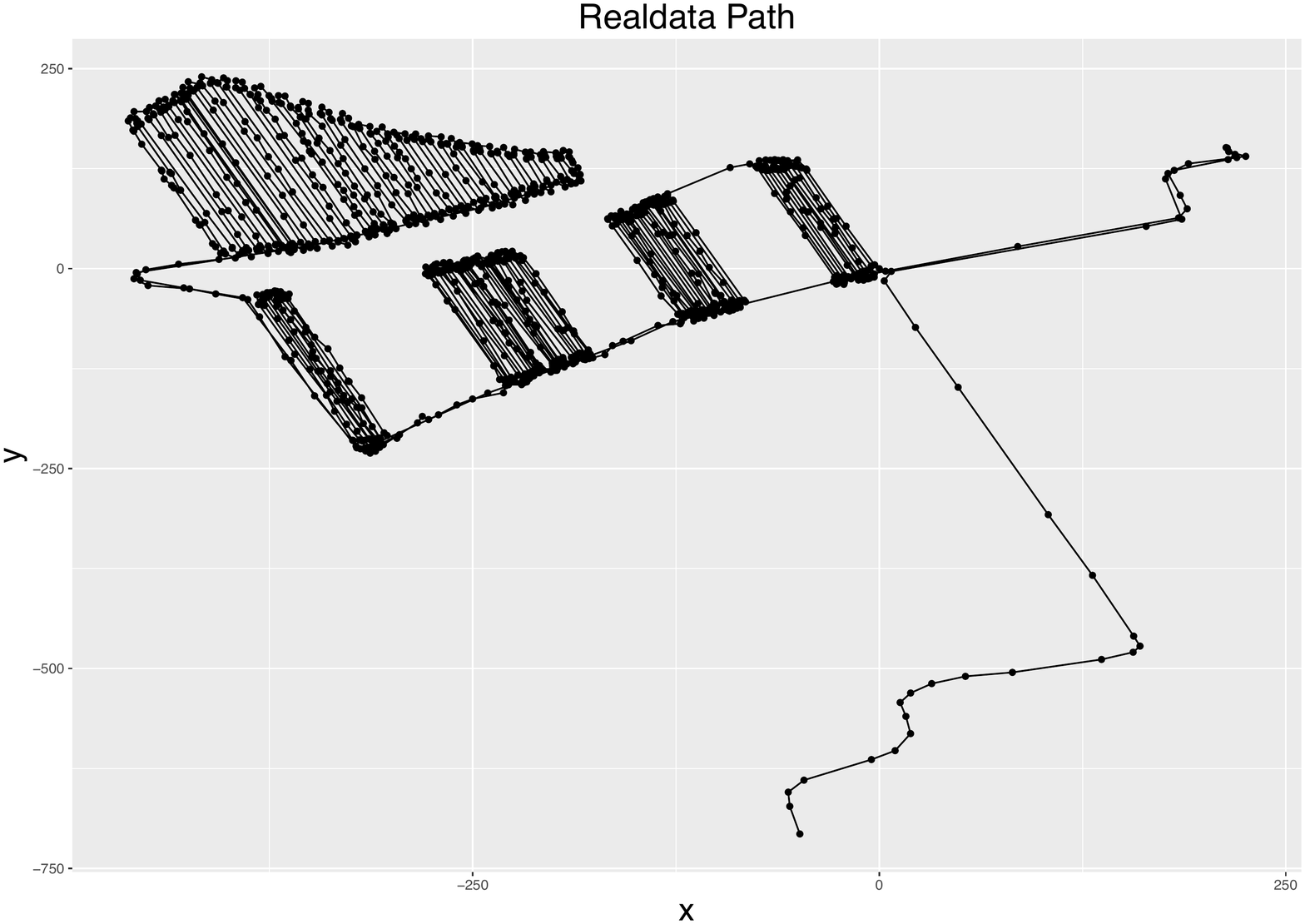}
\includegraphics[width=0.45\textwidth]{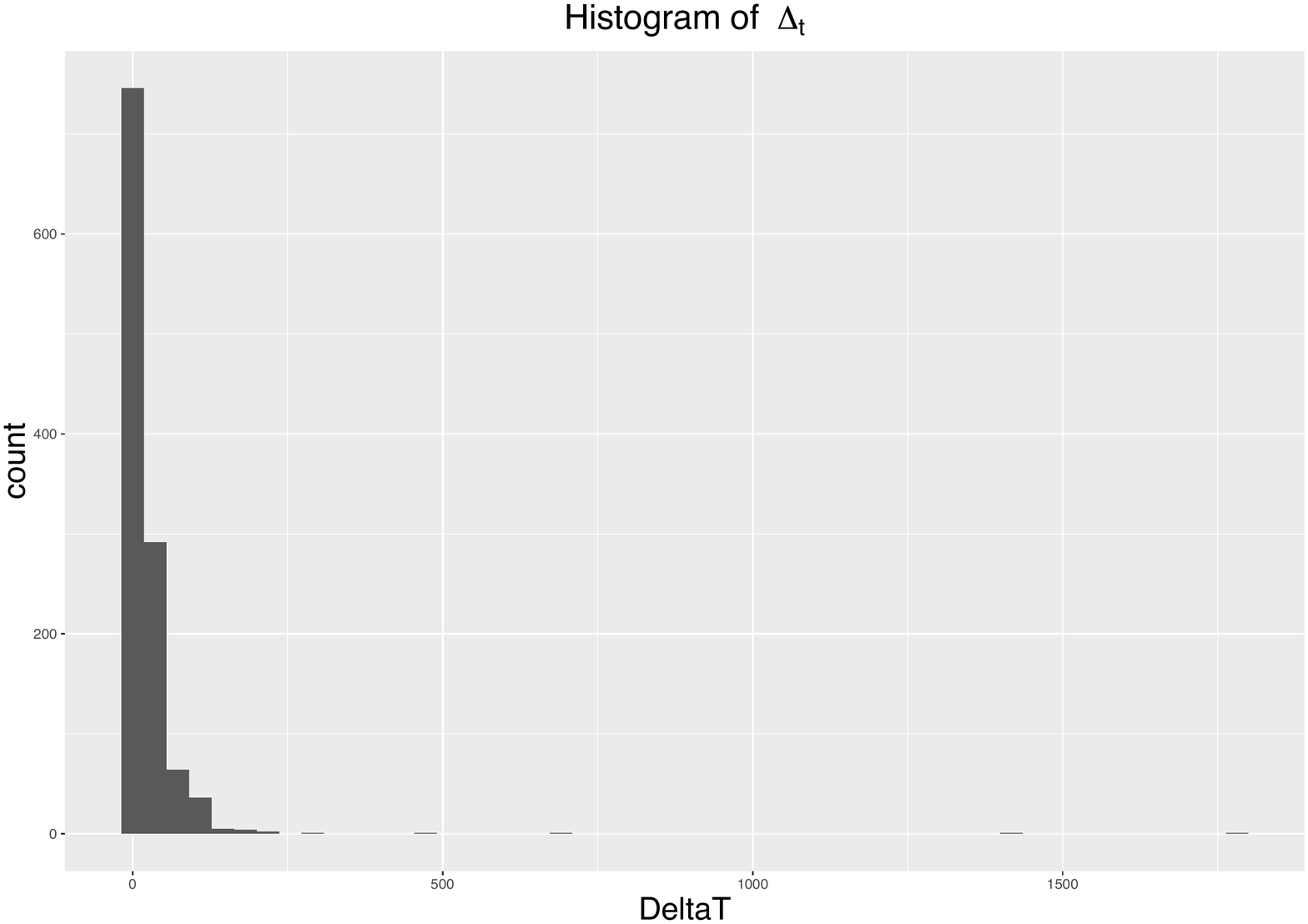}
\caption{The trajectory of a moving tractor. The time lags (right side figure) obtained from GPS units are irregular.}
\label{realdatareview}
\end{figure}

Therefore, in this section, we are introducing an OU-process model combing both position and velocity with the following equations  
\begin{equation}\label{OUprocess}
\begin{cases}
du_t = -\gamma u_t dt+ \lambda dW_t,\\
dx_t = u_t dt+\xi dW_t'.
\end{cases}
\end{equation}
The solution can be found by integrating $dt$ out, that gives us 
\begin{align}
\begin{cases}
u_t &=u_{t-1}e^{-\gamma t} +\int_{0}^{t} \lambda e^{-\gamma (t-s)}dW_s,\\
x_t &=x_{t-1} +\frac{u_{t-1}}{\gamma}(1- e^{-\gamma t}) + \int_{0}^{t} \frac{\lambda}{\gamma}e^{\gamma  s} \left(1-e^{-\gamma t}\right)dW_s + \int_{0}^{t}\xi dW_s'.
\end{cases}
\end{align}
As a result, the joint distribution is 
\begin{align}
\begin{bmatrix} x_t \\ u_t \end{bmatrix} &\sim N\left(
\begin{bmatrix}\mu_t^{(x)} \\ \mu_t^{(u)}  \end{bmatrix} , 
\begin{bmatrix}
\sigma_t^{(x)2} & \rho_t\sigma_t^{(x)} \sigma_t^{(u)} \\
\rho_t\sigma_t^{(x)} \sigma_t^{(u)} & \sigma_t^{(u)2}
\end{bmatrix} \right),
\end{align}
where $\mu_t^{(x)}$ and $\mu_t^{(u)} $ are from the forward map process 
\begin{align}
\begin{bmatrix}\mu_t^{(x)} \\ \mu_t^{(u)}  \end{bmatrix}  = 
\begin{bmatrix}
1 & \frac{1-e^{-\gamma \Delta_t}}{\gamma} \\ 0 &  e^{-\gamma \Delta_t}
\end{bmatrix}  \begin{bmatrix} x_{t-1}^{(x)} \\ u_{t-1}  \end{bmatrix} \triangleq \Phi \begin{bmatrix} x_{t-1}^{(x)} \\ u_{t-1}  \end{bmatrix},
\end{align}
and 
\begin{align*}
\begin{cases}
\sigma_t^{(x)2} &=\frac{\lambda^2 \left(e^{2 \gamma\Delta_t}-1\right) \left(1 -e^{-\gamma\Delta_t}\right)^2}{2 \gamma ^3 } + \xi^2\Delta_t\\
\sigma_t^{(u)2} &= \frac{\lambda ^2 \left(1- e^{-2 \gamma\Delta_t}\right)}{2 \gamma } \\
\rho_t\sigma_t^{(x)}\sigma_t^{(u)} & =\frac{\lambda ^2 \left(e^{\gamma\Delta_t} -1\right) \left(1-e^{-2\gamma\Delta_t}\right)}{2 \gamma ^2}
\end{cases}
\end{align*}
In the above equations $\Delta_t = T_t-T_{t-1}$ and initial values are $\Delta_1=0$, $x_0\sim N(0,L_x^2), u_0\sim N(0,L_u^2)$, $\rho_t^2 = 1-\frac{\xi^2 \Delta_t}{\sigma_t^{(x)^2}}$. To be useful, we are using $\frac{1}{1-\rho_t^2} =\frac{\sigma_t^{(x)^2}}{\xi^2 \Delta_t}$ instead in the calculation. 

Furthermore, the independent observation process is 
\begin{equation}\label{obseq}
\begin{cases} y_t=x_t+\epsilon_t,\\ v_t=u_t+\epsilon'_t, \end{cases} 
\end{equation}
where $\epsilon_t\sim N(0,\sigma),\epsilon'_t\sim N(0,\tau)$ are normally distributed independent errors. Thus, the joint distribution of observations is 
\begin{align}\label{obmodel}
\begin{bmatrix} y_t \\ v_t \end{bmatrix} &\sim N\left(
\begin{bmatrix}x_t \\ u_t \end{bmatrix} , 
\begin{bmatrix}
\sigma^2 & 0\\
0 & \tau^2
\end{bmatrix} \right).
\end{align}
Consequently, the parameter $\theta$ of an entire Ornstein-Uhlenbeck process is a set of five parameters from both hidden status and observation process, which is represented as $\theta = \{\gamma,\xi^2,\lambda^2,\sigma^2,\tau^2 \}$.

Starting from the joint distribution of $x_{0:t},u_{0:t}$ and $y_{1:t},v_{1:t}$ by given $\theta$, it can be found that
\begin{equation}\label{jointmatrix}
\begin{bmatrix} \begin{matrix} \tilde{X}\\ \tilde{Y}  \end{matrix} \biggr\rvert \theta \end{bmatrix}
\sim N\left(0, \tilde{\Sigma} \right),
\end{equation}
where $\tilde{X}$ represents for the hidden statues $\{x,u\}$, $\tilde{Y}$ represents for observed $\{y,v\}$, $\theta$ is the set of five parameters.  The inverse of the covariance matrix $\tilde{\Sigma}^{-1}$ is the procedure matrix in the form of
\begin{align*} \tilde{\Sigma}^{-1}=
\begin{bmatrix}
Q_{xx} & Q_{xu} & -\frac{1}{\sigma^2}I & 0\\
Q_{ux} & Q_{uu} & 0 &-\frac{1}{\tau^2}I \\
-\frac{1}{\sigma^2}I & 0 & \frac{1}{\sigma^2}I  & 0\\
 0  &  -\frac{1}{\tau^2}I  & 0 & \frac{1}{\tau^2}I 
\end{bmatrix}.
\end{align*}
To make the covariance matrix a more beautiful form and convenient computing, $\tilde{X}$, $\tilde{Y}$ and $\tilde{\Sigma}$ can be rearranged in a time series order, that makes $X_{1:t} = \{x_1,u_1,x_2,u_2,\cdots, x_t, u_t \}$, $Y_{1:t} = \{y_1,v_1,y_2,v_2,\cdots, y_t, v_t \}$ and the new procedure matrix $\Sigma^{-1}$ looks like 
\begin{align*} \Sigma^{-1} &=
\begin{bmatrix}
\sigma_{11}^{(x)2}+\frac{1}{\sigma^2} & \sigma_{11}^{(xu)2} & \cdots & \sigma_{1t}^{(x)2} & \sigma_{1t}^{(xu)2}  &  -\frac{1}{\sigma^2} & 0 & \cdots & 0 & 0\\
\sigma_{11}^{(ux)2}   & \sigma_{11}^{(u)2} +\frac{1}{\tau^2} & \cdots & \sigma_{1t}^{(ux)2} & \sigma_{1t}^{(x)2}  &  0 & -\frac{1}{\tau^2} & \cdots & 0 & 0 \\
\vdots & \vdots & \ddots & \vdots & \vdots & \vdots & \vdots &\ddots & \vdots & \vdots \\
\sigma_{t1}^{(x)2}   & \sigma_{t1}^{(xu)2} & \cdots & \sigma_{tt}^{(x)2} +\frac{1}{\sigma^2}  & \sigma_{tt}^{(xu)2}  &  0 & 0 & \cdots & -\frac{1}{\sigma^2} & 0 \\
\sigma_{t1}^{(ux)2}   & \sigma_{t1}^{(u)2} & \cdots & \sigma_{tt}^{(ux)2} & \sigma_{tt}^{(u)2} +\frac{1}{\tau^2}  &  0 & 0 & \cdots & 0 &-\frac{1}{\tau^2} \\
- \frac{1}{\sigma^2} & 0 & \cdots & 0 & 0 &  \frac{1}{\sigma^2} & 0 & \cdots & 0 & 0\\
0  & -\frac{1}{\tau^2}& \cdots & 0 & 0 &  0 &  \frac{1}{\tau^2} & \cdots & 0 & 0 \\
\vdots & \vdots & \ddots & \vdots & \vdots & \vdots & \vdots &\ddots & \vdots & \vdots \\
0 & 0& \cdots & -\frac{1}{\sigma^2}  &0&  0 & 0 & \cdots & \frac{1}{\sigma^2} & 0 \\
0 & 0 & \cdots & 0 & -\frac{1}{\tau^2}   &  0 & 0 & \cdots & 0 & \frac{1}{\tau^2}
\end{bmatrix} \\ 
& \triangleq \begin{bmatrix}
A_t& -B_t \\ -B_t^\top & B_t
\end{bmatrix},
\end{align*}
where $B_t$ is a $2t\times 2t$ diagonal matrix of observation errors at time $t$ in the form of $\begin{bmatrix}
\frac{1}{\sigma^2}& \cdot & \cdot &  \cdot  &  \cdot \\  \cdot & \frac{1}{\tau^2} & \cdot &  \cdot  &  \cdot  \\ 
\vdots & \vdots & \ddots & \vdots & \vdots \\
 \cdot  &  \cdot  & \cdot  & \frac{1}{\sigma^2}&  \cdot \\  \cdot  &  \cdot & \cdot  &  \cdot  & \frac{1}{\tau^2}
\end{bmatrix}$. 
In fact, the matrix $A_t$ is a $2t \times 2t$ bandwidth six sparse matrix at time $t$ in the process. For sake of simplicity, we are using $A$ and $B$ to represent the matrices  $A_t$ and $B_t$ here. Then we may find the covariance matrix by calculating the inverse of the procedure matrix as 
\begin{align*}
\Sigma &= \begin{bmatrix}
(A-B^\top B^{-1}B) ^{-1} & -(A-B^\top B^{-1}B)^{-1}B^\top B^{-1}\\
- B^{-1}B(A-B^\top B^{-1}B)^{-1} & (B-B^\top A^{-1}B) ^{-1}
\end{bmatrix} \\
&= \begin{bmatrix}
(A-B) ^{-1} & (A-B)^{-1}\\
(A-B)^{-1} & (I- A^{-1}B) ^{-1}B^{-1}
\end{bmatrix} \\
&\triangleq \begin{bmatrix}
\Sigma_{XX} & \Sigma_{XY} \\
\Sigma_{YX}  &\Sigma_{YY} 
\end{bmatrix}.
\end{align*}
A detailed structure of the covariance matrix $\Sigma_{XX} $ is presented in section \ref{covMatrixdetails}. 

\subsection{Approximations of The Parameters Posterior}

To find the log-posterior distribution of $X_{1:t}$ and $Y_{1:t}$, we shall start from the joint distribution. Similarly, the inverse of the covariance matrix is 
\begin{align*}
\Sigma_{YY}^{-1} &= B(I-A^{-1}B)= BA^{-1}\Sigma_{XX}^{-1}.
\end{align*}
By using Choleski decomposition and similar technical solution, second term in the integrated objective function is 
\begin{align*}
p(\theta \mid Y) &\propto p(Y\mid\theta)p(\theta) \propto e^{-\frac{1}{2} Y \Sigma_{YY}^{-1} Y } \sqrt{\det \Sigma_{YY}^{-1}} P(\theta).
\end{align*}
Then by taking natural logarithm on the posterior of $\theta$ and using the useful solutions in equations (\ref{sigmayy01}) and (\ref{sigmayy02}), we will have
\begin{align}\label{logL}
\ln L(\theta) &= -\frac{1}{2}Y^\top\Sigma_{YY}^{-1}Y+\frac{1}{2}\sum\ln\mbox{tr}(B)-\sum\ln\mbox{tr}(L)+\sum\ln\mbox{tr}(R).
\end{align}

\subsection{The Forecast Distribution}

It is known that 
\begin{align*}
p(Y_{1:t-1},\theta) &\sim N\left( 0,\Sigma_{YY}^{(t-1)} \right)\\
p(Y_{t},Y_{1:t-1},\theta) &\sim N\left( 0,\Sigma_{YY}^{(t)} \right)\\
p(Y_{t}\mid Y_{1:t},\theta) &\sim N\left( \bar{\mu}_{t},\bar{\Sigma}_{t} \right)
\end{align*}
where the covariance matrix of the joint distribution is $\Sigma_{YY}^{(t)} = (I_{t}-A_{t}^{-1}B_{t})^{-1}B_{t}^{-1}$. Then, by taking its inverse, we will get
\begin{align*}
\Sigma_{YY}^{(t) (-1)} = B_{t}(I_{t}-A_{t}^{-1}B_{t}).
\end{align*}
To be clear, the matrix $B_{t}$ is short for the matrix $B_{t}(\sigma^2,\tau^2)$, which is $2t\times 2t$ diagonal matrix with elements $\frac{1}{\sigma^2},\frac{1}{\tau^2}$ repeating for $t$ times on its diagonal. For instance, the very simple $B_1(\sigma^2,\tau^2) = 
\begin{bmatrix}
\frac{1}{\sigma^2} & 0  \\
0 & \frac{1}{\tau^2}
\end{bmatrix}_{2\times 2}$ is a $2\times 2$ matrix. 

Because of $A_t$ is symmetric and invertible, $B_t$ is the diagonal matrix defined as above, then they have the following property 
\begin{align*}
& A_tB=A_t^\top B_t^\top = (B_tA_t)^\top, \\
& A_t^{-1}B_t = A_t^{-\top}B_t^\top = (B_tA_t^{-1})^\top. 
\end{align*}
Followed up the form of $\Sigma_{YY}^{(t) (-1)}$, we can define that 
\begin{align*}
\Sigma_{YY}^{(t) (-1)} \triangleq \begin{bmatrix} 
B_{t-1} & 0 \\ 0 & B_1 \end{bmatrix}
\begin{bmatrix} 
Z_{t} & b_{t} \\
b_{t}^\top & K_{t}
\end{bmatrix} \begin{bmatrix} 
B_{t-1} & 0 \\ 0 & B_1\end{bmatrix}
\end{align*}
where $Z_{t}$ is a $2t \times 2t$ matrix, $ b_{t} $ is a $2t \times 2$ matrix and $K_{t}$ is a $2 \times 2$ matrix. Thus by taking its inverse again, we will get 
\begin{align*} \Sigma_{YY}^{(t)}= \left[ \begin{matrix}
B_{t-1}^{-1} (Z_{t}-b_{t}K_{t}^{-1}b_{t}^\top)^{-1}B_{t-1}^{-1}  & - B_{t-1}^{-1}  Z_{t}^{-1}b_{t}(K_{t}-b_{t}^\top Z_{t}^{-1}b_{t})^{-1}B_1^{-1} \\
-B_1^{-1}  K_{t}^{-1}b_{t}^\top (Z_{t}-b_{t}K_{t}^{-1}b_{t}^\top)^{-1}B_{t-1}^{-1}  & B_1^{-1}  (K_{t}-b_{t}^\top Z_{t}^{-1}b_{t})^{-1}B_1^{-1} 
\end{matrix}\right].
\end{align*}

It is easy to find the relationship between $A_{t}$ and  $A_{t}$ in the Sherman-Morrison-Woodbury form, which is 
\begin{align*} A_{t} = 
\begin{bmatrix}
A_{t-1} & \cdot & \cdot  \\ \cdot &\frac{1}{\sigma^2} &\cdot  \\ \cdot  & \cdot  & \frac{1}{\tau^2} 
\end{bmatrix} + U_{t}U_{t}^\top \triangleq M_{t}  + U_{t}U_{t}^\top,
\end{align*}
where, in fact, $M_{t} = \begin{bmatrix}
A_{t-1} & \cdot & \cdot  \\ \cdot &\frac{1}{\sigma^2} &\cdot  \\ \cdot  & \cdot  & \frac{1}{\tau^2}
\end{bmatrix}  = \begin{bmatrix}
A_{t-1} & 0 \\ 0 & B_1
\end{bmatrix}$ 
and its inverse is $M_{t}^{-1} =\begin{bmatrix}
A_{t-1}^{-1} & 0 \\ 0 & B_1^{-1}
\end{bmatrix}$. We may use Sherman-Morrison-Woodbury formula to find the inverse of $A_{t}$ in a recursive way, which is 
\begin{equation}
A_{t}^{-1} = (M_{t}+U_{t}U_{t}^\top)^{-1}= M_{t}^{-1}-M_{t}^{-1}U_{t}(I+U_{t}^\top M_{t}^{-1}U_{t})^{-1}U_{t}^\top M_{t}^{-1}.
\end{equation}
Consequently, with some calculations, we will get 
\begin{equation}\label{OUupdatingK}
K_{t} =B_1^{-1}D_{t} (I+ S_{t}^\top (B_1^{-1} - K_{t-1})  S_{t} +D_{t}^\top B_1^{-1}D_{t}  )^{-1}  D_{t}^\top B_1^{-1},
\end{equation}
and
\begin{align*}
b_{t} = \begin{bmatrix}
-b_{t-1} \\ B_1^{-1}-K_{t-1} 
\end{bmatrix}  S_{t} (I+ S_{t}^\top (B_1^{-1} - K_{t-1})  S_{t} +D_{t}^\top B_1^{-1}D_{t}  )^{-1} D_{t}^\top B_1^{-1}, 
\end{align*}
that are updating in a recursive way. Therefore, one can achieve the recursive updating formula for the mean and covariance matrix, which are  
\begin{align}
\begin{cases}
\bar{\mu}_{t}&=\Phi_{t} K_{t-1}B_1\bar{\mu}_{t-1} + \Phi_{t} (I-K_{t-1}B_1)Y_{t-1}\\
\bar{\Sigma}_{t}&=\left( B_1K_{t}B_1  \right)^{-1}
\end{cases}.
\end{align}
The matrix $K_{t}$ is updated via equation (\ref{OUupdatingK}), or updating its inverse in the following form makes the computation faster, that is 
\begin{align*}
\begin{cases}
K_{t}^{-1} &= B_1D_{t}^{-\top}D_{t}^{-1}B_1 + B_1\Phi_{t} (B_1^{-1} - K_{t-1}) \Phi_{t}^\top B_1+ B_1,\\
\bar{\Sigma}_{t}&= D_{t}^{-\top}D_{t}^{-1}+ \Phi_{t} (B_1^{-1} - K_{t-1}) \Phi_{t}^\top + B_1^{-1}
\end{cases}
\end{align*}
and $K_1 =B_1^{-1} - A_1^{-1} = \begin{bmatrix}
\frac{\sigma^4}{\sigma^2 +L_x^2} & 0 \\ 0 &\frac{\tau^4}{\tau^2 +L_u^2}
\end{bmatrix} $. For calculation details, readers can refer to section \ref{OUcalculation}. 

\subsection{The Estimation Distribution}

Because of the joint distribution (\ref{jointmatrix}), one can find the best estimation with a given $\theta$ by
\begin{equation*}
X_{1:t} \mid Y_{1:t},\theta \sim N(L^{-\top}W,L^{-\top}L^{-1}),
\end{equation*}
thus
\begin{align*}
\hat{X} _{1:t}= L^{-\top}(W+Z),
\end{align*}
where $Z \sim N(0, I(\sigma,\tau))$.

For $X_{t}$, the joint distribution with $Y_{1:t}$ updated to time $t$ is 
\begin{align*}
X_{t}, Y_{1:t} \mid \theta \sim N\left( 0, \begin{bmatrix}
C_{t}^\top(A_{t}-B_{t}) ^{-1}C_{t} & C_{t}^\top (A_{t}-B_{t})^{-1}\\
(A_{t}-B_{t})^{-1}C_{t} & (I- A_{t}^{-1}B_{t}) ^{-1}B_{t}^{-1}
\end{bmatrix} \right),
\end{align*}
where $C_{t}^\top=\begin{bmatrix}
0 & \cdots & 0 & 1 & 0 \\ 0 & \cdots & 0 & 0 & 1 
\end{bmatrix}$. Thus 
\begin{align*}
X_{t}\mid Y_{1:t},\theta \sim N(\mu_{t}^{(X)},\Sigma_{t}^{(X)}),
\end{align*}
where
\begin{align*}
\mu_{t}^{(X)} & = C_{t}^\top A^{-1}BY =C_{t}^\top L^{-\top}W,\\
\Sigma_{t}^{(X)} & =C_{t}^\top A^{-1}C_{t} =U_{t}^\top U_{t},
\end{align*}
and $U_{t} = L^{-1} C_{t}$.
The recursive updating formula is  
\begin{align}
\mu_{t}^{(X)}  &=  K_{t}B_1\bar{\mu}_{t} + (I - B_1K_{t})Y_{t}  \\
\Sigma_{t}^{(X)}  &=B_1^{-1}-K_{t}.
\end{align}

\subsection{Prior Distribution for Parameters}

The well known Hierarchical Linear Model, where the parameters vary at more than one level, was firstly introduced by Lindley and Smith in 1972 and 1973 \cite{lindley1972bayes} \cite{smith1973general}. Hierarchical Model can be used on data with many levels, although 2-level models are the most common ones. The state-space model in equations (\ref{obserY}) and (\ref{hiddX}) is one of Hierarchical Linear Model if $G_t$ and $F_t$ are linear, and non-linear model if $G_t$ and $F_t$ are non-linear processes. Researchers have made a few discussions and work on these both linear and non-linear models. In this section, we only discuss on the prior for parameters in these models. 

Various informative and non-informative prior distributions have been suggested for scale parameters in hierarchical models. Andrew Gelman gave a discussion on prior distributions for variance parameters in hierarchical models in 2006 \cite{gelman2006prior}. General considerations include using invariance \cite{jeffries1961theory}, maximum entropy \cite{jaynes1983papers} and agreement with classical estimators \cite{box2011bayesian}.  Regarding informative priors, Andrew suggests to distinguish them into three categories: The first one is traditional informative prior. A prior distribution giving numerical information is crucial to statistical modeling and it can be found from a literature review, an earlier data analysis or the property of the model itself. The second category is weakly informative prior. This genre prior is not supplying any controversial information but are strong enough to pull the data away from inappropriate inferences that are consistent with the likelihood. Some examples and brief discussions of weakly informative priors for logistic regression models are given in \cite{gelman2008weakly}. The last one is uniform prior, which allows the information from the likelihood to be interpreted probabilistically. 

Jonathan and Thomas in \cite{stroud2007sequential} have discussed a model, which is slightly different with a Gaussian state-space model from section one. The two errors $\omega_t$ and $\epsilon_t$ are assumed normally distributed as
\begin{align*}
\omega_t &\sim N(0,\alpha Q),\\
\epsilon_t &\sim N(0,\alpha R),
\end{align*}
where the two matrices $R$ and $Q$ are known and $\alpha$ is an unknown scale factor to be estimated. (Note that a perfect model is obtained by setting $Q= 0$.) Therefore, the density of Gaussian state-space model is
\begin{align*}
p(y_t\mid x_t,\alpha) &= N(F(x_t),\alpha R),\\
p(x_t\mid x_{t-1},\alpha) &= N(G(x_{t-1}),\alpha Q).
\end{align*}
The parameter $\alpha$ is assumed \textit{Inverse Gamma} distribution. 

For the priors of all the parameters in OU-process, shown in equation (\ref{OUprocess}) and (\ref{obseq}), firstly we should understand what meanings of these parameters are standing for. The reciprocal of $\gamma$ is typical velocity falling in the reasonable range of 0.1 to 100 $m/s$. $\xi$ is the error occurs in transition process, $\sigma$ and $\tau$ are errors in the forward map for position and velocity respectively. Generally, the error is a positive finite number. Considering prior distributions for these parameters, before looking at the data, we have an idea of ranges where these parameters are falling in. Conversely, we don't have any assumptions about the true value of $\lambda$, which means it could be anywhere. According to this assumption, the prior distributions are 
\begin{align*}
\gamma   &\sim IG(10,0.5),\\
\xi^2        &\sim IG(5,2.5),\\
\sigma^2 &\sim IG(5,2.5),
\end{align*}
where $IG(\alpha,\beta)$ represents the \textit{Inverse Gamma} distribution with two parameters $\alpha$ and $\beta$. 
\begin{figure}[h]
\centering
\includegraphics[width=0.45\textwidth]{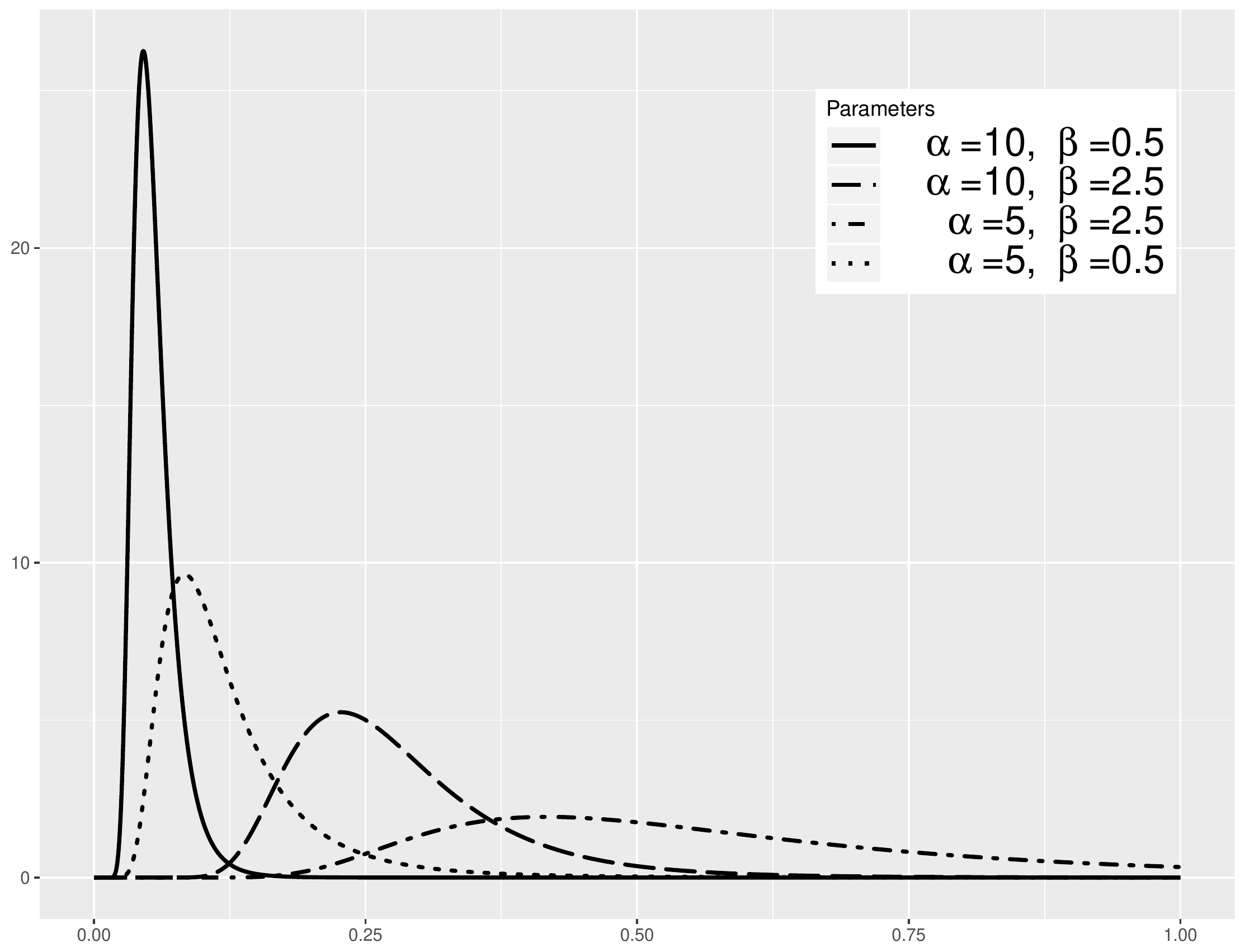}
\includegraphics[width=0.45\textwidth]{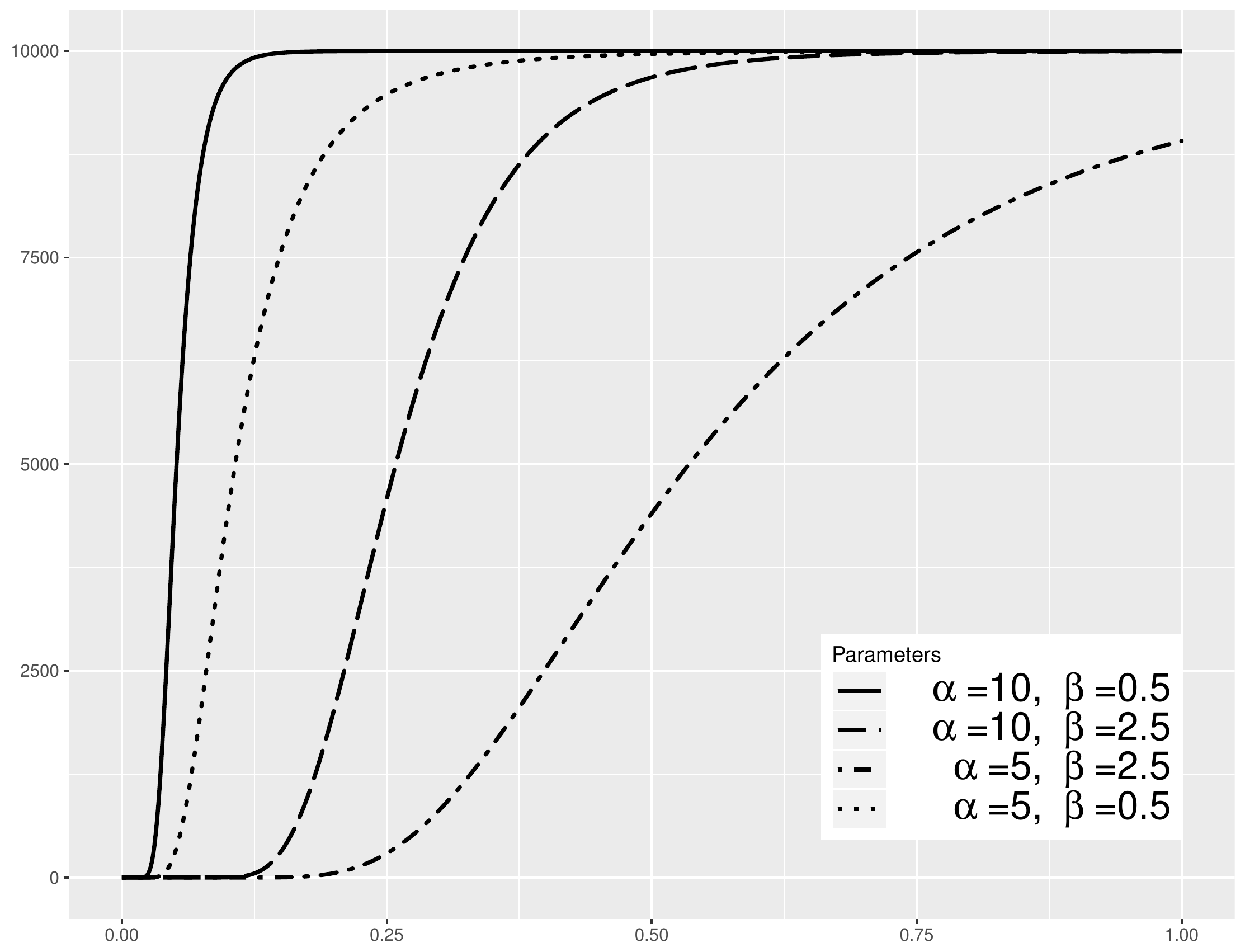}
\caption{Probability density function and cumulative distribution function of \textit{Inverse Gamma} with two parameters $\alpha$ and $\beta$. }
\label{IGPDFCDF}
\end{figure}

\subsection{Efficiency of DA-MH}

\begin{figure}[h]
\centering
\begin{subfigure}[t]{0.45\textwidth}
	\includegraphics[width=\textwidth]{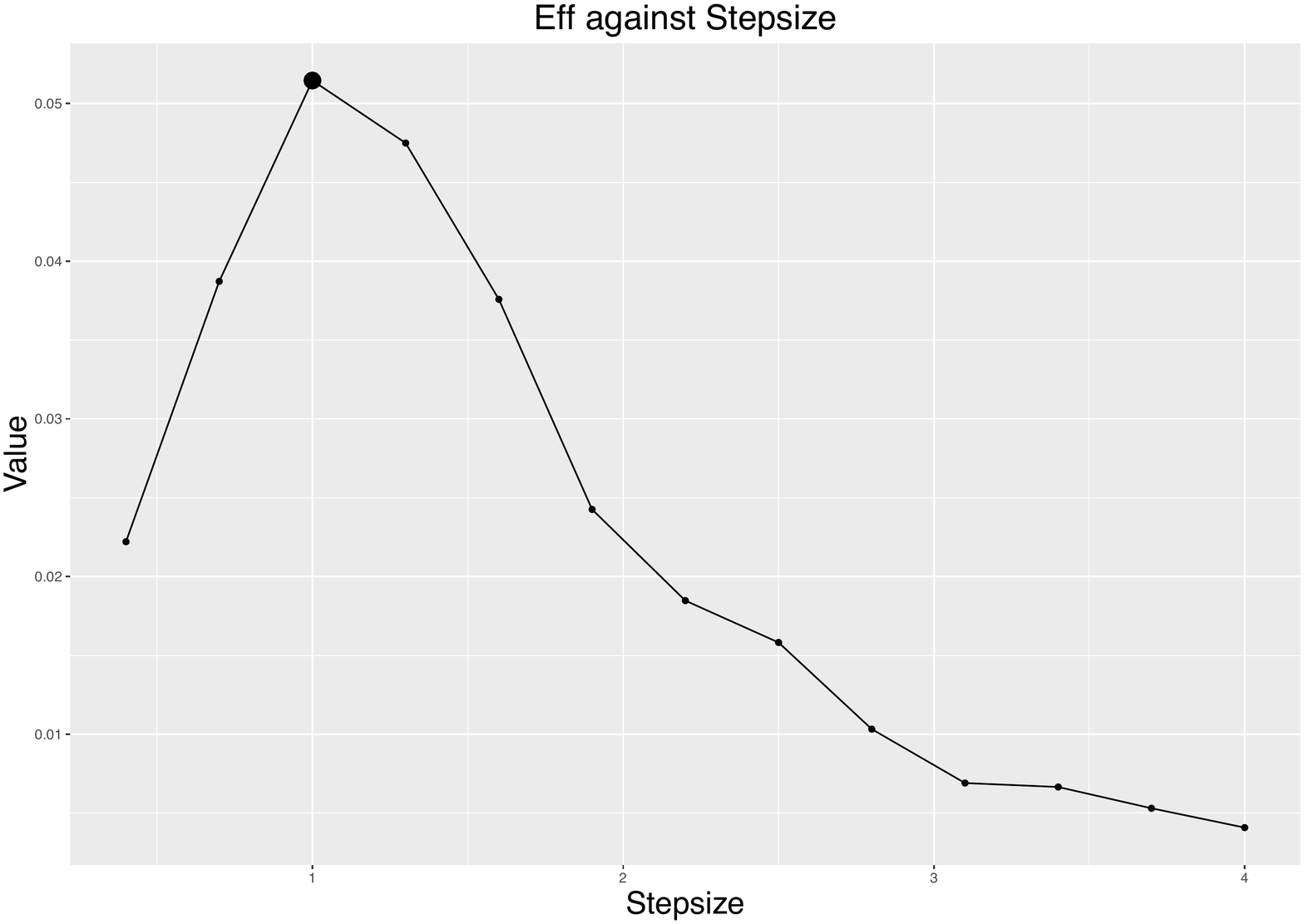}
	\caption{Efficiency against different step sizes}
\end{subfigure}
\begin{subfigure}[t]{0.45\textwidth}
	\includegraphics[width=\textwidth]{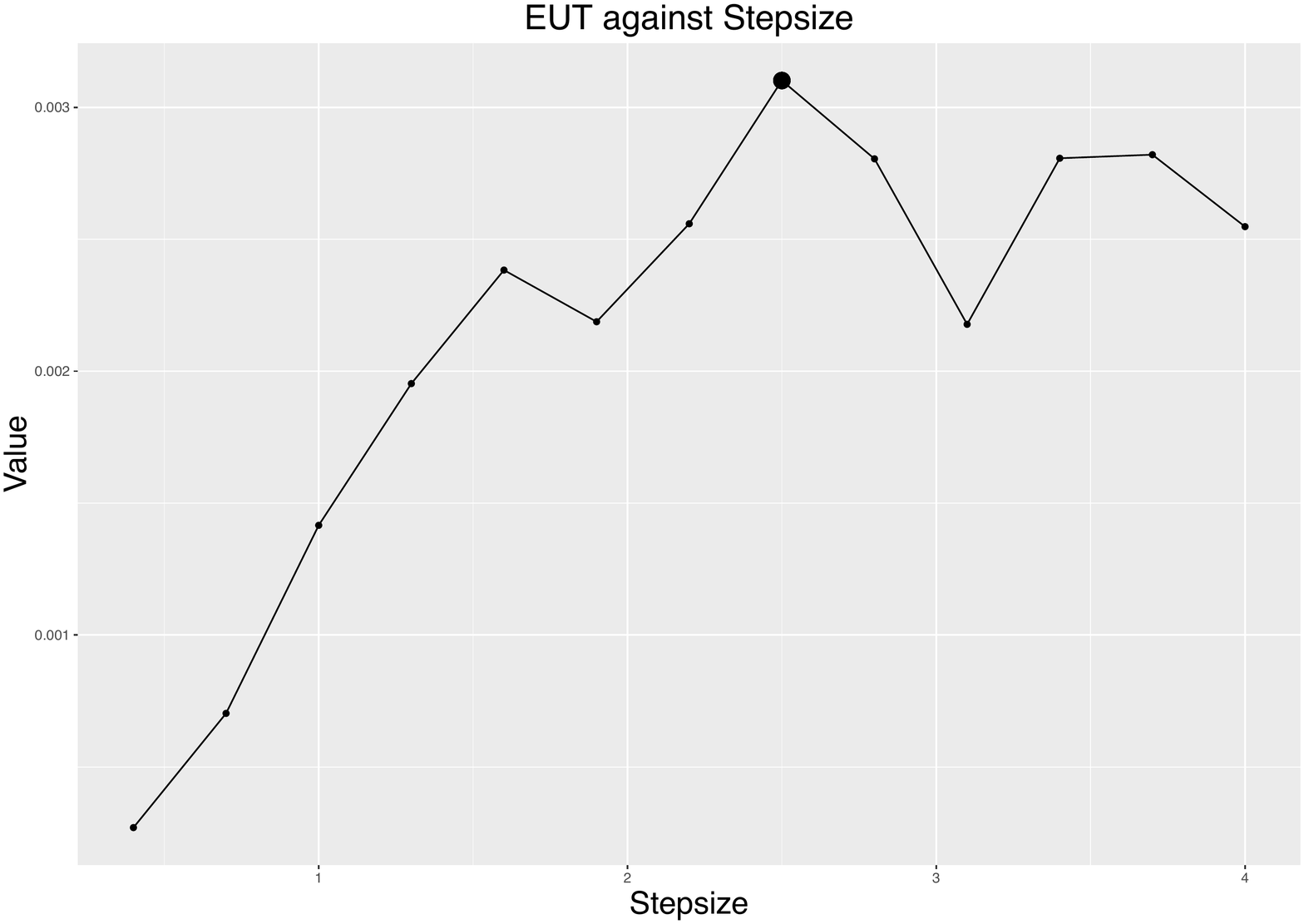}
	\caption{EffUT against different step sizes}
\end{subfigure}
\begin{subfigure}[t]{0.45\textwidth}
	\includegraphics[width=\textwidth]{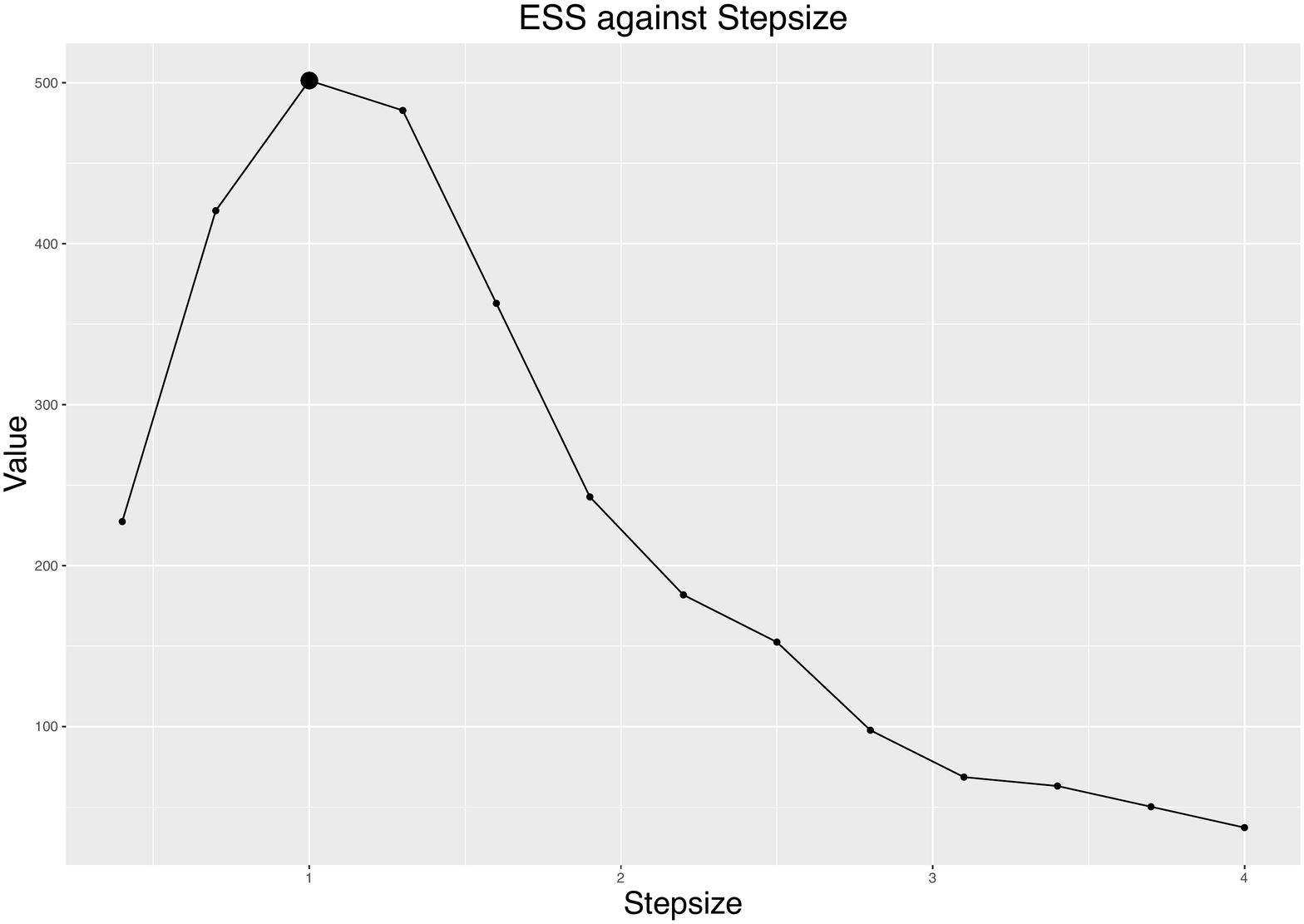}
	\caption{ESS against different step sizes}
\end{subfigure}
\begin{subfigure}[t]{0.45\textwidth}
	\includegraphics[width=\textwidth]{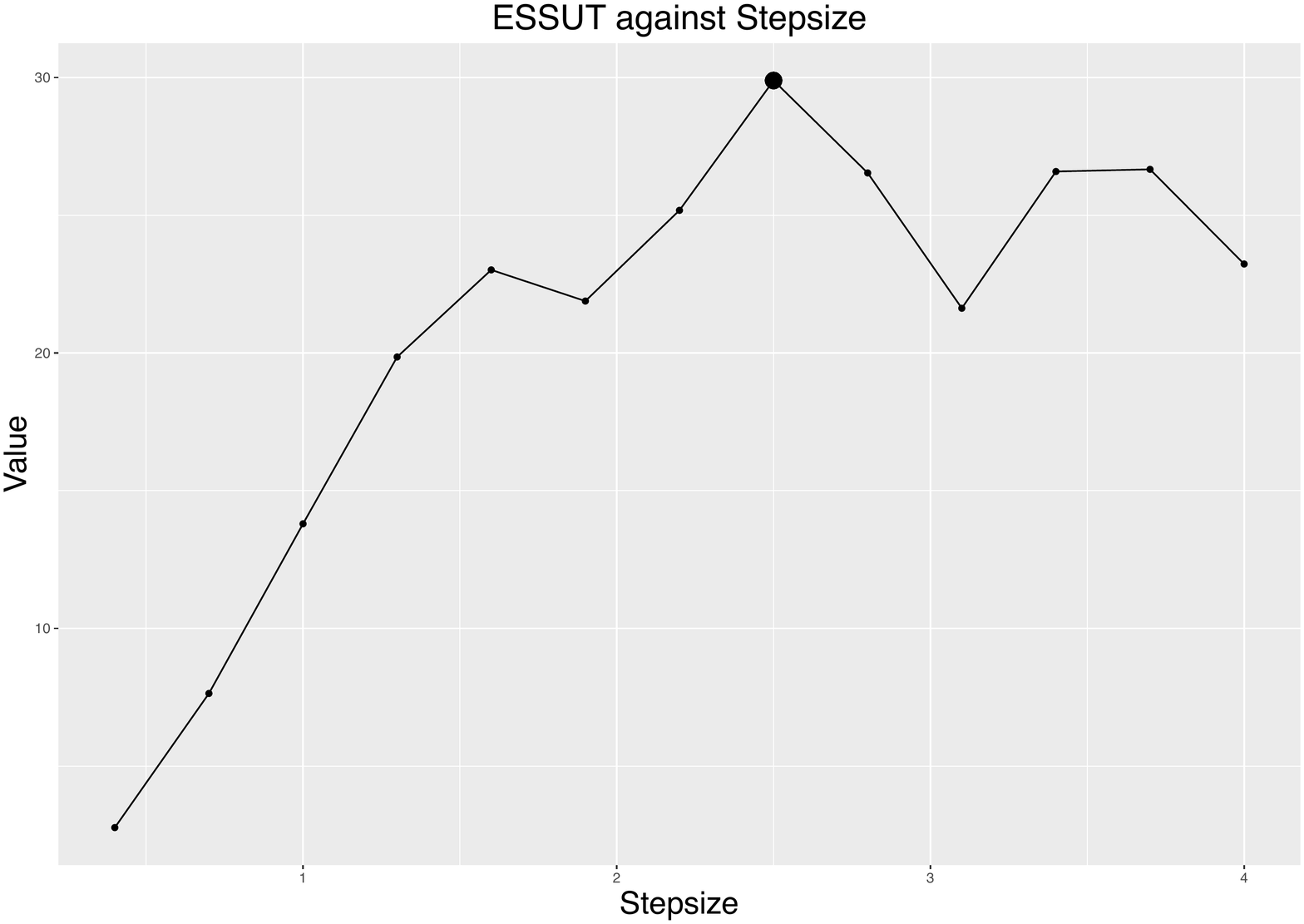}
	\caption{ESSUT against different step sizes}
\end{subfigure}
\caption{Influences of different step sizes on sampling efficiency (Eff), efficiency in unit time (EffUT), effective sample size (ESS) and effective sample size in unit time (ESSUT) found by using the same data}
\label{effeutessessutexamplefigure}
\end{figure}

We have discussed the efficiency of Delayed-Acceptance Metropolis-Hastings algorithm and how it is affected by the step size. To explain explicitly, here we give an example comparing Eff, EffUT, ESS and ESSUT, which are calculated by using the same dataset and running 10\,000 iterations of DA-MH. We are taking an 0.3-equal-spaced sequence $s=\{0.1,\dots,4\}$ from 0.1 to 4 and choosing each of them to calculate the criterion values. Table \ref{effeutessessutexampletable} and figure \ref{effeutessessutexamplefigure} show the results of this comparison. 

The best step size found by Eff is 1, which is as the same as it found by ESS. By using $s=1$ and running 1\,000 iterations, the DA-MH takes 36.35 seconds to get the Markov chain for $\theta$ and the acceptance rates $\alpha_1$ for approximate $\hat{\pi}(\cdot)$ and $\alpha_2$ for posterior distribution $\pi(\cdot)$ are 0.3097 and 0.8324 respectively. By using EffUT and ESSUT, the best step size is 2.5, which is bigger. The advantages of using this kind of step size are the computation time decreased to 5.10 seconds significantly. Because of the approximation $\hat{\pi}(\cdot)$ took bad proposals out and only approve good ones going to the next level, that can be seen from the lower rates $\alpha_1$ in table \ref{effeutessessutexampletable}. 


\begin{table}[h]
\centering
\begin{tabular}{|c|c|c|c|c|c|}
\hline
          & Values     & Time & Step Size & $\alpha_1$ & $\alpha_2$ \\ \hline
Eff      & 0.0515     & 36.35 & 1.0   & 0.3097    & 0.8324    \\ \hline
EffUT  & 0.0031     & 5.10   & 2.5   & 0.0360   & 0.7861   \\ \hline
ESS     & 501.4248 & 36.35 & 1.0   & 0.3097    & 0.8324     \\ \hline
ESSUT & 29.8912   & 5.10   & 2.5   & 0.0360   & 0.7861    \\ \hline
\end{tabular}
\caption{An example of Eff, EffUT, ESS and ESSUT found by running 10\,000 iterations with same data. The computation time is measured in seconds $s$. }
\label{effeutessessutexampletable}
\end{table}

\begin{figure}[h]
\centering
\includegraphics[width=8cm,height=5cm]{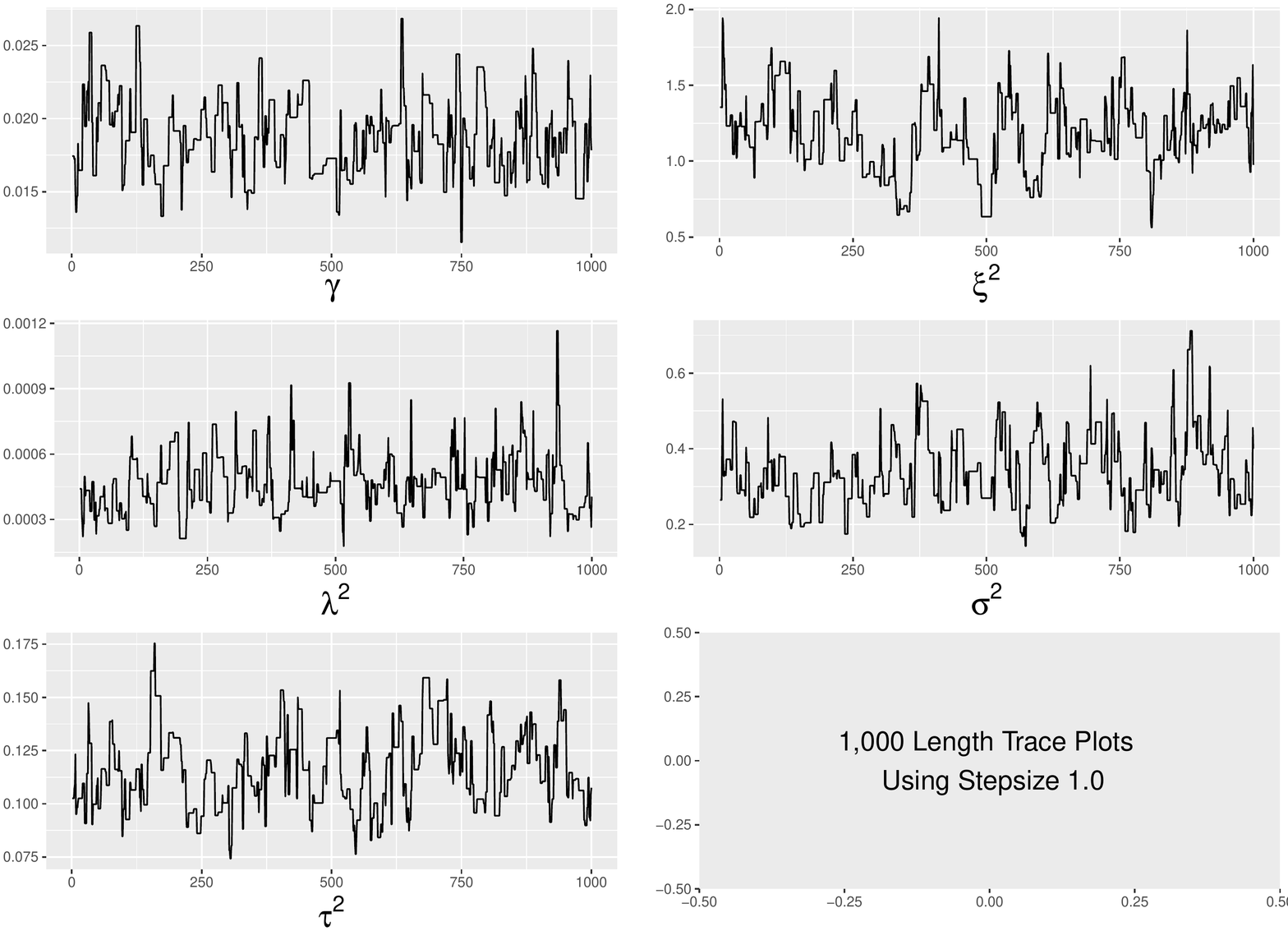}
\includegraphics[width=8cm,height=5cm]{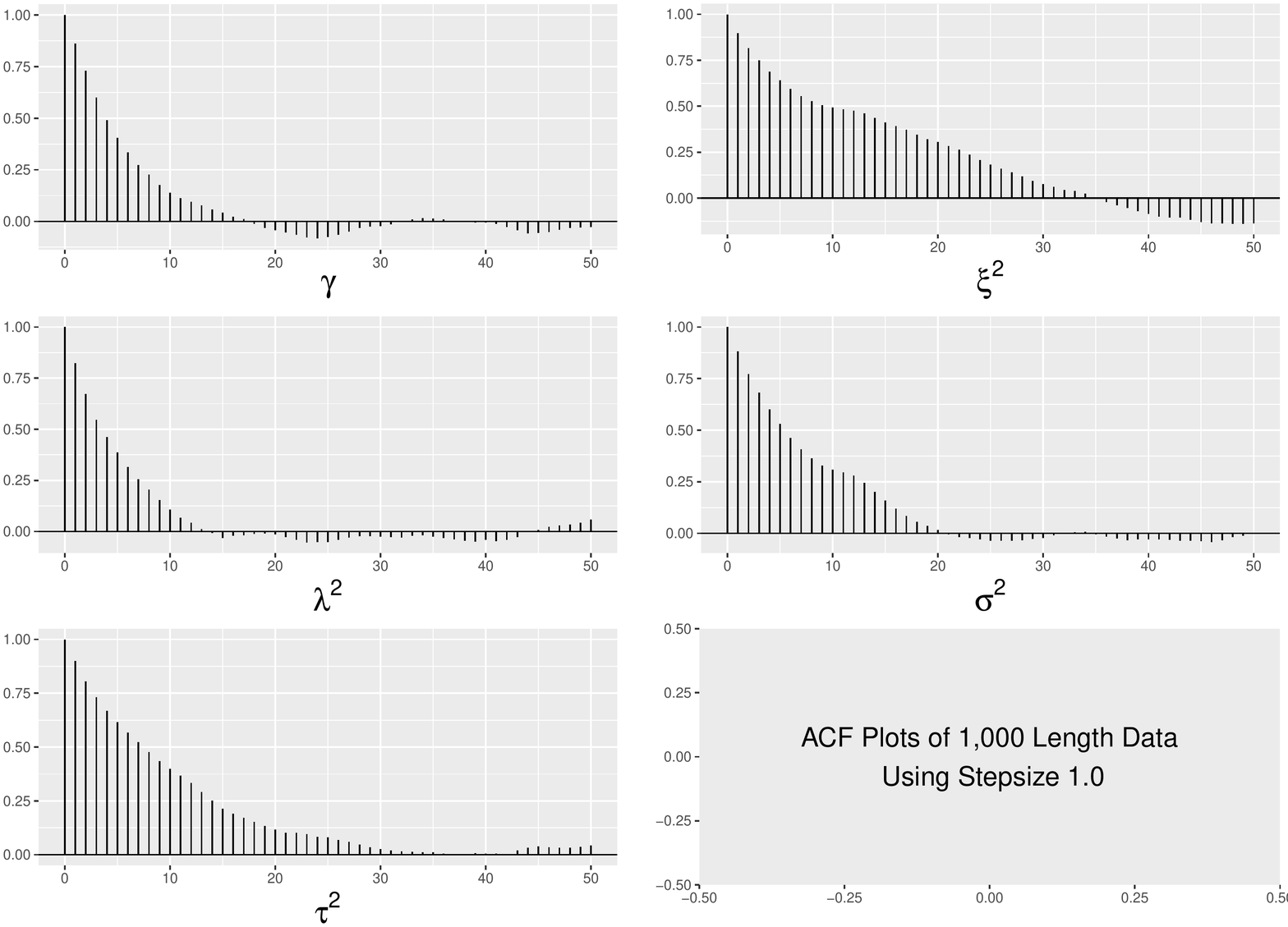}
\includegraphics[width=8cm,height=5cm]{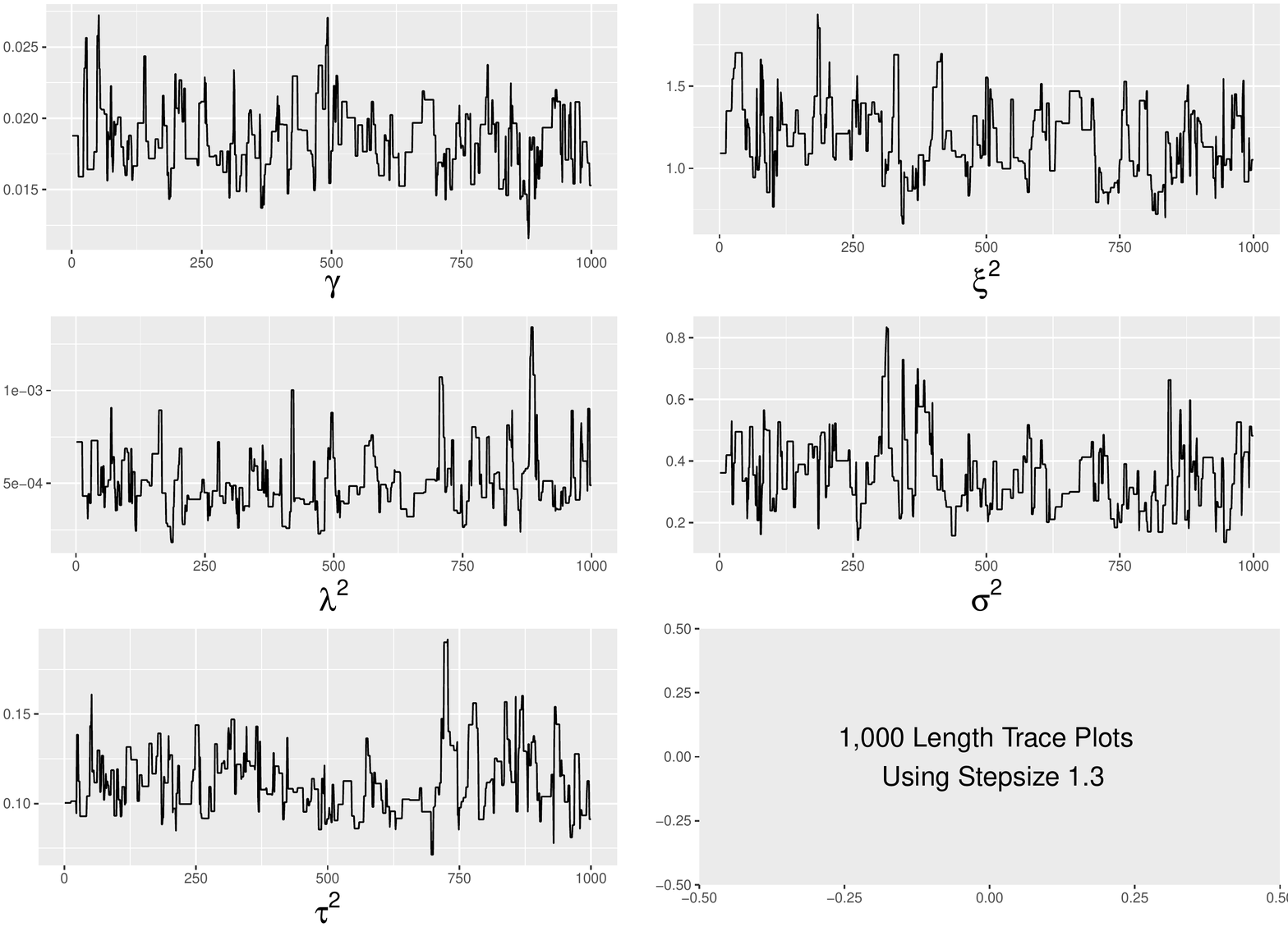}
\includegraphics[width=8cm,height=5cm]{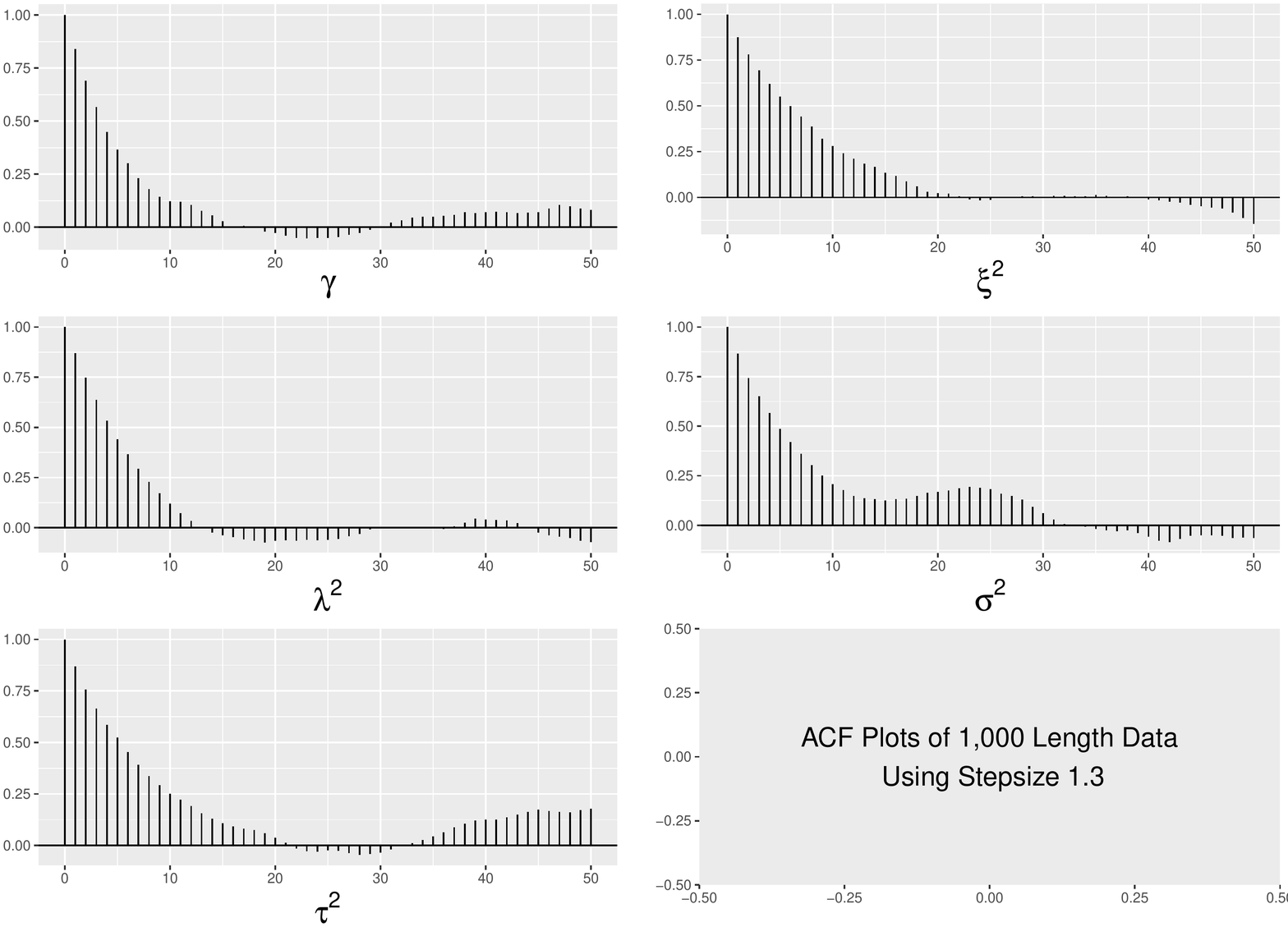}
\includegraphics[width=8cm,height=5cm]{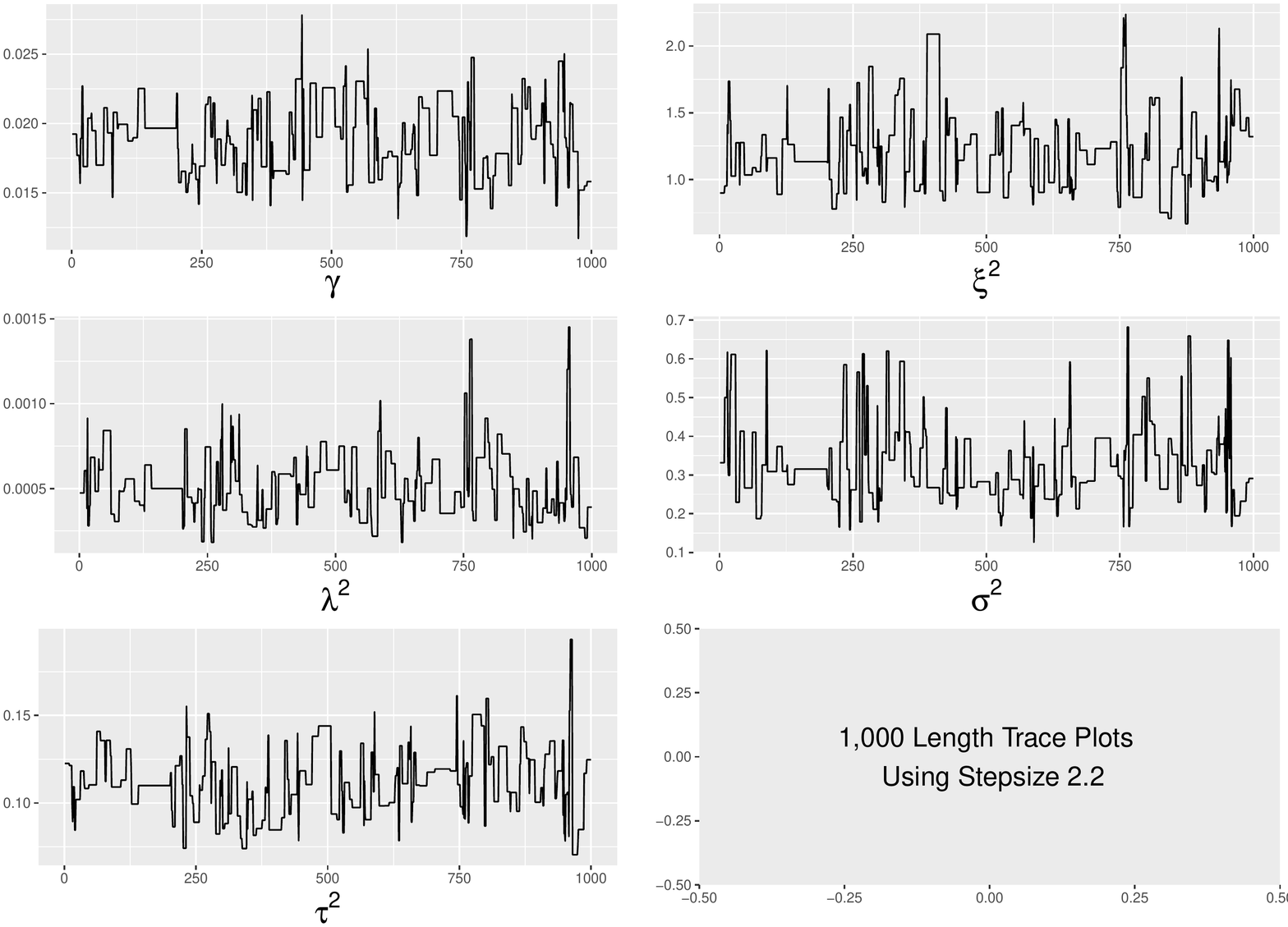}
\includegraphics[width=8cm,height=5cm]{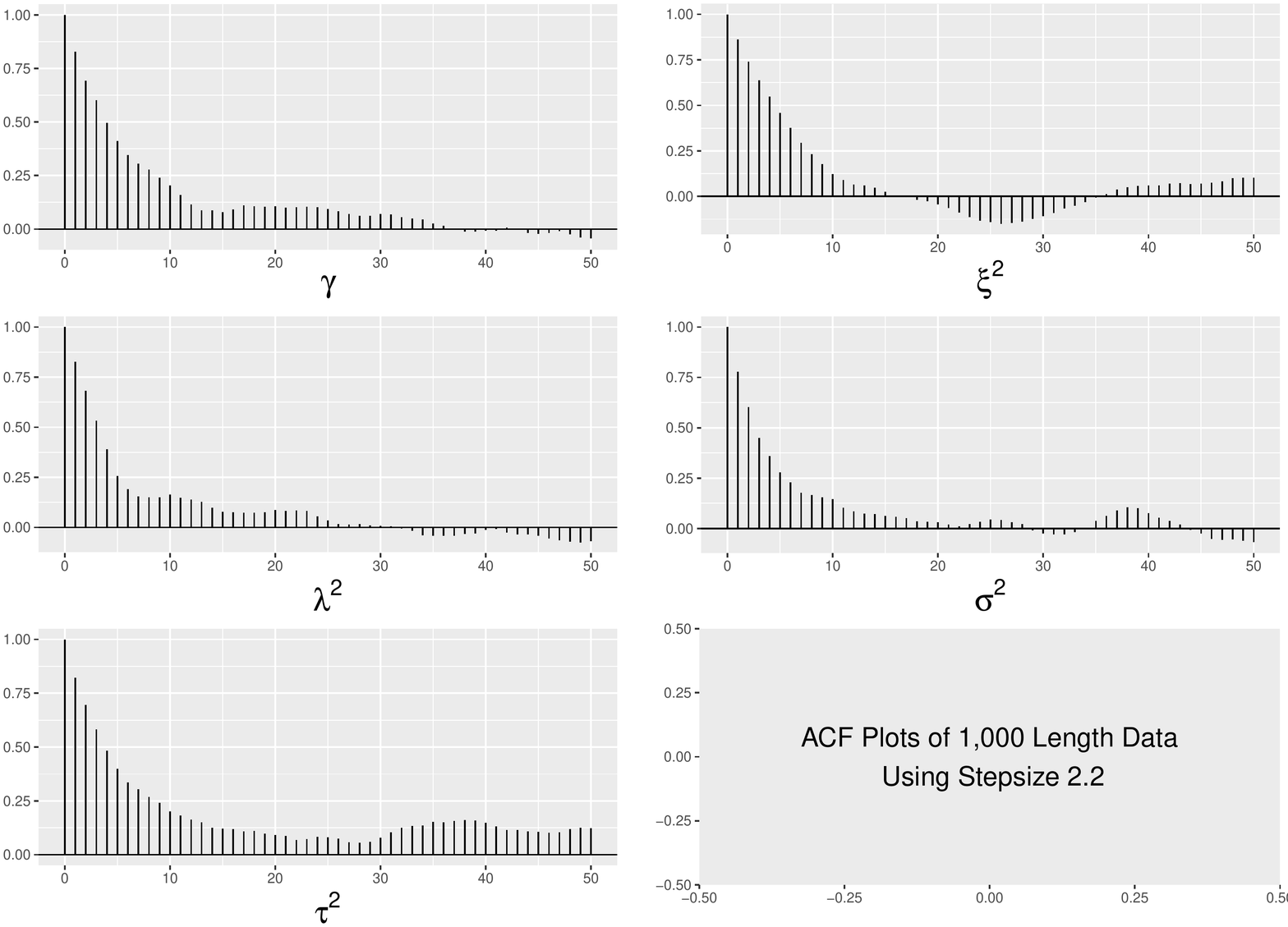}
\includegraphics[width=8cm,height=5cm]{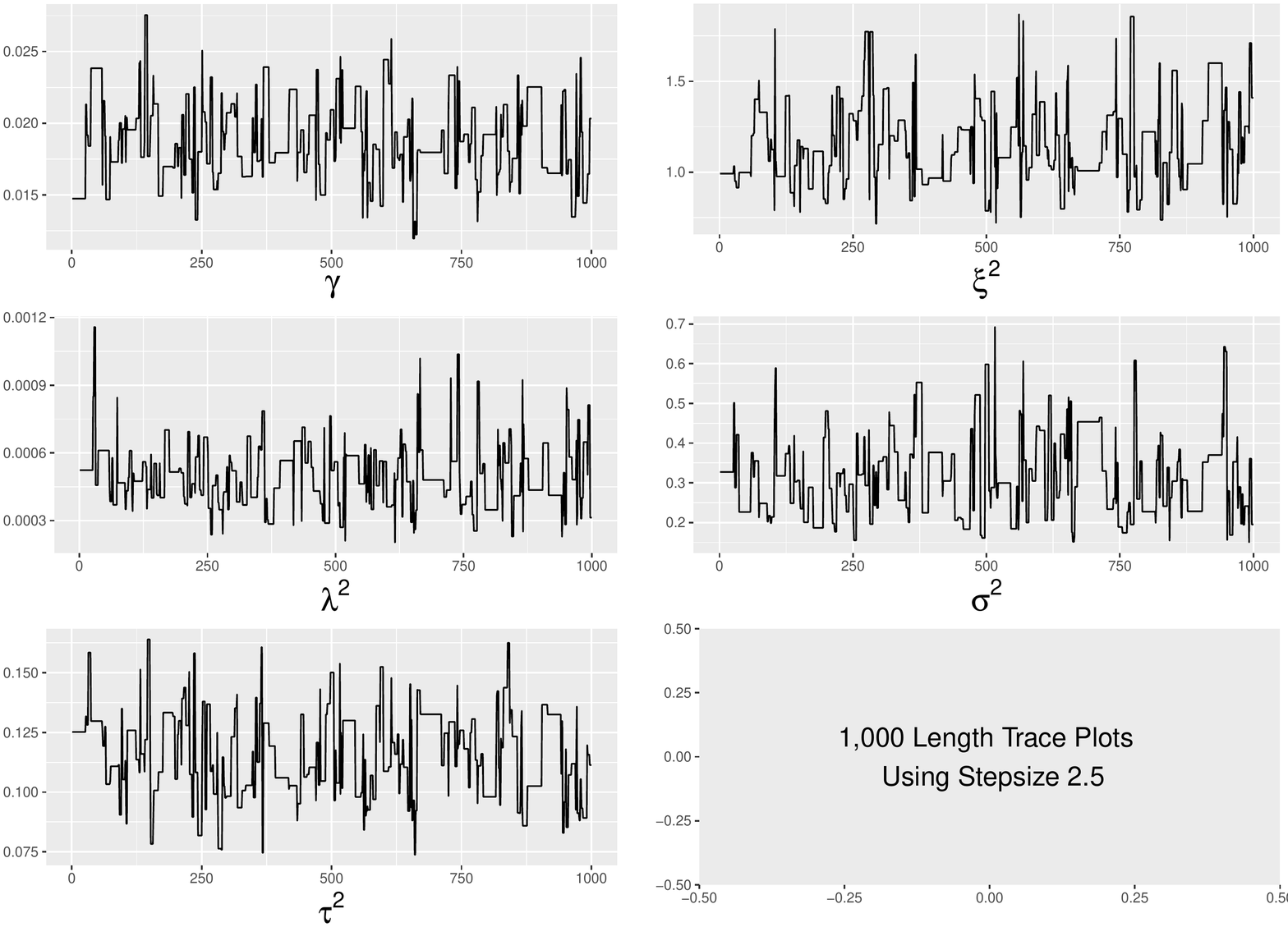}
\includegraphics[width=8cm,height=5cm]{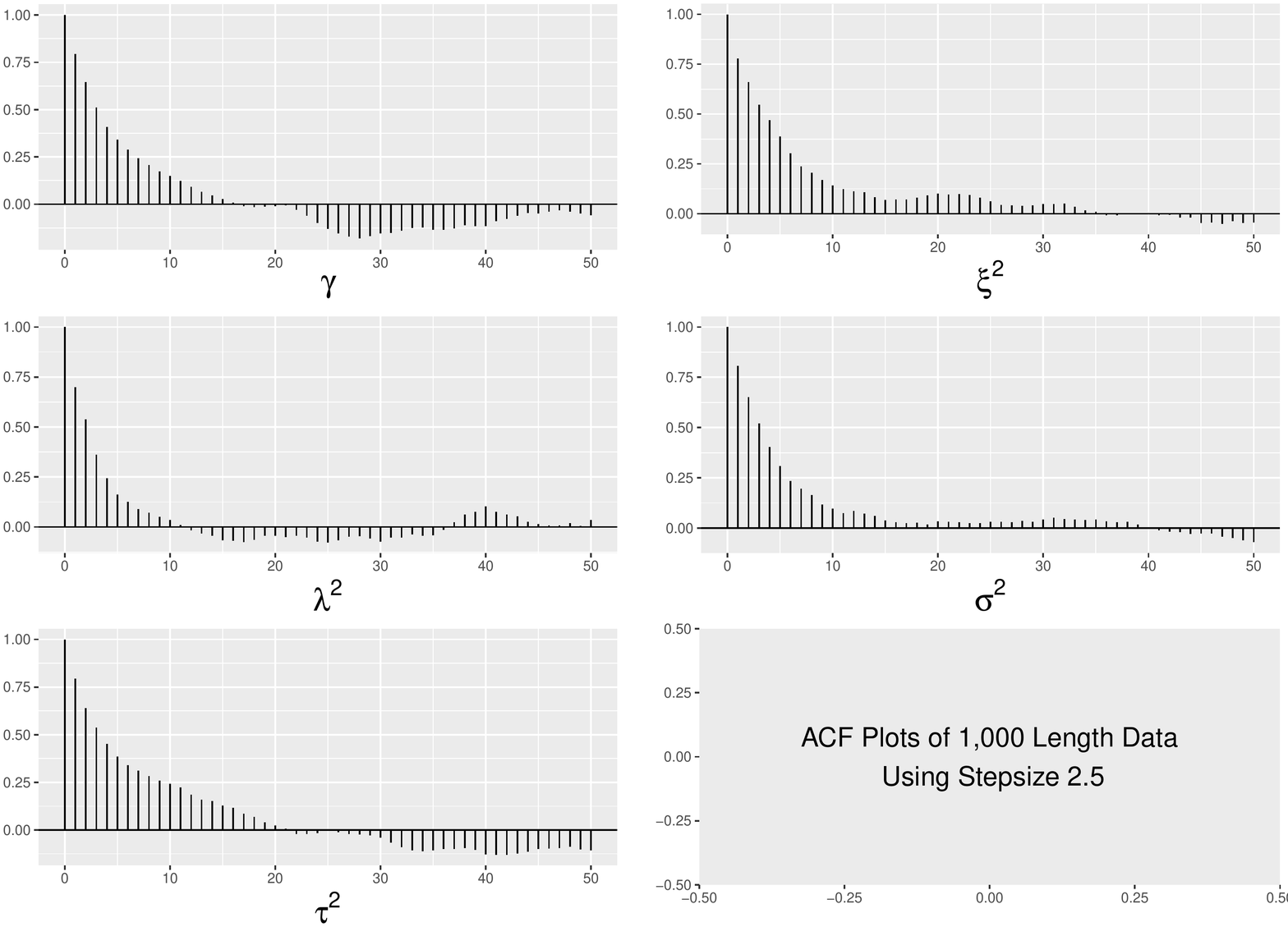}
\caption{Running the same amount of time and taking the same length of data, the step size $\epsilon=2.5$ returns the highest ESSUT value and generates more effective samples with a lower correlation. }\label{1koutof8kfigures}
\end{figure}

On the surface, a bigger step size causes lower acceptance rates $\alpha_1$ and it might not be a smart choice. However, on the other hand, one should notice the less time cost. To make it sensible, we are running the Delayed-Acceptance MH with different step sizes, as presented in table \ref{effeutessessutexampletable},  for the same (or similar) amount of time. Because of the bigger step size takes less time than smaller one, so we achieve a longer chain. To be more clear, we take 1\,000 samples out from a longer chain, such as 8\,500, and calculate Eff, EffUT, ESS and ESSUT separately using the embedded function \textsf{IAT}, \cite{christen2010general}, and \textsf{ESS} of the package \textsf{LaplacesDemon} in \textit{R} and the above formulas . As we can see from the outcomes, by running the similar amount of time, the Markov chain using a bigger step size has a higher efficiency and effective sample size in unit time. More intuitively, the advantage of using larger step size is the sampling algorithm generates more representative samples per second. Figure (\ref{1koutof8kfigures}) is comparing different $\theta$ chains found by using different step size but running the same amount of time. As we can see that $\theta$ with the optimal step size has a lower correlated relationship. 
\begin{table}[h]
\centering
\begin{tabular}{|c|c|c|c|c|c|c|}
\hline
Step Size& Length & Time & Eff   & EffUT & ESS & ESSUT \\ \hline
1.0    &   1\,000        & 3.48   & 0.0619 & 0.0178   &  69.4549     & 19.9583   \\ \hline
1.3    &   1\,400        & 3.40   & 0.0547 & 0.0161   &  75.3706   & 22.1678 \\ \hline
1.3    &   $1\,000^\star$ & 3.40 & 0.0813 & 0.0239  & 72.5370  & 21.3344   \\ \hline
2.2    &   5\,000          &  3.31 & 0.0201 &  0.0061  &  96.6623    & 29.2031   \\ \hline
2.2    &   $1\,000^\star$ & 3.31  &  0.0941 & 0.0284 & 94.2254 &  28.4669 \\ \hline
2.5    &   7\,000          &  3.62  & 0.0161 &0.0044  & 112.3134   &  31.0258    \\ \hline
2.5    &   $1\,000^\star$ &  3.62 & \textbf{0.1095} &  \textbf{0.0302}  &  \textbf{113.4063} & \textbf{31.3277} \\ \hline
\end{tabular}
\caption{Comparing Eff, EffUT, ESS and ESSUT values using different step size. The $1000^\star$ means taking 1\,000 samples from a longer chain, like 1\,000 out of 5\,000 sample chain. The computation time is measured in seconds $s$.}
\label{stepsizecompare}
\end{table}

\subsection{Sliding Window State and Parameters Estimation}

The length of data used in the algorithm really affects the computation time. The forecast distribution $p(Y_{t}\mid Y_{1:t-1},\theta)$ and estimation distribution $p(X_{t}\mid Y_{1:t},\theta)$ require finding the inverse of the covariance $\Sigma_{YY}^{(t+1)}$, however, which is time consuming if the sample size is big to generate a large sparse matrix. For a moving vehicle, one is more willing to get the estimation and moving status instantly rather than being delayed. Therefore, a compromise solution is using fixed-length sliding window sequential filtering. A fixed-lag sequential parameter learning method was proposed in \cite{polson2008practical} and named as \textit{Practical Filtering}. The authors rely on the approximation of 
\begin{equation*}
p(x_{0:n-L},\theta\mid y_{0:n-1}) \approx p(x_{0:n-L},\theta \mid y_{0:n})
\end{equation*}
for large $L$. The new observations coming after the $n$th data has little influence on $x_{0:n-L}$. 

Being inspired, we are not using the first $0$ to $n-1$ date and ignoring the latest $n$th, but using all the latest with truncating the first few history ones. Suppose we are given a fixed-length $L$, up to time $t$, which should be greater than $L$, we are estimating $x_t$ by using all the retrospective observations to the point at $t-L+1$. In another word, the estimation distribution for the current state is 
\begin{equation}
p(X_{t}\mid Y_{t-L+1:t},\theta),
\end{equation}
where $t>L$. We name this method \textit{Sliding Window Sequential Parameter Learning Filter}. 

\begin{figure}[h]
\centering
\begin{subfigure}[t]{0.45\textwidth}
    \includegraphics[width=\textwidth]{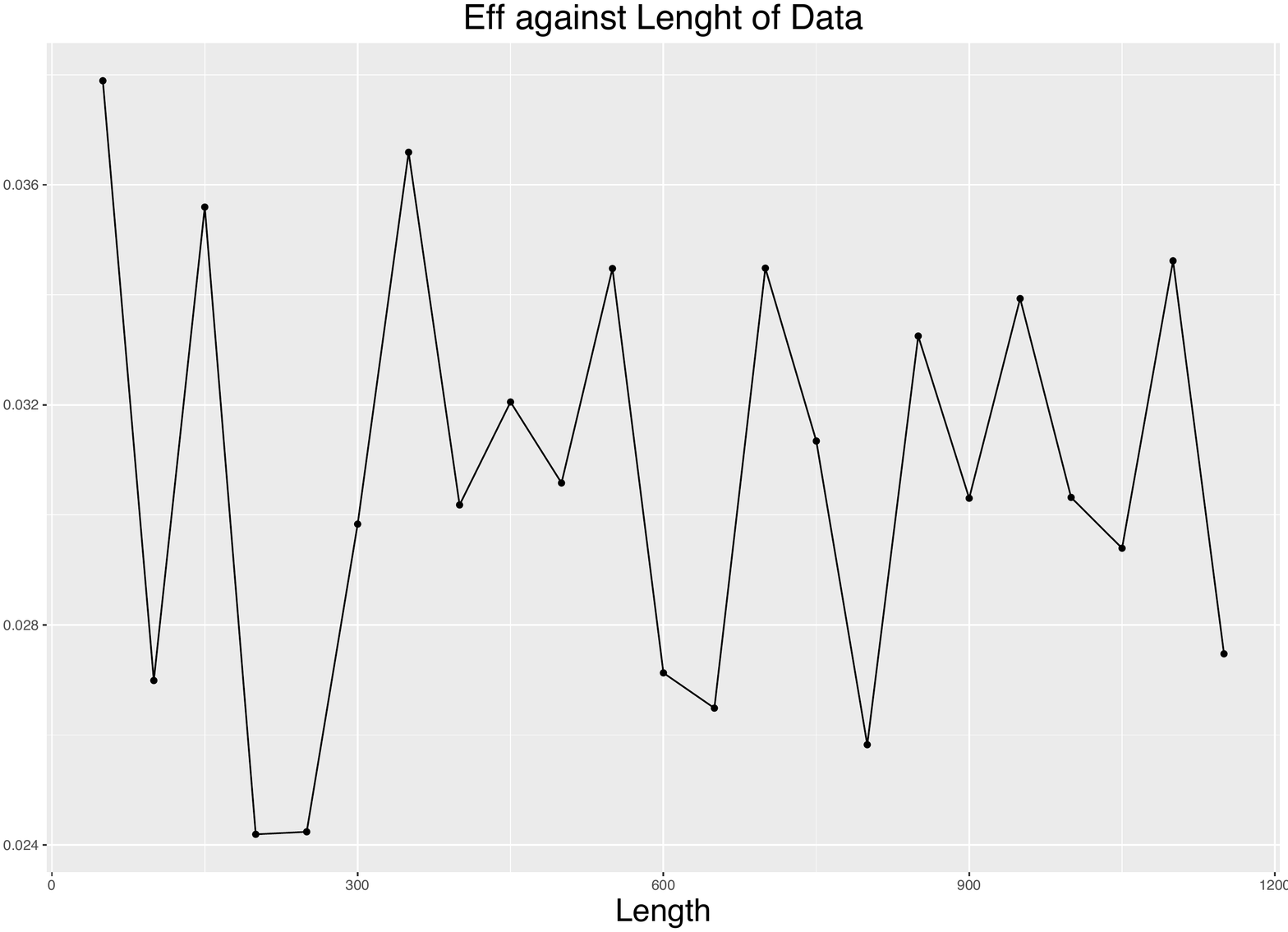}
    \caption{Efficiency against data length}
\end{subfigure}
\begin{subfigure}[t]{0.45\textwidth}
    \includegraphics[width=\textwidth]{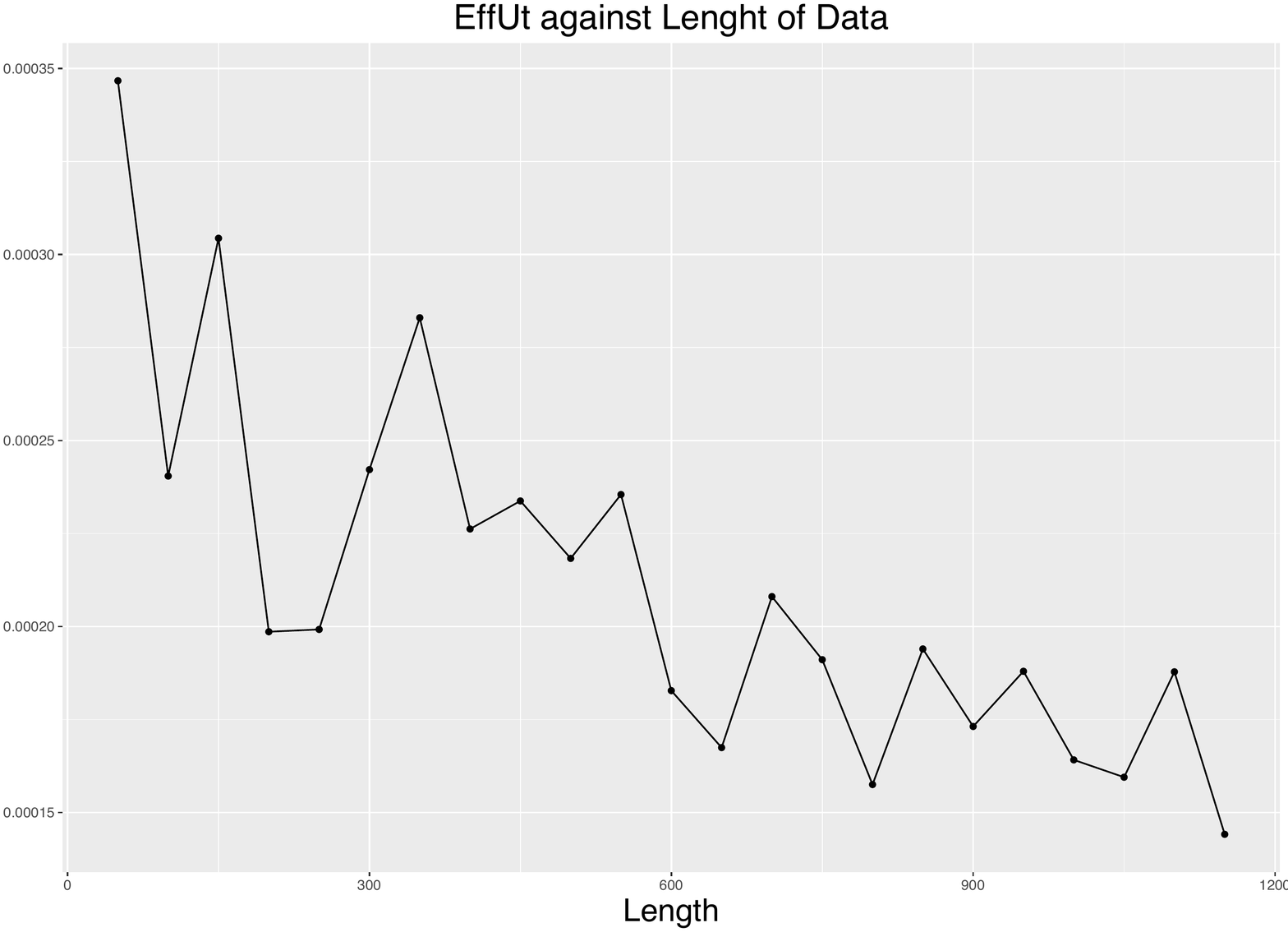}
    \caption{EffUT against data length}
\end{subfigure}
\begin{subfigure}[t]{0.45\textwidth}
    \includegraphics[width=\textwidth]{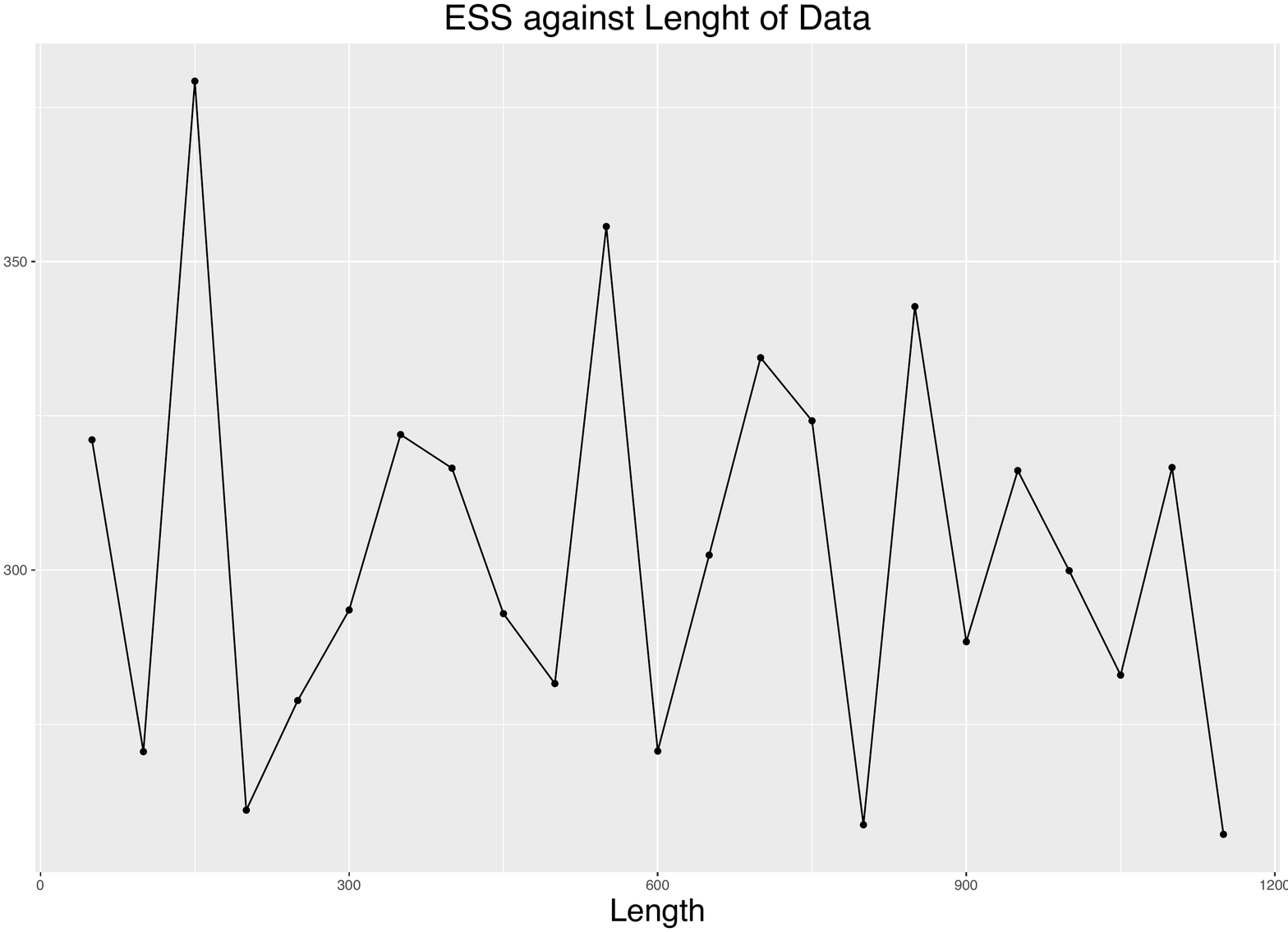}
    \caption{ESS against data length}
\end{subfigure}
\begin{subfigure}[t]{0.45\textwidth}
    \includegraphics[width=\textwidth]{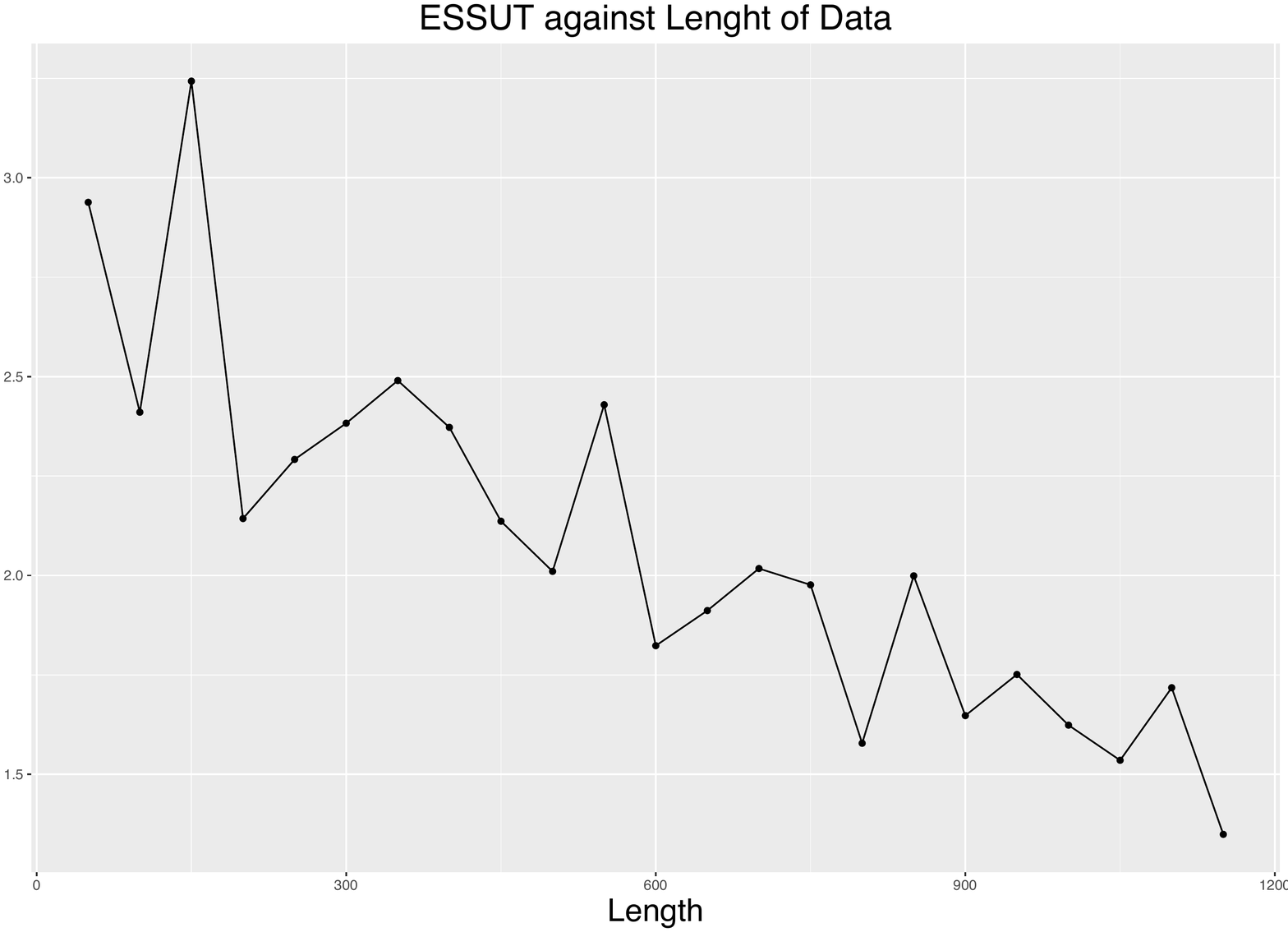}
    \caption{ESSUT against data length}
\end{subfigure}
\caption{Comparison efficiency (Eff), efficiency in unit time (EffUT), effective sample size (ESS) and effective sample size in unit time (ESSUT) against the different length of data. Increasing data length doesn't significantly improve the efficiency and ESSUT.}\label{compareLengthData}
\end{figure}

The next question is how to choose an appropriate $L$. The length of data used in MH and DA-MH algorithms has an influence on the efficiency and accuracy of parameter learning and state estimation. Being tested on real data set, there is no doubt that the more data be in use, the more accurate the estimation is, and lower efficient is in computation. In table \ref{lengthofdatacompare}, one can see the pattern of parameters $\gamma,\xi,\tau$ follow the same trend with the choice of $L$ and $\sigma$ increases when $L$ decreases. Since estimation bias is inevitable, we are indeed to keep the bias as small as possible, and in the meantime, the higher efficiency and larger effective sample size are bonus items. In figure \ref{compareLengthData}, we can see that the efficiency and effective sample size is not varying along the sample size used in sampling algorithm, but in unit time, they are decreasing rapidly as data size increases. 
In addition, from a practical point of view, the observation error $\sigma$ should be kept at a reasonable level, let's say $50cm$, and the computation time should be as less as possible. To reach that level, $L=100$ is an appropriate choice. For a one-dimensional linear model, $L$ can be chosen larger and that doesn't change too much. If the data up to time $t$ is less than or equal to the chosen $L$, the whole data set is used in learning $\theta$ and estimating $X_t$.

\begin{figure}[h]
\centering
\includegraphics[width=0.4\textwidth,height=0.2\textheight]{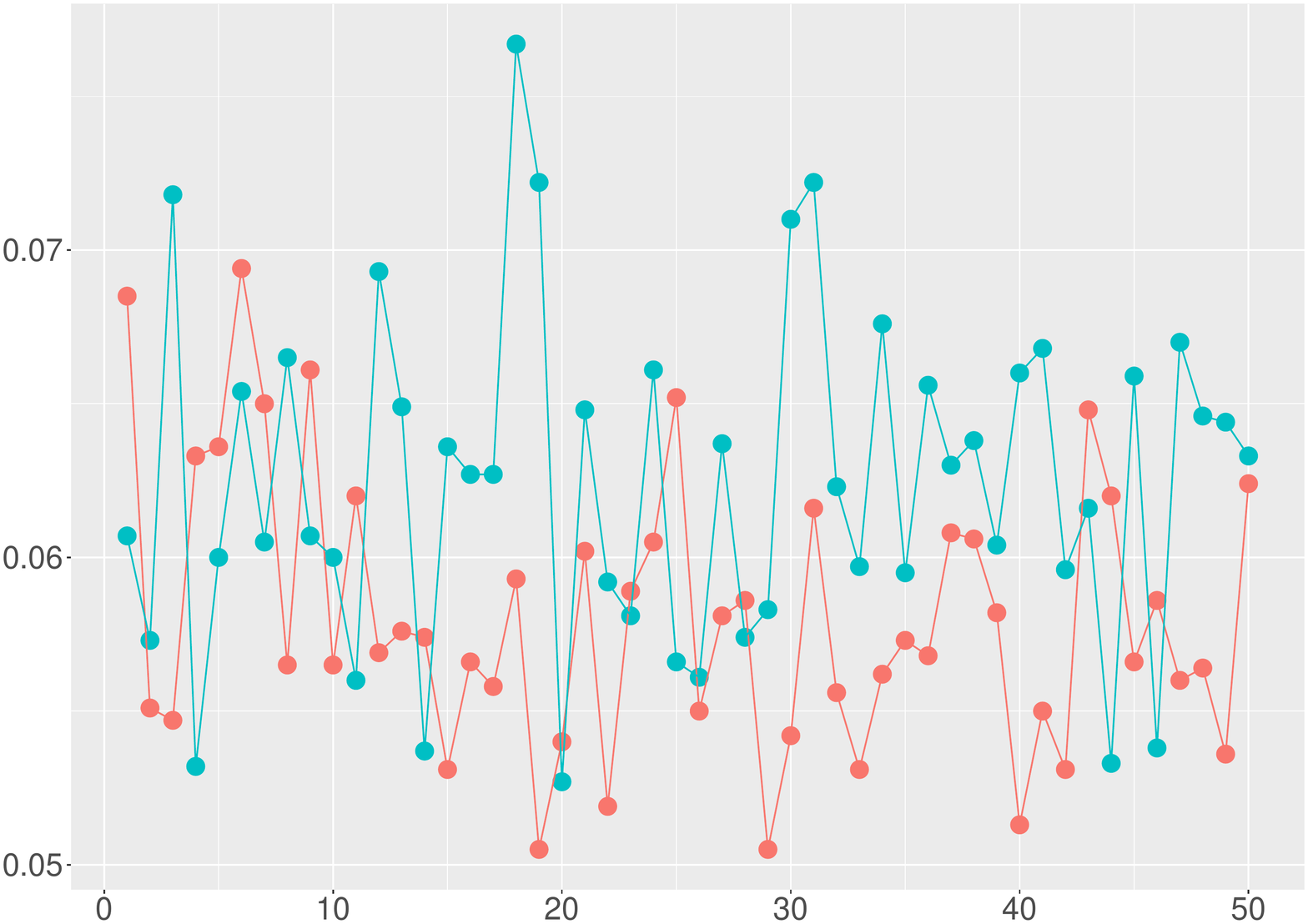}
\includegraphics[width=0.4\textwidth,height=0.2\textheight]{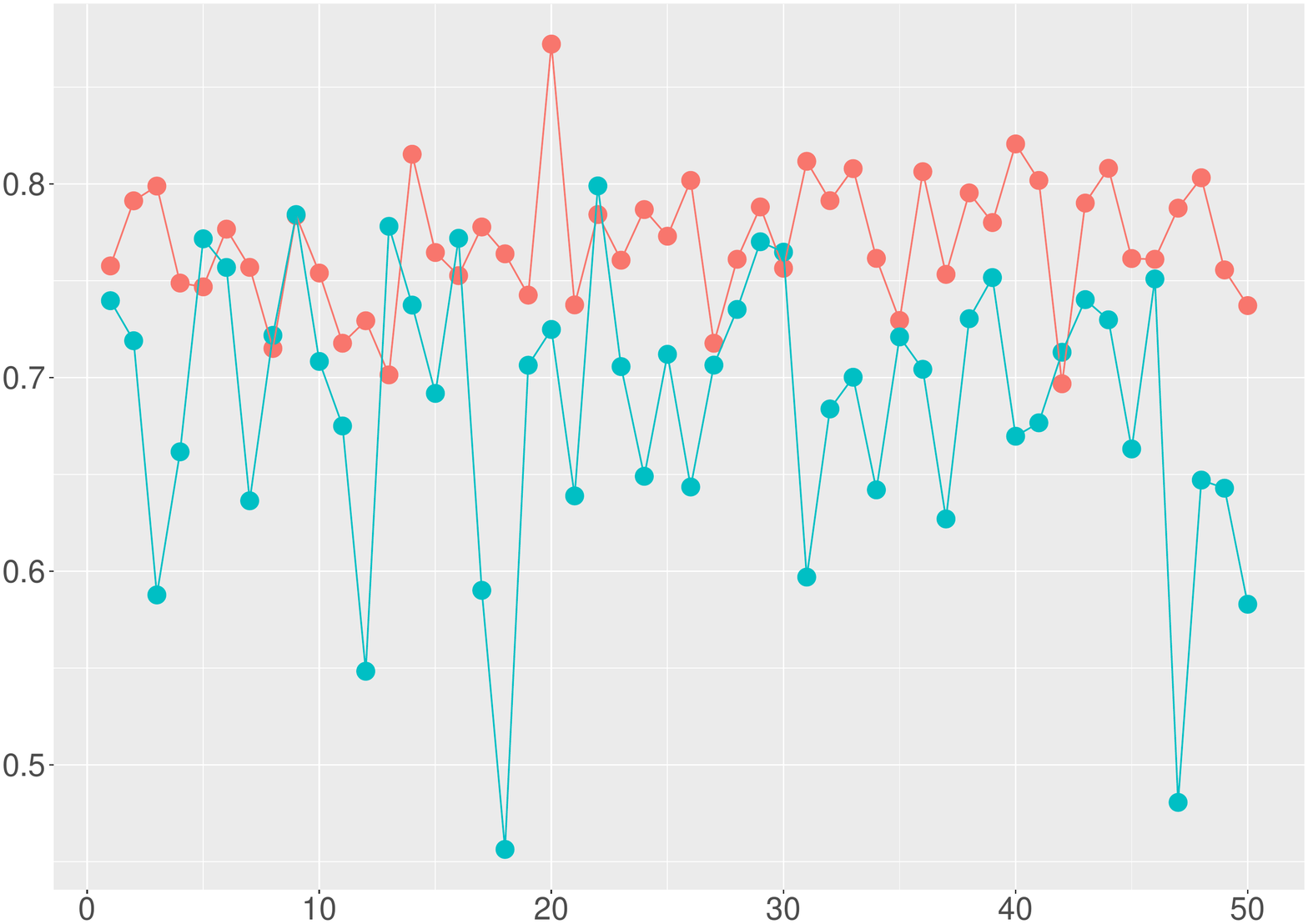}
\includegraphics[width=0.4\textwidth,height=0.2\textheight]{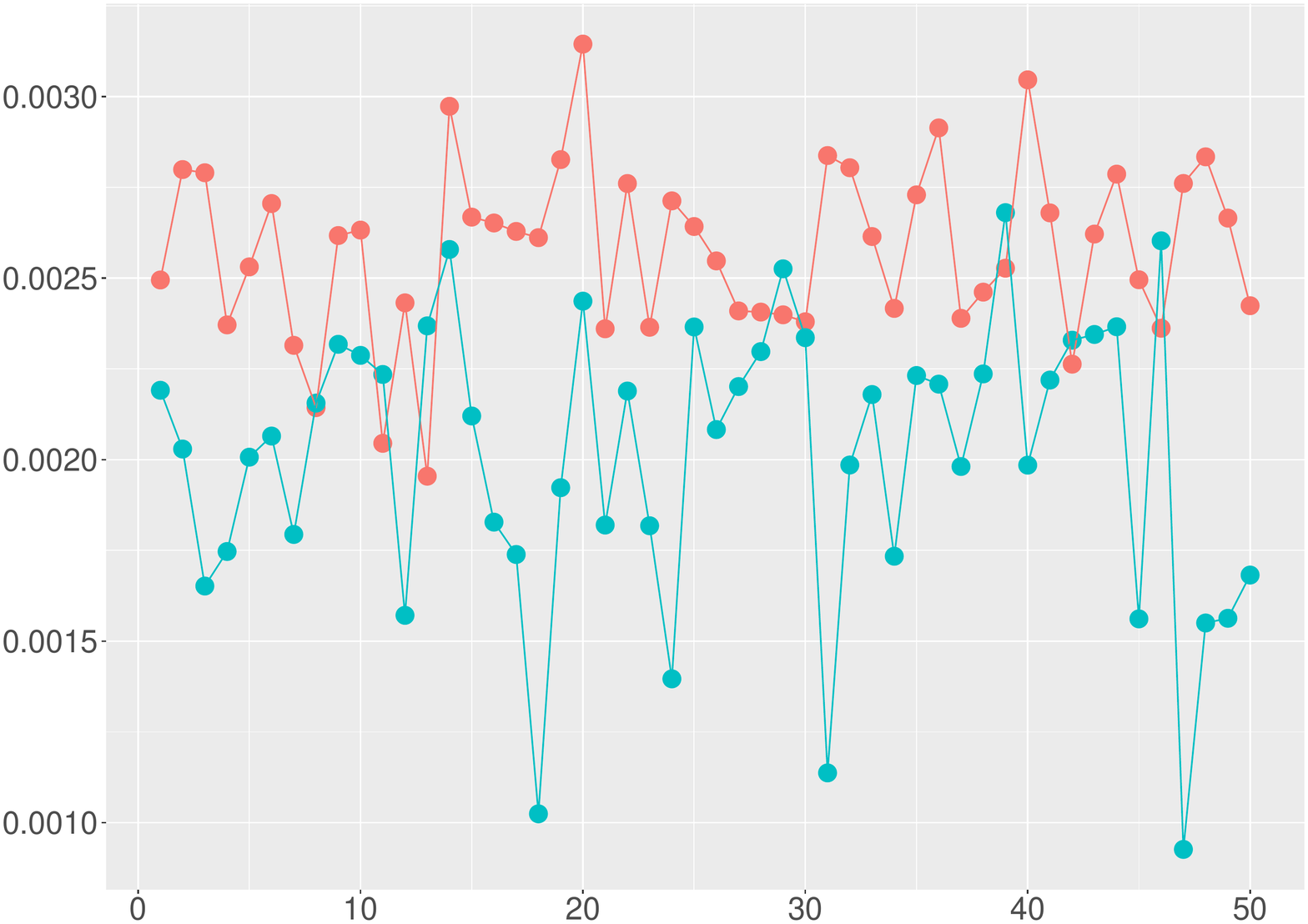}
\includegraphics[width=0.4\textwidth,height=0.2\textheight]{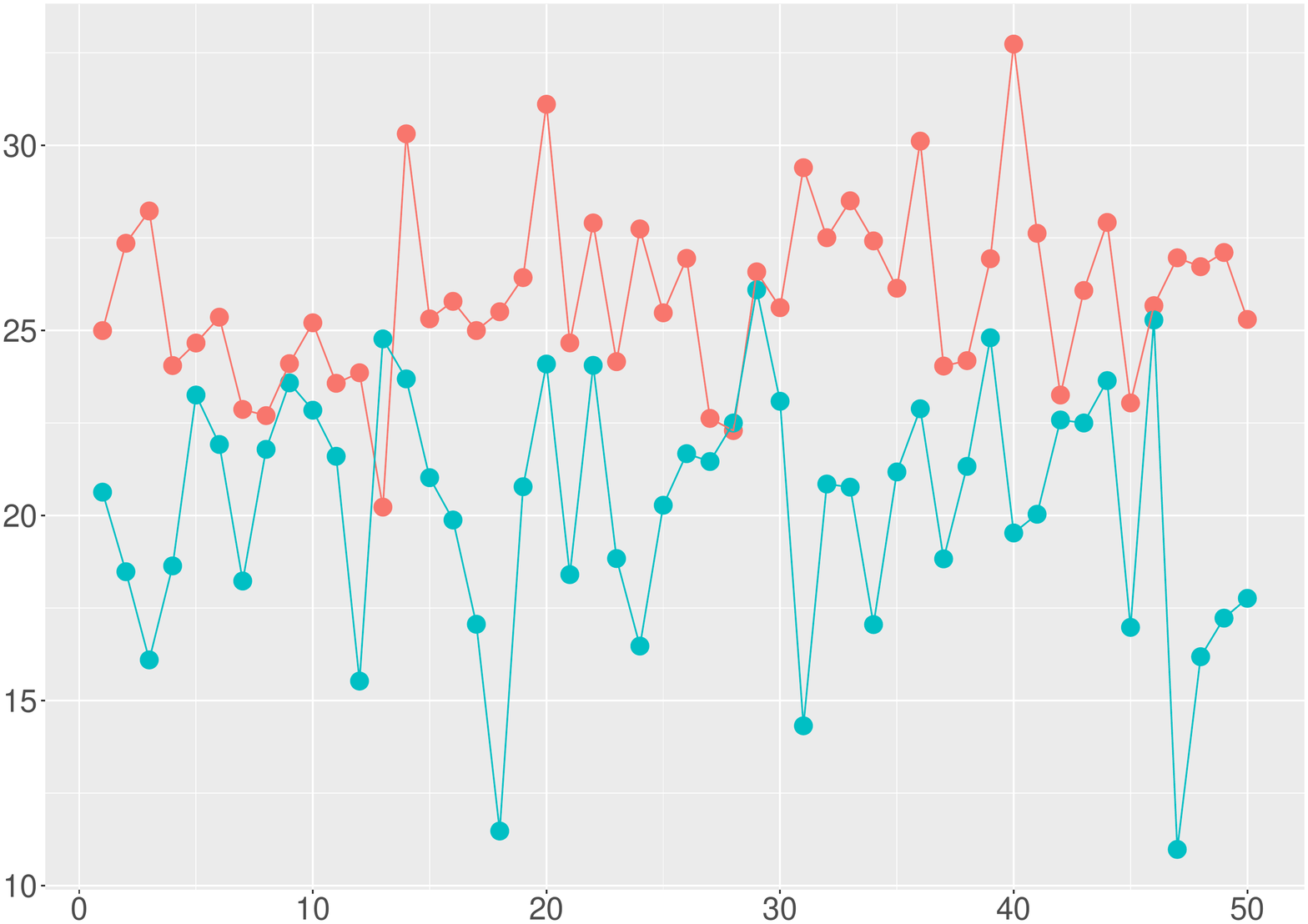}
\caption{Key features comparison. }\label{batchwindowkeyfeature}
\end{figure}

\begin{figure}[h]
\centering
\includegraphics[width=0.4\textwidth,height=0.2\textheight]{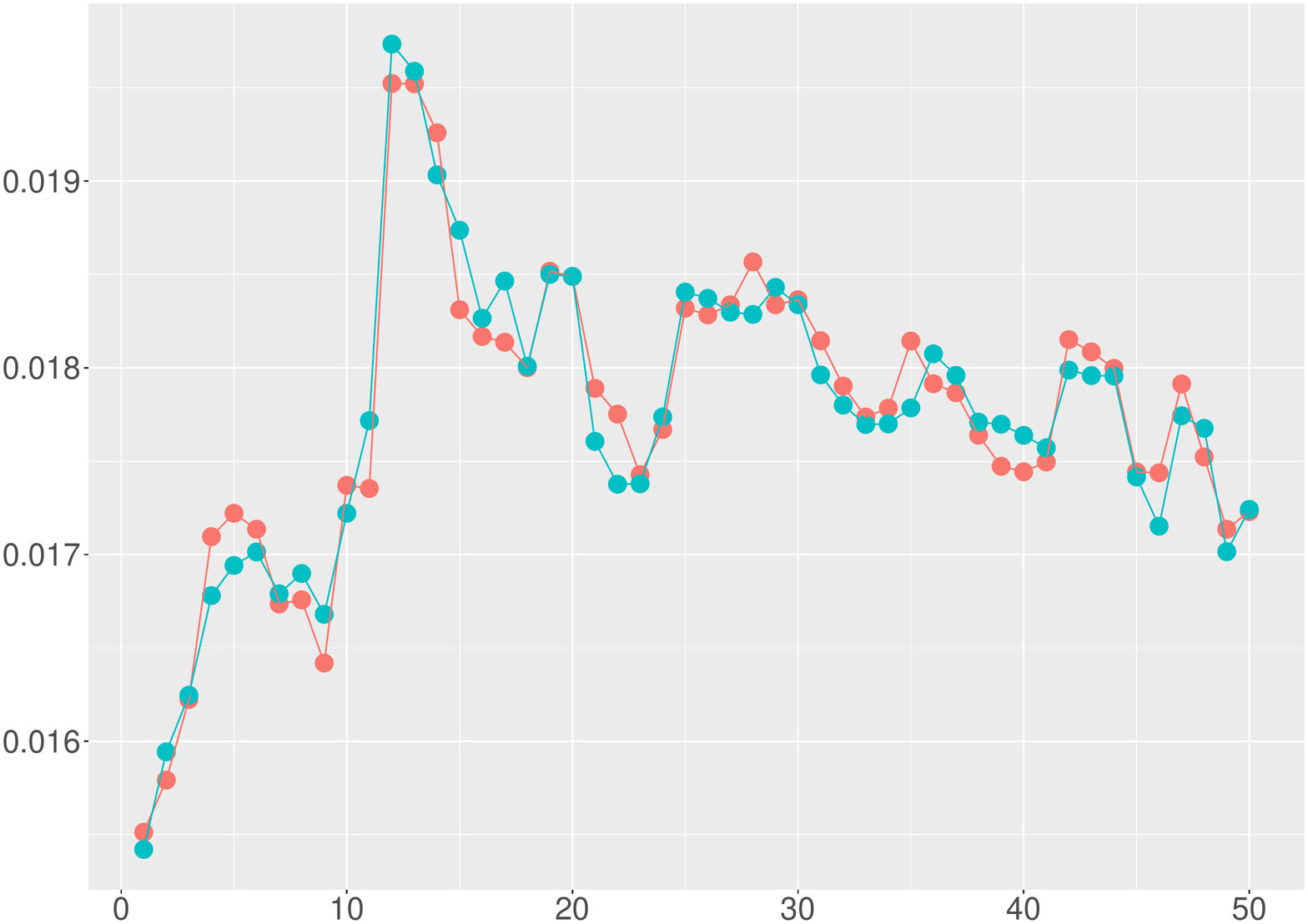}
\includegraphics[width=0.4\textwidth,height=0.2\textheight]{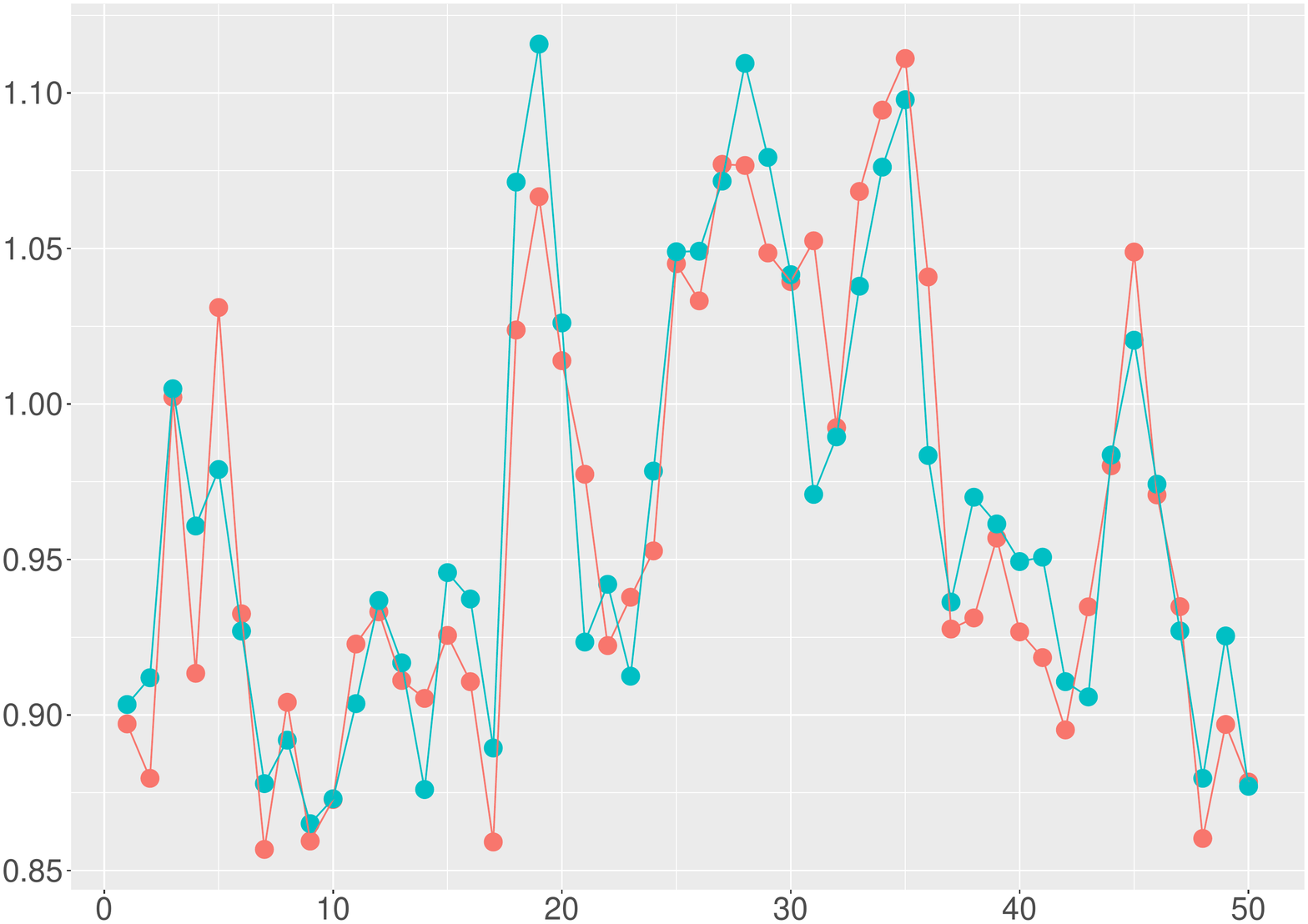}
\includegraphics[width=0.4\textwidth,height=0.2\textheight]{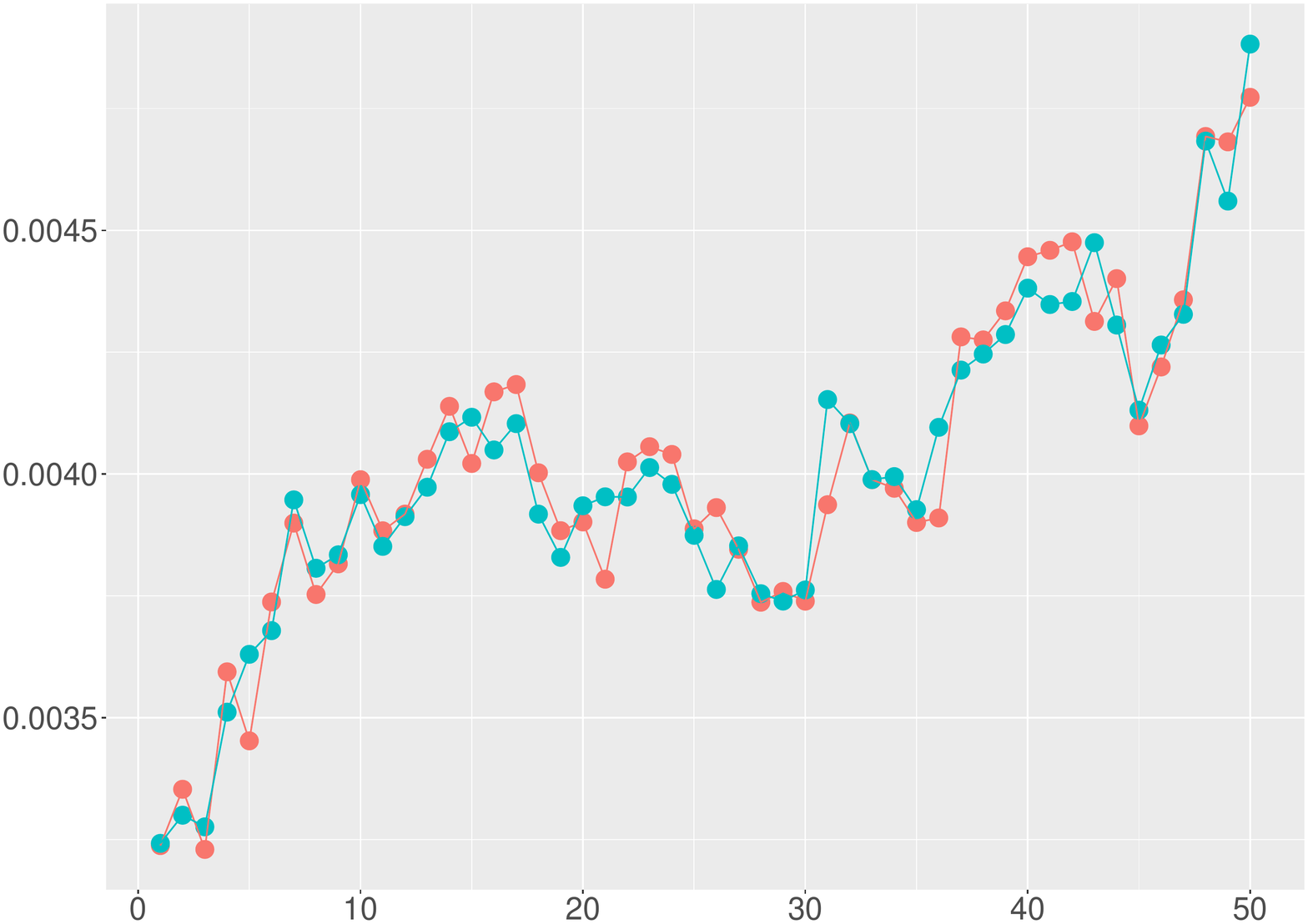}
\includegraphics[width=0.4\textwidth,height=0.2\textheight]{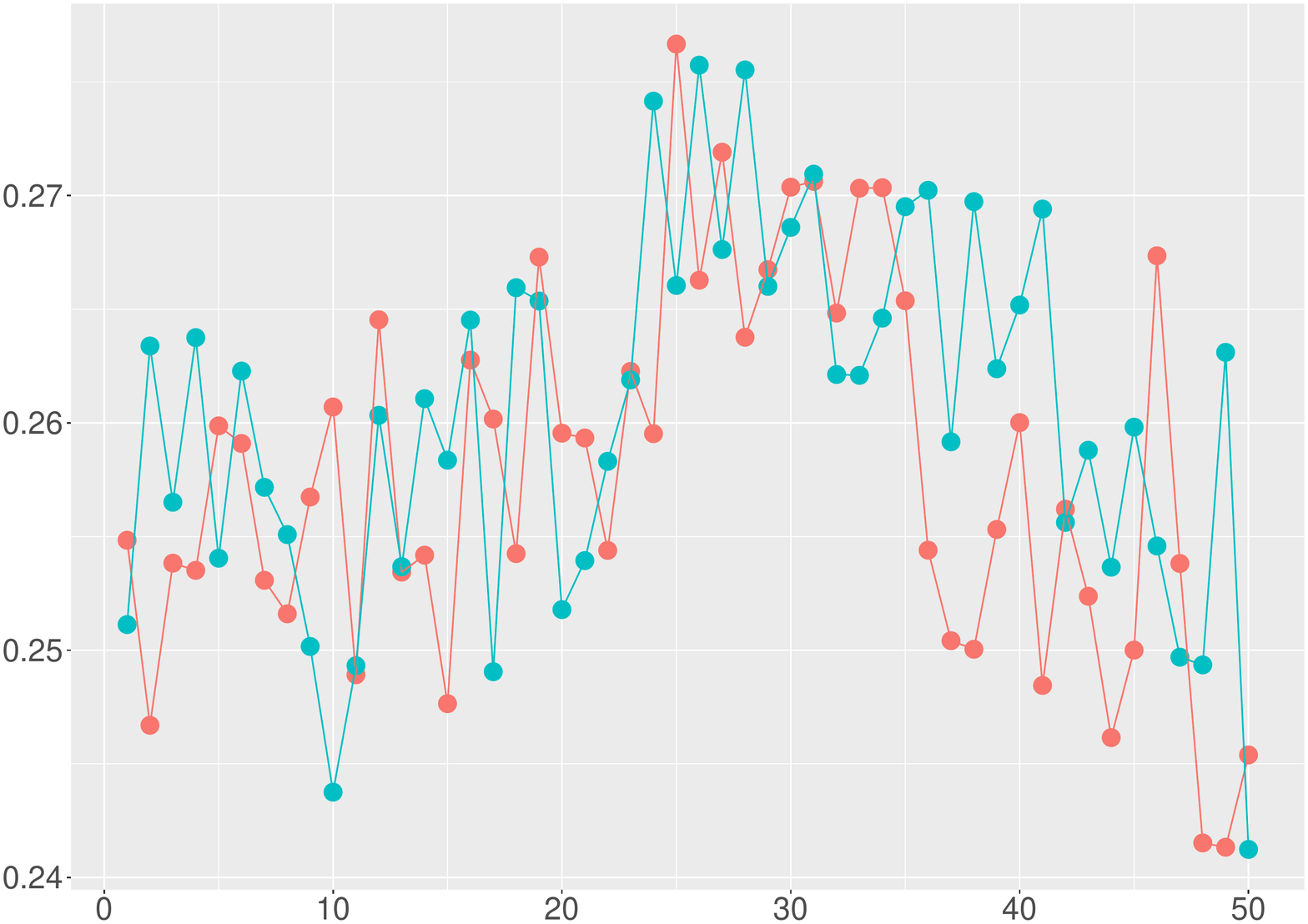}
\includegraphics[width=0.4\textwidth,height=0.2\textheight]{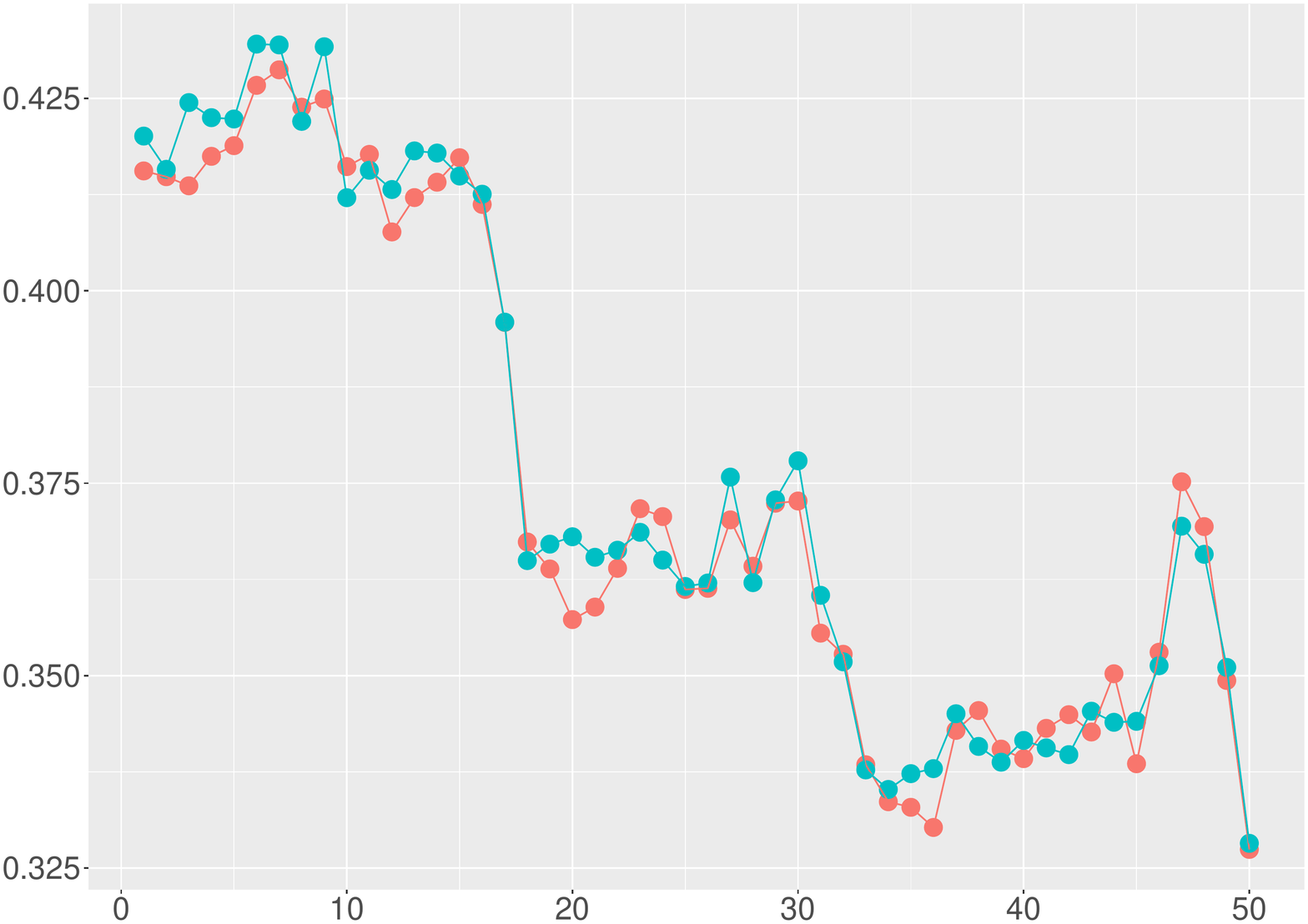}
\caption{Parameter Comparison. }\label{batchwindowparameter}
\end{figure}

For the true posterior, the algorithm requires a cheap estimation $\hat{\pi}(\cdot)$, which is found by one-variable-at-a-time Metropolis-Hastings algorithm. The advantage is getting a precise estimation of the parameter structure, and disadvantage is, obviously, lower efficiency. Luckily, we find that it is not necessary to run this MH every time when estimate a new state from $x_{t-1}$ to $x_t$. In fact, in the DA-MH process, the cheap $\hat{\pi}$ doesn't vary too much in the filtering process with new data coming into the dataset. We may use this property in the algorithm. At first, we use all available data from $1$ to $t$ with length up to $L$ to learn the structure of $\theta$ and find out the cheap approximation $\hat{\pi}$. Then, use DA-MH to estimate the true posterior $\pi$ for $\theta$ and $x_t$. After that, extend dataset to $1:t+1$ if $t\leq L$ or shift the data window to $2:t+1$ if $t>L$ and run DA-MH again to estimate $\theta$ and $x_{t+1}$. From figures \ref{batchwindowkeyfeature} and \ref{batchwindowparameter}, we can see that the main features and parameters in the estimating process between using batch and sliding window methods have not significant differences.

\begin{figure}[h]
\centering
\begin{subfigure}[t]{0.45\textwidth}
	\includegraphics[width=\textwidth]{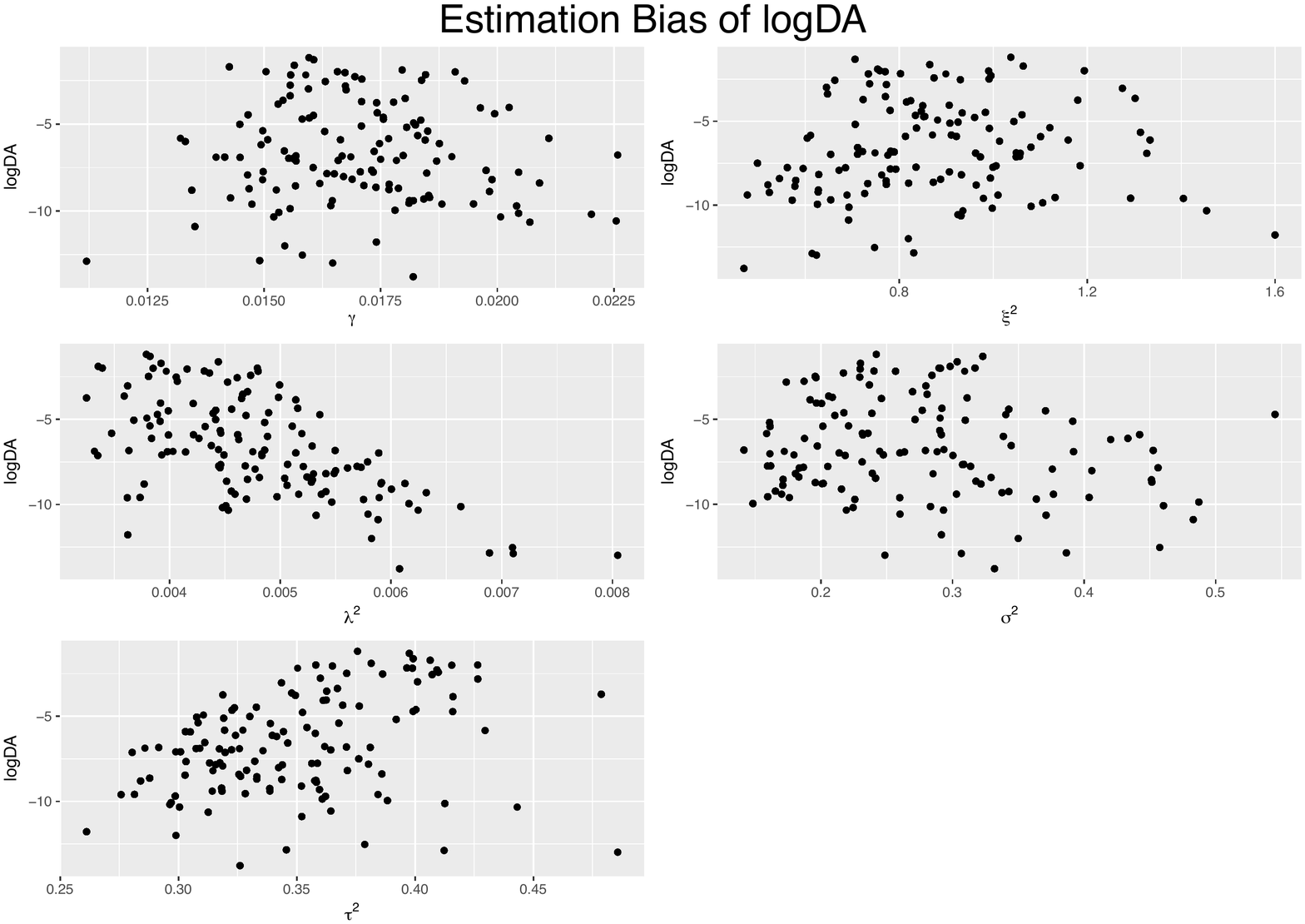}
	\caption{$\ln DA$ surfaces of not-updating-mean}
\end{subfigure}
\begin{subfigure}[t]{0.45\textwidth}
	\includegraphics[width=\textwidth]{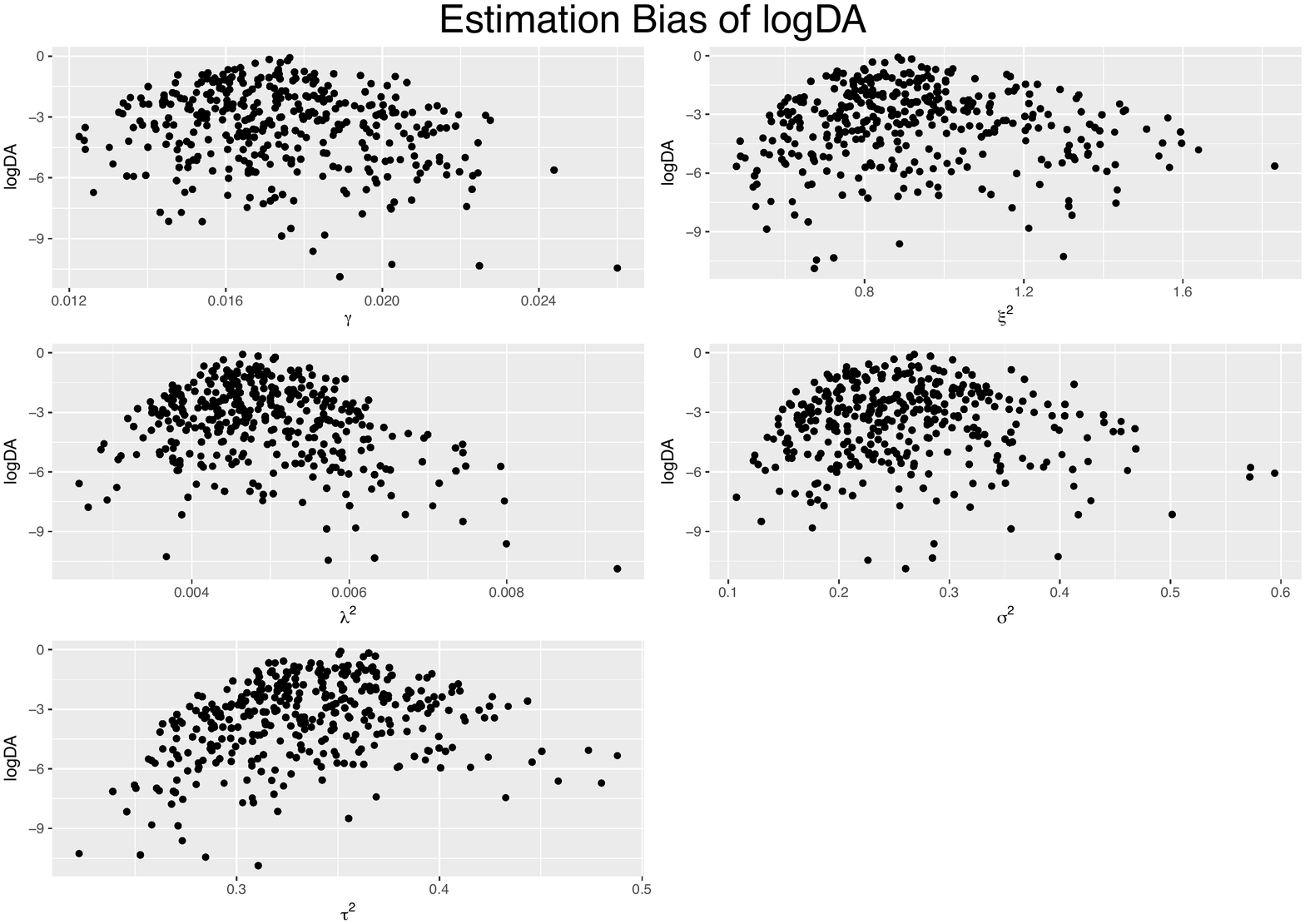}
	\caption{$\ln DA$ surfaces of updating-mean}
\end{subfigure}
\begin{subfigure}[t]{0.45\textwidth}
	\includegraphics[width=\textwidth]{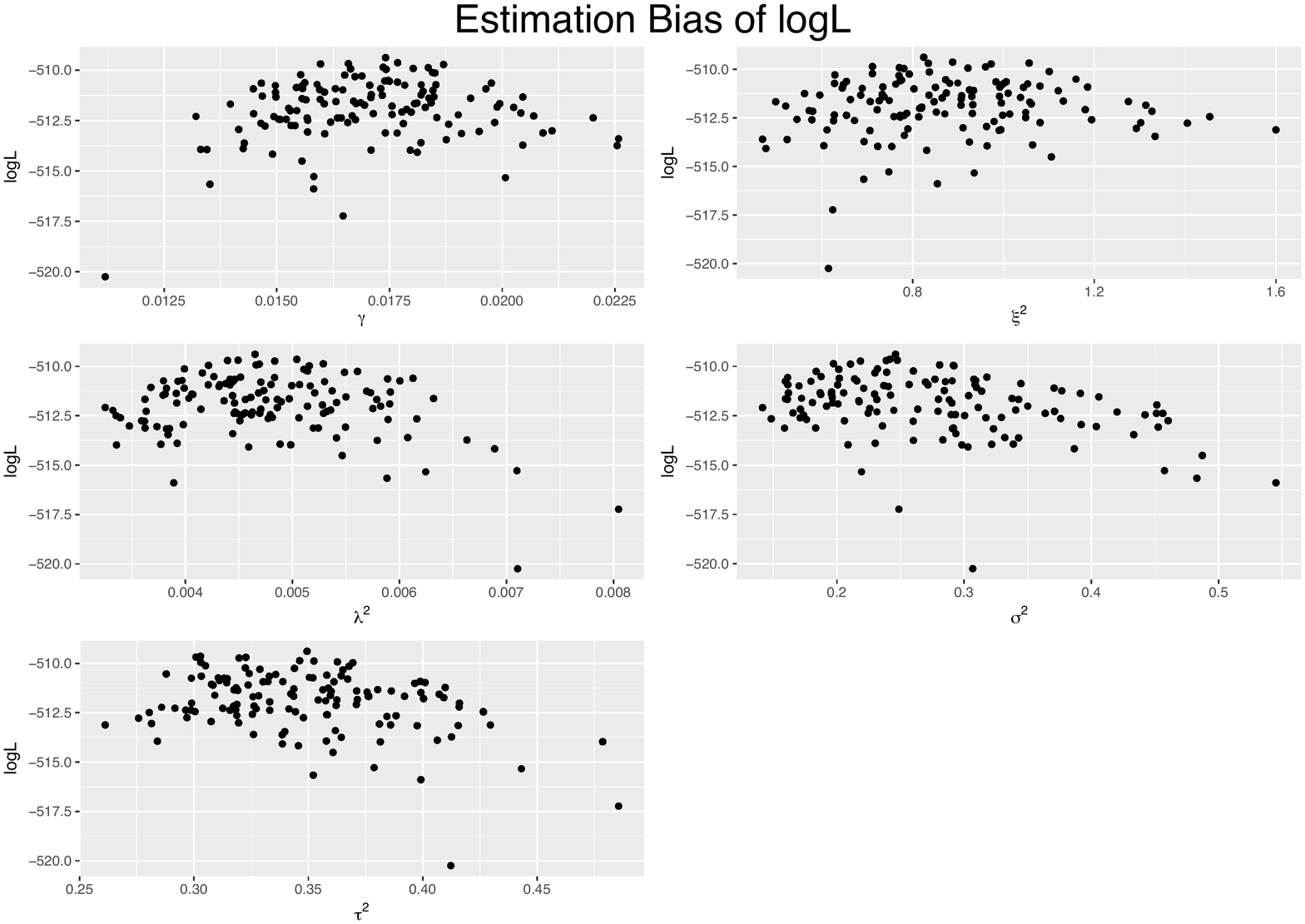}
	\caption{$\ln L$ surfaces of not-updating-mean}
\end{subfigure}
\begin{subfigure}[t]{0.45\textwidth}
	\includegraphics[width=\textwidth]{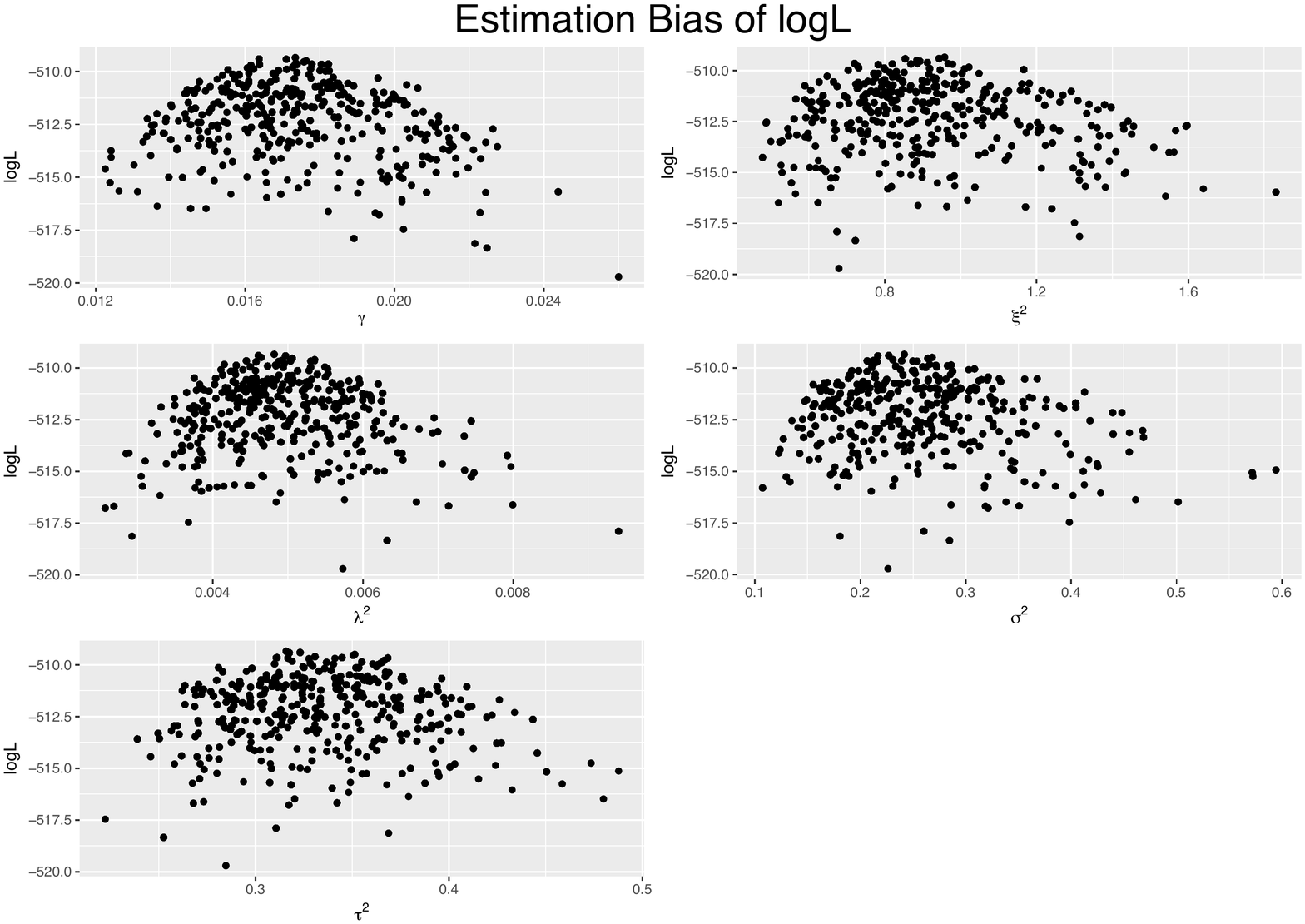}
	\caption{$\ln L$ surfaces of updating-mean}
\end{subfigure}
\caption{Comparison $\ln DA$ and $\ln L$ surfaces between not-updating-mean and updating-mean methods. It is obviously that the updating-mean method has dense log-surfaces indicating more effective samples.} \label{comparenotanupDAL}
\end{figure}

\begin{figure}[h]
\centering
\begin{subfigure}[t]{0.45\textwidth}
\includegraphics[width=\textwidth]{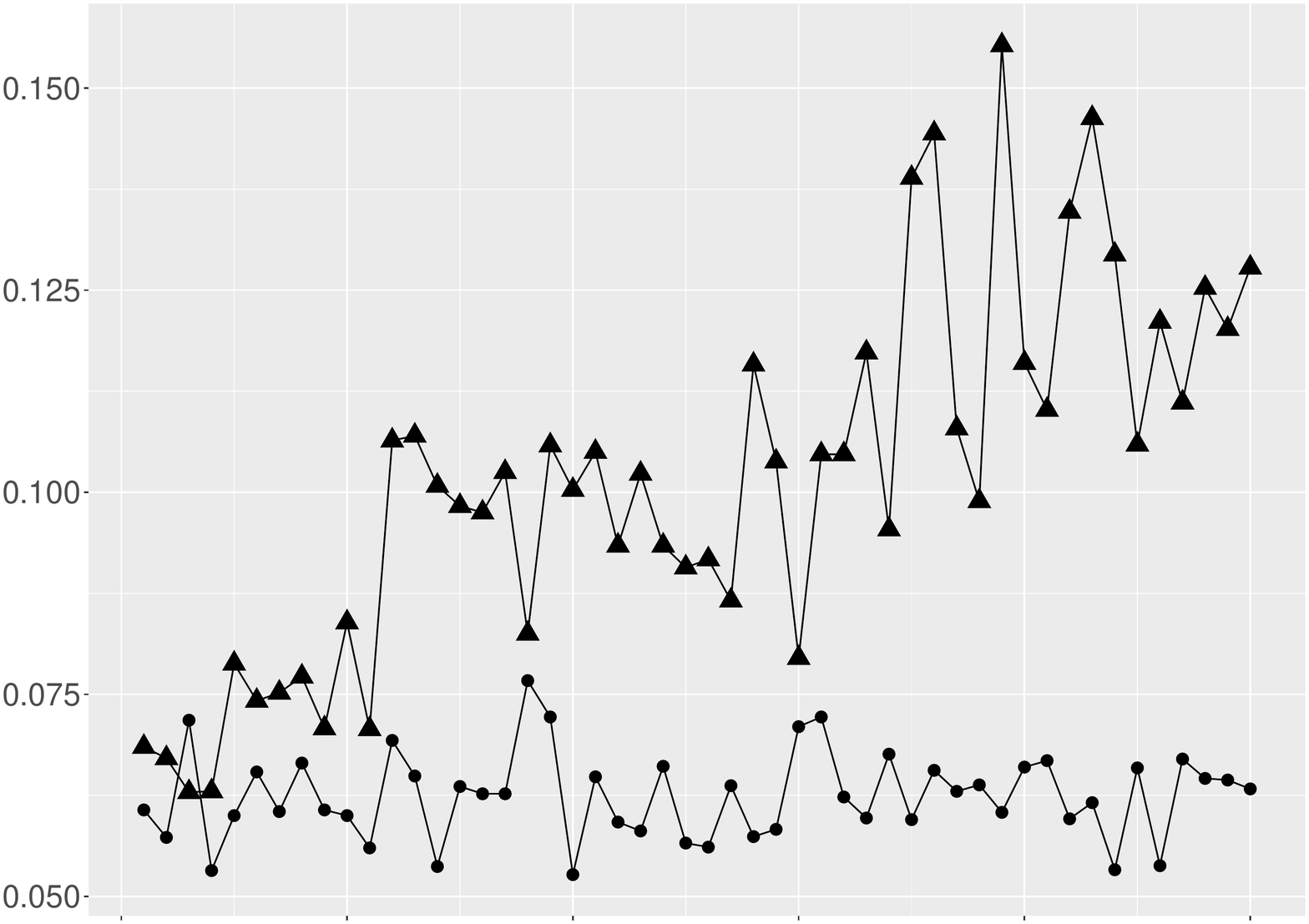}
   \caption{Comparing $\alpha_1$}
\end{subfigure}
\begin{subfigure}[t]{0.45\textwidth}
\includegraphics[width=\textwidth]{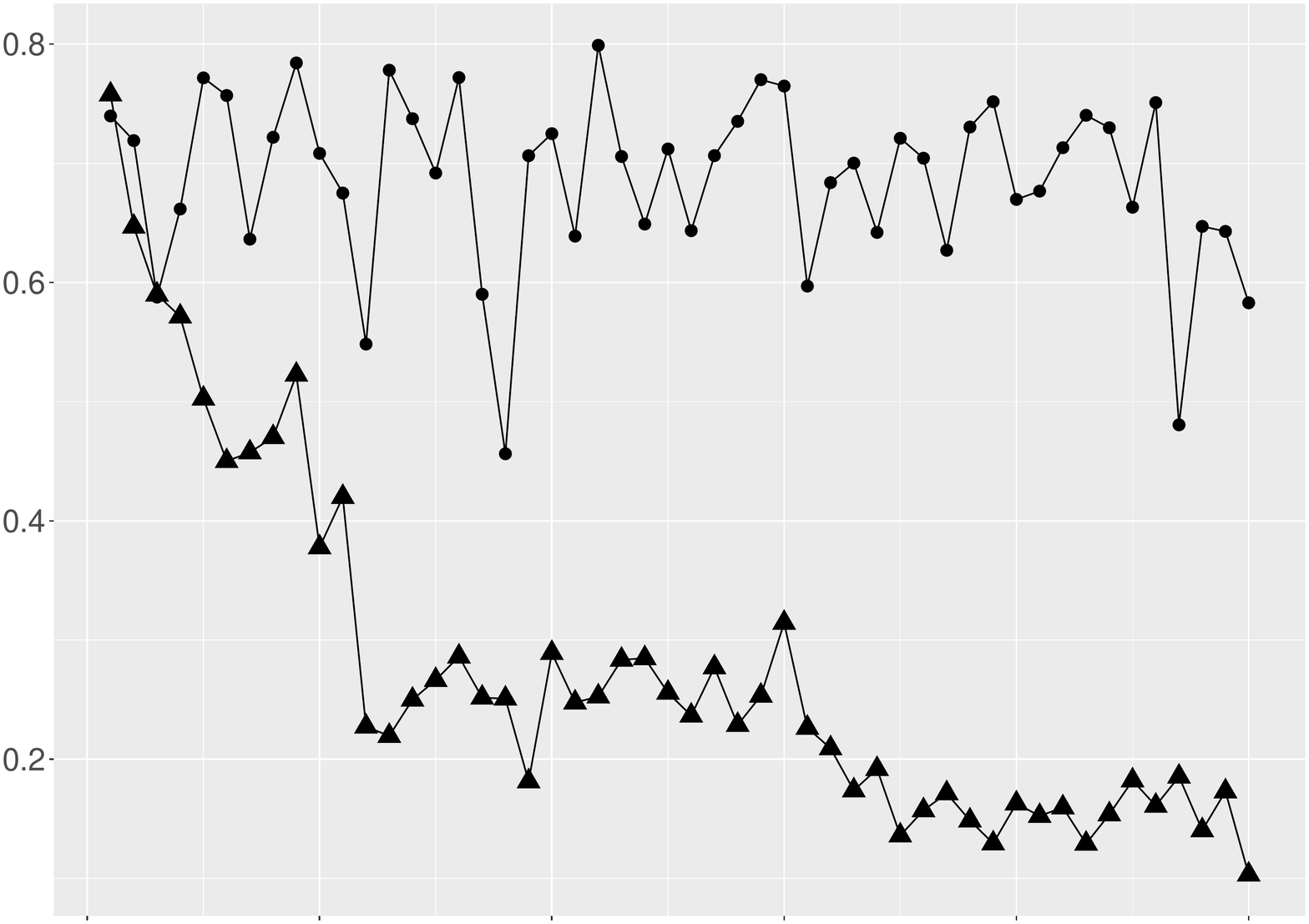}
   \caption{Comparing $\alpha_2$}
\end{subfigure}
\begin{subfigure}[t]{0.45\textwidth}
\includegraphics[width=\textwidth]{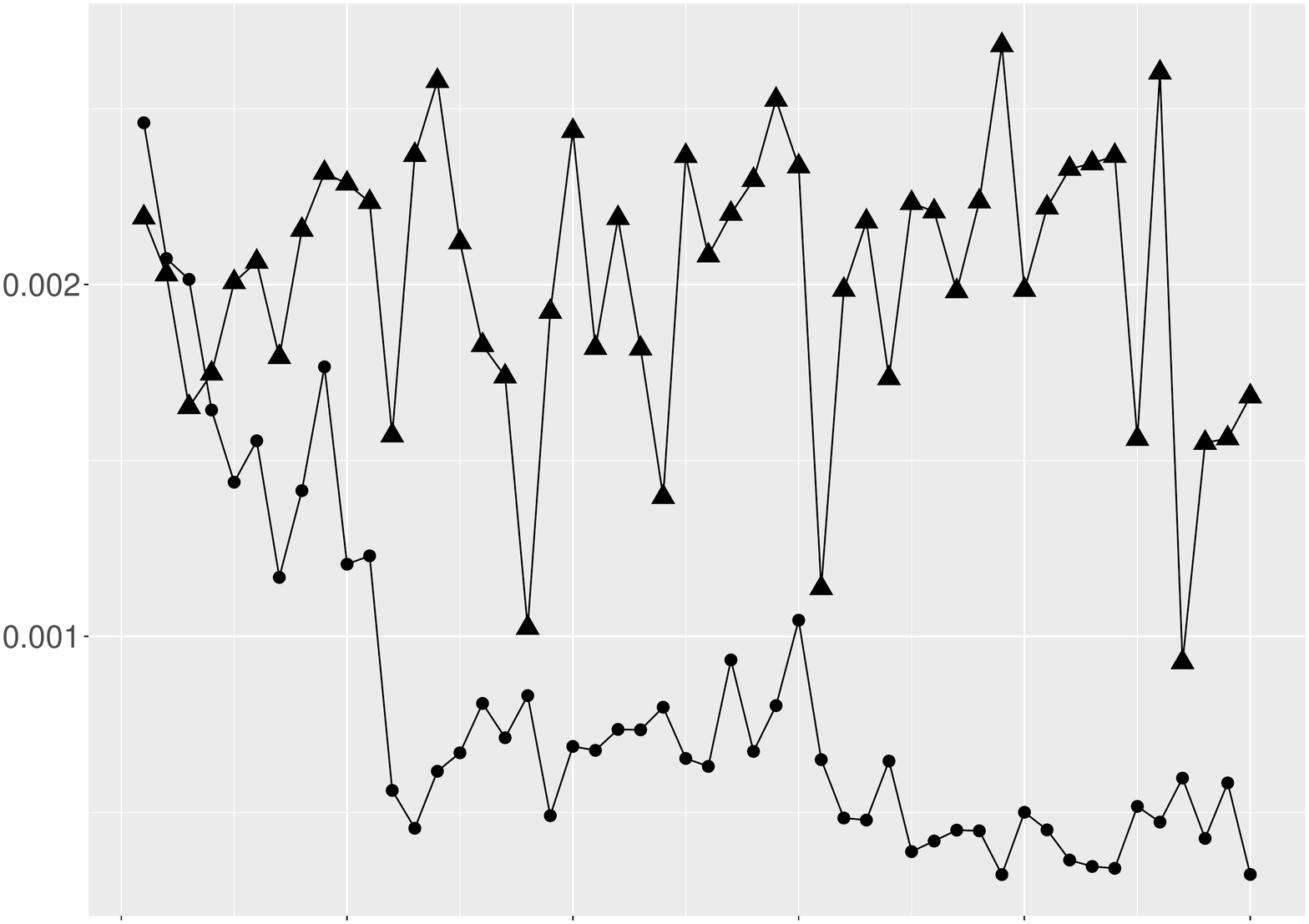}
   \caption{Comparing EffUT}
\end{subfigure}
\begin{subfigure}[t]{0.45\textwidth}
\includegraphics[width=\textwidth]{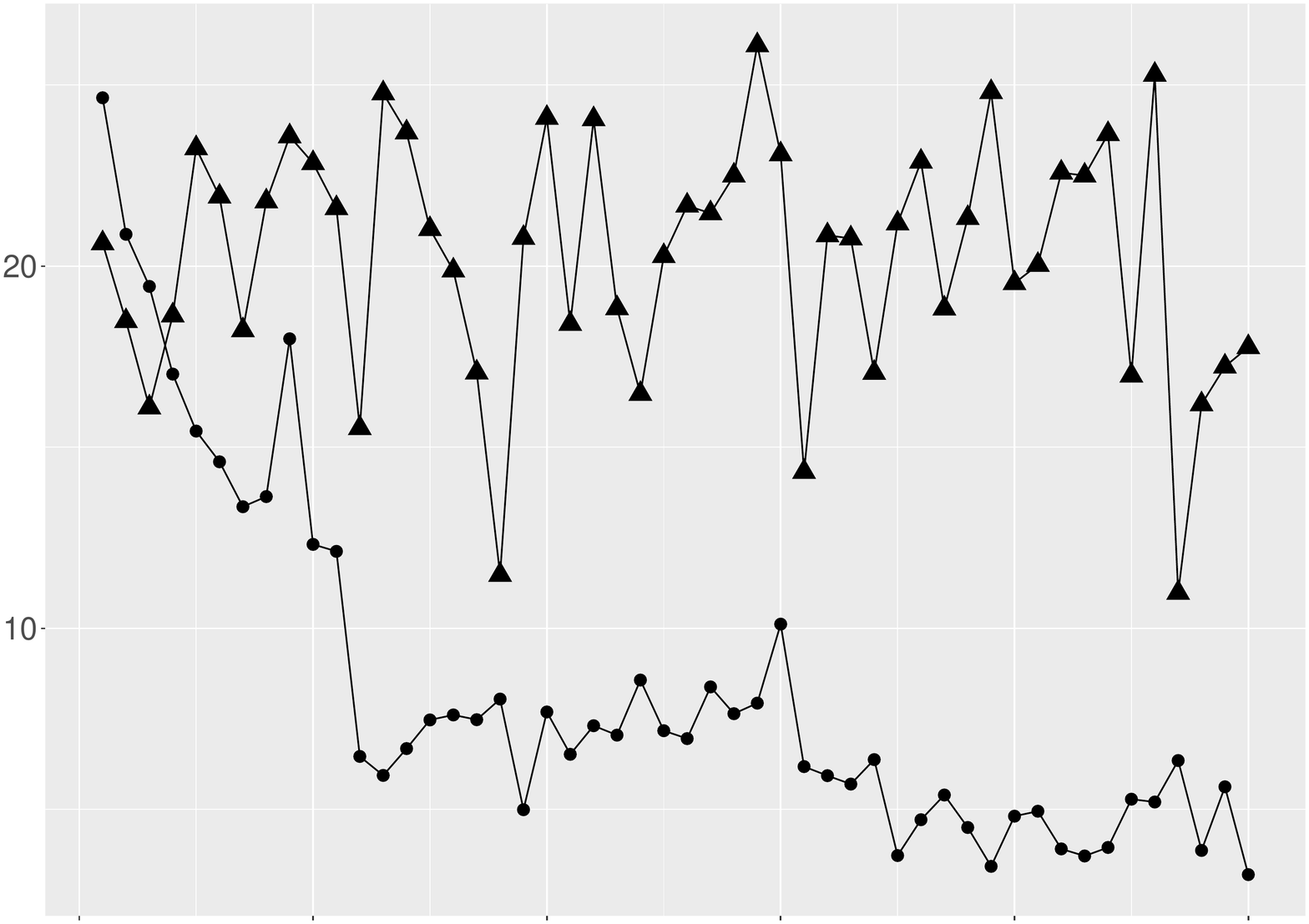}
   \caption{Comparing ESSUT}
\end{subfigure}
\caption{Comparison of acceptance rates $\alpha_1$, $\alpha_2$, EffUT and ESSUT between not-updating-mean and updating-mean methods. Black solid dots $\bullet$ indicate values obtained from not-updating-mean method and black solid triangular $\blacktriangle$ indicate values obtained from updating-mean method. The acceptance rates of the updating-mean method are more stable and effective samples are larger in unit computation time. }\label{comparenotanupfeatures}
\end{figure}

To avoid estimation bias in the algorithm, we are introducing \textit{threshold} and \textit{cut off} processes. \textit{threshold} means when a bias occurs in the algorithm, the cheap $\hat{\pi}$ may not be appropriate and a new one is needed. Thus, we have to update $\hat{\pi}$ with a latest data we have. A \textit{cut off} process stops the algorithm when a large $\Delta_t$ happens. A large time gap indicates the vehicle stops at some time point and it causes irregularity and bias. A smart way is stopping the process and waiting for new data coming in. By running testings on real data, the \textit{threshold} is chosen $\alpha_2<0.7$ and \textit{cut off} is $\Delta_t\geq 300$ seconds. These two values are on researchers' choice. From figures \ref{comparenotanupDAL} and  \ref{comparenotanupfeatures}, we can see that by using the \textit{threshold}, we are efficiently avoiding bias and getting more effective samples.

 
So far, the complete algorithm is summarized in the following algorithm \ref{algorithmslidingwindow}: 

\begin{algorithm}[h]
\SetAlgoLined 
Initialization: Set up $L$, \textit{threshold} and  \textit{cut off} criteria. \\
Learning process: Estimate $\theta$ with $p(\theta\mid Y_{1:\min \{t,L\} } ) \propto p(Y_{1:\min \{t,L\} } \mid \theta )p(\theta )$ by one-variable-at-a-time Random Walk Metropolis-Hastings algorithm gaining the target acceptance rates and find out the structure of $\theta\sim N(\mu,\Sigma)$ and the approximation $\hat{\pi}(\cdot)$. \label{algorithmlearningsurface}\\
Estimate $X_{ \max\{1,t-L+1 \} :\min \{t,L\} }$ with $Y_{ \max\{1,t-L+1 \} :\min \{t,L\} }$: \For{$i$ from 1 to $N$}{ \label{algorithmestimaiton}
Propose $\theta_i^*$ from $N(\theta_i\mid\mu,\Sigma)$, accept it with probability $\alpha_1=\min\left\lbrace  1,\frac{\hat{\pi}(\theta_i^*)q(\theta_i, \theta_i^*)}{\hat{\pi}(\theta_i)q(\theta_i^*, \theta_i)}  \right\rbrace$ and go to next step; otherwise go to step \ref{algorithmDA}.\label{algorithmDA}\\
Accept $\theta_i^*$ with probability $\alpha_2=\min \left\lbrace  1,\frac{\pi(\theta_i^*)\hat{\pi}(\theta_i) }{\pi(\theta_i)\hat{\pi}(\theta_i^*)} \right\rbrace$ and go to next step; otherwise go to step \ref{algorithmDA}. \\
Calculate $\mu_i^{(t)},\Sigma_i^{(t)}$ for $X_t$ and $\mu_i^{(t+s)},\Sigma_i^{(t+s)}$ for $X_{t+s}$.\\
}
Calculate $\mu_X^{(t)} = \frac{1}{N} \sum_i \mu_i^{(t)}$, $\Var(X^{(t)}) = \frac{1}{N} \sum_i (\mu_i^{(t)} \mu_i^{(t)\top} +\Sigma_i) -\frac{1}{N^2} (\sum_i  \mu_i^{(t)}) (\sum_i \mu_i^{(t)})^\top$ and $\mu_X^{(t+s)}$, $\Var(X^{(t+s)})$ in the same formula.  \\
Check \textit{threshold} and  \textit{cut off} criteria. \uIf{\textit{threshold} is TRUE}{Update $\theta\sim N(\mu,\Sigma)$}\uElseIf{ \textit{cut off} is TRUE}{Stop process. }
\Else{ Go to next step.\\}
Shift the window by setting $t = t+1$ and go back to step \ref{algorithmestimaiton}.
 \caption{Sliding Window MCMC.}\label{algorithmslidingwindow}
\end{algorithm}

\subsection{Implementation}

\begin{figure}[h]
\centering
\includegraphics[width=0.7\textwidth,height=8cm]{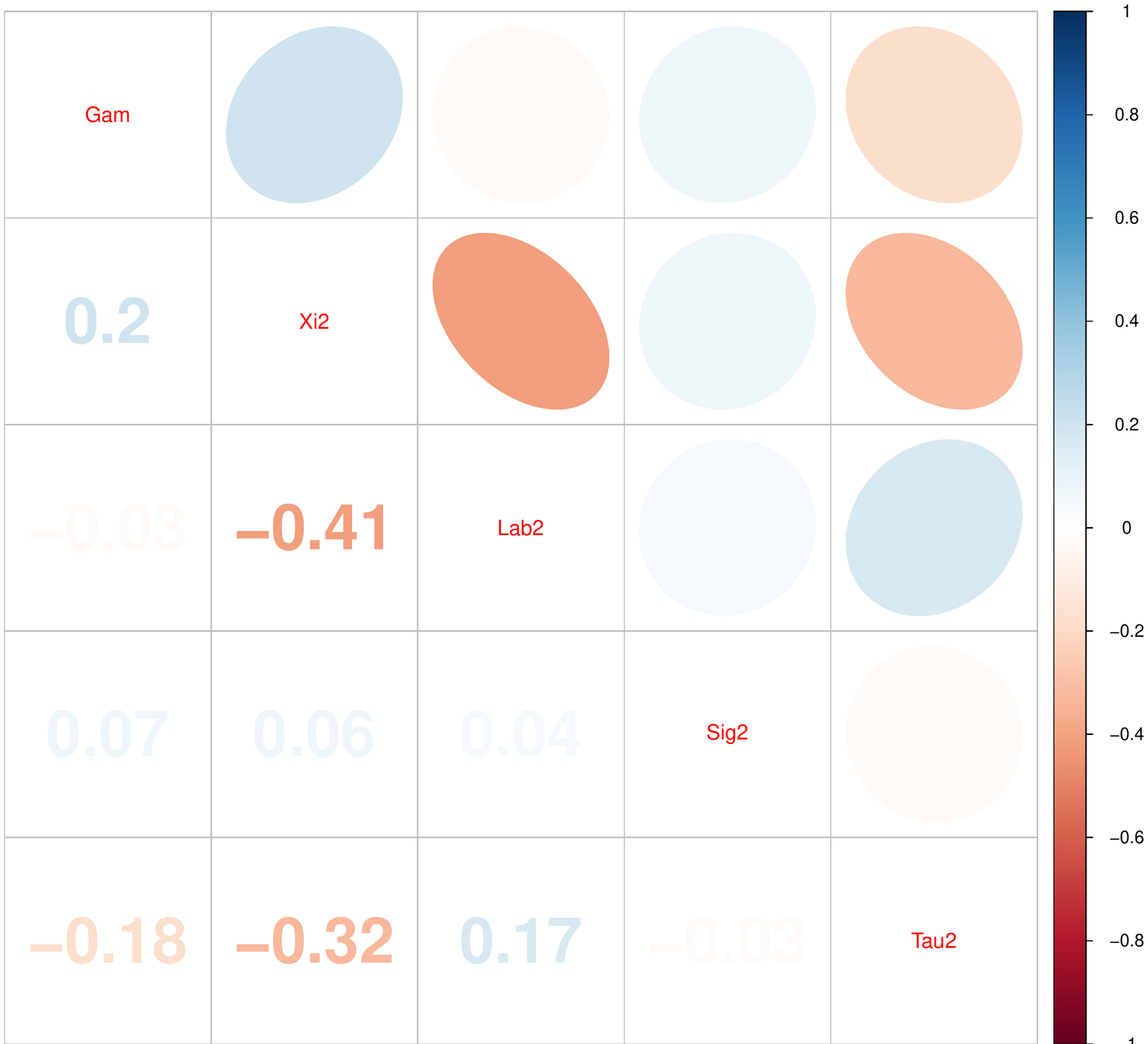}
\caption{Visualization of the parameters correlation matrix, which is found in learning phase. }\label{realdatacorMatrix}
\end{figure}

To implement the algorithm \ref{algorithmslidingwindow}, firstly we should get an idea of how the hyper parameter space looks like by running step \ref{algorithmlearningsurface} of the algorithm with some observed data. By setting $L=100$ and running 5\,000 iterations, we can find the whole $\theta$ samples in 59 seconds. For each parameter of $\theta$, we take 1\,000 sub-samples out of 5\,000 as new sequences. The new $\theta^*$ is representative for the hyper parameter space. Then the traces and correlation is derived from $\theta^*$. Meanwhile, the acceptance rates for each parameter are $\alpha_\gamma = 0.453,\alpha_{\xi^2}=0.433, \alpha_{\lambda^2}=0.435, \alpha_{\sigma^2}=0.414, \alpha_{\tau^2}=0.4490$ respectively. Hence, the structure of  $\hat{\theta}\sim N\left( m_t,C_t\right)$ is achieved. That can be seen in figure \ref{realdatacorMatrix}. 

\begin{figure}[h]
\centering
\begin{subfigure}[t]{0.45\textwidth}
	\includegraphics[width=\textwidth]{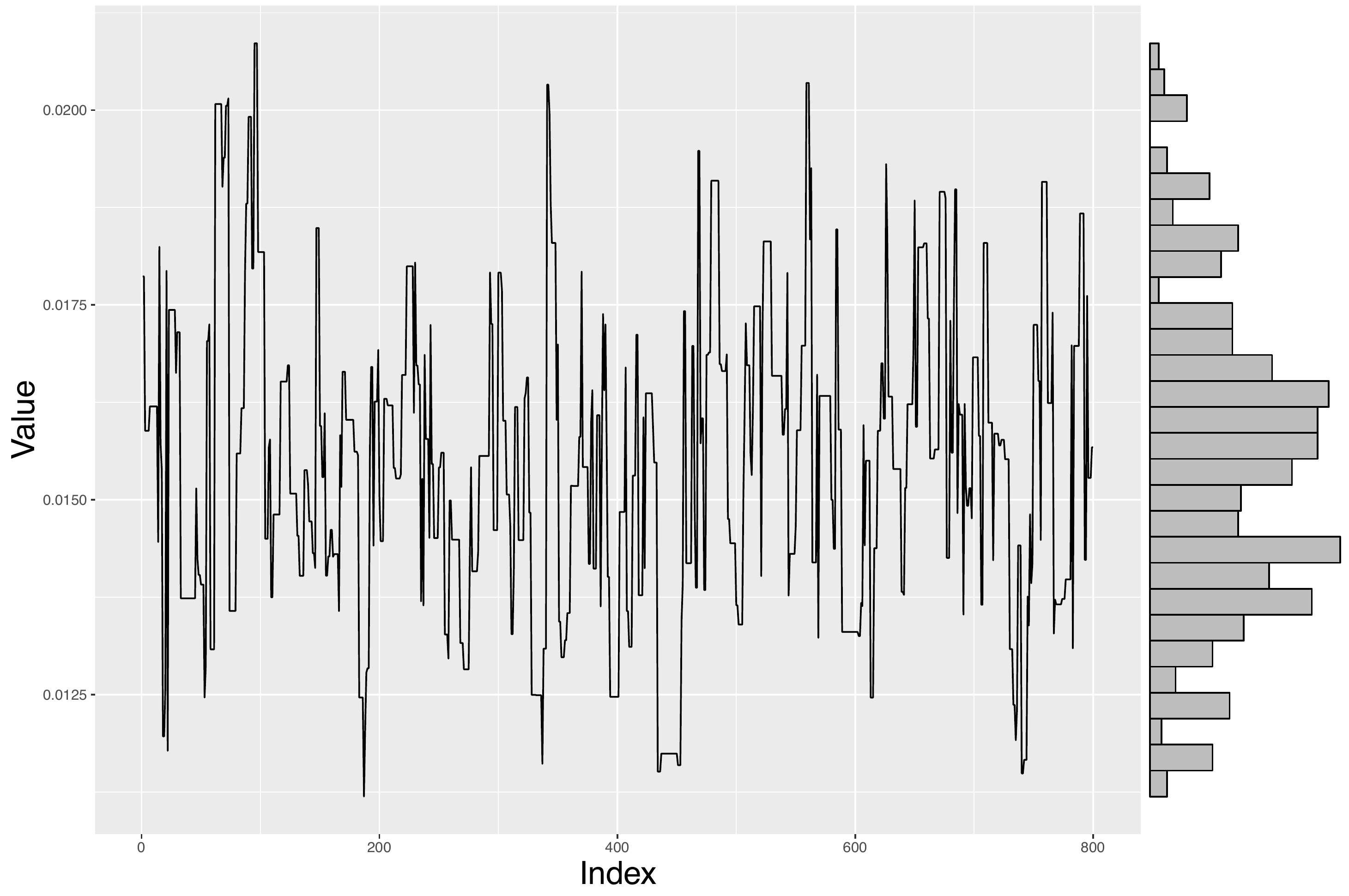}
    \caption{Trace plot of $\gamma$}
\end{subfigure}
\begin{subfigure}[t]{0.45\textwidth}
	\includegraphics[width=\linewidth]{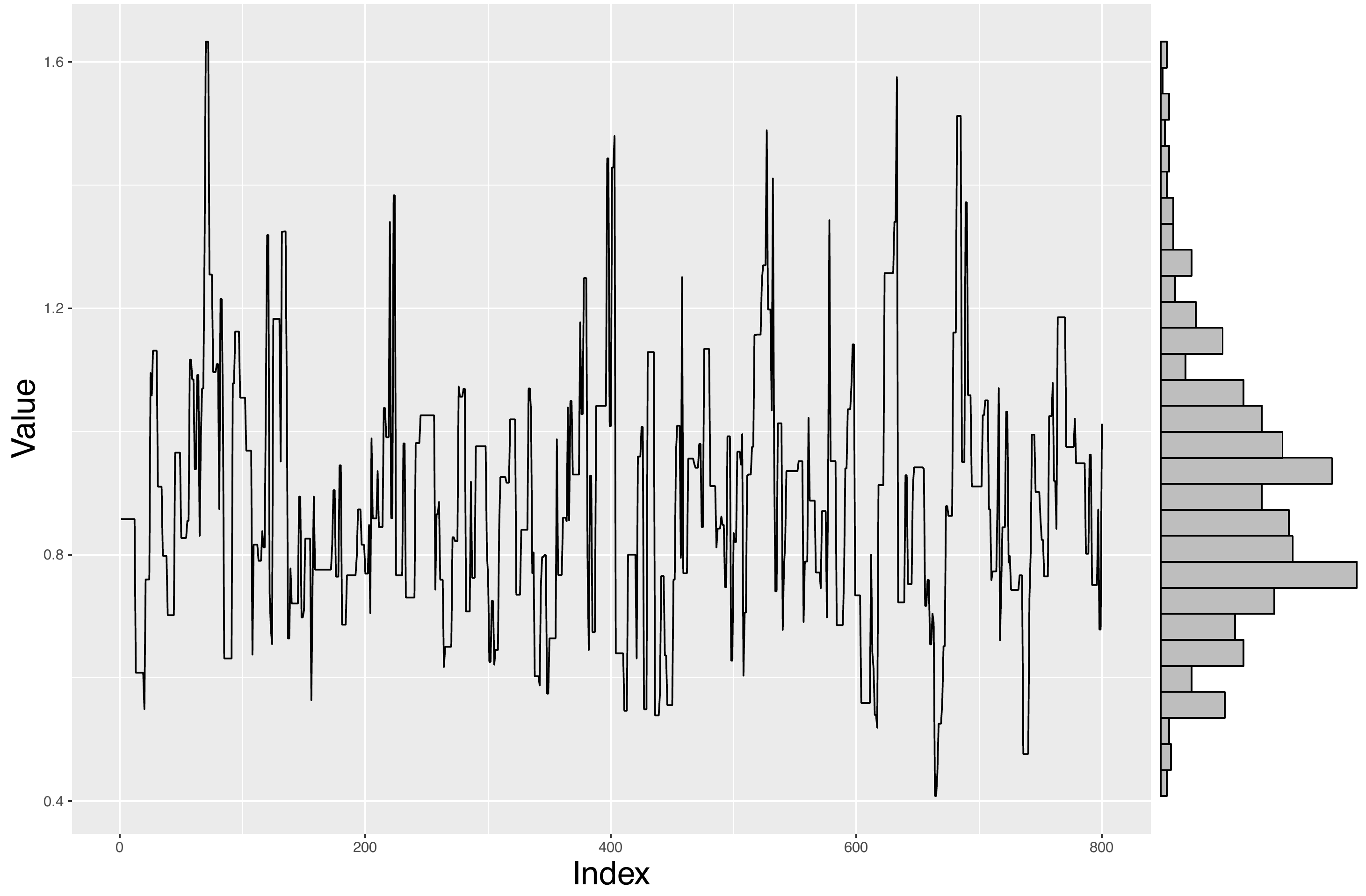}
 	\caption{Trace plot of $\xi^2$}
\end{subfigure}
\begin{subfigure}[t]{0.45\textwidth}
	\includegraphics[width=\linewidth]{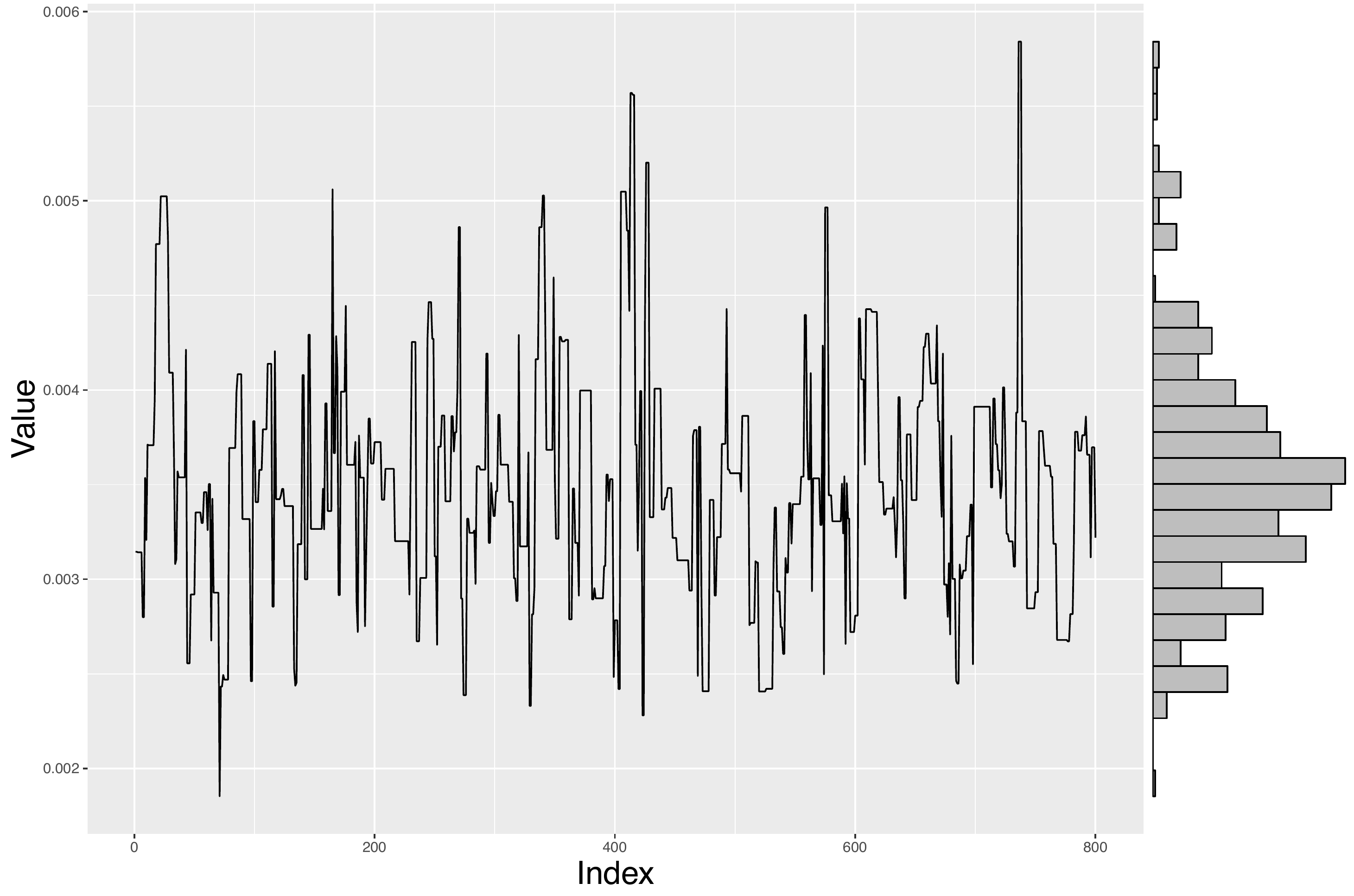}
	\caption{Trace plot of $\lambda^2$}
\end{subfigure}
\begin{subfigure}[t]{0.45\textwidth}
	\includegraphics[width=\linewidth]{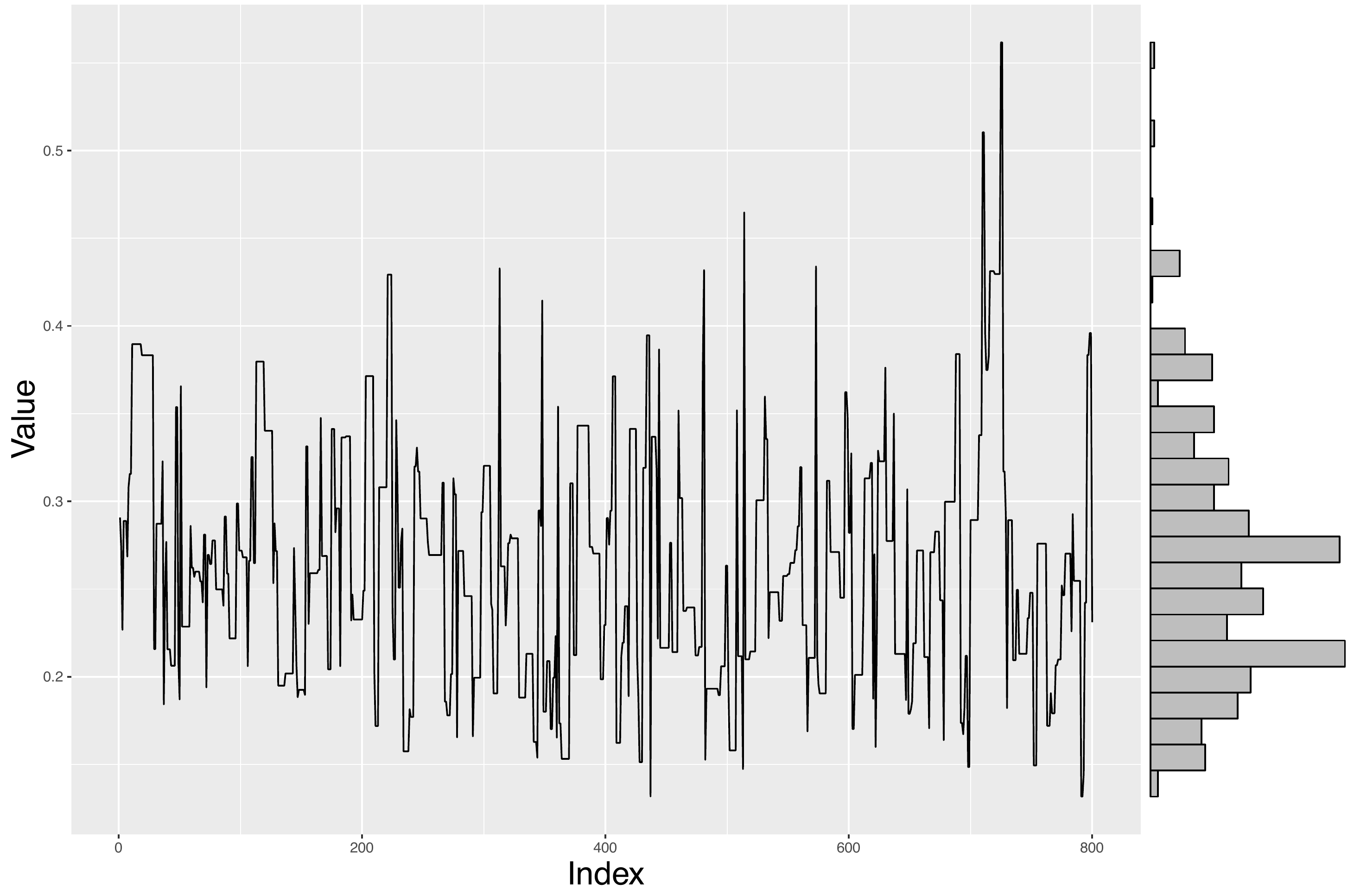}
	\caption{Trace plot of $\sigma^2$}
\end{subfigure}
\begin{subfigure}[t]{0.45\textwidth}
	\includegraphics[width=\linewidth]{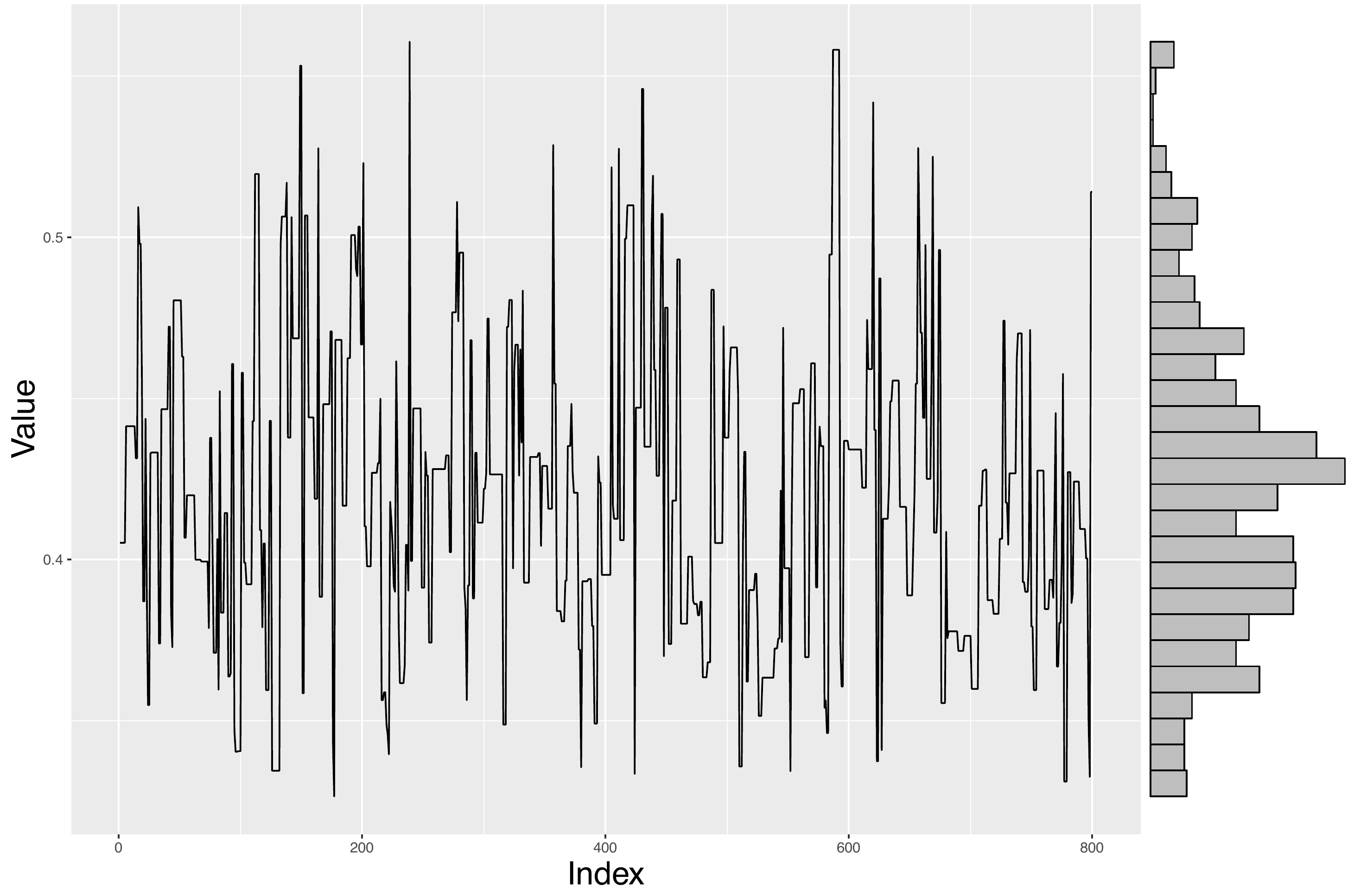}
	\caption{Trace plot of $\tau^2$}
\end{subfigure}
\caption{Trace plots of $\theta$ from learning phase after taking 1\,000 burn-in samples out from 5\,000. }
\end{figure}

Since a cheap surrogate $\hat{\pi}(\cdot)$ for the true $\pi(\cdot)$ is found in step \ref{algorithmlearningsurface}, it is time to move to the next step. Algorithm \ref{algorithmslidingwindow} takes fixed $L$ length data from $Y_{1:L}$ to $Y_{t-L+1:t}$ until an irregular large time lag meets the \textit{cut off} criterion. In the implementation, the first \textit{cut off} occurs at $t = 648$th data point. The first estimated $\hat{X}_{1:L}$ was found by the batch method and $\hat{X}_{L+1}$ to $\hat{X}_{t}$ were found sequentially around 9 seconds with 10\,000 iterations each time. 

\begin{figure}[h]
\centering
\includegraphics[width=0.7\textwidth]{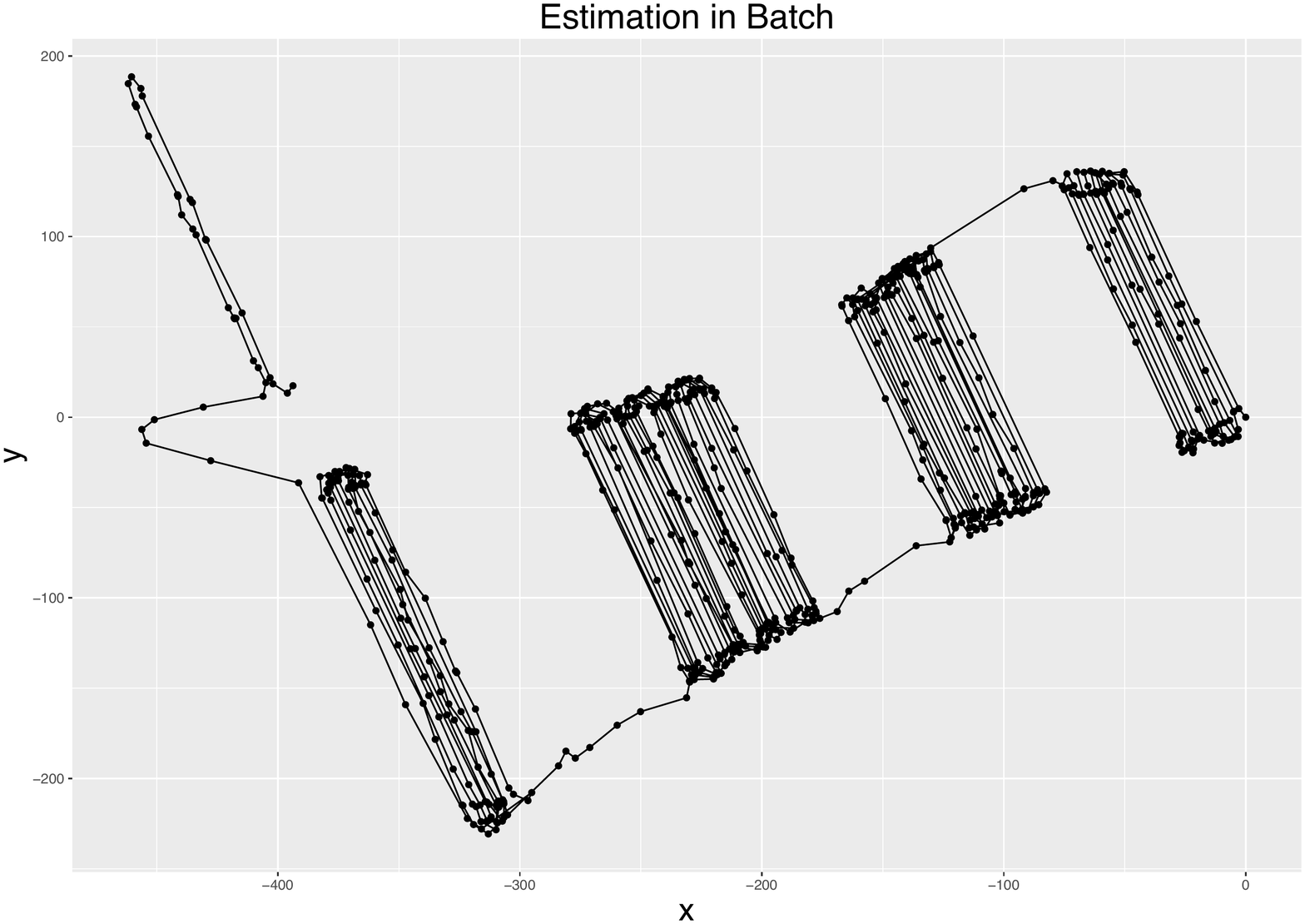}
\includegraphics[width=0.7\textwidth]{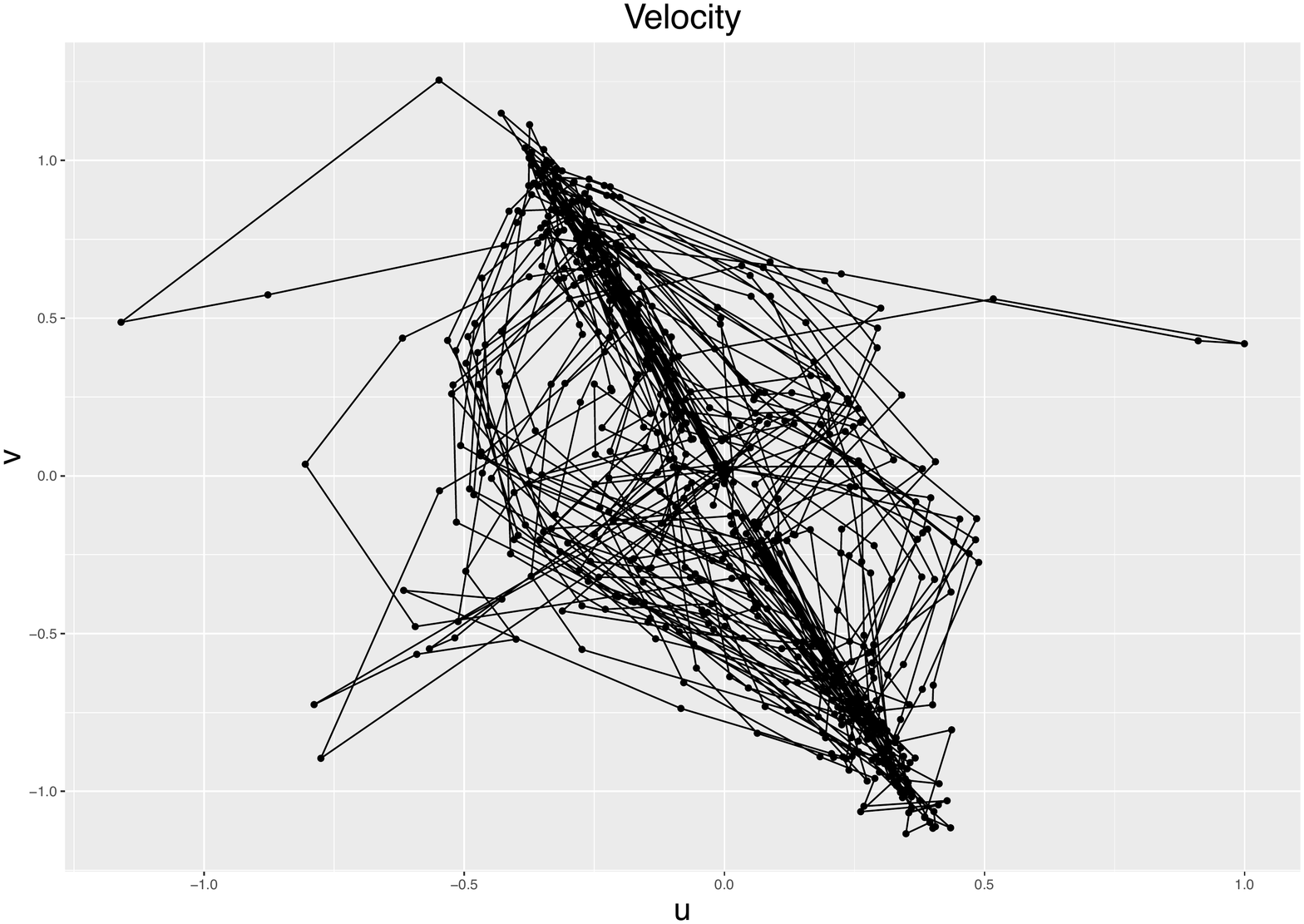}
\caption{Position and velocity for $X$ and $Y$ found by combined batch and sequential methods. }
\end{figure}

\begin{figure}[h]
\centering
\includegraphics[width=0.45\textwidth]{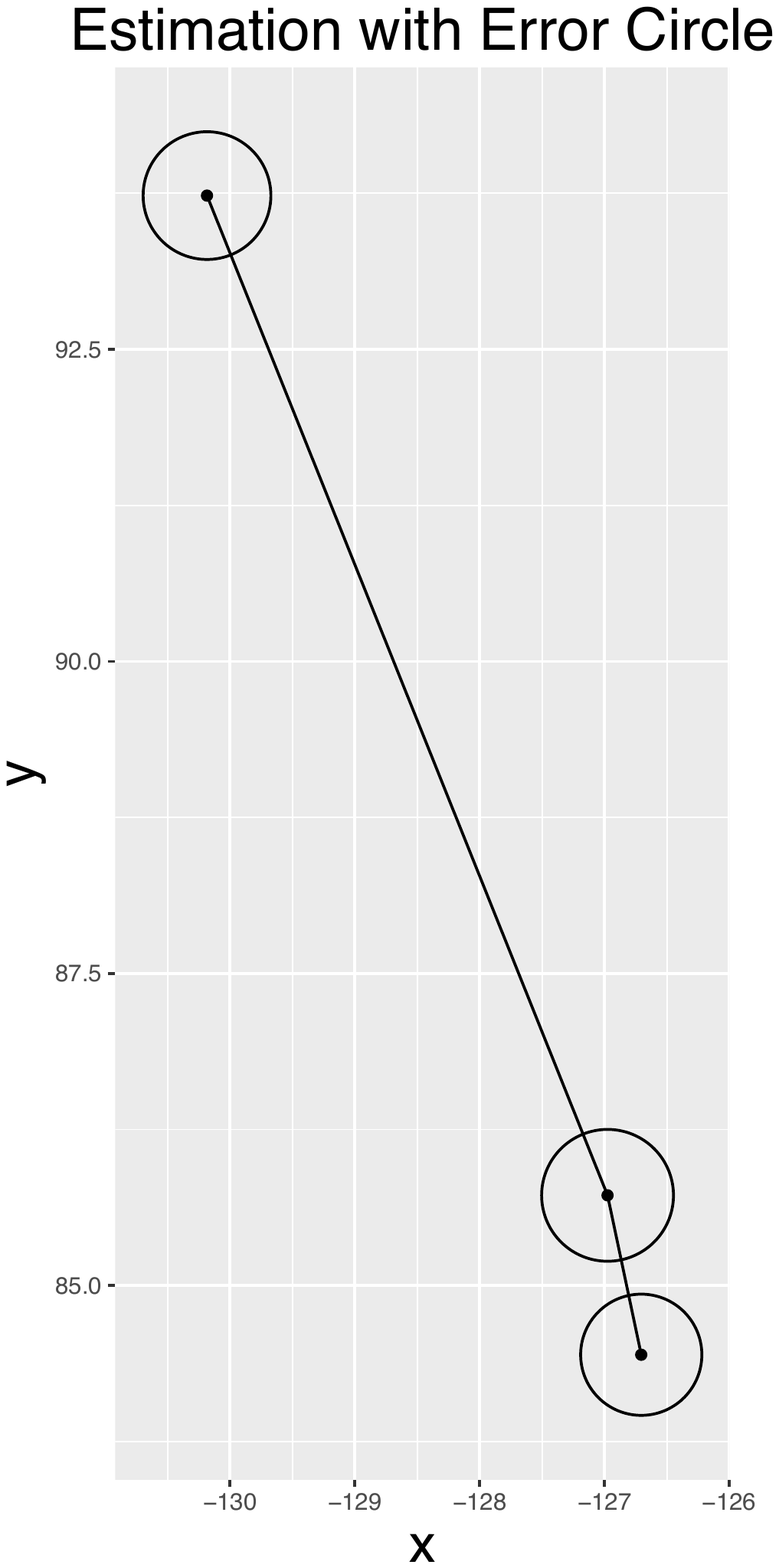}
\includegraphics[width=0.45\textwidth]{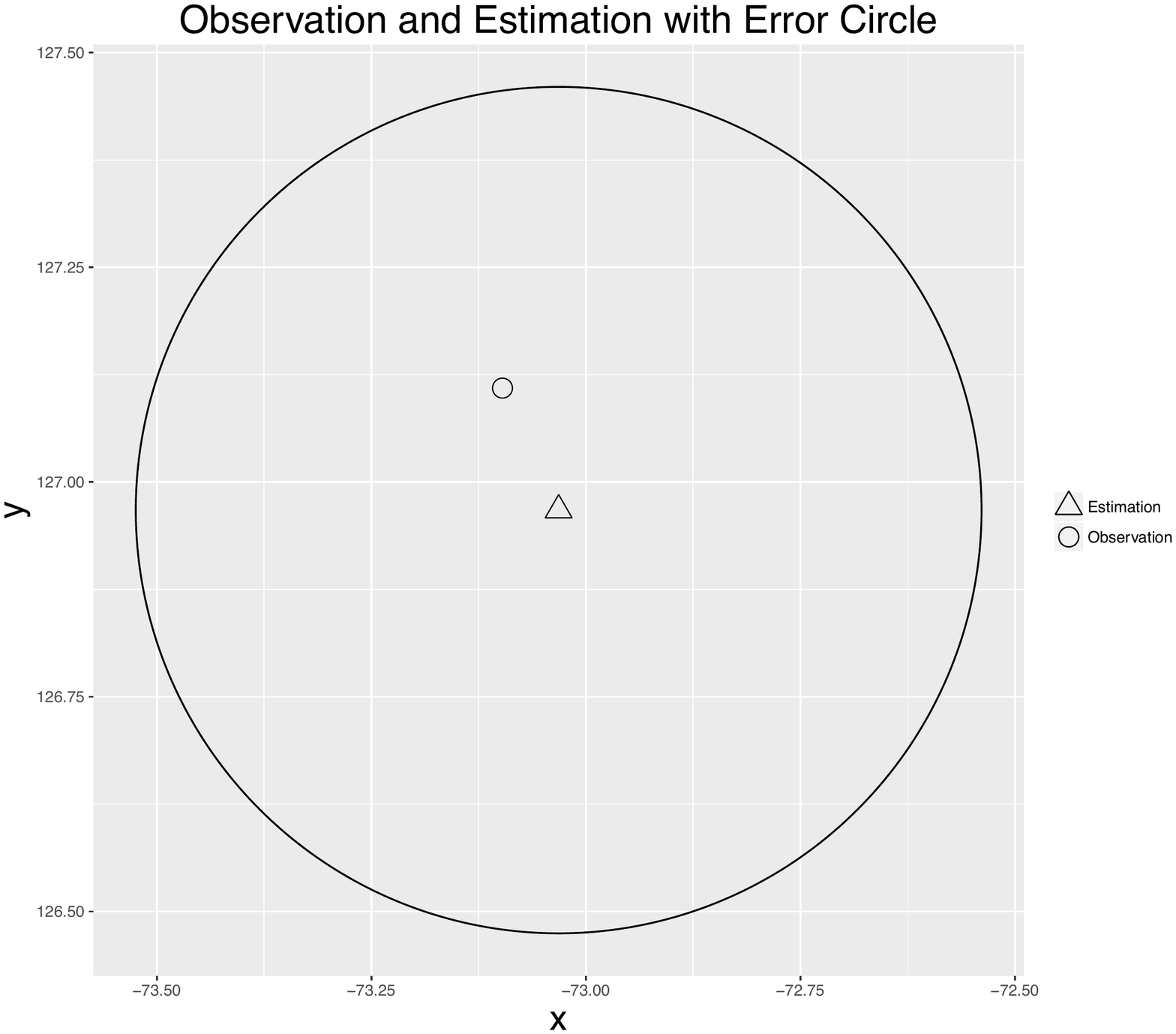}
\caption{Zoom in on estimations. For each estimation $\hat{X}_i (i=1,\dots,t)$, there is a error circle around it. }
\end{figure}



\clearpage


\section{Discussion and Future Work}

In this paper, we are using the a self-tuning one-variable-at-a-time Metropolis-Hastings Random Walk to learn the parameter hyper space for a linear state-space model.  Starting from the joint covariance and distribution of $X$ and $Y$, we have a recursive way to update the mean and covariance sequentially. After getting the cheap approximation posterior distribution, Delayed-Acceptance Metropolis-Hastings algorithm accelerates the estimating process.  The advantage of this algorithm is that it is easily to understand and implement in practice. Contrast, Particle Learning algorithm is high efficient but the sufficient statistics are not available all the time . 

Some future work can be done on inferring state from precious movement with other kinetic information, not just with diffusive  velocity. Besides, I am more interested in increase the efficiency and accuracy of MCMC method.

\section{Appendices}

\subsection{Linear Simulation Calculations}\label{linearcalculation}

\subsubsection*{Forecast} 

Calculating the log-posterior of parameters requires finding out the forecast distribution of $p(y_{1:t}\mid y_{1:t-1},\theta)$. A general way is using the joint distribution of $y_{t}$ and $y_{1:t-1}$, which is $p(y_{1:t}\mid \theta)\sim N(0,\Sigma_{YY})$ and following the procedure in section \ref{sectionforecast} to work out the inverse matrix of a multivariate normal distribution. For example, one may find the inverse of the covariance matrix 
\begin{align*}
\Sigma_{YY}^{-1} = B(I-A^{-1}B) =\frac{1}{\sigma^4}(\sigma^2 I-A^{-1}) \triangleq \frac{1}{\sigma^4} \left[\begin{matrix} 
Z_{t} & b_{t} \\
b_{t}^\top & K_{t}
\end{matrix} \right].
\end{align*}
Therefore, the original form of this covariance is 
\begin{align*} \Sigma_{YY} =\sigma^4 \left[ \begin{matrix}
(Z-bK^{-1}b^\top)^{-1} & -Z^{-1}b(K-b^\top Z^{-1}b)^{-1}\\
-K^{-1}b^\top (Z-bK^{-1}b^\top)^{-1} & (K-b^\top Z^{-1}b)^{-1}
\end{matrix}\right].
\end{align*}
For sake of simplicity, here we are using $Z$ to represent the $t\times t$ matrix $Z_{t}$, $b$ to represent the $t \times 1$ vector  $b_{t}$  and $K$ to represent the $1\times 1$ constant $K_{t}$. By denoting $C_{t}^\top = \begin{bmatrix} 0 & \cdots & 0 & 1\end{bmatrix}$ and post-multiplying $\Sigma_{YY}^{-1}$, it gives us 
\begin{equation}
\Sigma_{YY}^{-1} C_{t}= \frac{1}{\sigma^4}(\sigma^2 I-A^{-1}) C_{t}= \frac{1}{\sigma^4} \left[\begin{matrix} b_{t} \\ K_{t} \end{matrix} \right].
\end{equation} 

By using the formula, one can find a recursive way to update $K_{t}$ and $b_{t-1}$, which are 
\begin{align}
K_{t}  &=\frac{\sigma^4}{\tau^2+\sigma^2+\phi^2(\sigma^2-K_{t-1})},\\
b_{t} &= \begin{bmatrix}
\frac{b_{t-1}\phi K_{t}}{\sigma^2} \\ \frac{K_{t}(\sigma^2+\tau^2)-\sigma^4 }{\phi\sigma^2}
\end{bmatrix}. 
\end{align}
With the above formula, the recursive way of updating the mean and covariance are 
\begin{align}
\bar{\mu}_{t}       & = \frac{\phi}{\sigma^2}K_{t-1}\bar{\mu}_{t-1} + \phi (1 - \frac{K_{t-1}}{\sigma^2})y_{t-1}, \\
\bar{\Sigma}_{t}  &= \sigma^4K_{t}^{-1},
\end{align}

where $K_1=\frac{\sigma^4}{\sigma^2+\tau^2+L^2\phi^2}$. For calculation details, readers can refer to appendices (\ref{linearcalculation}).

By using the formula, one term of equation (\ref{beforeSMformula}) becomes 
\begin{equation}
A_{t}^{-1}C_{t} = \left( I - \frac{M_{t}^{-1}u_{t}u_{t}^\top }{1+u_{t}^\top M_{t}^{-1} u_{t}} \right)M_{t}^{-1}C_{t},
\end{equation}
in which
\begin{align*}
M_{t}^{-1}C_{t}    &=\left[ \begin{array}{cc} A_{t-1}^{-1} & 0 \\ 0 & \sigma^2 \end{array} \right]C_{t}=\sigma^2 C_{t},\\
u_{t}^\top C_{t} & = \left[ \begin{array}{ccccc} 0 & \cdots & 0 &\frac{-\phi}{\tau} & \frac{1}{\tau} \end{array} \right] \left[ \begin{array}{c} 0 \\ \vdots \\ 0\\ 1 \end{array} \right]= \frac{1}{\tau}.
\end{align*}
Then the above equation becomes
\begin{equation}
A_{t}^{-1}C_{t} = \sigma^2 C_{t}-\frac{M_{t}^{-1} u_{t} \frac{\sigma^2}{\tau}}{1+u^\top M_{t}^{-1} u}.
\end{equation}
Moreover,
\begin{align*}
M_{t}^{-1} u_{t} &=\left[ \begin{array}{cc} A_{t-1}^{-1} & 0 \\ 0 & \sigma^2 \end{array} \right] \left[ \begin{array}{c}  0 \\ \vdots \\0 \\ -\frac{\phi}{\tau} \\\frac{1}{\tau} \end{array} \right] =
\left[ \begin{array}{cc} A_{t-1}^{-1} & 0 \\ 0 & \sigma^2 \end{array} \right] \left[ \begin{array}{c} -\frac{\phi}{\tau} C_{t-1} \\ \frac{1}{\tau}\end{array} \right] = \left[ \begin{array}{c} -\frac{\phi}{\tau} A_{t-1}^{-1}C_{t-1} \\ \frac{\sigma^2}{\tau}\end{array} \right] ,\\
u^\top M_{t}^{-1}  u &=\left[ \begin{array}{ccccc}  0 & \cdots & 0 & -\frac{\phi}{\tau} & \frac{1}{\tau} \end{array} \right] \left[ \begin{array}{c} -\frac{\phi}{\tau} A_{t-1}^{-1}C_{t-1} \\ \frac{\sigma^2}{\tau} \end{array} \right] =  \left[ \begin{array}{cc} -\frac{\phi}{\tau} C_{t-1}^\top & \frac{1}{\tau}\end{array} \right]  \left[ \begin{array}{c} -\frac{\phi}{\tau} A_{t-1}^{-1}C_{t-1} \\ \frac{\sigma^2}{\tau}\end{array} \right] =\frac{\phi^2}{\tau^2} C_{t-1}^\top A_{t-1}^{-1}C_{t-1}+\frac{\sigma^2}{\tau^2}.
\end{align*}
Thus
\begin{align}
\begin{split}
A_{t}^{-1}C_{t} &= \left[ \begin{array}{c} -b_{t} \\ \sigma^2-K_{t}\end{array} \right] = \sigma^2C_{t}-\frac{1}{1+\frac{\phi^2}{\tau^2} C_{t-1}^\top A_{t-1}^{-1}C_{t-1}+\frac{\sigma^2}{\tau^2}} \left[\begin{array}{c} -\frac{\phi\sigma^2}{\tau^2} A_{t-1}^{-1}C_{t-1} \\\frac{\sigma^4}{\tau^2} \end{array}\right] \\
&= \sigma^2C_{t}-\frac{1}{\tau^2+\phi^2C_{t-1}^\top A_{t-1}^{-1}C_{t-1}+\sigma^2} \left[\begin{array}{c} -\phi\sigma^2 A_{t-1}^{-1}C_{t-1} \\ \sigma^4 \end{array}\right]
\end{split}
\end{align}
and
\begin{equation*}
\sigma^2-K_{t} = \sigma^2 - \frac{\sigma^4}{\tau^2+\phi^2C_{t-1}^\top A_{t-1}^{-1}C_{t-1}+\sigma^2} = \sigma^2 - \frac{\sigma^4}{\tau^2+\sigma^2+\phi^2(\sigma^2-K_{t-1})},
\end{equation*}
therefore
\begin{align}
K_{t}  &=\frac{\sigma^4}{\tau^2+\sigma^2+\phi^2(\sigma^2-K_{t-1})},
\end{align}
and
\begin{align*}b_{t} = 
\begin{bmatrix}
\frac{b_{t-1}\phi K_{t}}{\sigma^2} \\ \frac{K_{t}(\sigma^2+\tau^2)-\sigma^4 }{\phi\sigma^2}
\end{bmatrix},
\end{align*}

\begin{align*}
\bar{\mu}_{t}      &= 0-\sigma^4 K_t^{-1}b_t^\top (Z-b_tK_t^{-1}b_t^\top)^{-1} \sigma^{-4} (Z-b_tK_t^{-1}b_t^\top) y_{1:t-1} \\
					 & =-K_t^{-1}b_t^\top y_{1:t-1} \\
					 & = \frac{\phi}{\sigma^2}K_{t-1}\bar{\mu}_{t-1} + \phi (1 - \frac{K_{t-1}}{\sigma^2})y_{t-1}, \\
\bar{\Sigma}_{t} &= \sigma^4(K_t-b_t^\top Z^{-1}b_t)^{-1}- \sigma^4K_t^{-1}b_t^\top (Z_t-b_tK_t^{-1}b_t^\top)^{-1} (Z_t-b_tK_t^{-1}b_t^\top)Z_t^{-1}b_t(K_t-b_t^\top Z_t^{-1}b_t)^{-1}\\
                     & = \sigma^4(I-K_t^{-1}b_t^\top Z_t^{-1}b_t)(K_t-b_t^\top Z_t^{-1}b_t)^{-1} \\
                     & = \sigma^4K_{t}^{-1},
\end{align*}
where $K_1=\frac{\sigma^4}{\sigma^2+\tau^2+L^2\phi^2}$.

\subsubsection*{Estimation}

As introduced before, $p(x_t \mid y_{1:t})$ is a mixture Gaussian distribution with given $\theta$ and its mean and variance can be found by 
\begin{align}
\mu_x^{(t)} &= \frac{1}{N} \sum_i \mu_i\\
\Var(x)^{(t)}  &= \E(\Var(x\mid y,\theta)) + \Var(\E(x\mid y,\theta)) = \frac{1}{N} \sum_i (\mu_i \mu_i^\top +\Sigma_i) -\frac{1}{N^2} (\sum_i  \mu_i) (\sum_i \mu_i)^\top. 
\end{align}

To find $\mu_i$ and $\Sigma_i$, we will use the joint distribution of $x_{t}$ and $y_{1:t}$, which is $p(x_{t}, y_{1:t}  \mid  \theta)\sim N(0,\Gamma)$ and 
\begin{equation*}
\Gamma=\begin{bmatrix} C_{t}^\top(A-B)^{-1}C_{t} & C_{t}^\top(A-B)^{-1}\\(A-B)^{-1}C_{t} & (I-A^{-1}B)^{-1}B^{-1} \end{bmatrix}.
\end{equation*}
Because of 
\begin{align*}
C_{t}^\top A_{t}^{-1} = \left[\begin{matrix} - b_{t}^\top & \sigma^2- K_{t} \end{matrix} \right],
\end{align*}
thus, for any given $\theta_i$, $x_{t}\mid y_{1:t},\theta_i \sim N(\mu_{t}^{(x)},\sigma_{t}^{(x)2})$, where
\begin{align*}
\mu_i  &= \phi \hat{x}_{t-1} +  C_{t}^\top (A-B)^{-1}B (I-A^{-1}B)y_{1:t}\\
                      &= \phi \hat{x}_{t-1} +  C_{t}^\top A^{-1}B y_{1:t} \\ &= \phi \hat{x}_{t-1} +  \frac{1}{\sigma^2}C_{t}^\top A^{-1} y_{1:t}\\
                      &=0+  \frac{1}{\sigma^2}\left[\begin{matrix} - b_{t}^\top & \sigma^2- K_{t} \end{matrix} \right]  \left[\begin{matrix} y_{1:t-1} \\ y_{t} \end{matrix} \right] \\
                      &= - \frac{1}{\sigma^2}b_{t-1}^\top y_{1:t-1}+(1-\frac{K_{t}}{\sigma^2})y_{t}\\
                      &=\frac{K_{t}\bar{\mu}_{t}}{\sigma^2}+(1-\frac{K_{t}}{\sigma^2})y_{t} \\
\Sigma_i&=C_{t}^\top(A-B)^{-1}C_{t}-  C_{t}^\top(A-B)^{-1}  B(I-A^{-1}B) (A-B)^{-1}C_{t}\\
                      &= C_{t}^\top(A-B)^{-1}C_{t} -  C_{t}^\top A^{-1}B(A-B)^{-1}C_{t}\\
                      &= C_{t}^\top A^{-1}C_{t} \\ &= \sigma^2-K_{t}.
\end{align*}
By substituting them into the equation (\ref{linearmu}) and (\ref{linearsigma}), the estimated $\hat{x}_t$ is easily got.

\subsection{OU process calculation}\label{OUcalculation}

\subsubsection*{Forecast}
We are now using the capital letter $Y$ to represent the joint $\{y,v\}$ and $Y_{1:t} = \{y_1,v_1,y_2,v_2,\cdots, y_t, v_t \}$, $Y_{t+1} = \{y_{t+1}, v_{t+1}\}$. It is known that 
\begin{align*}
p(Y_{1:t},\theta) &\sim N\left( 0,\Sigma_{YY}^{(t)} \right)\\
p(Y_{t+1},Y_{1:t},\theta) &\sim N\left( 0,\Sigma_{YY}^{(t+1)} \right)\\
p(Y_{t+1}\mid Y_{1:t},\theta) &\sim N\left( \bar{\mu}_{t+1},\bar{\Sigma}_{t+1} \right)
\end{align*}
where the covariance matrix of the joint distribution is $\Sigma_{YY}^{(t+1)} = (I_{t+1}-A_{t+1}^{-1}B_{t+1})^{-1}B_{t+1}^{-1}$. Then, by taking its inverse, we will get
\begin{align*}
\Sigma_{YY}^{(t+1) (-1)} = B_{t+1}(I_{t+1}-A_{t+1}^{-1}B_{t+1}).
\end{align*}
To be clear, the matrix $B_t$ is short for the matrix $B_t(\sigma^2,\tau^2)$, which is $2t\times 2t$ diagonal matrix with elements $\frac{1}{\sigma^2},\frac{1}{\tau^2}$ repeating for $t$ times on its diagonal. For instance, the very simple $B_1(\sigma^2,\tau^2) = 
\begin{bmatrix}
\frac{1}{\sigma^2} & 0  \\
0 & \frac{1}{\tau^2}
\end{bmatrix}_{2\times 2}$ is a $2\times 2$ matrix. 

Because of $A$ is symmetric and invertible, $B$ is the diagonal matrix defined as above, then they have the following property 
\begin{align*}
& AB=A^\top B^\top = (BA)^\top, \\
& A^{-1}B = A^{-\top}B^\top = (BA^{-1})^\top. 
\end{align*}
Followed up the form of $\Sigma_{YY}^{(t+1) (-1)}$, we can find out that 
\begin{align*}
\Sigma_{YY}^{(t+1) (-1)} &= B_{t+1}(I_{t+1}-A_{t+1}^{-1}B_{t+1}) \\
&= B_{t+1}(B_{t+1}^{-1}-A_{t+1}^{-1})B_{t+1} \\
&\triangleq \begin{bmatrix} 
B_t & 0 \\ 0 & B_1 \end{bmatrix}
\begin{bmatrix} 
Z_{t+1} & b_{t+1} \\
b_{t+1}^\top & K_{t+1}
\end{bmatrix} \begin{bmatrix} 
B_t & 0 \\ 0 & B_1\end{bmatrix}
\end{align*}
where $Z_{t+1}$ is a $2t \times 2t$ matrix, $ b_{t+1} $ is a $2t \times 2$ matrix and $K_{t+1}$ is a $2 \times 2$ matrix. Thus by taking its inverse again, we will get 
\begin{align*} \Sigma_{YY}^{(t+1)}= \left[ \begin{matrix}
B_t^{-1} (Z_{t+1}-b_{t+1}K_{t+1}^{-1}b_{t+1}^\top)^{-1}B_t^{-1}  & - B_t^{-1}  Z_{t+1}^{-1}b_{t+1}(K_{t+1}-b_{t+1}^\top Z_{t+1}^{-1}b_{t+1})^{-1}B_1^{-1} \\
-B_1^{-1}  K_{t+1}^{-1}b_{t+1}^\top (Z_{t+1}-b_{t+1}K_{t+1}^{-1}b_{t+1}^\top)^{-1}B_t^{-1}  & B_1^{-1}  (K_{t+1}-b_{t+1}^\top Z_{t+1}^{-1}b_{t+1})^{-1}B_1^{-1} 
\end{matrix}\right].
\end{align*}

It is easy to find the relationship between $A_{t+1}$ and  $A_{t}$ in the Sherman-Morrison-Woodbury form, which is 
\begin{align*} A_{t+1} = 
\begin{bmatrix}
A_t & \cdot & \cdot  \\ \cdot &\frac{1}{\sigma^2} &\cdot  \\ \cdot  & \cdot  & \frac{1}{\tau^2} 
\end{bmatrix} + U_{t+1}U_{t+1}^\top \triangleq M_{t+1}  + U_{t+1}U_{t+1}^\top,
\end{align*}
where $M_{t+1} = \begin{bmatrix}
A_t & \cdot & \cdot  \\ \cdot &\frac{1}{\sigma^2} &\cdot  \\ \cdot  & \cdot  & \frac{1}{\tau^2}
\end{bmatrix}  = \begin{bmatrix}
A_t & 0 \\ 0 & B_1
\end{bmatrix}$ 
and its inverse is $M_{t+1}^{-1} =\begin{bmatrix}
A_t^{-1} & 0 \\ 0 & B_1^{-1}
\end{bmatrix}$. Additionallly, $U$ is a $2t+2 \times 2$ matrix in the following form 
\begin{align*}
U_{t+1} = \frac{1}{\sqrt{ 1-\rho_{t+1}^2} } \begin{bmatrix}
\mathbf{0}_{2t-2} & \mathbf{0}_{2t-2}  \\ \frac{1}{\sigma_{t+1}^{(x)}}& 0 \\
\frac{1-e^{-\gamma \Delta_{t+1}}}{\gamma \sigma_{t+1}^{(x)}}-\frac{\rho_{t+1} e^{-\gamma\Delta_{t+1}}}{\sigma_{t+1}^{(u)}} & \frac{\sqrt{1-\rho_{t+1}^2}e^{-\gamma\Delta_{t+1}}}{\sigma_{t+1}^{(u)}} \\
-\frac{1}{\sigma_{t+1}^{(x)}} & 0 \\
\frac{\rho_{t+1}}{\sigma_{t+1}^{(u)}} & -\frac{\sqrt{1-\rho_{t+1}^2}}{\sigma_{t+1}^{(u)}}
\end{bmatrix} \triangleq  \begin{bmatrix}
C_t S_{t+1} \\ D_{t+1}
\end{bmatrix},
\end{align*}
denoted by $S_{t+1} = \frac{1}{\sqrt{ 1-\rho_{t+1}^2} } \begin{bmatrix}
\frac{1}{\sigma_{t+1}^{(x)}}& 0 \\
\frac{1-e^{-\gamma \Delta_{t+1}}}{\gamma \sigma_{t+1}^{(x)}}-\frac{\rho_{t+1} e^{-\gamma\Delta_{t+1}}}{\sigma_{t+1}^{(u)}} & \frac{\sqrt{1-\rho_{t+1}^2}e^{-\gamma\Delta_{t+1}}}{\sigma_{t+1}^{(u)}}
\end{bmatrix}$, $D_{t+1} =  \frac{1}{\sqrt{ 1-\rho_{t+1}^2} }\begin{bmatrix}
-\frac{1}{\sigma_{t+1}^{(x)}} & 0 \\
\frac{\rho_{t+1}}{\sigma_{t+1}^{(u)}} & -\frac{\sqrt{1-\rho_{t+1}^2}}{\sigma_{t+1}^{(u)}}
\end{bmatrix}$ and $C_{t+1} = \begin{bmatrix} 0 & 0 \\ \vdots & \vdots \\ 0 & 0 \\ 1 & 0 \\ 0 & 1 \end{bmatrix} = \begin{bmatrix}
\mathbf{0}_t \\ I_{2} \end{bmatrix}$.

By post-multiplying $\Sigma_{YY}^{(t+1)(-1)}$ with $C_{t+1}$, it gives us
\begin{align*}
\Sigma_{YY}^{(t+1)(-1)} C_{t+1} &=  B_{t+1} (I_{t+1} -A_{t+1} ^{-1}B_{t+1} ) C_{t+1}  \\ 
&= \begin{bmatrix} B_t & 0 \\ 0 & B_1 \end{bmatrix} \left( \begin{bmatrix} B_t^{-1} & 0 \\ 0 & B_1^{-1} \end{bmatrix}  -A_{t+1}^{-1} \right) \begin{bmatrix} B_t & 0 \\ 0 & B_1 \end{bmatrix}   C_{t+1}\\
&= \begin{bmatrix} B_t & 0 \\ 0 & B_1 \end{bmatrix}\begin{bmatrix} Z_{t+1} & b_{t+1} \\ b_{t+1}^\top  & K_{t+1} \end{bmatrix}  \begin{bmatrix} B_t & 0 \\ 0 & B_1 \end{bmatrix}   C_{t+1}\\
& = \begin{bmatrix} B_t & 0 \\ 0 & B_1 \end{bmatrix}\begin{bmatrix} b_{t+1}B_1\\ K_{t+1}B_1 \end{bmatrix}.
\end{align*}
and the property of $A_{t+1}^{-1}$ is 
\begin{equation*}
A_{t+1}^{-1}C_{t+1} = \begin{bmatrix}
-b_{t+1} \\ B_1^{-1} - K_{t+1}
\end{bmatrix}.
\end{equation*}
Moreover, by pre-multiplying $C_{t+1}^\top$ on the left side of the above equation, we will have 
\begin{align}
C_{t+1}^\top A_{t+1}^{-1}C_{t+1}  &= B_1^{-1} -K_{t+1},\\
K_{t+1} &= B_1^{-1} - C_{t+1}^\top A_{t+1}^{-1}C_{t+1}.\label{OUKtp1}
\end{align}

We may use Sherman-Morrison-Woodbury formula to find the inverse of $A_{t+1}$ in a recursive way, which is 
\begin{equation}
A_{t+1}^{-1} = (M_{t+1}+U_{t+1}U_{t+1}^\top)^{-1}= M_{t+1}^{-1}-M_{t+1}^{-1}U_{t+1}(I+U_{t+1}^\top M_{t+1}^{-1}U_{t+1})^{-1}U_{t+1}^\top M_{t+1}^{-1}.
\end{equation}
Consequently, it is easy to find that $M_{t+1}^{-1}C_{t+1} =\begin{bmatrix} 0 \\ B_1^{-1} \end{bmatrix} $ and 
\begin{align*}
A_{t+1}^{-1}C_{t+1} &= \begin{bmatrix} 0 \\ B_1^{-1} \end{bmatrix}  - \begin{bmatrix}
A_t^{-1} & 0 \\ 0 & B_1^{-1} \end{bmatrix} \begin{bmatrix} C_tS_{t+1} \\ D \end{bmatrix}
(I+U_{t+1}^\top M_{t+1}^{-1}U_{t+1})^{-1}  \begin{bmatrix}
S_{t+1}^\top C_t^\top & D_{t+1}^\top \end{bmatrix} \begin{bmatrix} 0 \\ B_1^{-1} \end{bmatrix}  \\ 
& = \begin{bmatrix} 0 \\ B_1^{-1} \end{bmatrix}  - \begin{bmatrix}
A_t^{-1} C_tS_{t+1} \\B_1^{-1}D_{t+1} \end{bmatrix} 
(I+U_{t+1}^\top M_{t+1}^{-1}U_{t+1})^{-1}  D_{t+1}^\top B_1^{-1} \\ 
& = \begin{bmatrix} 0 \\ B_1^{-1} \end{bmatrix}  - \begin{bmatrix}
A_t^{-1} C_tS_{t+1} \\B_1^{-1}D_{t+1} \end{bmatrix} 
(I+ S_{t+1}^\top C_t^\top A_t^{-1} C_t S_{t+1} +D_{t+1}^\top B_1^{-1}D_{t+1}  )^{-1}  D_{t+1}^\top B_1^{-1} \\ 
& = \begin{bmatrix} 0 \\ B_1^{-1} \end{bmatrix}  - \begin{bmatrix}
A_t^{-1} C_tS_{t+1} \\B_1^{-1}D_{t+1} \end{bmatrix} 
(I+ S_{t+1}^\top (B_1^{-1} - K_t)  S_{t+1} +D_{t+1}^\top B_1^{-1}D_{t+1}  )^{-1}  D_{t+1}^\top B_1^{-1}.
\end{align*}
Thus, by using the equation (\ref{OUKtp1}), we will get 
\begin{equation}\label{recursiveKp1}
K_{t+1} =B_1^{-1}D_{t+1} (I+ S_{t+1}^\top (B_1^{-1} - K_t)  S_{t+1} +D_{t+1}^\top B_1^{-1}D_{t+1}  )^{-1}  D_{t+1}^\top B_1^{-1},
\end{equation}
and
\begin{align*}
b_{t+1} &= A_t^{-1}C_t S_{t+1} (I+ S_{t+1}^\top (B_1^{-1} - K_t)  S_{t+1} +D_{t+1}^\top B_1^{-1}D_{t+1}  )^{-1} D_{t+1}^\top B_1^{-1} \\
&= \begin{bmatrix}
-b_t \\ B_1^{-1}-K_t 
\end{bmatrix}  S_{t+1} (I+ S_{t+1}^\top (B_1^{-1} - K_t)  S_{t+1} +D_{t+1}^\top B_1^{-1}D_{t+1}  )^{-1} D_{t+1}^\top B_1^{-1}.
\end{align*}
To achieve the recursive updating formula, firstly we need to find the form of $b_{t+1}^\top B_t^2 Y_{1:t}$. In fact, it is 
\begin{align*}
b_{t+1}^\top B_t Y_{1:t} &= B_1^{-\top}D_{t+1}  (I+ S_{t+1}^\top  (B_1^{-1} - K_t)  S_{t+1} +D_{t+1}^\top B_1^{-1}D_{t+1}  )^{-\top} S_{t+1}^\top 
\begin{bmatrix}
-b_t^\top  & B_1^{-1}-K_t 
\end{bmatrix} B_t \begin{bmatrix}
Y_{1:t-1} \\ Y_t
\end{bmatrix}\\
&= B_1^{-\top}D_{t+1}  (I+ S_{t+1}^\top  (B_1^{-1} - K_t)  S_{t+1} +D_{t+1}^\top B_1^{-1}D_{t+1}  )^{-\top} S_{t+1}^\top 
\left(  -b_t^\top  B_{t-1}  Y_{1:t-1} + (B_1^{-1}-K_t )  B_1  Y_t      \right) \\ 
&= B_1^{-\top}D_{t+1}  (I+ S_{t+1}^\top  (B_1^{-1} - K_t)  S_{t+1} +D_{t+1}^\top B_1^{-1}D_{t+1}  )^{-\top} S_{t+1}^\top 
\left(  K_t B_1\bar{\mu}_t+ (I-K_t B_1)  Y_t      \right), \\
\end{align*}
By using equation (\ref{recursiveKp1}) and simplifying the above equation, one can achieve a recursive updating form of the mean, which is 
\begin{align*}
\bar{\mu}_{t+1} &= -B_1K_{t+1}^{-1}b_{t+1}^\top B_t Y_{1:t} \\
&= -D_{t+1}^{-\top}S_{t+1}^\top (K_tB_1\bar{\mu}_t + (I-K_tB_1)Y_t) \\ 
&= -D_{t+1}^{-\top}S_{t+1}^\top ( Y_t +   K_tB_1(\bar{\mu}_t-Y_t)),
\end{align*}
where by simplifying $D^{-\top}S^\top$, one may find 
\begin{align*}
D_{t+1}^{-\top}S_{t+1}^\top = \begin{bmatrix}
-1 & -\frac{1-e^{-\gamma \Delta_{t+1}}}{\gamma} \\ 0 & - e^{-\gamma \Delta_{t+1}}
\end{bmatrix} = -\Phi_{t+1},
\end{align*}
which is the negative of forward process. Then the final form of recursive updating formula are 
\begin{align}
\begin{cases}
\bar{\mu}_{t+1}&=\Phi_{t+1} K_tB_1\bar{\mu}_t + \Phi_{t+1} (I-K_tB_1)Y_t\\
\bar{\Sigma}_{t+1}&=\left( B_1K_{t+1}B_1  \right)^{-1}
\end{cases}.
\end{align}
The matrix $K_{t+1}$ is updated via 
\begin{equation}
K_{t+1} =B_1^{-1}D_{t+1} (I+ S_{t+1}^\top (B_1^{-1} - K_t)  S_{t+1} +D_{t+1}^\top B_1^{-1}D_{t+1}  )^{-1}  D_{t+1}^\top B_1^{-1},
\end{equation}
or updating its inverse in the following form makes the computation faster, that is 
\begin{align*}
\begin{cases}
K_{t+1}^{-1} &= B_1D_{t+1}^{-\top}D_{t+1}^{-1}B_1 + B_1\Phi_{t+1} (B_1^{-1} - K_t) \Phi_{t+1}^\top B_1+ B_1,\\
\bar{\Sigma}_{t+1}&= D_{t+1}^{-\top}D_{t+1}^{-1}+ \Phi_{t+1} (B_1^{-1} - K_t) \Phi_{t+1}^\top + B_1^{-1}
\end{cases}
\end{align*}
and $K_1 =B_1^{-1} - A_1^{-1} = \begin{bmatrix}
\frac{\sigma^4}{\sigma^2 +L_x^2} & 0 \\ 0 &\frac{\tau^4}{\tau^2 +L_u^2}
\end{bmatrix} $.

\subsubsection*{Estimation}

Because of the joint distribution (\ref{jointmatrix}), one can find the best estimation with a given $\theta$ by
\begin{align*}
X \mid Y,\theta &\sim N \left( A^{-1}BY, A^{-1} \right) \\
&\sim N(L^{-\top}L^{-1}BY,L^{-\top}L^{-1})\\
&\sim N(L^{-\top}W,L^{-\top}L^{-1}),
\end{align*}
thus
\begin{align*}
\hat{X} = L^{-\top}(W+Z),
\end{align*}
where $Z \sim N(0, I(\sigma,\tau))$.

For $X_{t+1}$, the joint distribution with $Y$ updated to stage $t+1$ is 
\begin{align*}
X_{t+1}, Y\mid \theta \sim N\left( 0, \begin{bmatrix}
C_{t+1}^\top(A-B) ^{-1}C_{t+1} & C_{t+1}^\top (A-B)^{-1}\\
(A-B)^{-1}C_{t+1} & (I- A^{-1}B) ^{-1}B^{-1}
\end{bmatrix} \right),
\end{align*}
where $C_{t+1}^\top = \begin{bmatrix}
0 & \cdots & 0 & 1 & 0 \\
0 & \cdots & 0 & 0 & 1
\end{bmatrix}$ is a $2 \times 2(t+1)$ matriX. Thus
\begin{align*}
X_{t+1}\mid Y,\theta \sim N(\bar{\mu}_{t+1}^{(X)},\bar{\Sigma}_{t+1}^{(X)}),
\end{align*}
where
\begin{align*}
\bar{\mu}_{t+1}^{(X)} & = C_{t+1}^\top A^{-1}BY =C_{t+1}^\top L^{-\top}W,\\
\bar{\Sigma}_{t+1}^{(X)} & =C_{t+1}^\top A^{-1}C_{t+1} =U_{t+1}^\top U_{t+1},
\end{align*}
and $U_{t+1} = L^{-1} C_{t+1} = \mbox{solve}(L,C_{t+1})$.

The filtering distribution of the state given parameters is $p(X_t\mid Y_{1:t}, \theta )$. To find its form, one can use the joint distribution of $X_{t+1}$ and $Y_{1:t+1}$, which is $p(X_{t+1}, Y_{1:t+1}  \mid  \theta)\sim N(0,\Gamma)$, where
\begin{equation*}
\Gamma=\begin{bmatrix} C_{t+1}^\top(A-B)^{-1}C_{t+1} & C_{t+1}^\top(A-B)^{-1}\\(A-B)^{-1}C_{t+1} & (I-A^{-1}B)^{-1}B^{-1} \end{bmatrix}.
\end{equation*}
Because of 
\begin{align*}
C_{t+1}^\top A_{t+1}^{-1} = \left[\begin{matrix} - b_{t+1}^\top & B_1^{-1}- K_{t+1} \end{matrix} \right],
\end{align*}
then $X_{t+1}\mid Y_{1:t+1},\theta \sim N(\bar{\mu}_{t+1}^{(X)},\bar{\sigma}_{t+1}^{(X)2})$, where
\begin{align*}
\bar{\mu}_{t+1}^{(X)}       &= \Phi \hat{x}_{t} +  C_{t+1}^\top (A-B)^{-1}B (I-A^{-1}B)Y_{1:t+1}\\
                      &= \Phi \hat{x}_{t} +  C_{t+1}^\top A^{-1}B Y_{1:t+1} \\ 
                      &=0+ \begin{bmatrix} - b_{t+1}^\top & B_1^{-1}-K_{t+1} \end{bmatrix}\begin{bmatrix} B_t & 0 \\ 0 & B_1 \end{bmatrix} \begin{bmatrix} Y_{1:t} \\ Y_{y+1} \end{bmatrix} \\
                      &=   -b^\top B_t Y_{1:t} + (I - B_1K_{t+1})Y_{t+1} \\
                      & =  K_{t+1}B_1\bar{\mu}_{t+1 } + (I - B_1K_{t+1})Y_{t+1}  \\
\bar{\sigma}_{t+1}^{(X)2}&=C_{t+1}^\top(A-B)^{-1}C_{t+1}-  C_{t+1}^\top(A-B)^{-1}  B(I-A^{-1}B) (A-B)^{-1}C_{t+1}\\
                      &= C_{t+1}^\top(A-B)^{-1}C_{t+1} -  C_{t+1}^\top A^{-1}B(A-B)^{-1}C_{t+1}\\
                      &= C_{t+1}^\top A^{-1}C_{t+1} \\ &=B_1^{-1}-K_{t+1}.
\end{align*}

\begin{landscape}
\subsubsection*{Covariance Matrix in Details}\label{covMatrixdetails}
\begin{align*}
\Sigma_t &= \begin{bmatrix}
\sigma_t^{(x)2} & \rho_{t} \sigma_t^{(x)} \sigma_t^{(u)}\\
 \rho_{t} \sigma_t^{(x)} \sigma_t^{(u)} & \sigma_t^{(u)2} 
\end{bmatrix}\\
\Sigma_t^{-1} &= \frac{1}{1-\rho_{t}^2} \begin{bmatrix} \frac{1}{\sigma_t^{(x)2}} & -\frac{\rho_{t}}{\sigma_t^{(x)} \sigma_t^{(u)} }\\
-\frac{\rho_{t}}{\sigma_t^{(x)} \sigma_t^{(u)} } & \frac{1}{\sigma_t^{(u)2}} \end{bmatrix}\\
M_t^\top &= \begin{bmatrix}
1 & 0 \\
\frac{1-e^{-\gamma \Delta_t}}{\gamma} & e^{-\gamma \Delta_t}
\end{bmatrix} \\
z_t &= \begin{bmatrix} x_t \\ u_t \end{bmatrix}
\end{align*}
\begin{align*}
M_t^\top \Sigma_t^{-1} &= \frac{1}{1-\rho_{t}^2} 
\begin{bmatrix}
\frac{1}{\sigma_t^{(x)2}}  & -\frac{\rho_{t}}{\sigma_t^{(x)} \sigma_t^{(u)} }\\
\frac{1-e^{-\gamma \Delta_t}}{\gamma\sigma_t^{(x)2}} -\frac{\rho_{t} e^{-\gamma \Delta_t}}{\sigma_t^{(x)} \sigma_t^{(u)} }  & 
 -\frac{\rho_{t} (1-e^{-\gamma \Delta_t}) }{ \gamma \sigma_t^{(x)} \sigma_t^{(u)} } + \frac{  e^{-\gamma\Delta_t} }{   \sigma_t^{(u)2} }
\end{bmatrix} \\
M_t^\top \Sigma_t^{-1} M_t&= \frac{1}{1-\rho_{t}^2} 
\begin{bmatrix}
\frac{1}{\sigma_t^{(x)2}}  &  \frac{1-e^{-\gamma\Delta_t}}{\gamma\sigma_t^{(x)2}} -\frac{\rho_{t} e^{-\gamma\Delta_t}}{\sigma_t^{(x)}\sigma_t^{(u)}} \\
\frac{1-e^{-\gamma \Delta_t}}{\gamma\sigma_t^{(x)2}} -\frac{\rho_{t} e^{-\gamma \Delta_t}}{\sigma_t^{(x)} \sigma_t^{(u)} }  & 
\frac{ (1-e^{-\gamma \Delta_t})^2}{\gamma^2\sigma_t^{(x)2}} - \frac{2 \rho_{t} e^{-\gamma \Delta_t} (1-e^{-\gamma\Delta_t}) }{\gamma\sigma_t^{(x)} \sigma_t^{(u)} } +\frac{e^{-2\gamma\Delta_t}}{\sigma_t^{(u)2}}
\end{bmatrix} 
\end{align*}

\begin{align*}
\Sigma_{4}^{(X)-1}&=\frac{1}{1-\rho_{t}^2}\begin{bmatrix}
 \frac{1-\rho_{t}^2}{\sigma_1^{(x)2}}  + \frac{1}{\sigma_2^{(x)2}}  &  - \frac{1}{\sigma_2^{(x)2}}  &   0 &   0  &   \frac{1-e^{- \gamma\Delta_2}}{\gamma\sigma_2^{(x)2}} -\frac{\rho_{t} e^{-\gamma\Delta_2}}{\sigma_2^{(x)}\sigma_2^{(u)}} & \frac{\rho_{t}}{\sigma_2^{(x)}\sigma_2^{(u)}} & 0 & 0 \\
 - \frac{1}{\sigma_2^{(x)2}}    &  \frac{1}{\sigma_2^{(x)2}}  + \frac{1}{\sigma_3^{(x)2}}  &  - \frac{1}{\sigma_3^{(x)2}}   & 0 & -\frac{1-e^{- \gamma\Delta_2}}{\gamma\sigma_2^{(x)2}} +\frac{\rho_{t} e^{-\gamma\Delta_2}}{\sigma_2^{(x)}\sigma_2^{(u)}} & \frac{1-e^{- \gamma\Delta_3}}{\gamma\sigma_3^{(x)2}} -\frac{\rho_{t} e^{-\gamma\Delta_3}}{\sigma_3^{(x)}\sigma_3^{(u)}} -\frac{\rho_{t}}{\sigma_2^{(x)}\sigma_2^{(u)}}    & \frac{\rho_{t}}{\sigma_3^{(x)}\sigma_3^{(u)}}  & 0 \\
 0 & - \frac{1}{\sigma_3^{(x)2}}   & \frac{1}{\sigma_3^{(x)2}}  + \frac{1}{\sigma_4^{(x)2}}   &  - \frac{1}{\sigma_4^{(x)2}}   & 0 & -\frac{1-e^{- \gamma\Delta_3}}{\gamma\sigma_3^{(x)2}} +\frac{\rho_{t} e^{-\gamma\Delta_3}}{\sigma_3^{(x)}\sigma_3^{(u)}} & \frac{1-e^{- \gamma\Delta_4}}{\gamma\sigma_4^{(x)2}} -\frac{\rho_{t} e^{-\gamma\Delta_4}}{\sigma_4^{(x)}\sigma_4^{(u)}} -\frac{\rho_{t}}{\sigma_3^{(x)}\sigma_3^{(u)}}  & \frac{\rho_{t}}{\sigma_4^{(x)}\sigma_4^{(u)}}  \\
 0 & 0 & - \frac{1}{\sigma_4^{(x)2}}   &  \frac{1}{\sigma_4^{(x)2}}   & 0 & 0 & -\frac{1-e^{- \gamma\Delta_4}}{\gamma\sigma_4^{(x)2}} +\frac{\rho_{t} e^{-\gamma\Delta_4}}{\sigma_4^{(x)}\sigma_4^{(u)}}& -\frac{\rho_{t}}{\sigma_4^{(x)}\sigma_4^{(u)}}  \\
\top &  \top & 0 & 0 & \frac{1-\rho_{t}^2}{\sigma_1^{(u)2}}+C_2 & -\frac{e^{-\gamma\Delta_2}}{\sigma_2^{(u)2}}+\frac{\rho_{t}(1-e^{-\gamma\Delta_2})}{\gamma\sigma_2^{(x)}\sigma_2^{(u)}} & 0 & 0 \\
  \top &  \top & \top& 0 & -\frac{e^{-\gamma\Delta_2}}{\sigma_2^{(u)2}}+\frac{\rho_{t}(1-e^{-\gamma\Delta_2})}{\gamma\sigma_2^{(x)}\sigma_2^{(u)}}   & \frac{1}{\sigma_2^{(u)2}}+C_3 & -\frac{e^{-\gamma\Delta_3}}{\sigma_3^{(u)2}}+\frac{\rho_{t}(1-e^{-\gamma\Delta_3})}{\gamma\sigma_3^{(x)2}\sigma_3^{(u)}} & 0 \\
  0 &\top&\top & \top& 0 & -\frac{e^{-\gamma\Delta_3}}{\sigma_3^{(u)2}}+\frac{\rho_{t}(1-e^{-\gamma\Delta_3})}{\gamma\sigma_3^{(x)}\sigma_3^{(u)}}  & \frac{1}{\sigma_3^{(u)2}}+C_4 & -\frac{e^{-\gamma\Delta_4}}{\sigma_4^{(u)2}}+\frac{\rho_{t}(1-e^{-\gamma\Delta_4})}{\gamma\sigma_4^{(x)}\sigma_4^{(u)}}  \\
  0 & 0 & \top&\top& 0 & 0 & -\frac{e^{-\gamma\Delta_4}}{\sigma_4^{(u)2}}+\frac{\rho_{t}(1-e^{-\gamma\Delta_4})}{\gamma\sigma_4^{(x)}\sigma_4^{(u)}}  & \frac{1}{\sigma_4^{(u)2}}
\end{bmatrix}
\end{align*}
where $C_t = \frac{ (1-e^{-\gamma\Delta_t})^2 }{\gamma^2\sigma_t^{(x)2}}  + \frac{ e^{-2\gamma\Delta_t}}{\sigma_t^{(u)2}} - \frac{2\rho_{t} e^{-\gamma\Delta_t} (1-e^{-\gamma\Delta_t}) }{ \gamma \sigma_t^{(x)}\sigma_t^{(u)}}$, $\Delta_t = T_t-T_{t-1}$.
\end{landscape}

\subsection{Real Data Implementation}

\subsubsection*{Efficiency Plots}


\subsubsection*{Comparing Estimation with Different Length of Data }

\begin{figure}[h]
\centering
\includegraphics[width=\textwidth,height=0.5\textheight]{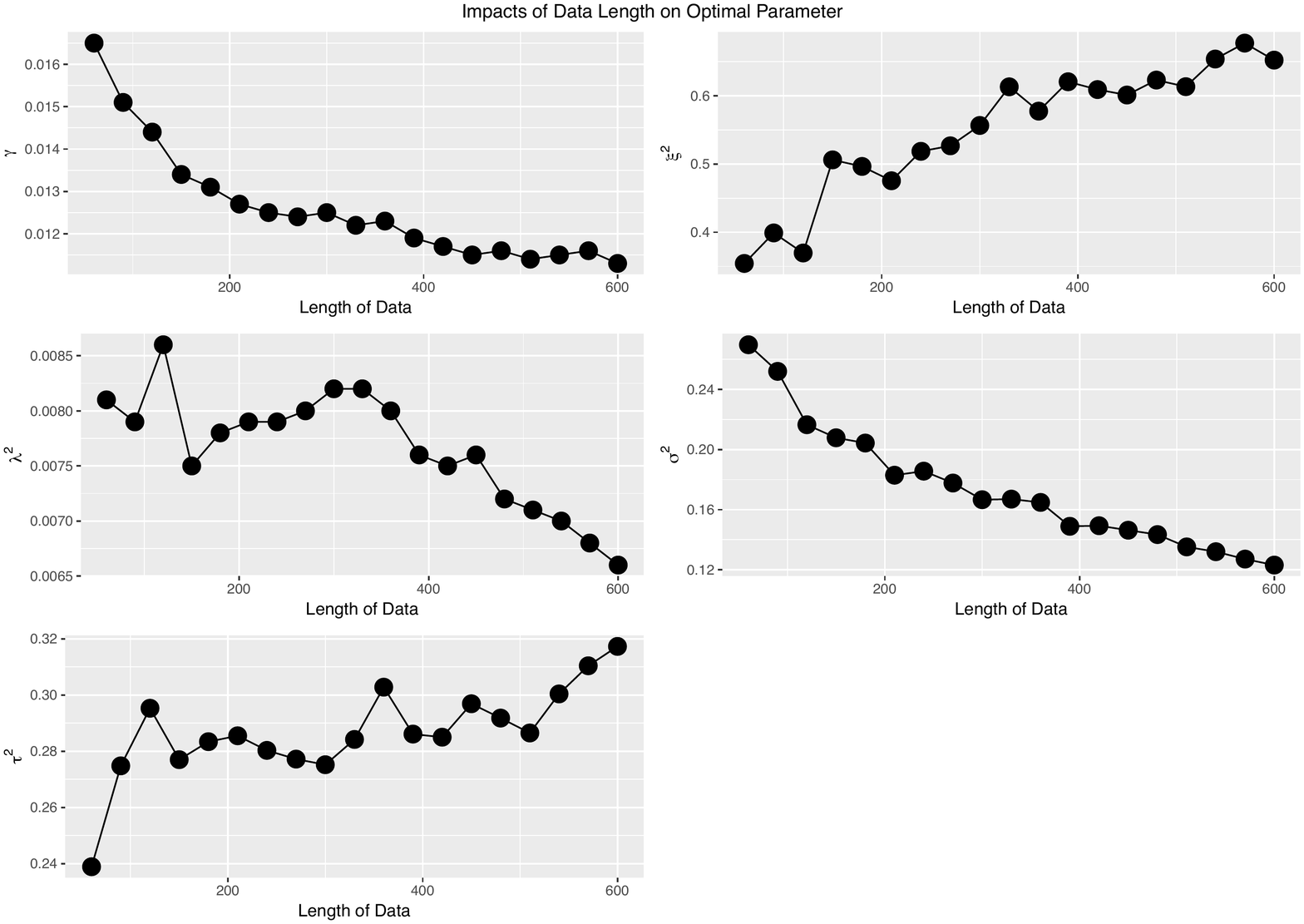}
\caption{Impacts of data length on optimal parameter. }
\end{figure}

\begin{landscape}

\begin{table}[h]
\centering
\caption{Parameter estimation by running the whole surface learning and DA-MH processes with different length of data}
\begin{tabular}{|c|c|c|c|c|c|c|c|c|c|c|c|c|}
\hline
\textbf{Length} & \textbf{Time} & $\gamma$ & $\xi^2$ & $\lambda^2$ & $\sigma^2$ & $\tau^2$ & $\alpha_1$ & $\alpha_2$ &$x$ & $v_x$ &$y$&$v_y$\\ \hline
\textbf{Obs}    & -             & -              & -            & -              & -               & -             & -           & -           & -339.0569  & 0.0413      & -100.2065  & 1.1825      \\ \hline
\textbf{600}    & 85.96         & 0.0113         & 0.6521       & 0.0066         & 0.1231          & 0.3173        & 0.0536      & 0.7873      & -339.0868  & 0.4331      & -100.1498  & -0.7498     \\ \hline
\textbf{570}    & 85.72         & 0.0116         & 0.6770       & 0.0068         & 0.1271          & 0.3104        & 0.0542      & 0.7638      & -339.0872  & 0.4292      & -100.1476  & -0.7356     \\ \hline
\textbf{540}    & 84.25         & 0.0115         & 0.6537       & 0.0070         & 0.1320          & 0.3004        & 0.0662      & 0.7553      & -339.0889  & 0.4326      & -100.1435  & -0.7375     \\ \hline
\textbf{510}    & 85.13         & 0.0114         & 0.6132       & 0.0071         & 0.1352          & 0.2865        & 0.0684      & 0.7310      & -339.0907  & 0.4376      & -100.1387  & -0.7425     \\ \hline
\textbf{480}    & 81.23         & 0.0116         & 0.6229       & 0.0072         & 0.1434          & 0.2918        & 0.0534      & 0.8127      & -339.0921  & 0.4368      & -100.1359  & -0.7408     \\ \hline
\textbf{450}    & 81.57         & 0.0115         & 0.6010       & 0.0076         & 0.1463          & 0.2969        & 0.0580      & 0.7931      & -339.0924  & 0.4432      & -100.1348  & -0.7521     \\ \hline
\textbf{420}    & 80.31         & 0.0117         & 0.6090       & 0.0075         & 0.1493          & 0.2850        & 0.0626      & 0.7636      & -339.0938  & 0.4392      & -100.1310  & -0.7397     \\ \hline
\textbf{390}    & 78.84         & 0.0119         & 0.6204       & 0.0076         & 0.1489          & 0.2861        & 0.0620      & 0.7581      & -339.0931  & 0.4373      & -100.1320  & -0.7354     \\ \hline
\textbf{360}    & 76.66         & 0.0123         & 0.5774       & 0.0080         & 0.1648          & 0.3028        & 0.0554      & 0.7762      & -339.0971  & 0.4457      & -100.1248  & -0.7563     \\ \hline
\textbf{330}    & 76.38         & 0.0122         & 0.6130       & 0.0082         & 0.1670          & 0.2842        & 0.0636      & 0.7830      & -339.0969  & 0.4403      & -100.1220  & -0.7336     \\ \hline
\textbf{300}    & 73.27         & 0.0125         & 0.5564       & 0.0082         & 0.1666          & 0.2752        & 0.0548      & 0.8212      & -339.0989  & 0.4457      & -100.1174  & -0.7443     \\ \hline
\textbf{270}    & 73.68         & 0.0124         & 0.5266       & 0.0080         & 0.1777          & 0.2772        & 0.0636      & 0.6698      & -339.1027  & 0.4489      & -100.1104  & -0.7546     \\ \hline
\textbf{240}    & 71.85         & 0.0125         & 0.5185       & 0.0079         & 0.1856          & 0.2803        & 0.0548      & 0.7336      & -339.1050  & 0.4495      & -100.1067  & -0.7590     \\ \hline
\textbf{210}    & 71.26         & 0.0127         & 0.4754       & 0.0079         & 0.1829          & 0.2855        & 0.0656      & 0.7561      & -339.1057  & 0.4559      & -100.1065  & -0.7754     \\ \hline
\textbf{180}    & 70.25         & 0.0131         & 0.4964       & 0.0078         & 0.2043          & 0.2834        & 0.0566      & 0.7880      & -339.1107  & 0.4498      & -100.0955  & -0.7620     \\ \hline
\textbf{150}    & 70.87         & 0.0134         & 0.5060       & 0.0075         & 0.2078          & 0.2770        & 0.0582      & 0.7801      & -339.1129  & 0.4436      & -100.0916  & -0.7507     \\ \hline
\textbf{120}    & 68.38         & 0.0144         & 0.3696       & 0.0086         & 0.2165          & 0.2953        & 0.0570      & 0.7754      & -339.1168  & 0.4705      & -100.0825  & -0.8057     \\ \hline
\textbf{90}     & 65.73         & 0.0151         & 0.3990       & 0.0079         & 0.2520          & 0.2748        & 0.0552      & 0.8188      & -339.1296  & 0.4550      & -100.0556  & -0.7740     \\ \hline
\textbf{60}     & 68.81         & 0.0165         & 0.3543       & 0.0081         & 0.2697          & 0.2389        & 0.0694      & 0.7176      & -339.1412  & 0.4527      & -100.0204  & -0.7573     \\ \hline
\end{tabular}
\end{table}\label{lengthofdatacompare}
\end{landscape}

\subsubsection{Comparison Between Batch and Sliding Window Methods}

\subsubsection{Parameter Evolution Visualization}

\begin{figure}[h]
\centering
\includegraphics[width=0.3\textwidth,height=0.18\textheight]{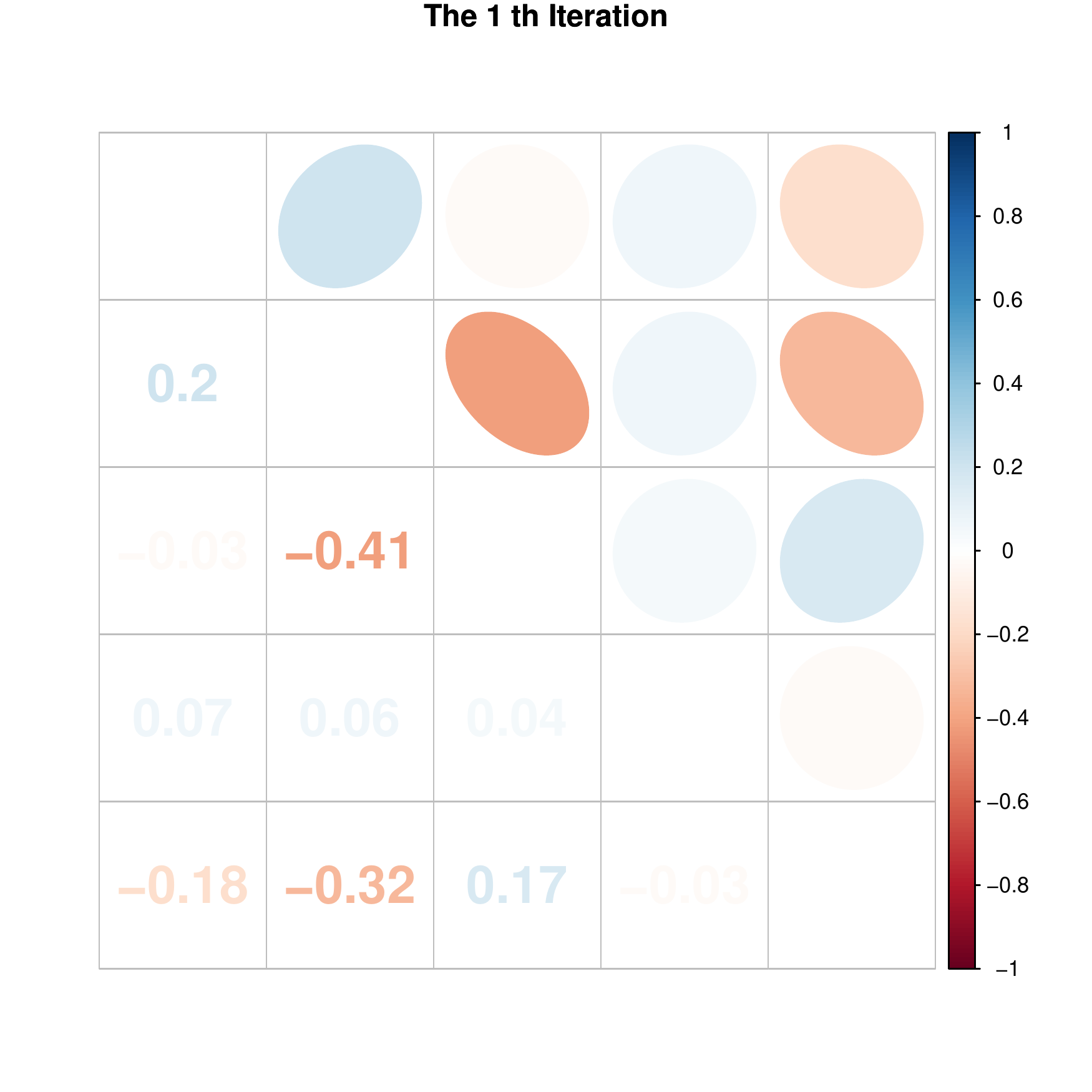}
\includegraphics[width=0.3\textwidth,height=0.18\textheight]{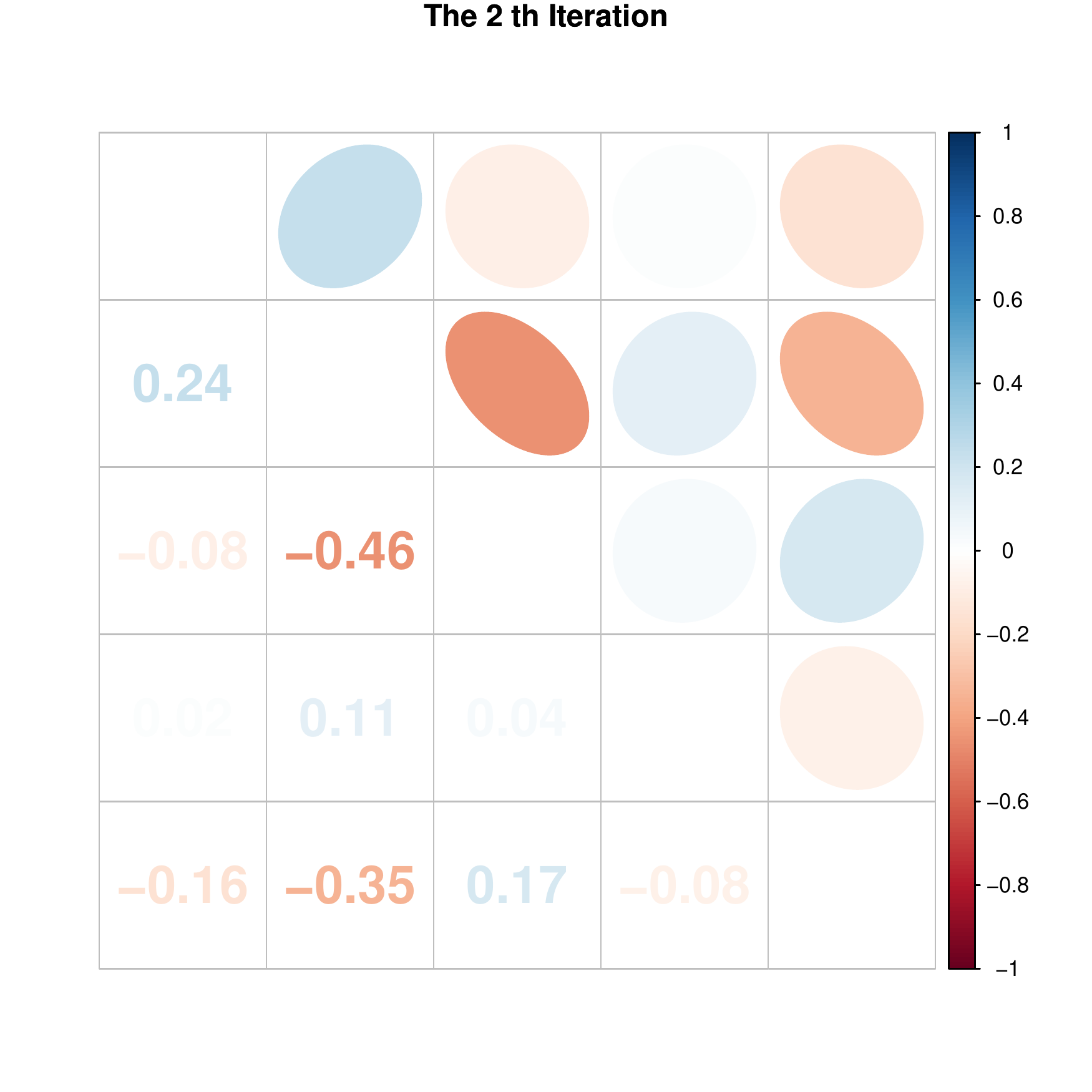}
\includegraphics[width=0.3\textwidth,height=0.18\textheight]{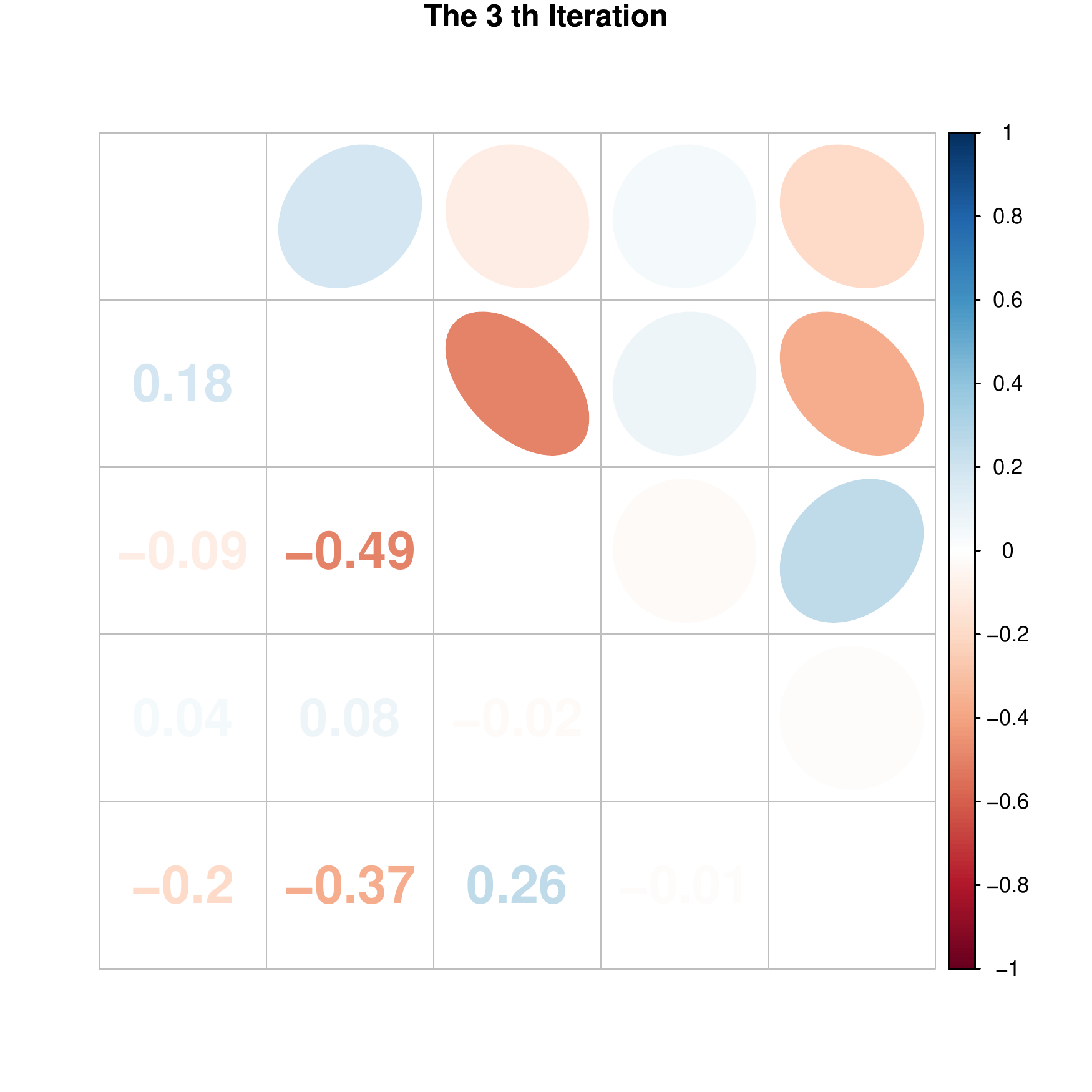}
\includegraphics[width=0.3\textwidth,height=0.18\textheight]{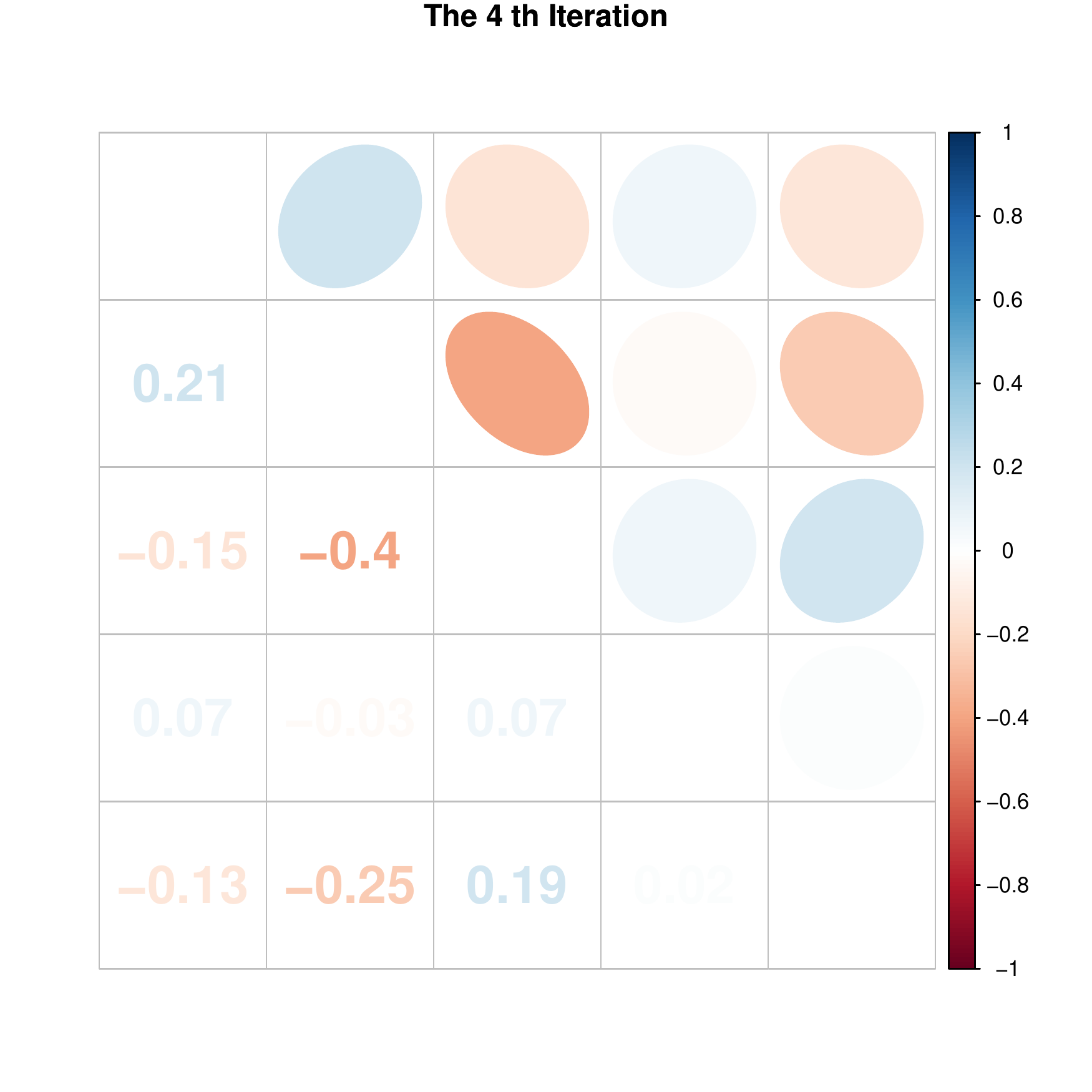}
\includegraphics[width=0.3\textwidth,height=0.18\textheight]{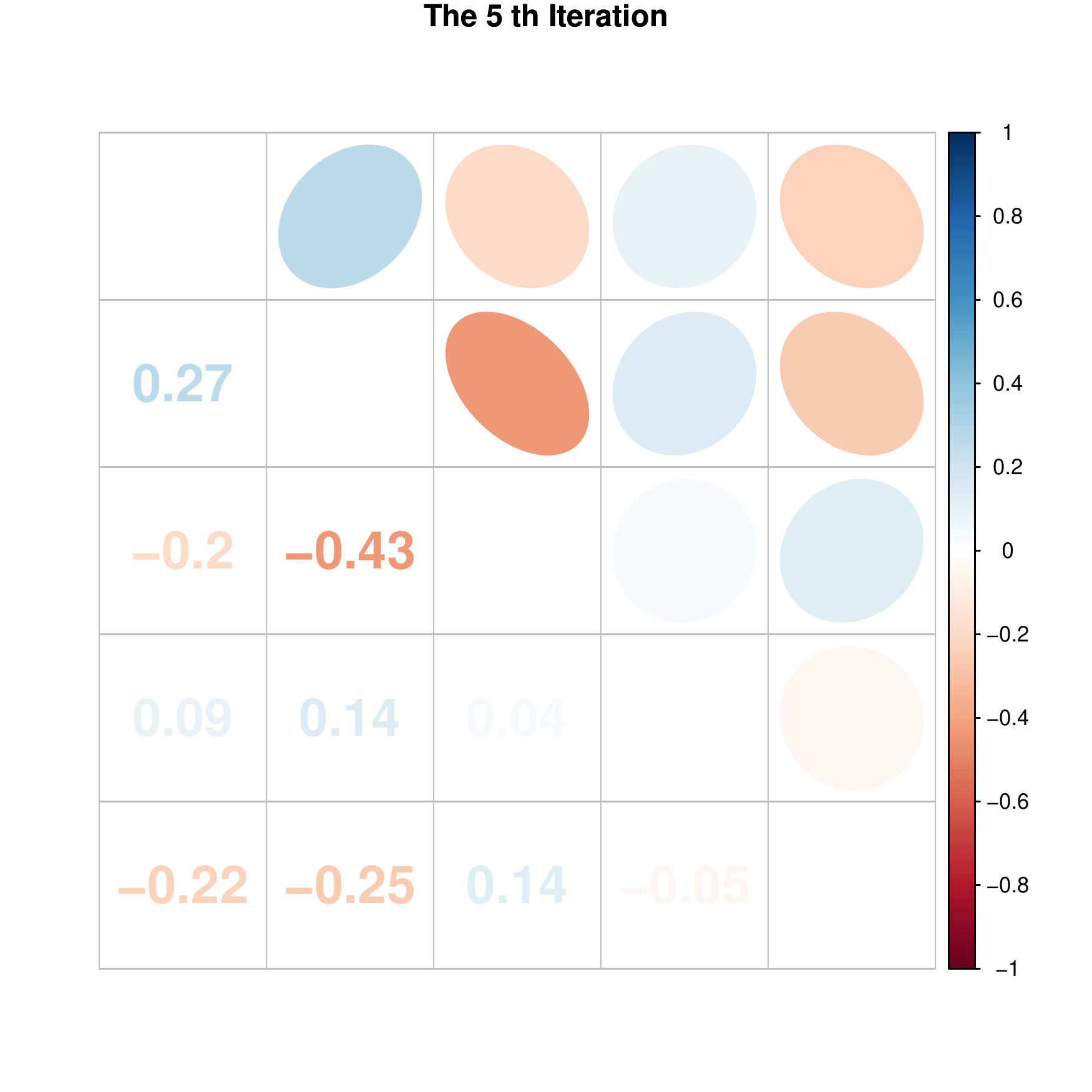}
\includegraphics[width=0.3\textwidth,height=0.18\textheight]{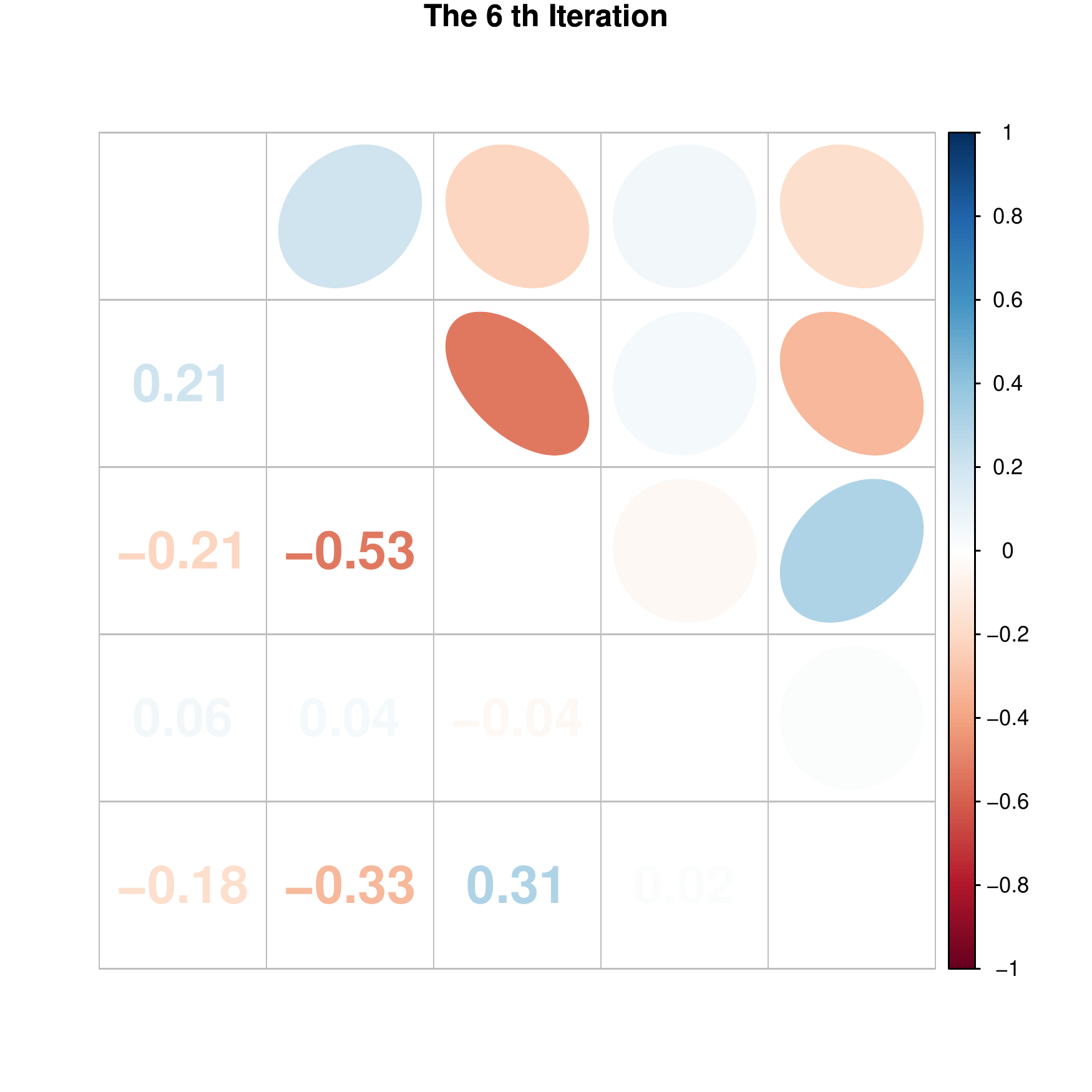}
\includegraphics[width=0.3\textwidth,height=0.18\textheight]{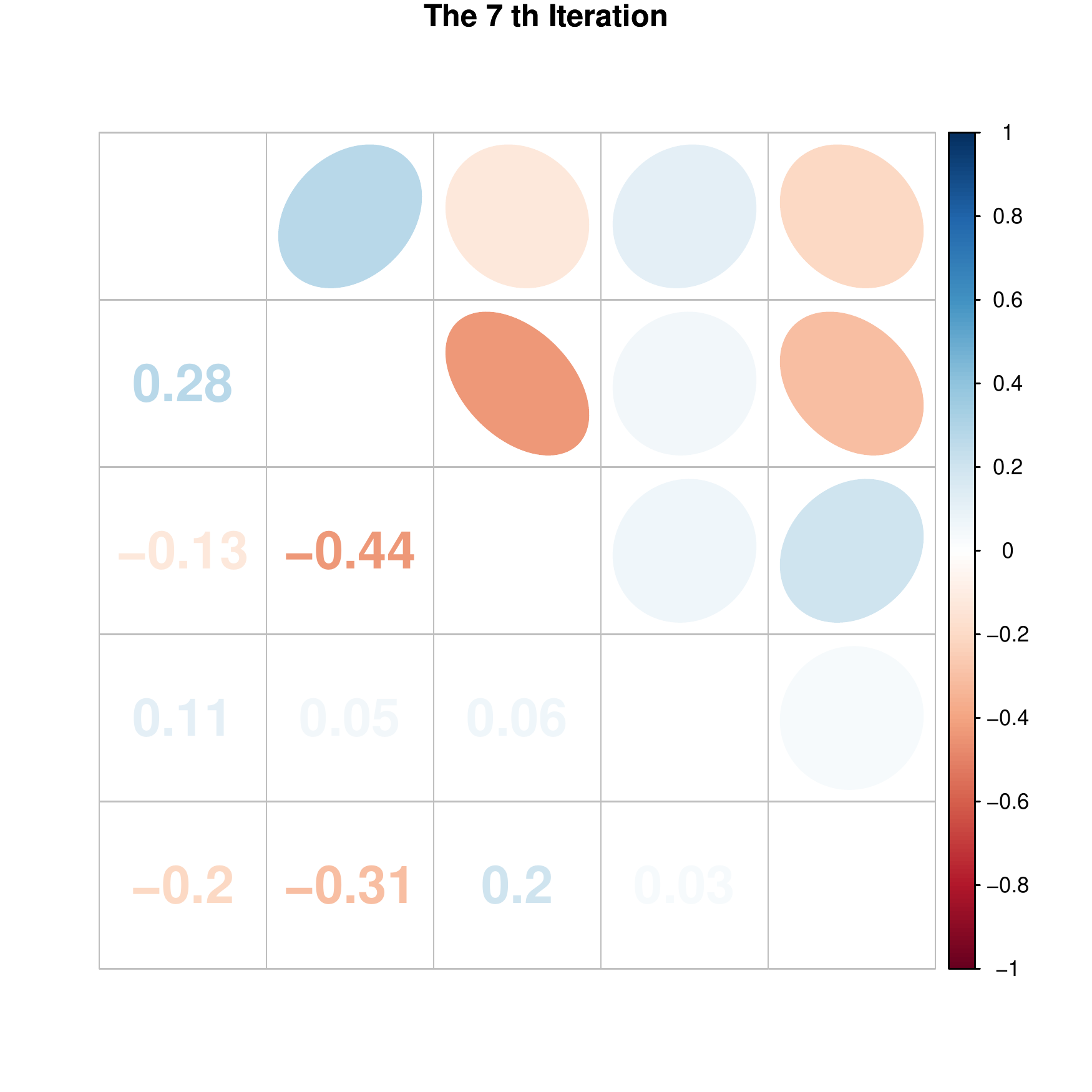}
\includegraphics[width=0.3\textwidth,height=0.18\textheight]{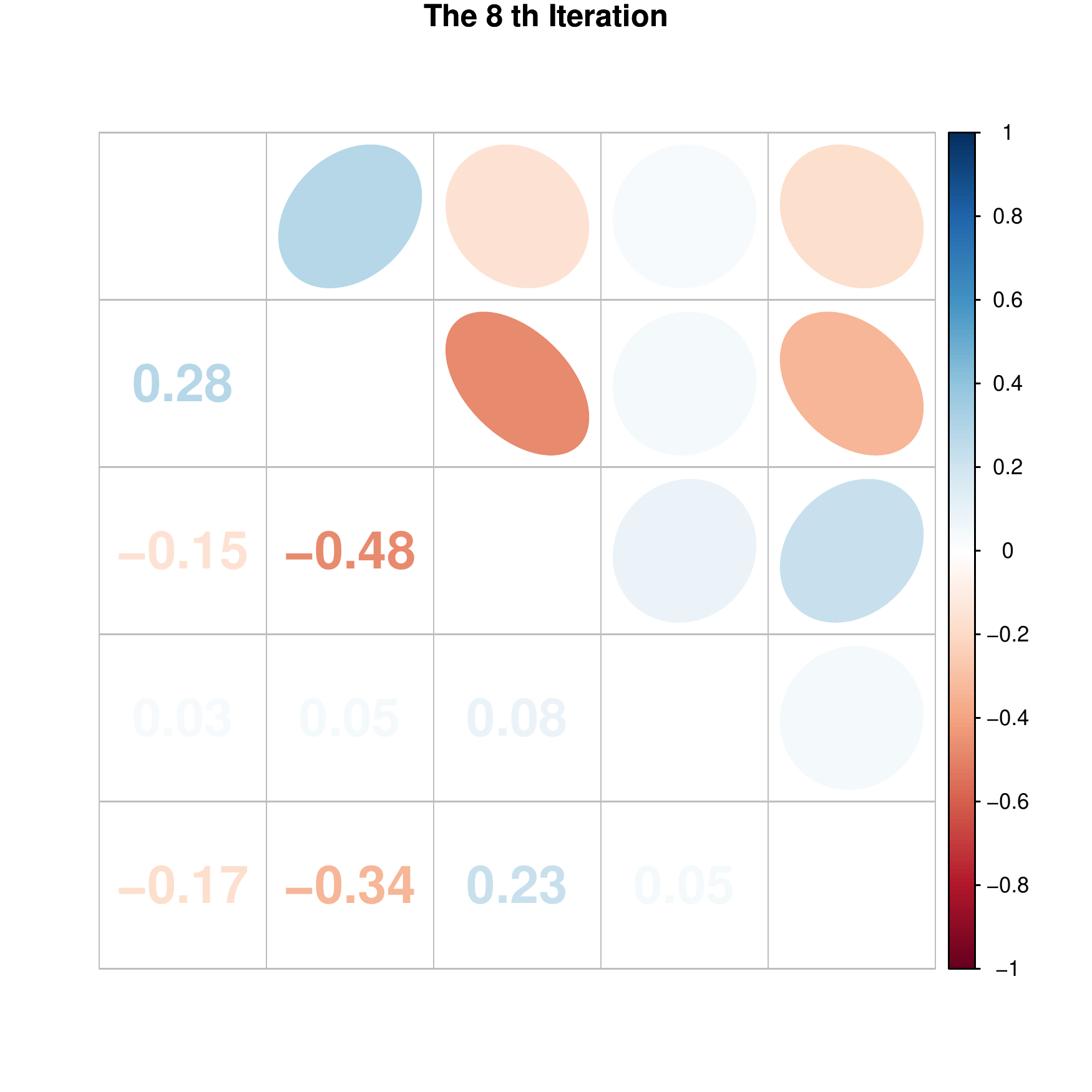}
\includegraphics[width=0.3\textwidth,height=0.18\textheight]{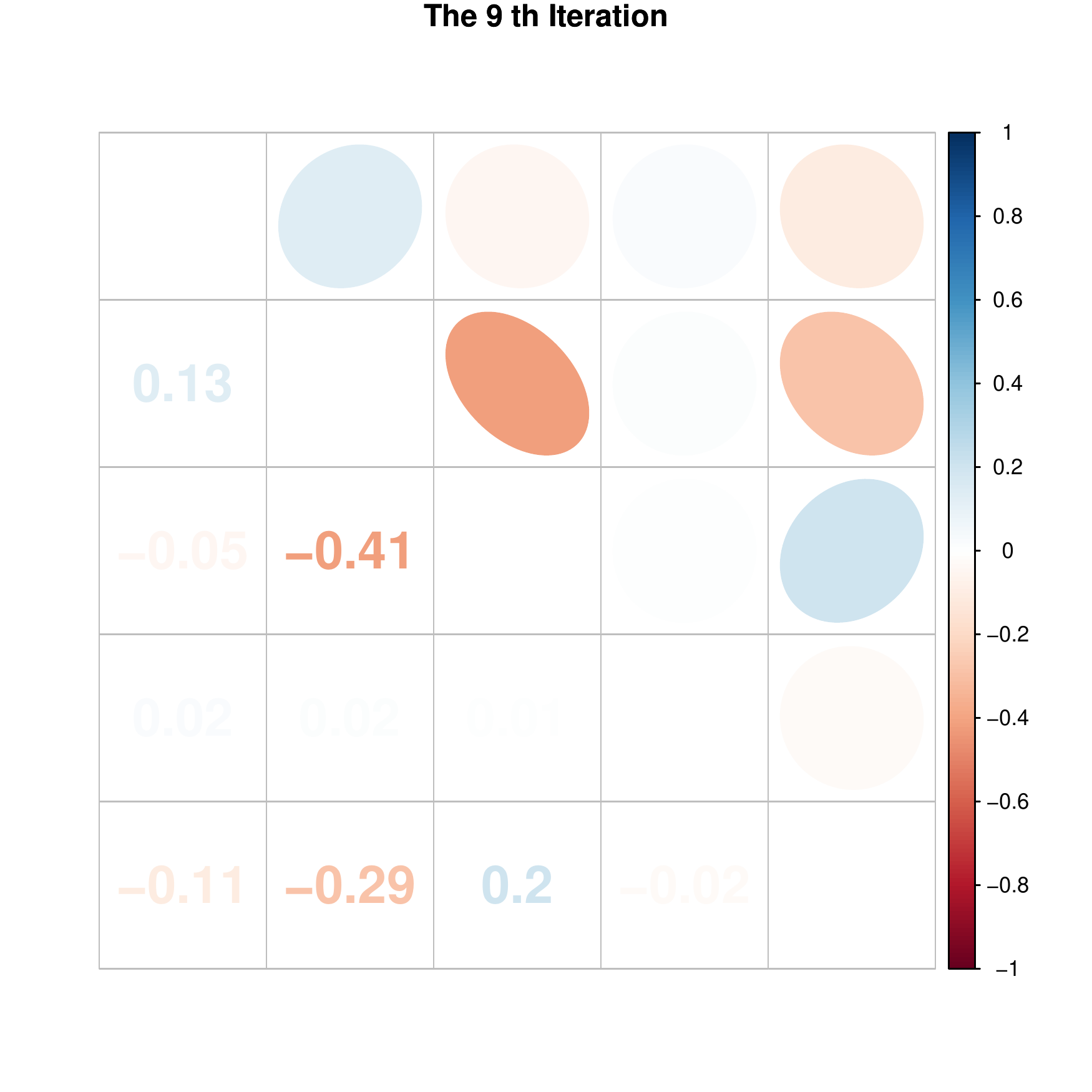}
\includegraphics[width=0.3\textwidth,height=0.18\textheight]{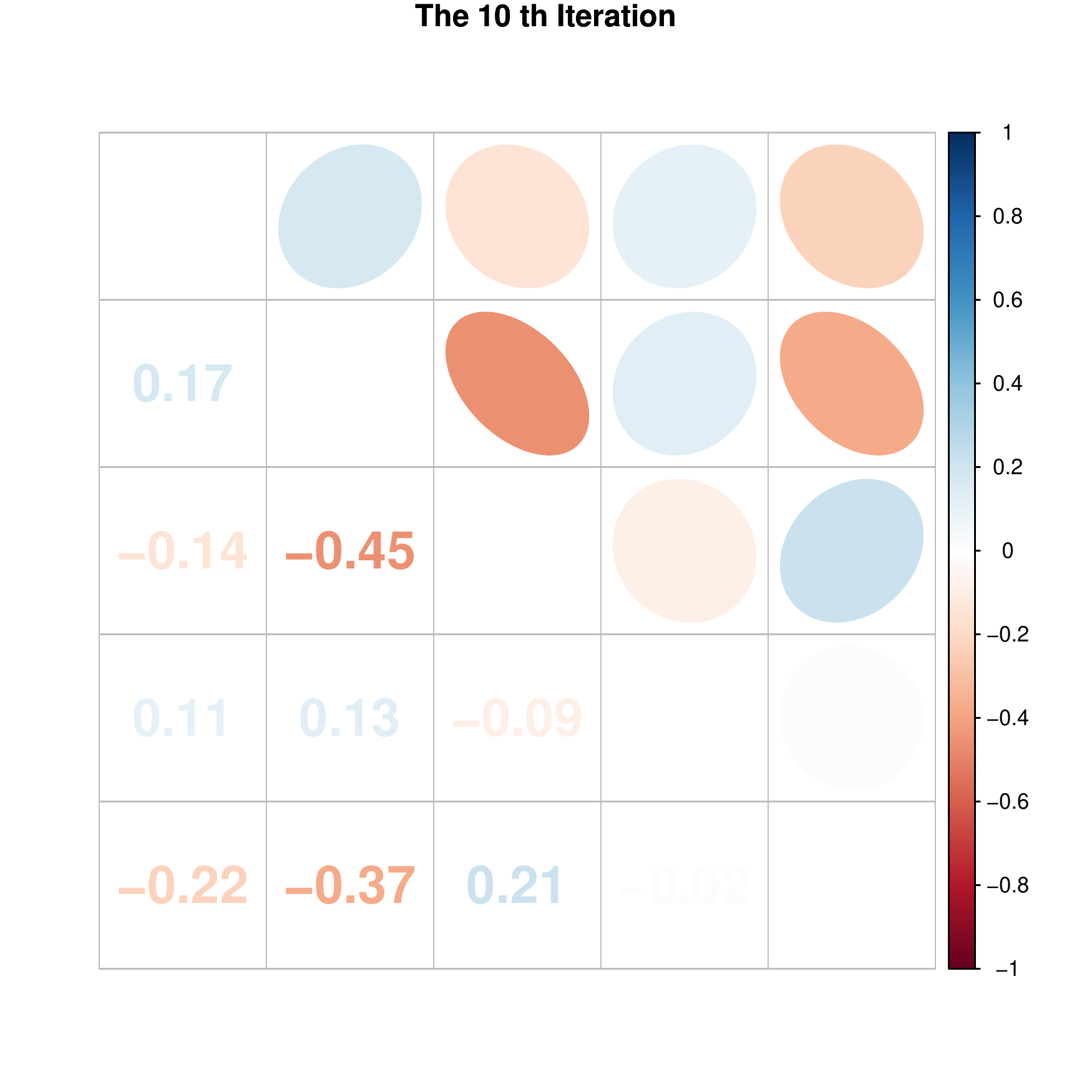}
\includegraphics[width=0.3\textwidth,height=0.18\textheight]{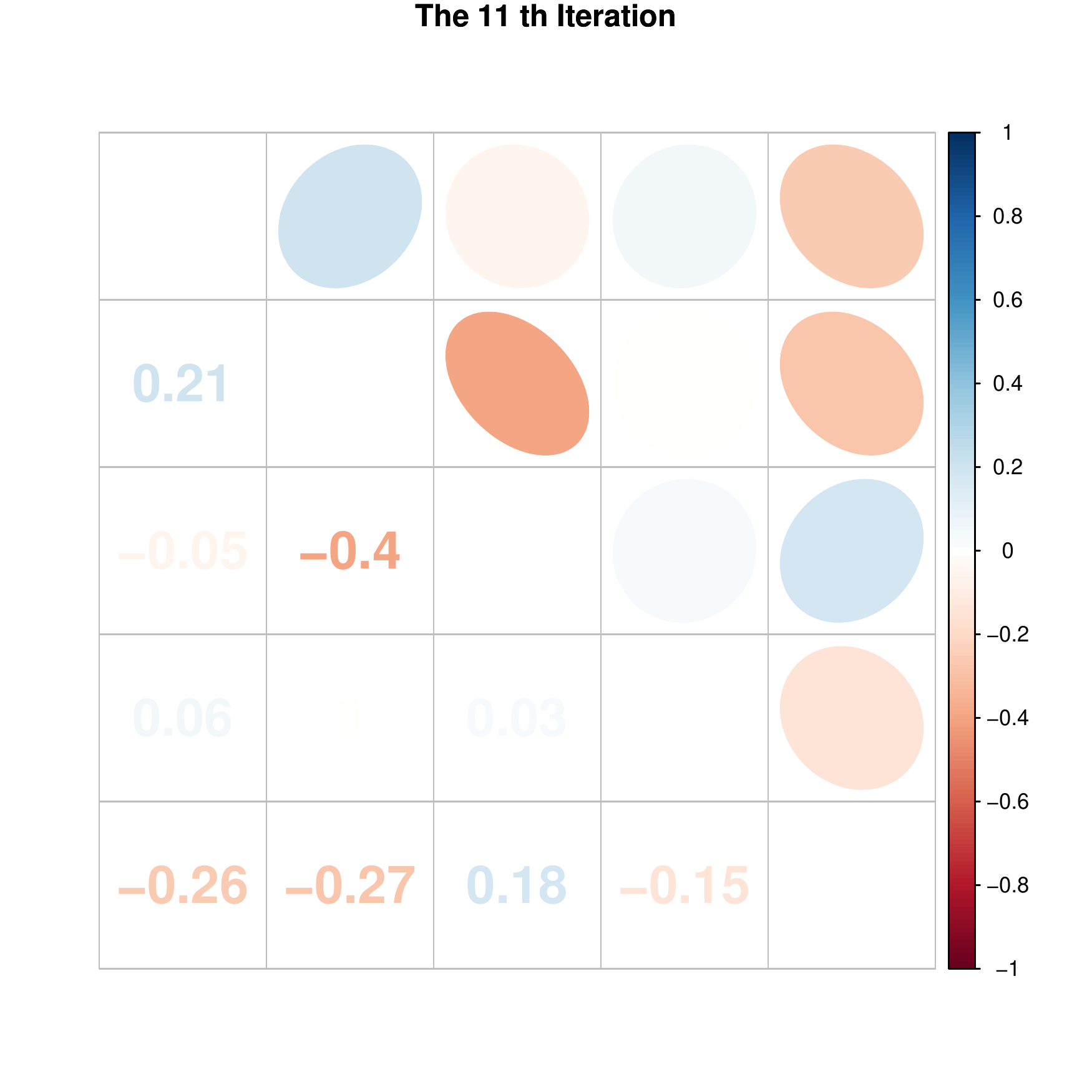}
\includegraphics[width=0.3\textwidth,height=0.18\textheight]{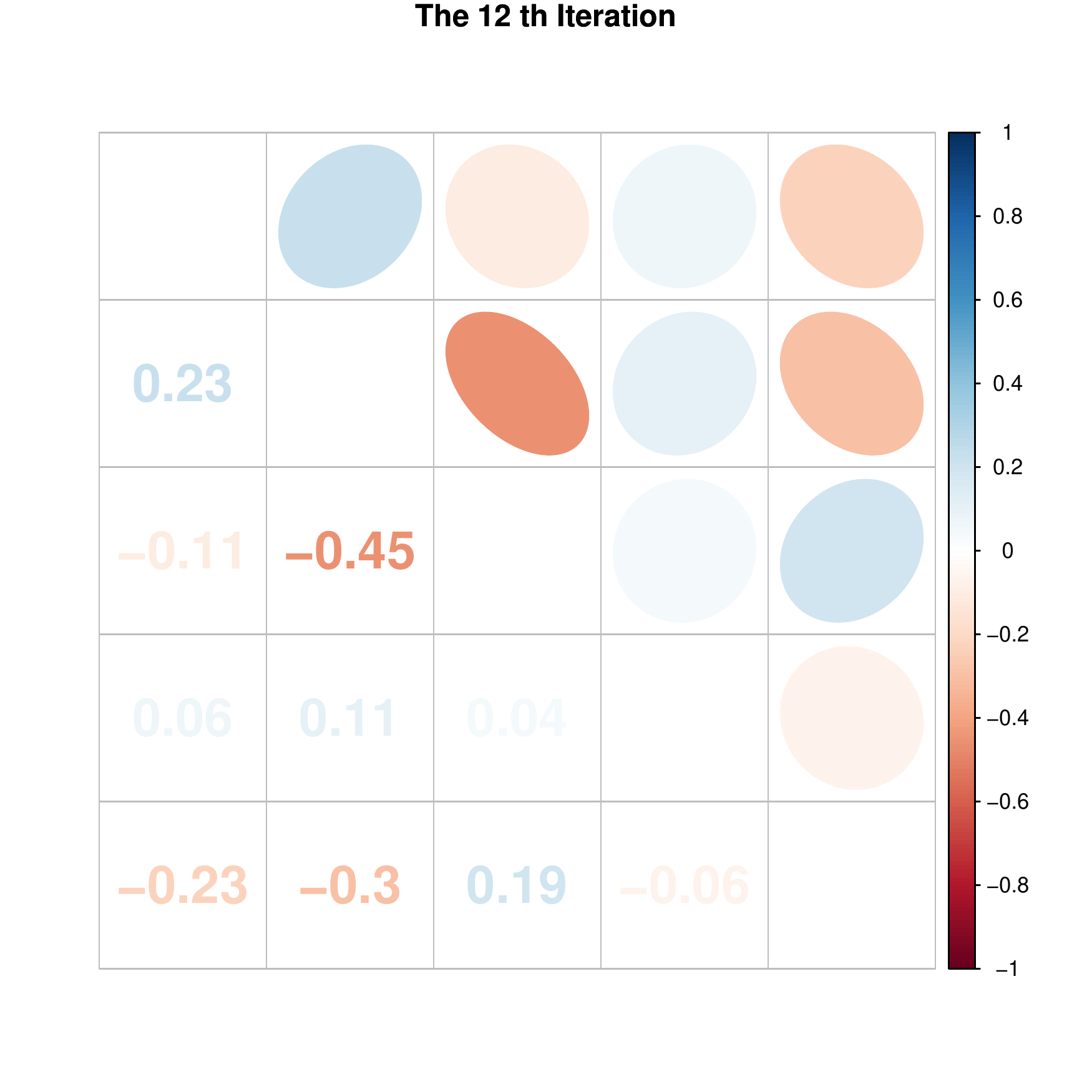}
\caption{Parameter Evolution Visualization. }
\end{figure}

\clearpage
\begin{figure}[h]
\centering
\includegraphics[width=0.3\textwidth,height=0.2\textheight]{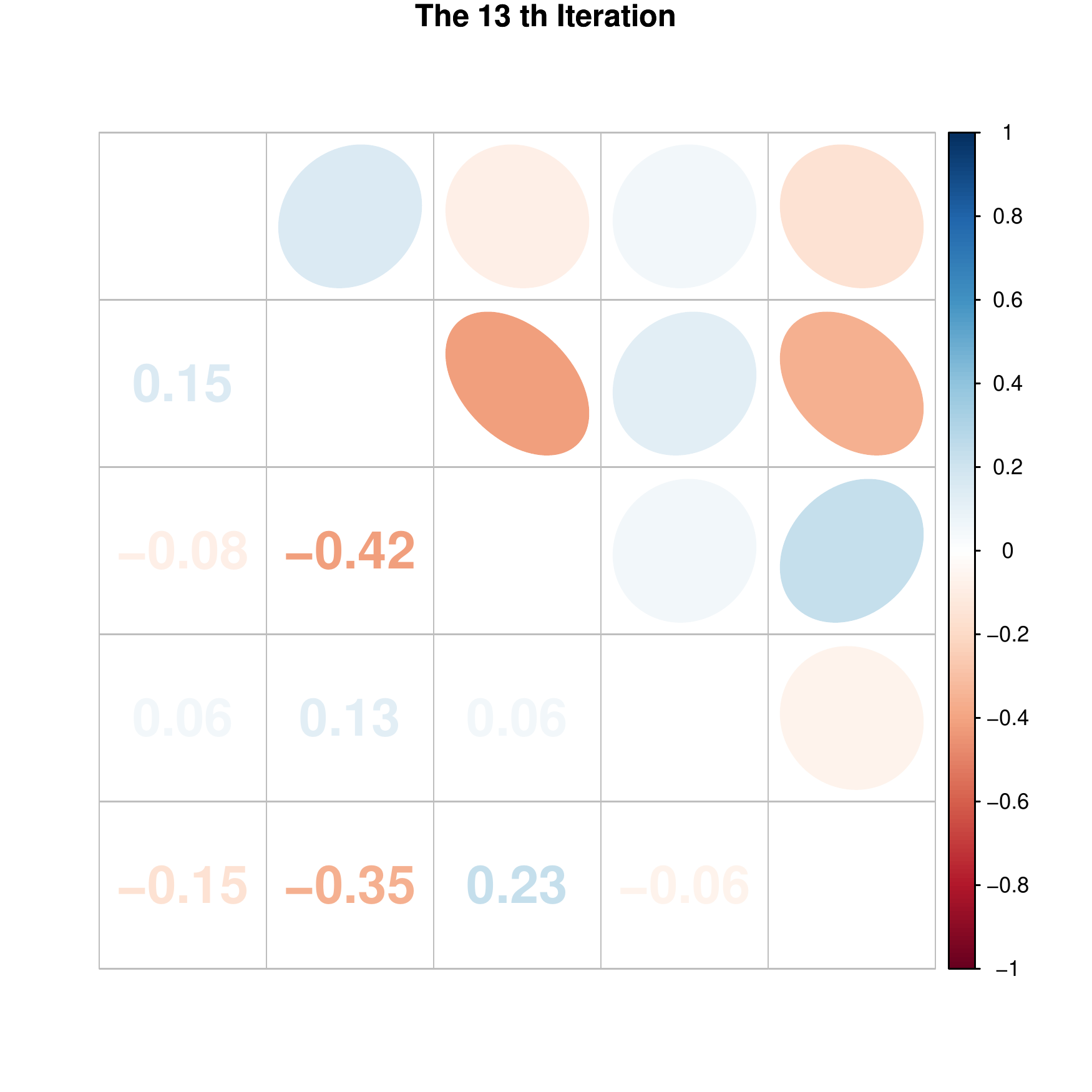}
\includegraphics[width=0.3\textwidth,height=0.2\textheight]{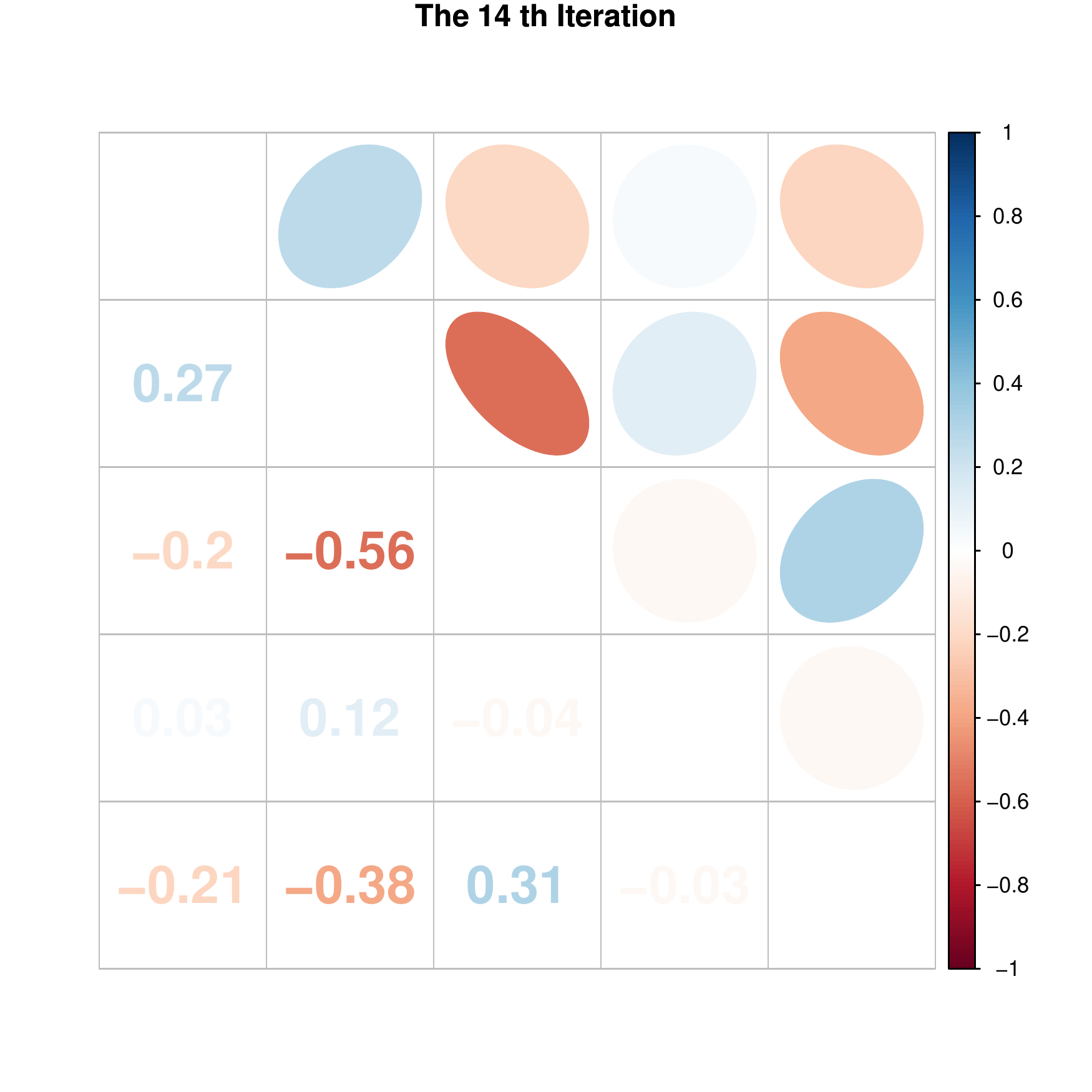}
\includegraphics[width=0.3\textwidth,height=0.2\textheight]{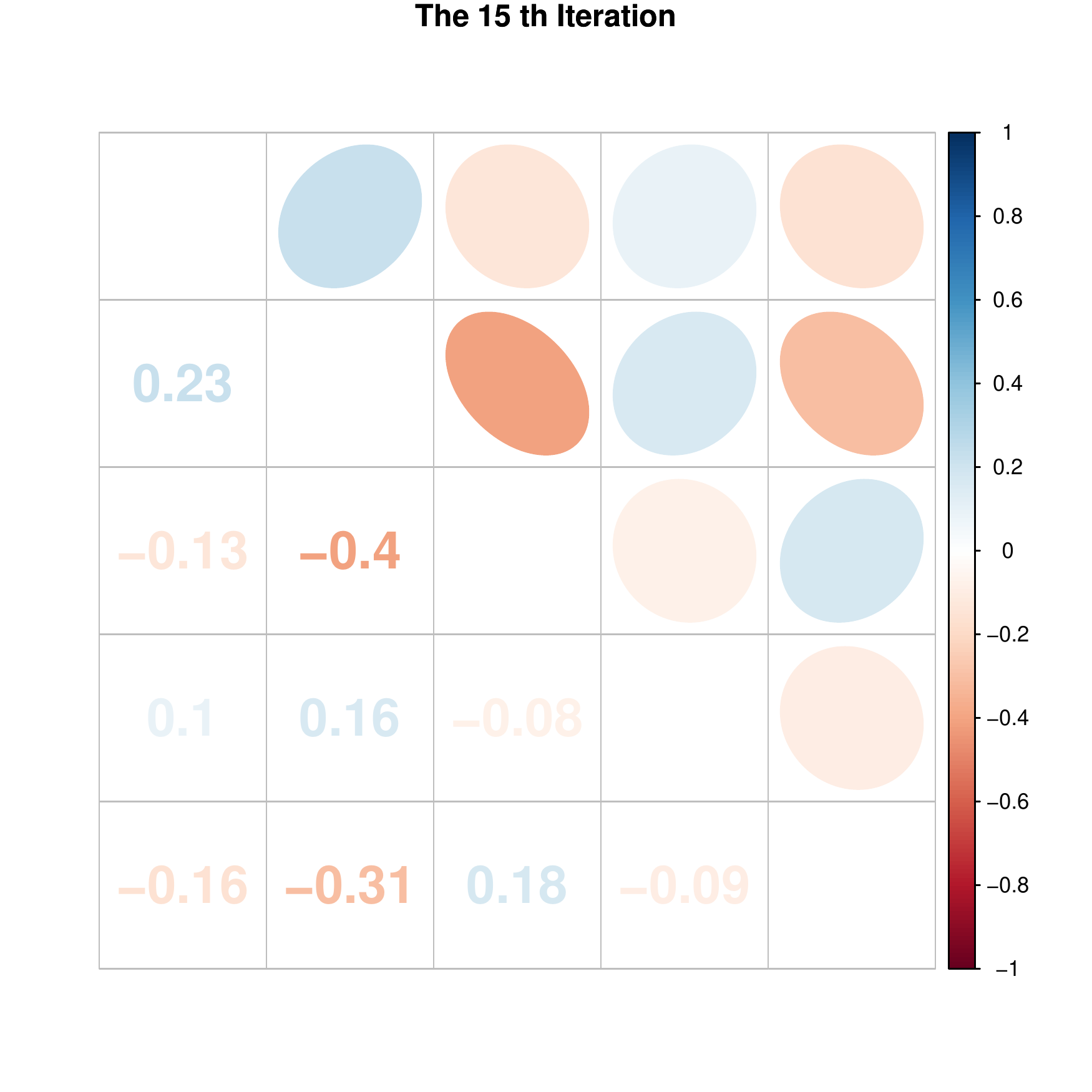}
\includegraphics[width=0.3\textwidth,height=0.2\textheight]{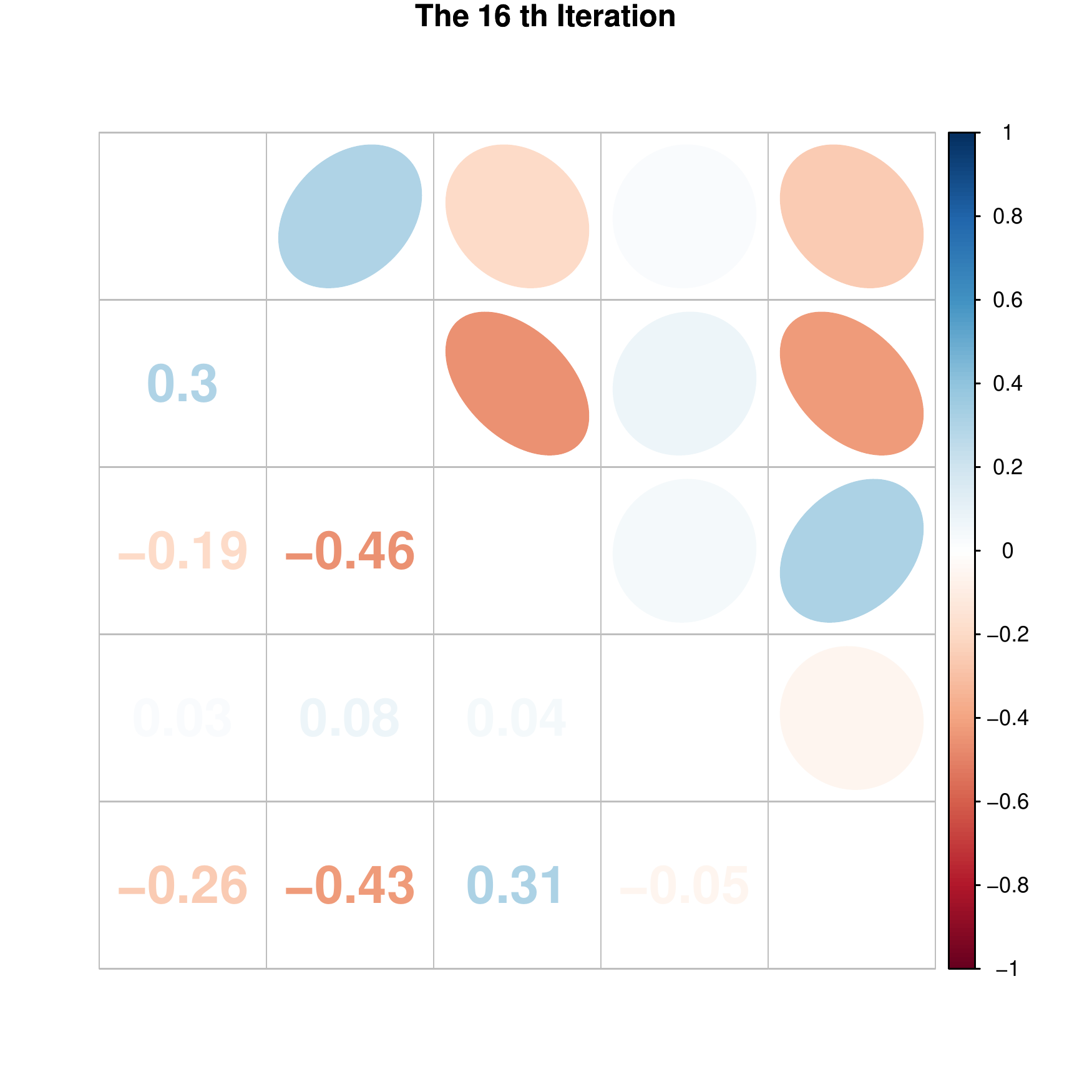}
\includegraphics[width=0.3\textwidth,height=0.2\textheight]{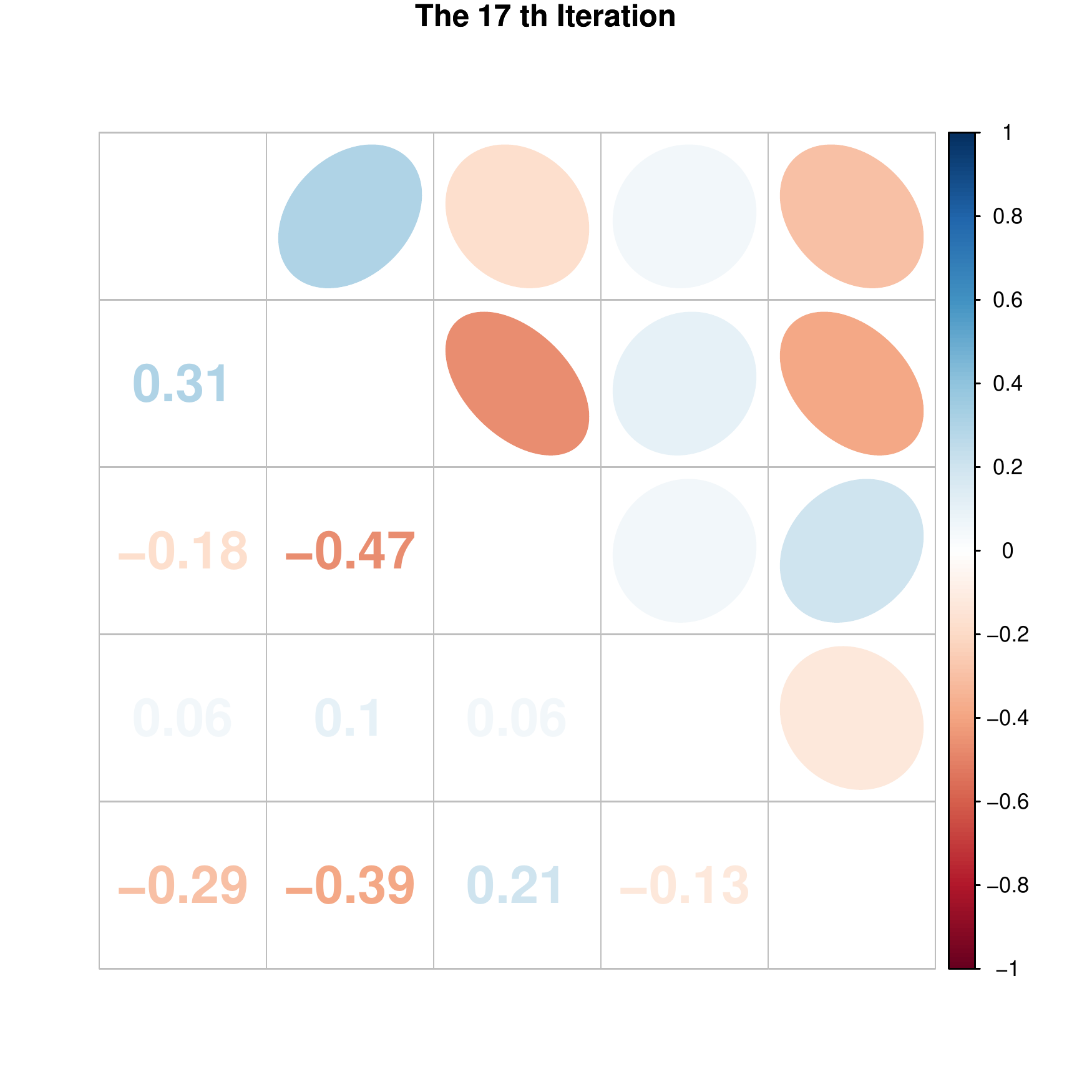}
\includegraphics[width=0.3\textwidth,height=0.2\textheight]{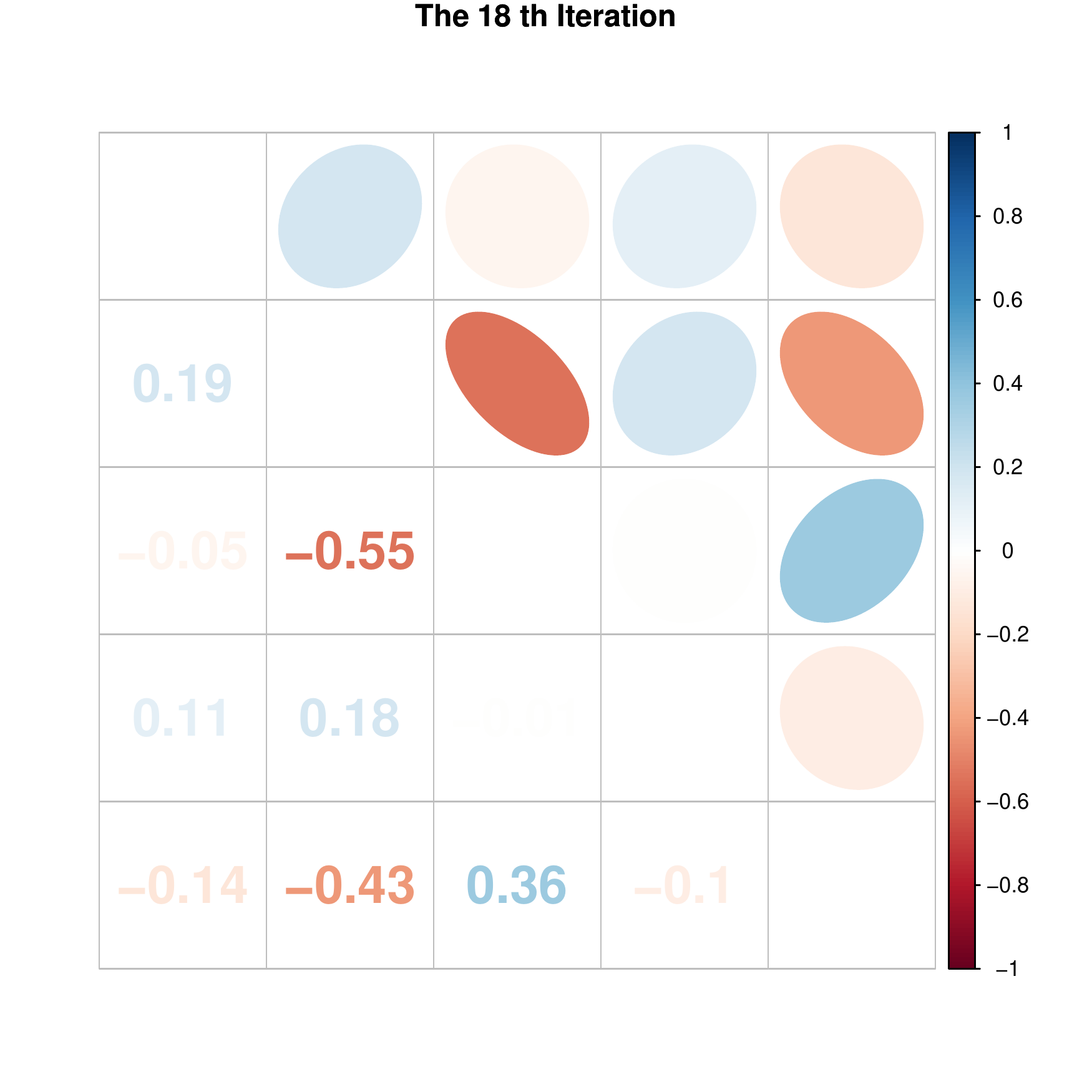}
\includegraphics[width=0.3\textwidth,height=0.2\textheight]{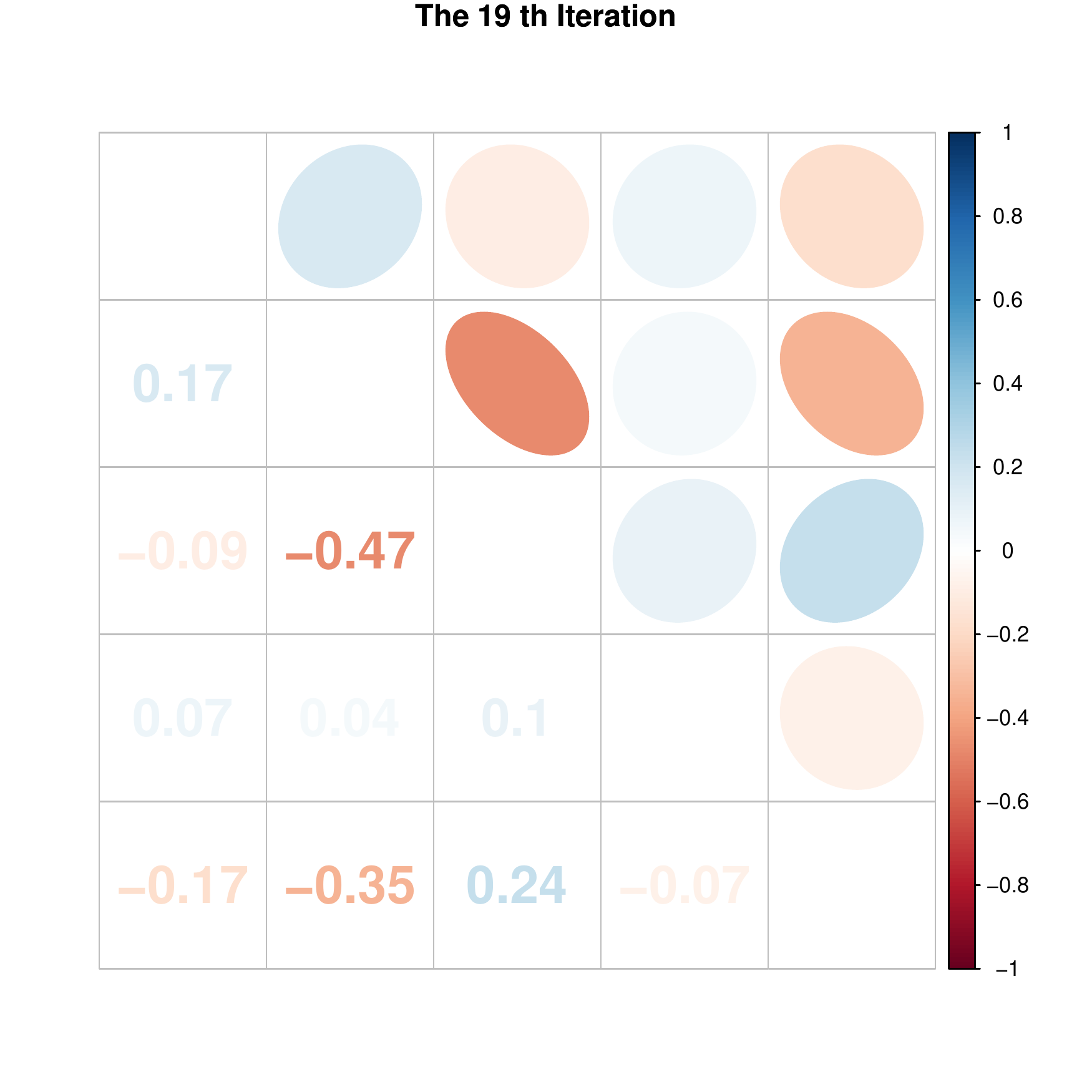}
\includegraphics[width=0.3\textwidth,height=0.2\textheight]{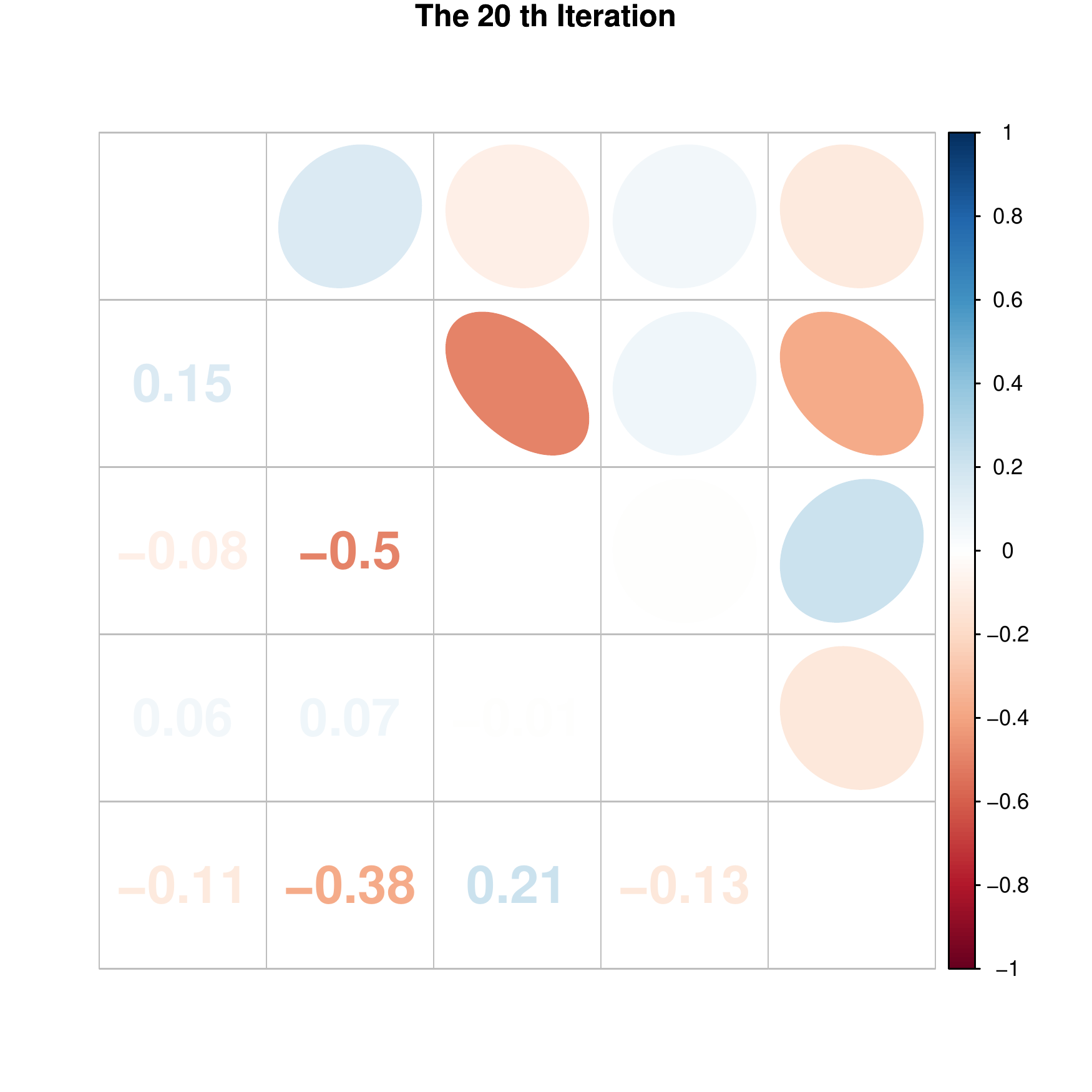}
\includegraphics[width=0.3\textwidth,height=0.2\textheight]{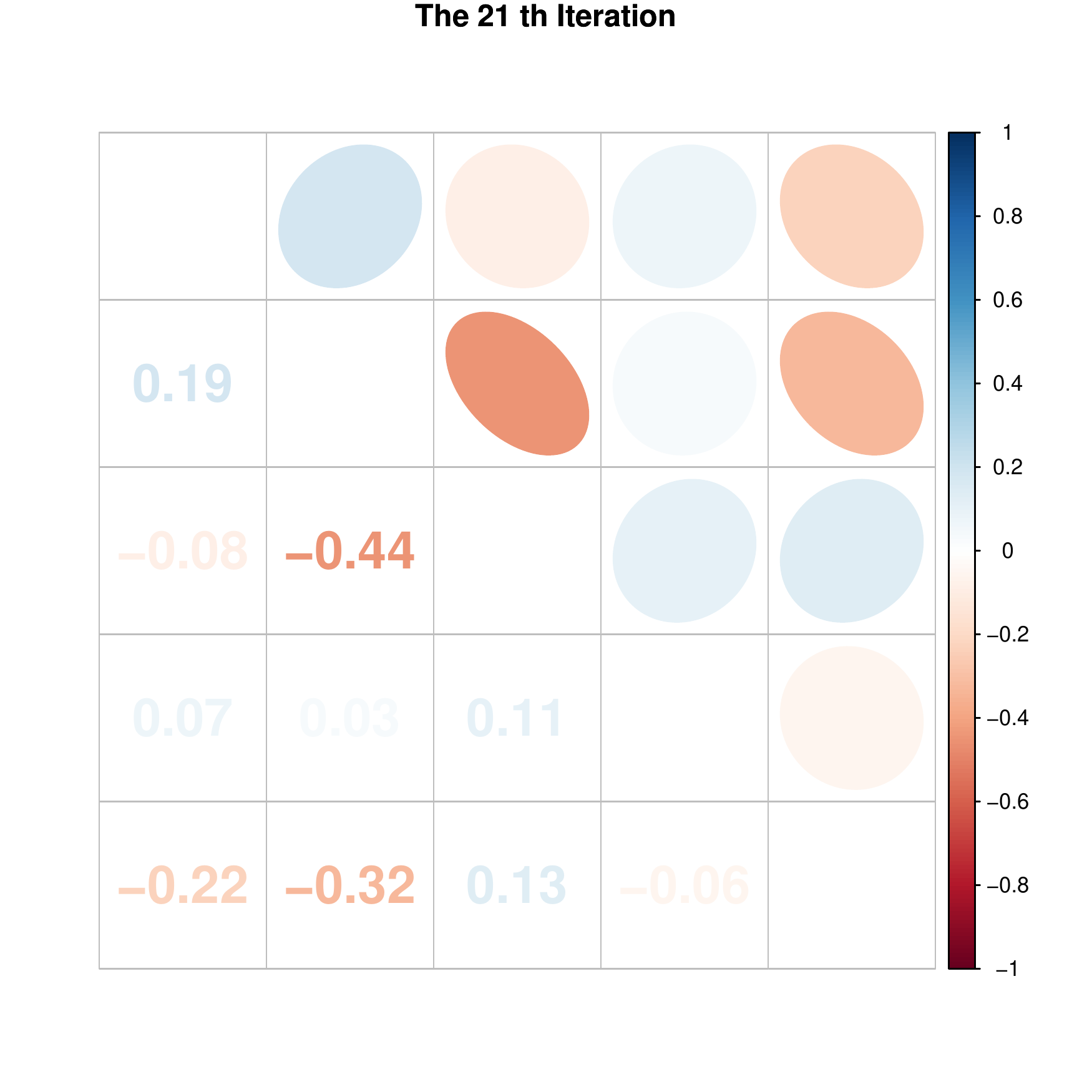}
\includegraphics[width=0.3\textwidth,height=0.2\textheight]{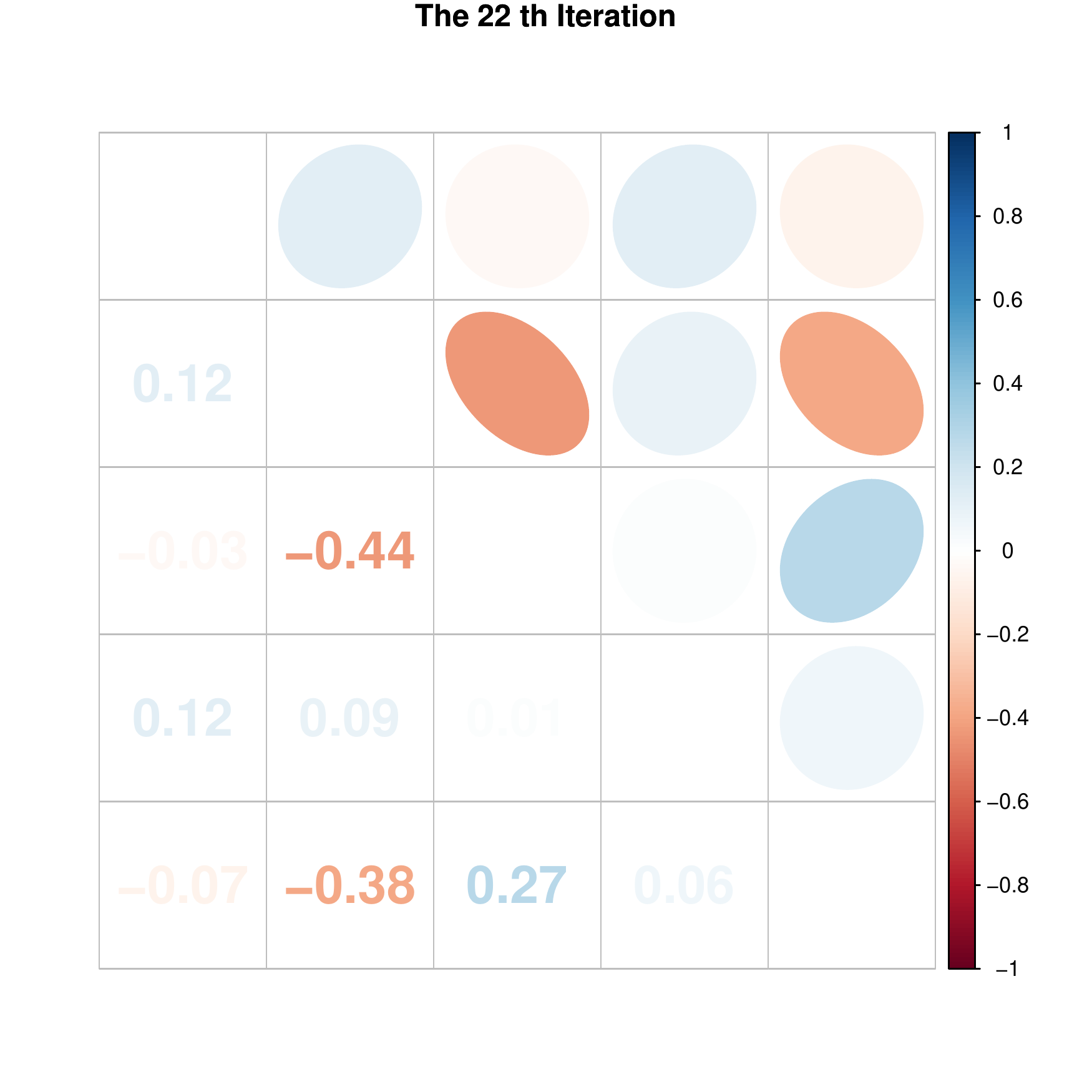}
\includegraphics[width=0.3\textwidth,height=0.2\textheight]{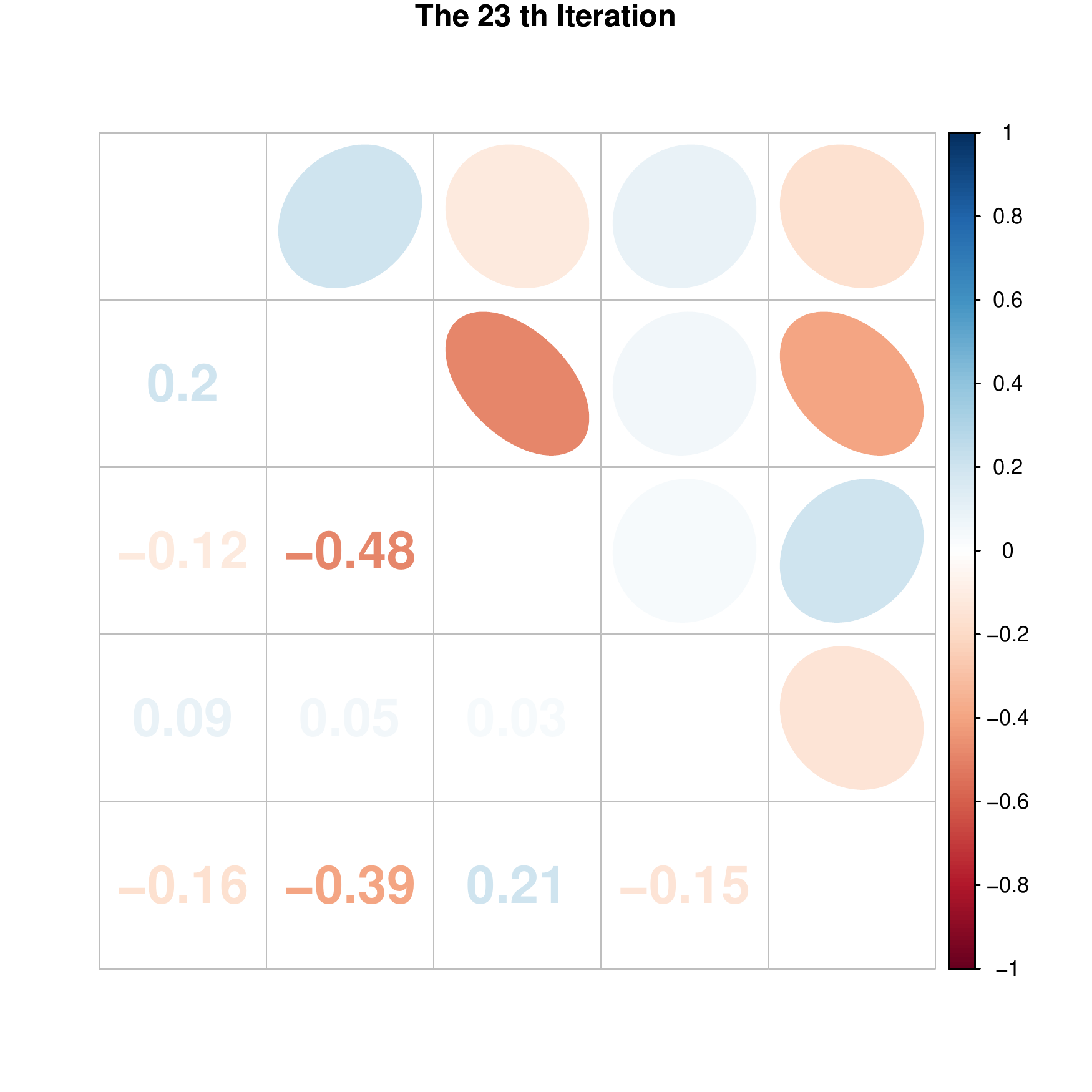}
\includegraphics[width=0.3\textwidth,height=0.2\textheight]{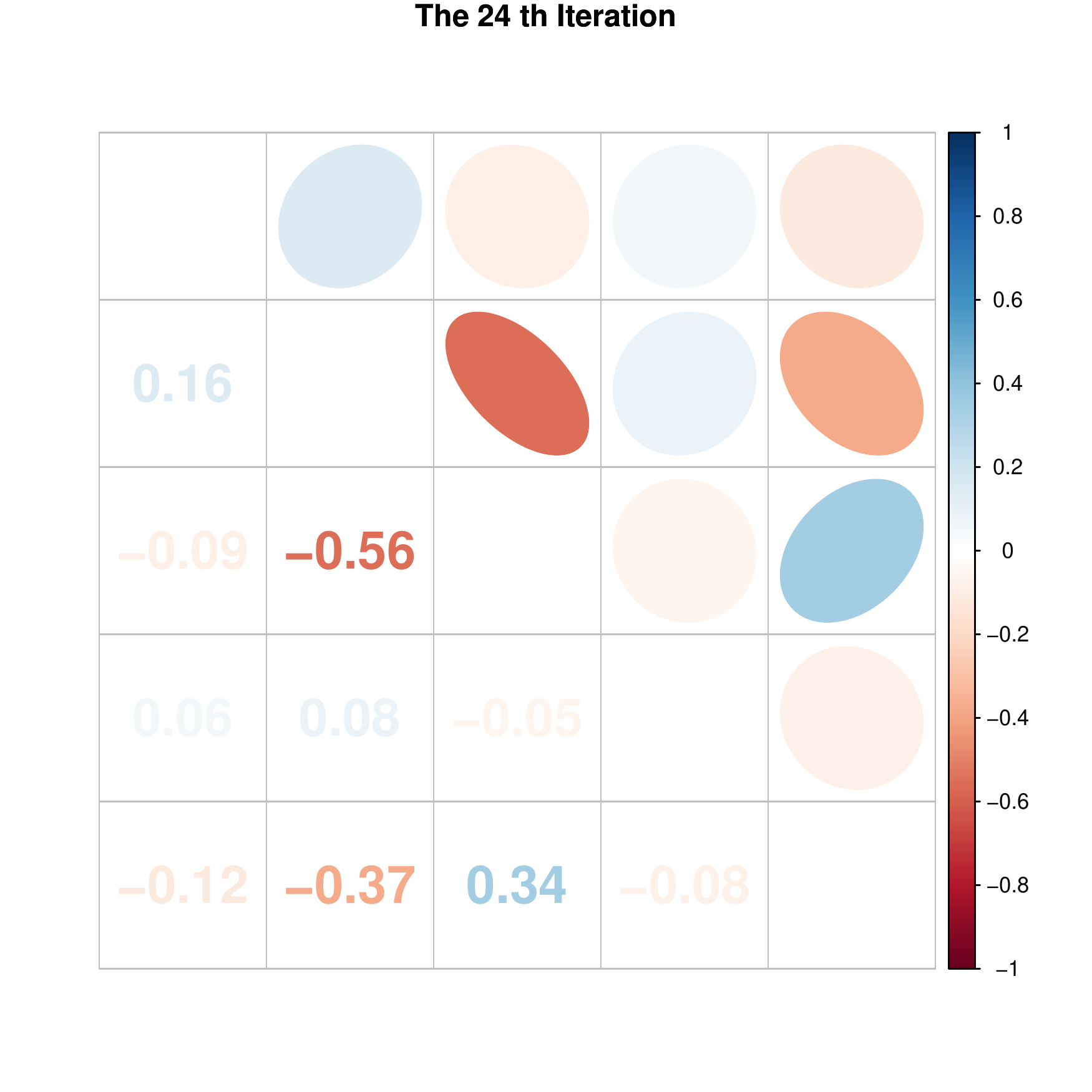}
\caption{Parameter Evolution Visualization. }
\end{figure}

\clearpage
\begin{figure}[h]
\centering
\includegraphics[width=0.3\textwidth,height=0.2\textheight]{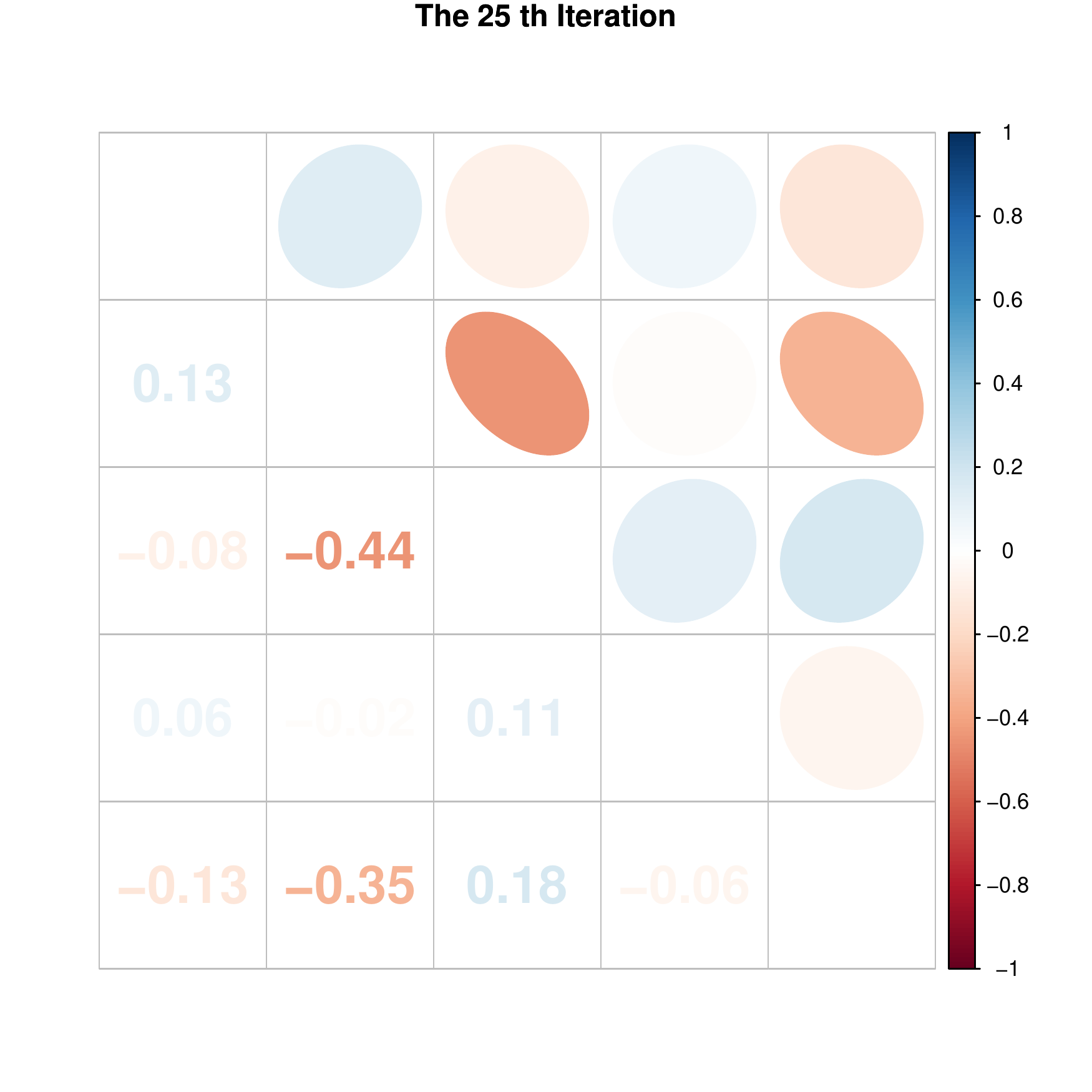}
\includegraphics[width=0.3\textwidth,height=0.2\textheight]{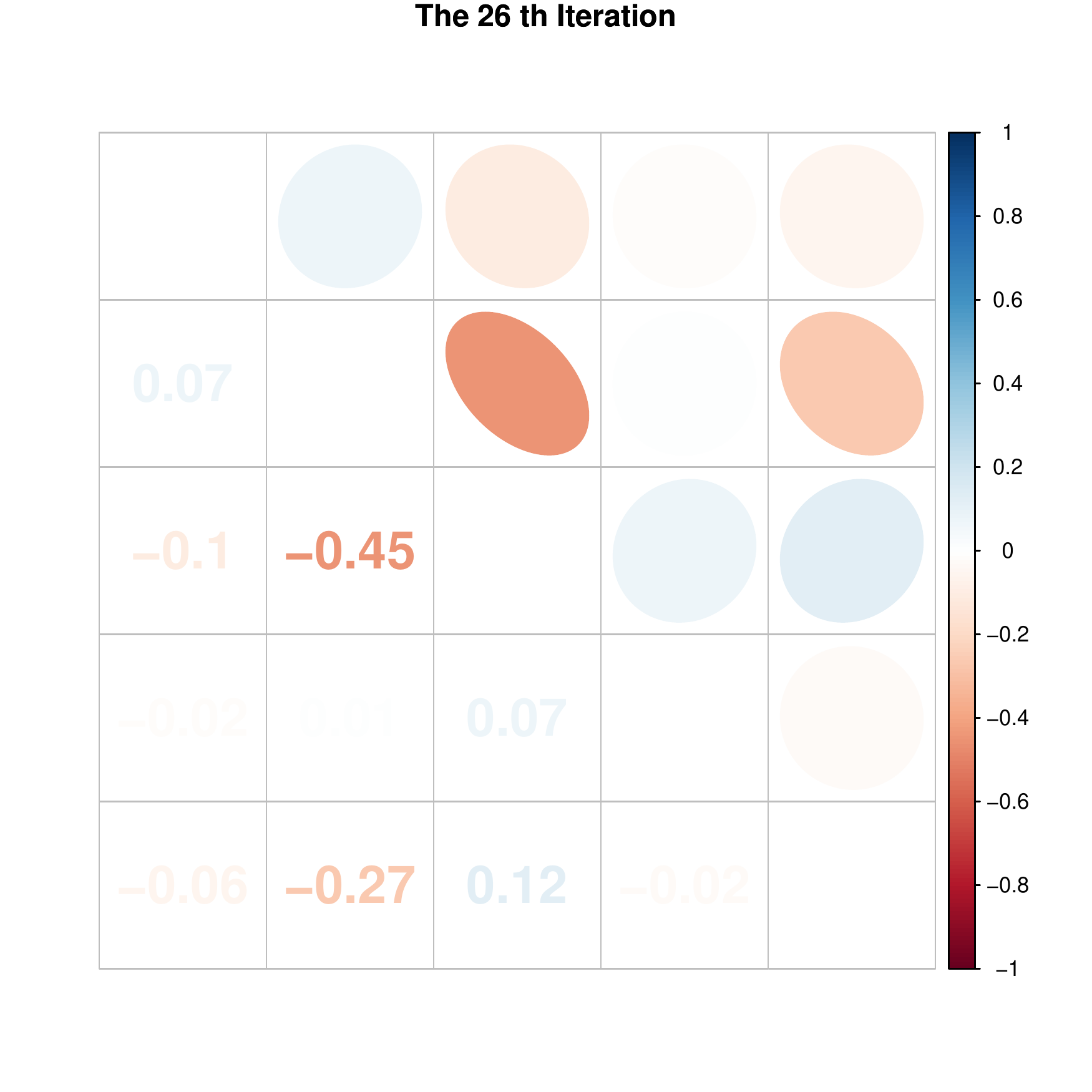}
\includegraphics[width=0.3\textwidth,height=0.2\textheight]{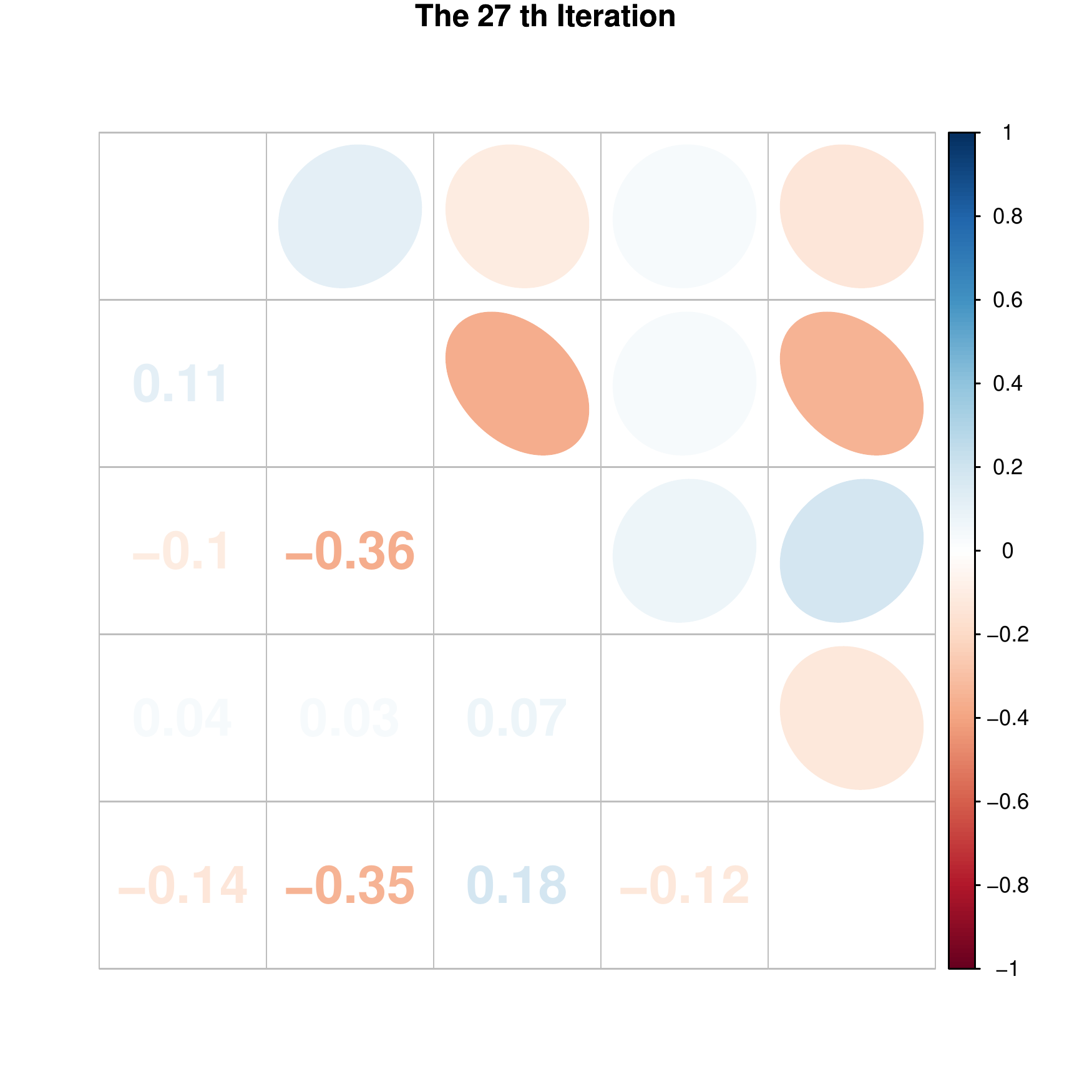}
\includegraphics[width=0.3\textwidth,height=0.2\textheight]{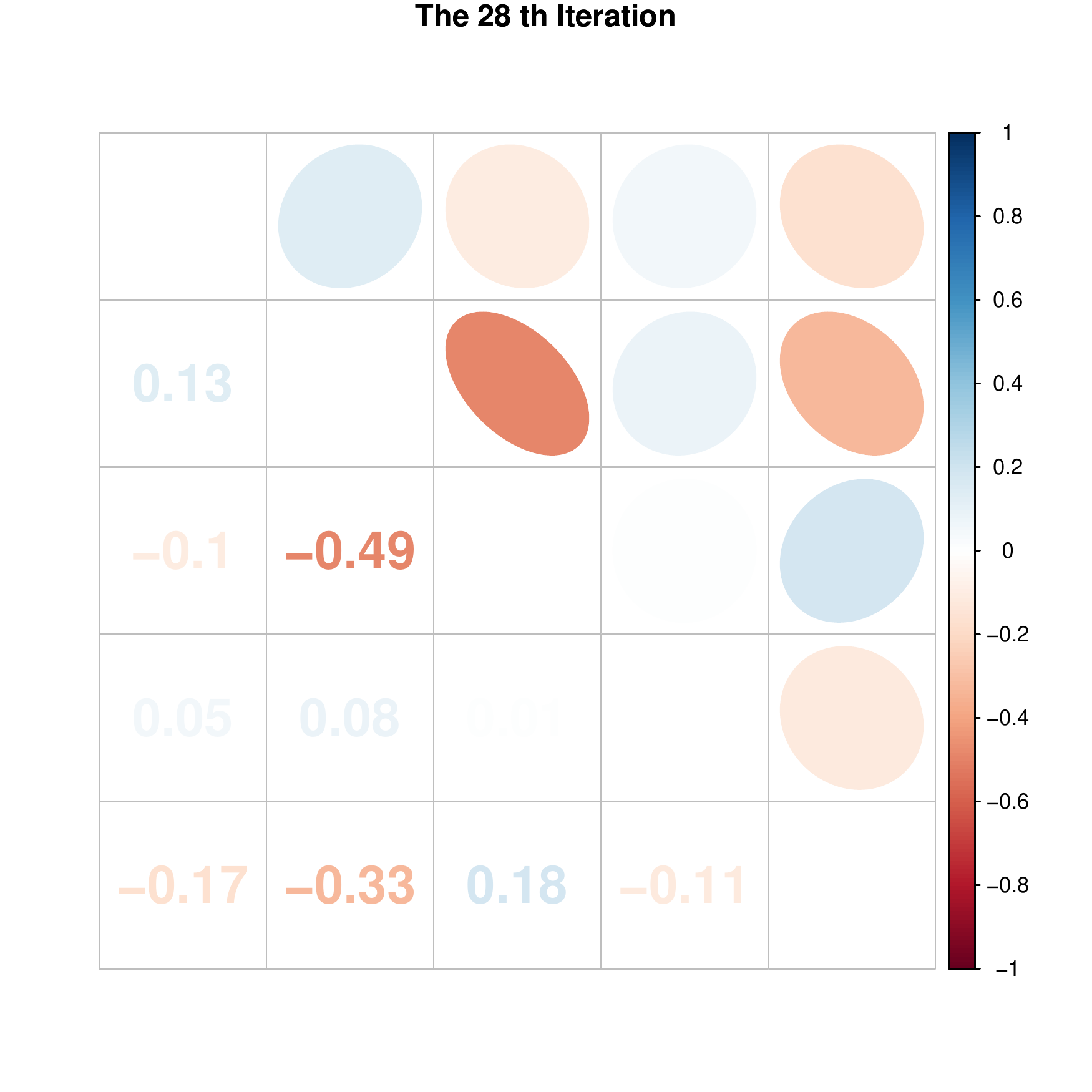}
\includegraphics[width=0.3\textwidth,height=0.2\textheight]{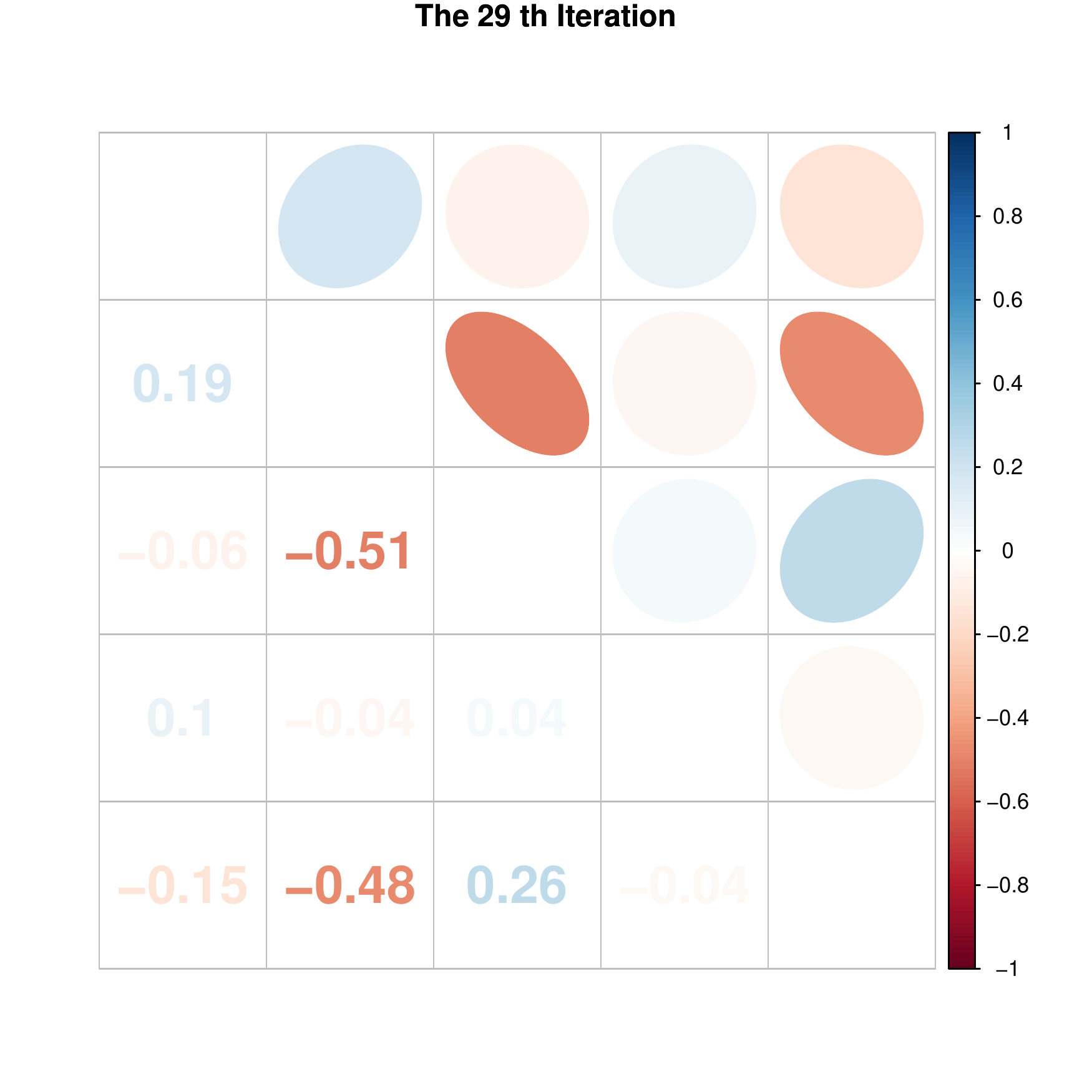}
\includegraphics[width=0.3\textwidth,height=0.2\textheight]{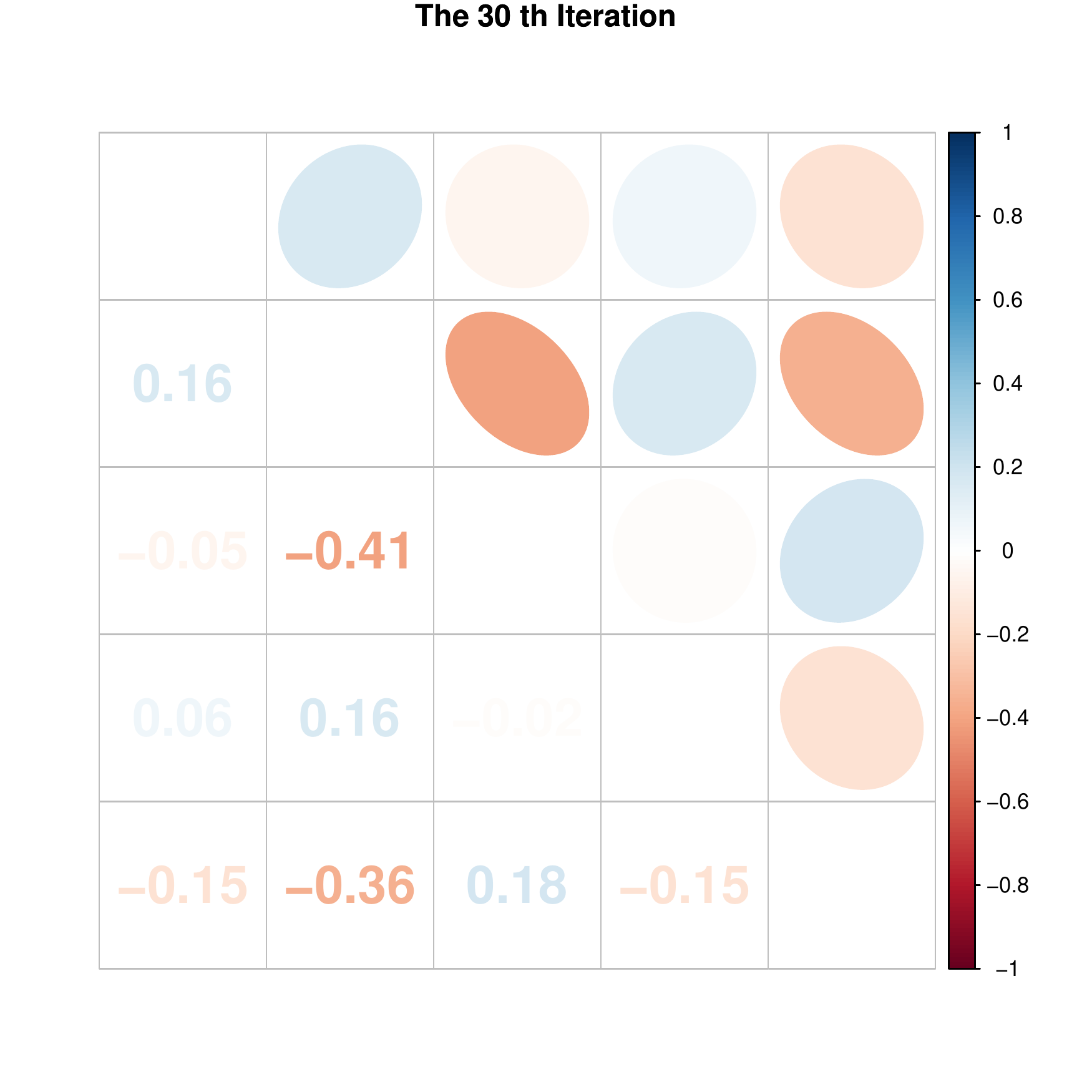}
\includegraphics[width=0.3\textwidth,height=0.2\textheight]{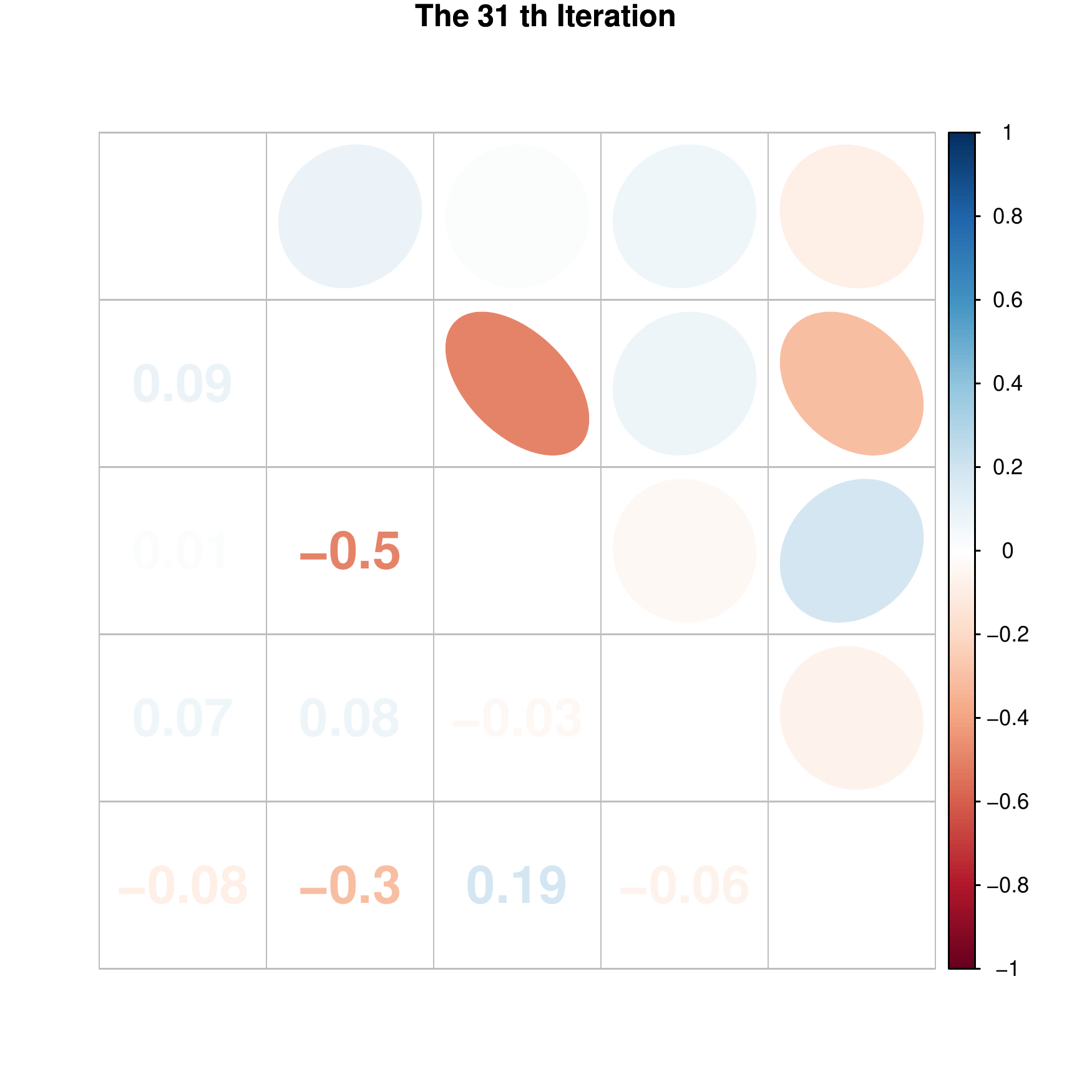}
\includegraphics[width=0.3\textwidth,height=0.2\textheight]{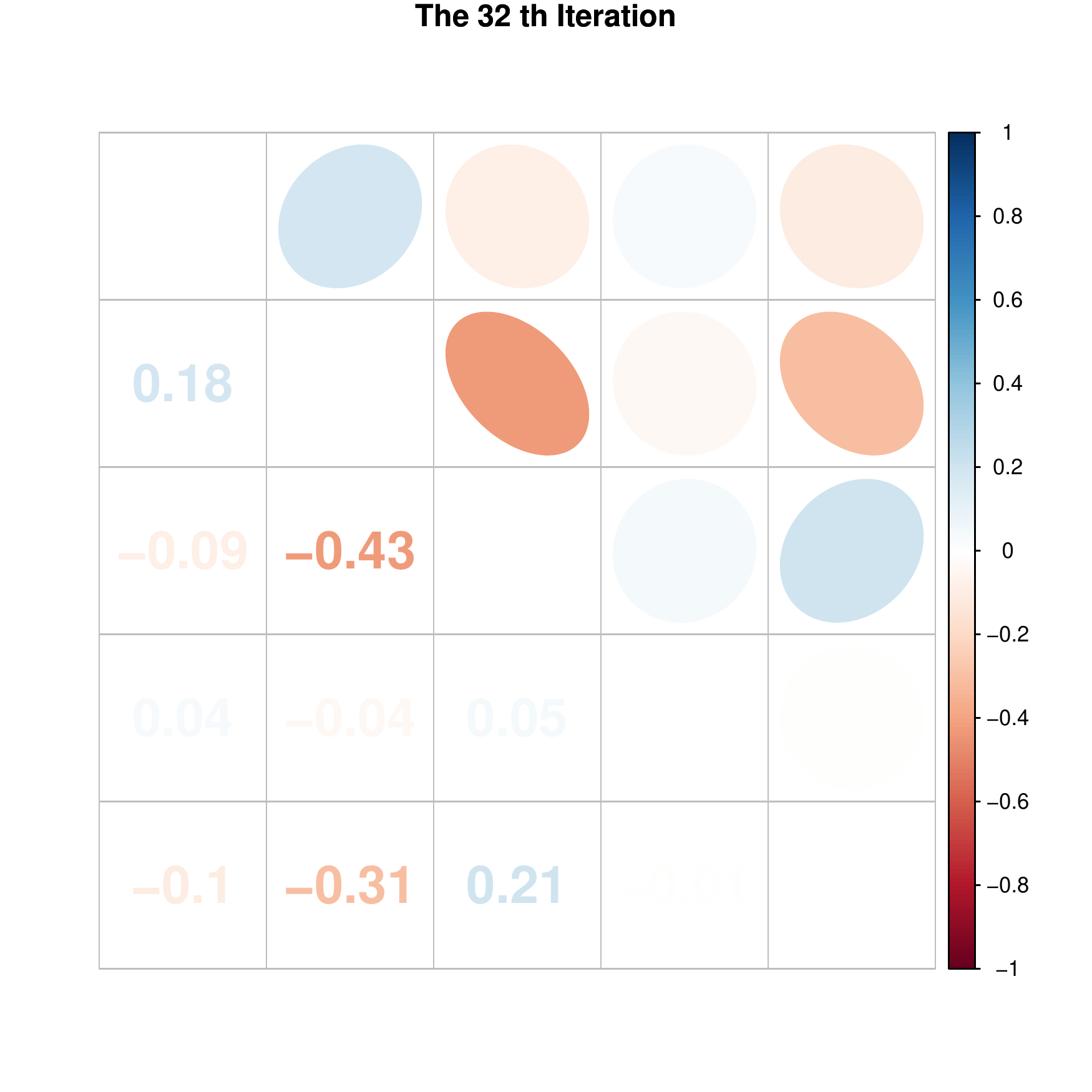}
\includegraphics[width=0.3\textwidth,height=0.2\textheight]{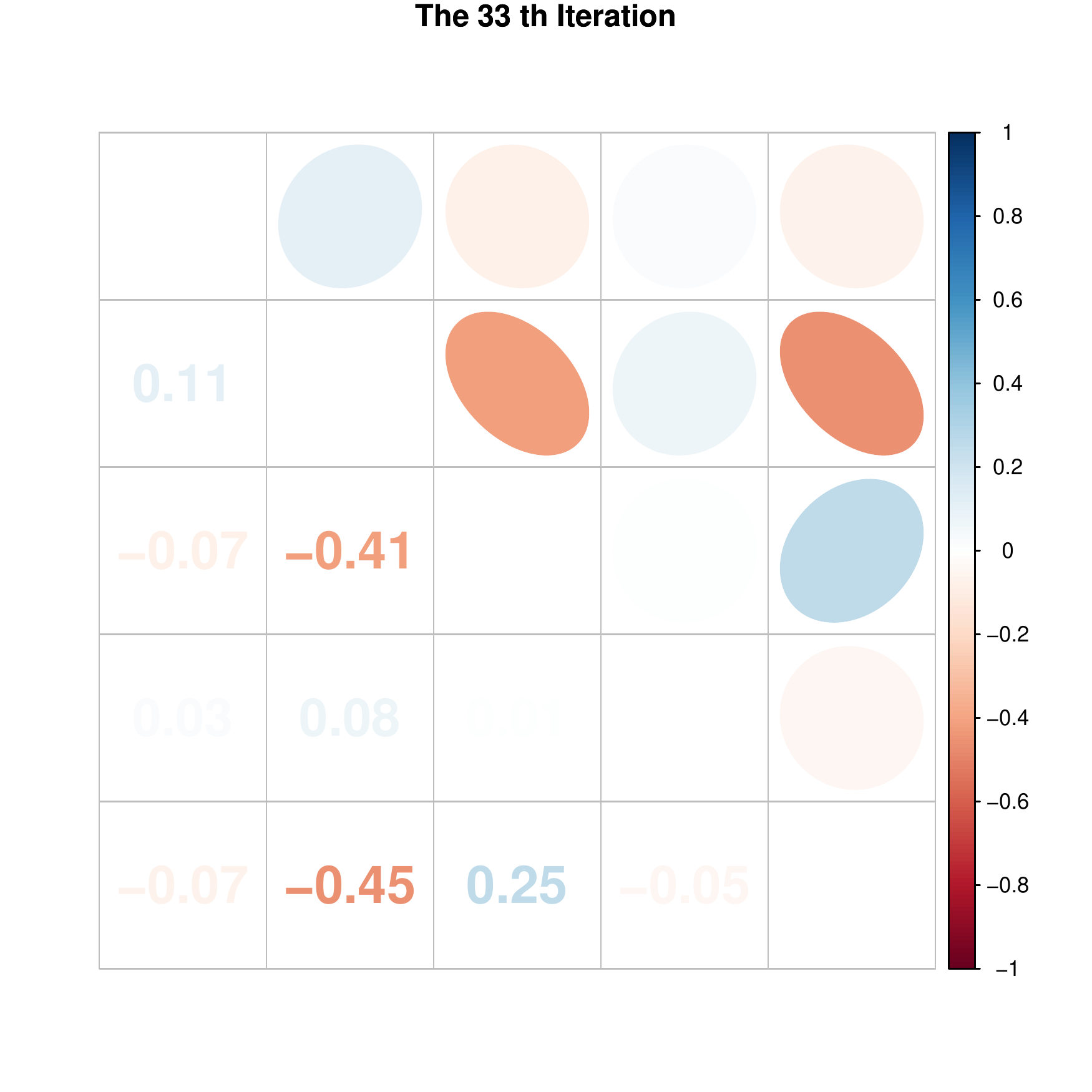}
\includegraphics[width=0.3\textwidth,height=0.2\textheight]{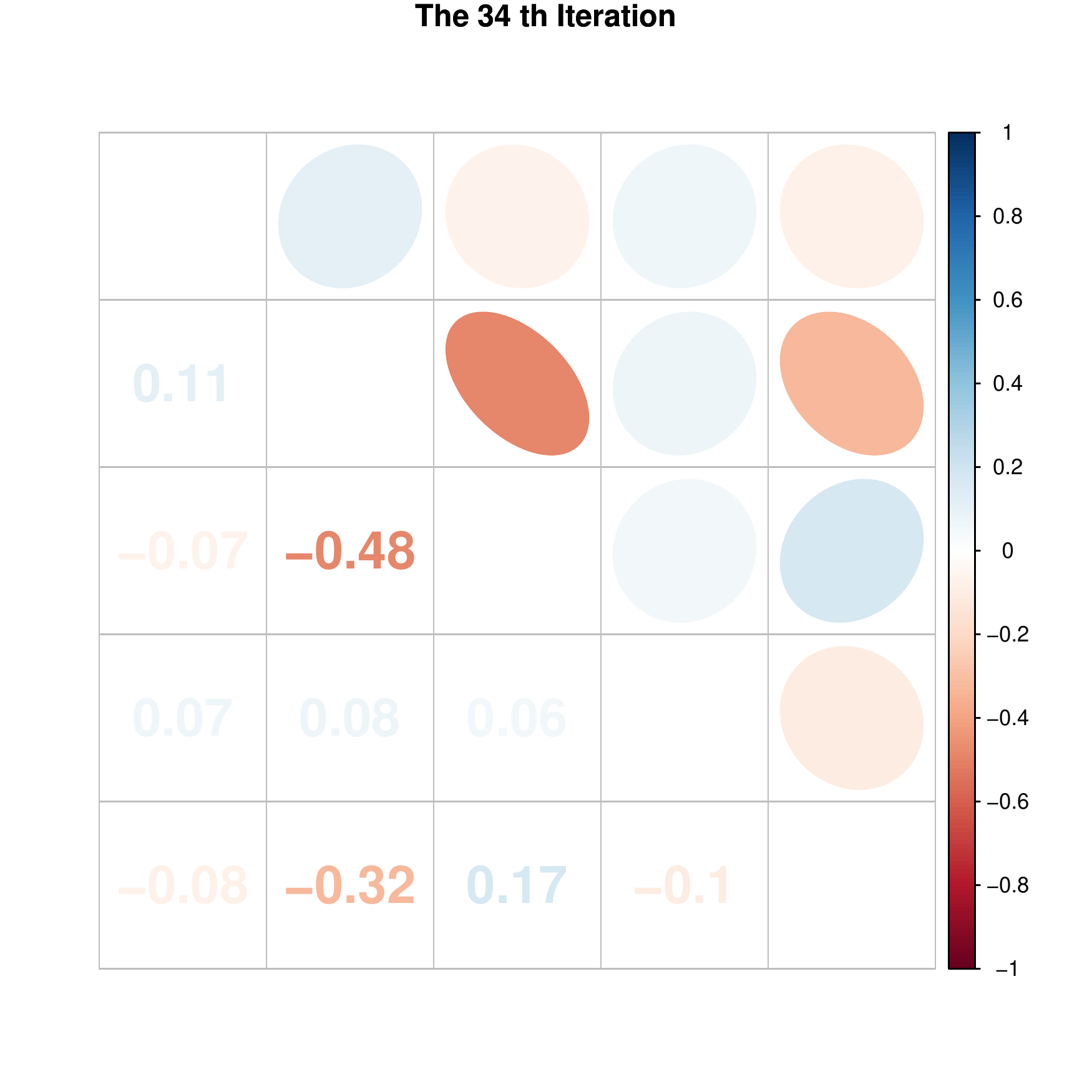}
\includegraphics[width=0.3\textwidth,height=0.2\textheight]{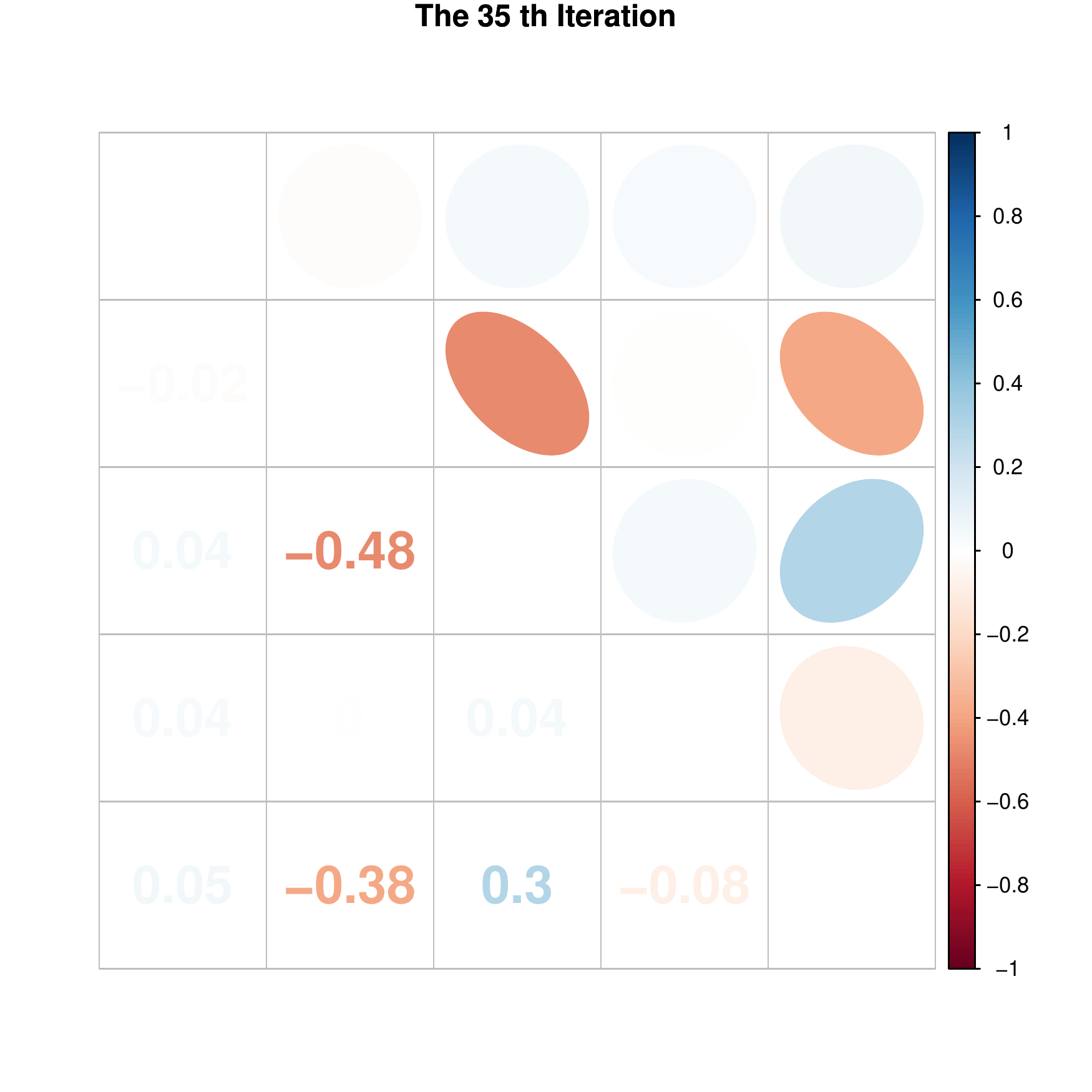}
\includegraphics[width=0.3\textwidth,height=0.2\textheight]{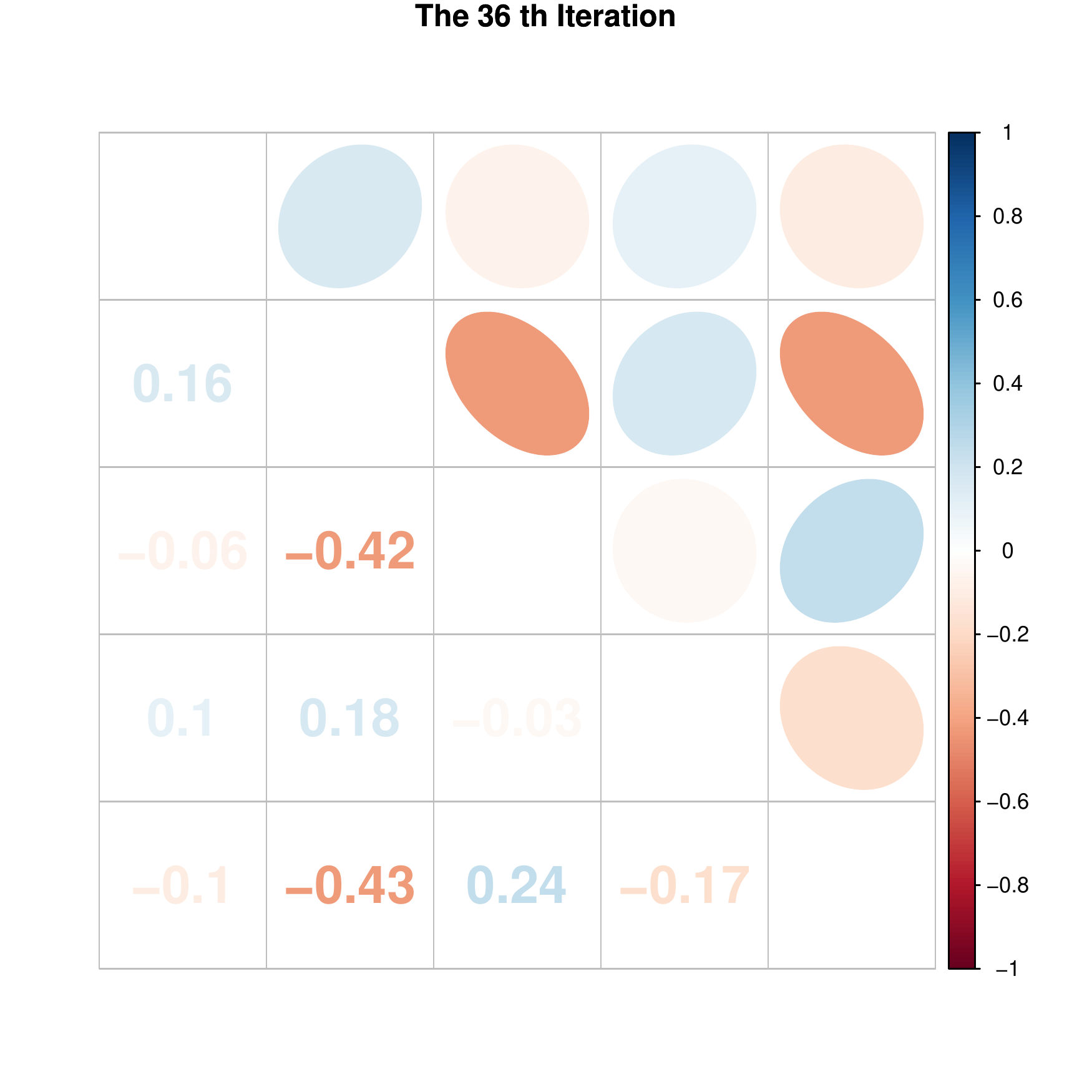}
\caption{Parameter Evolution Visualization. }
\end{figure}

\clearpage

\begin{figure}[h]
\centering
\includegraphics[width=0.3\textwidth,height=0.2\textheight]{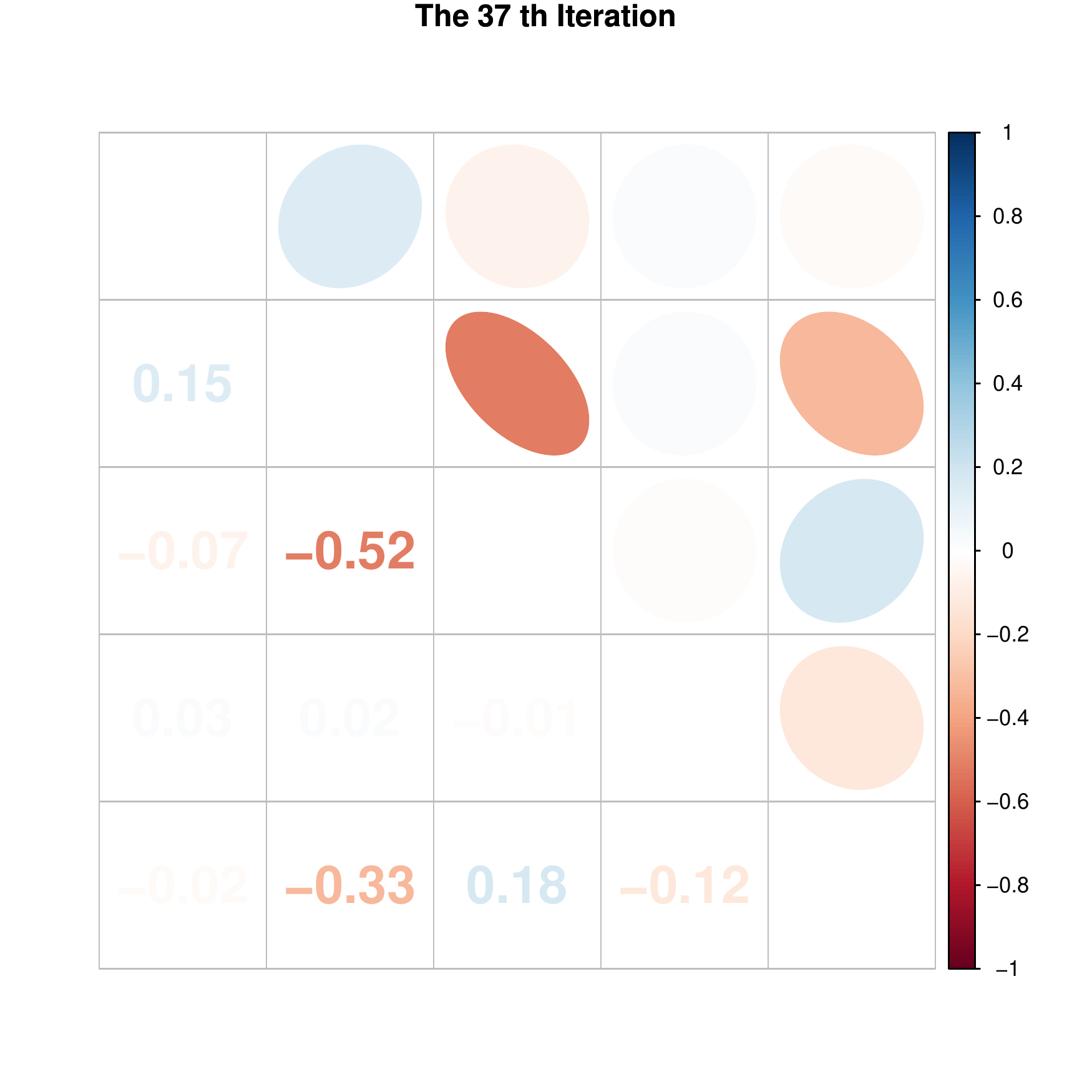}
\includegraphics[width=0.3\textwidth,height=0.2\textheight]{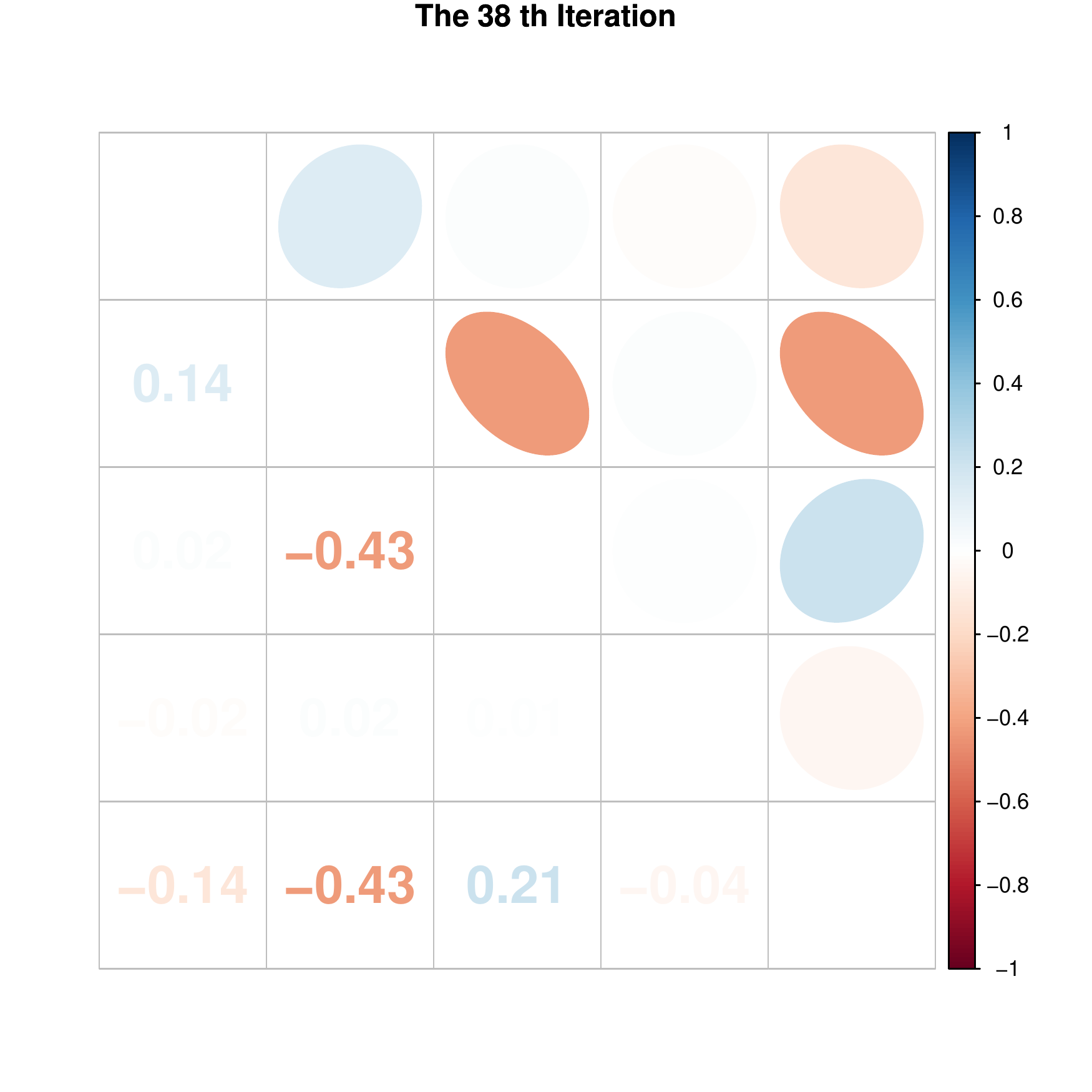}
\includegraphics[width=0.3\textwidth,height=0.2\textheight]{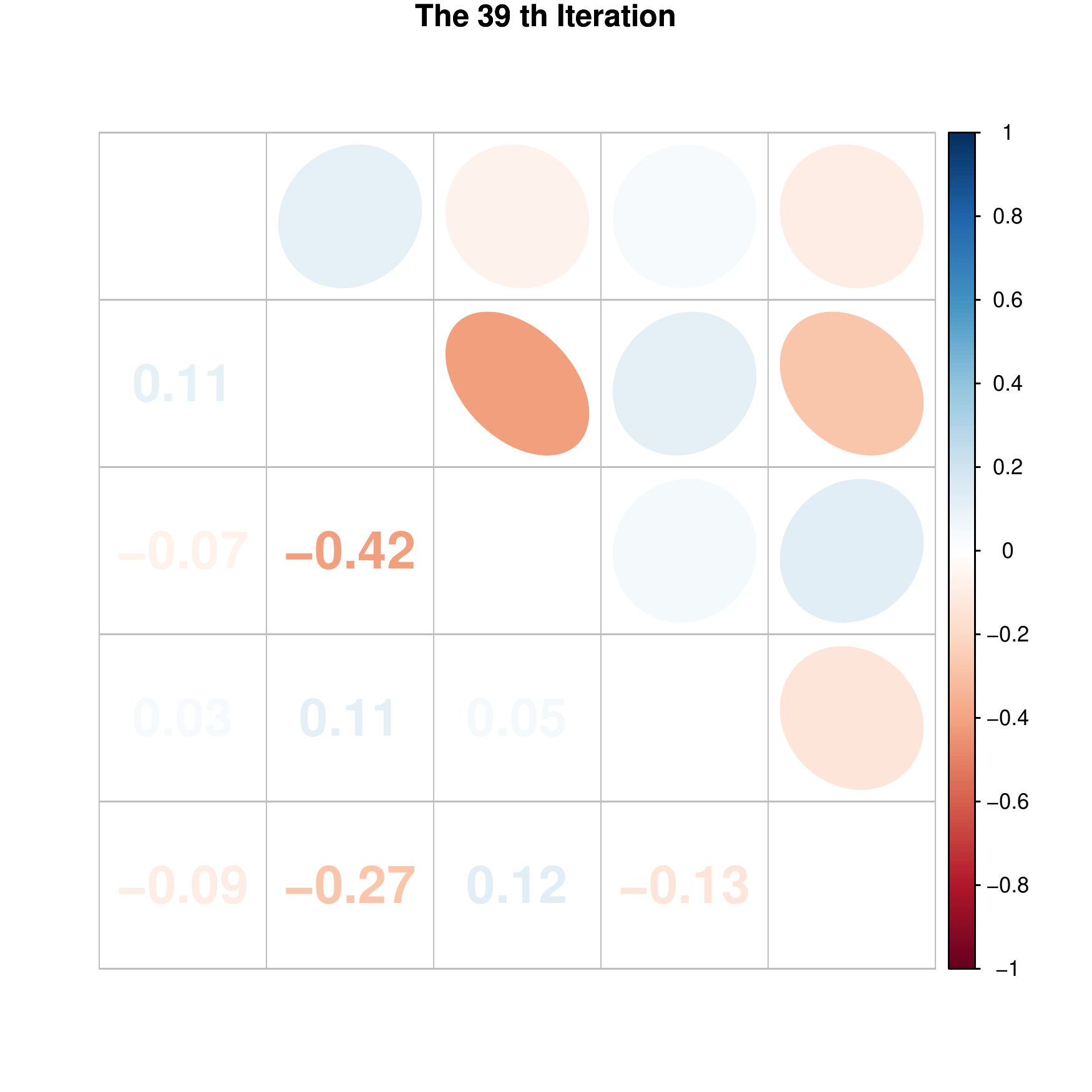}
\includegraphics[width=0.3\textwidth,height=0.2\textheight]{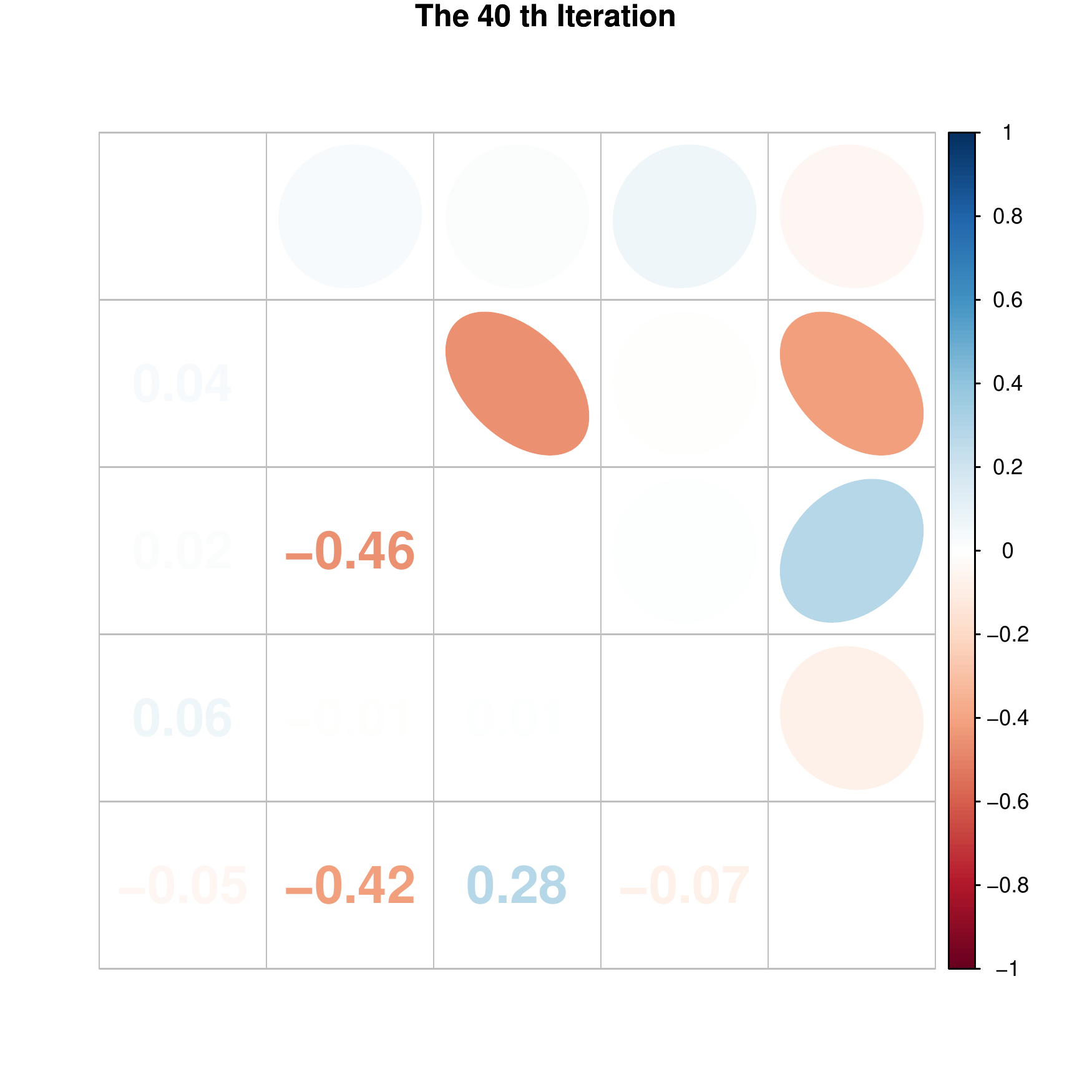}
\includegraphics[width=0.3\textwidth,height=0.2\textheight]{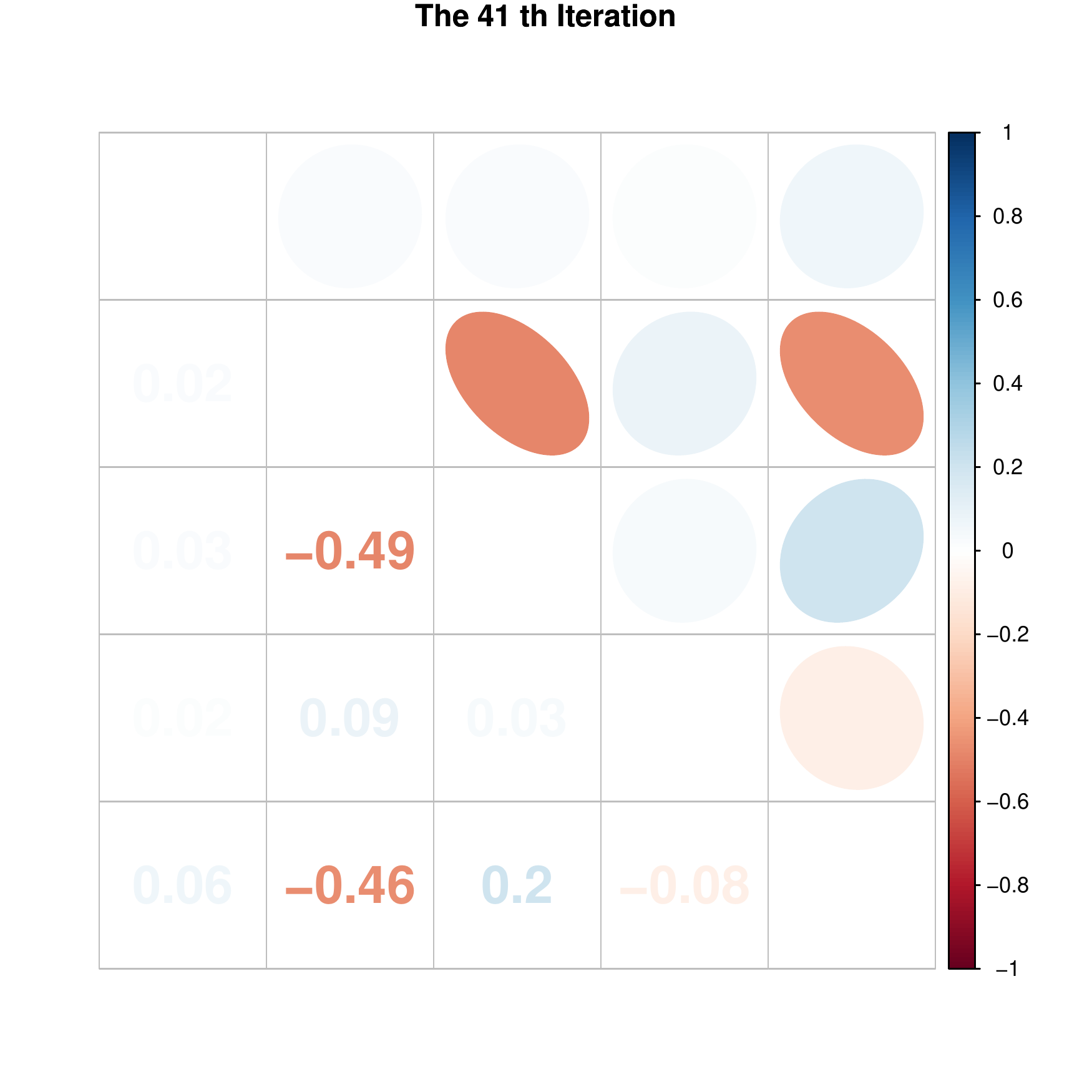}
\includegraphics[width=0.3\textwidth,height=0.2\textheight]{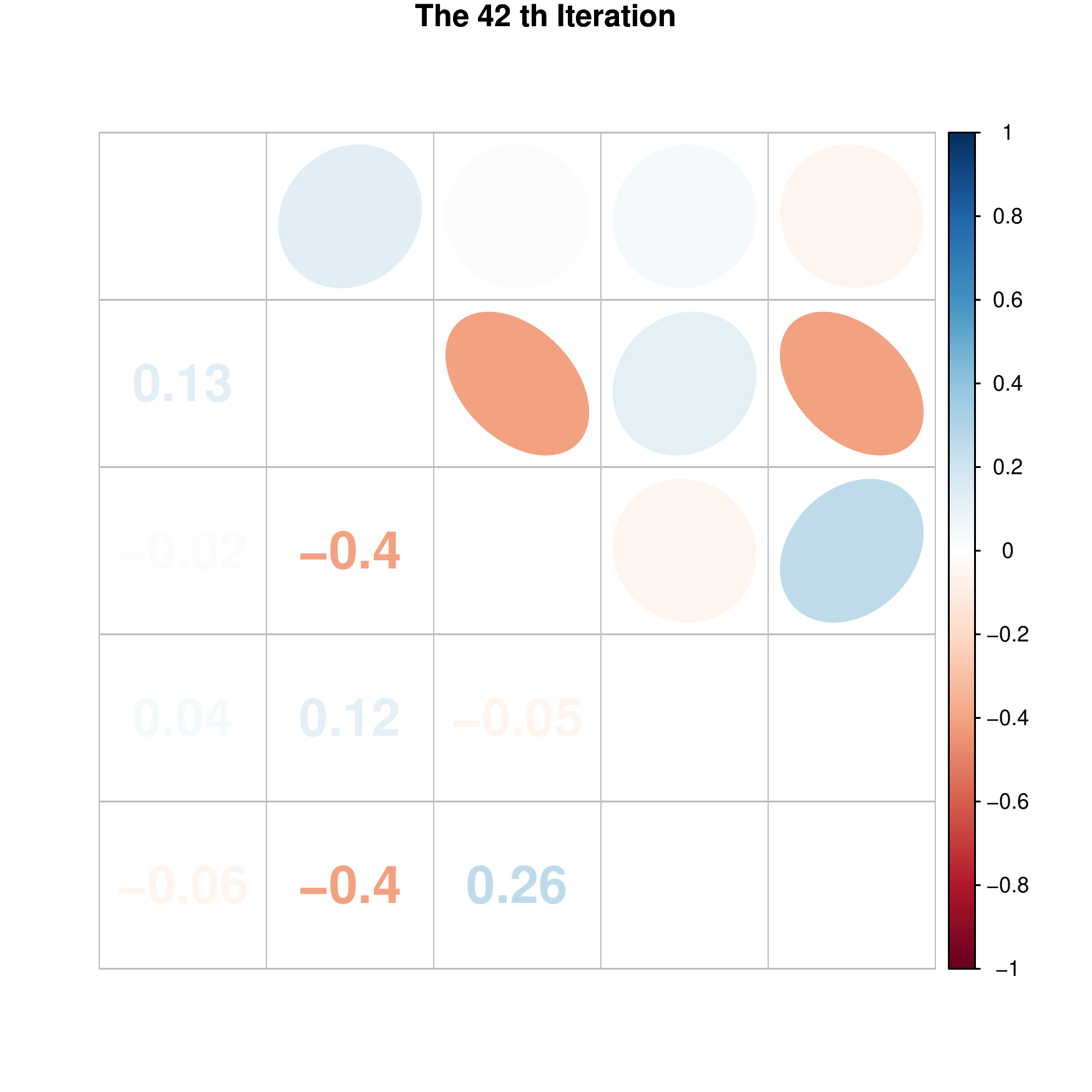}
\includegraphics[width=0.3\textwidth,height=0.2\textheight]{ggplots/paraEvolution/corMatrix42.pdf}
\includegraphics[width=0.3\textwidth,height=0.2\textheight]{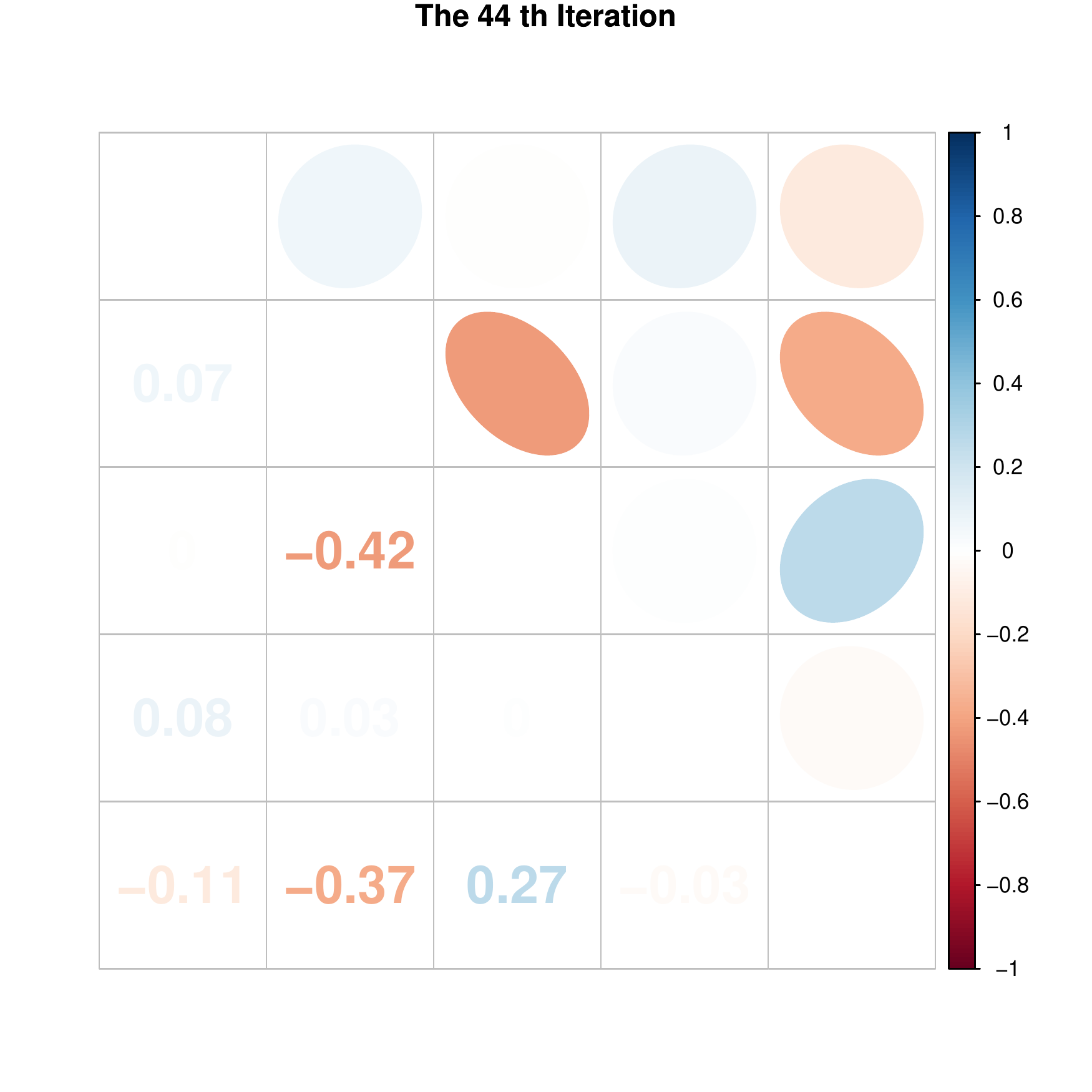}
\includegraphics[width=0.3\textwidth,height=0.2\textheight]{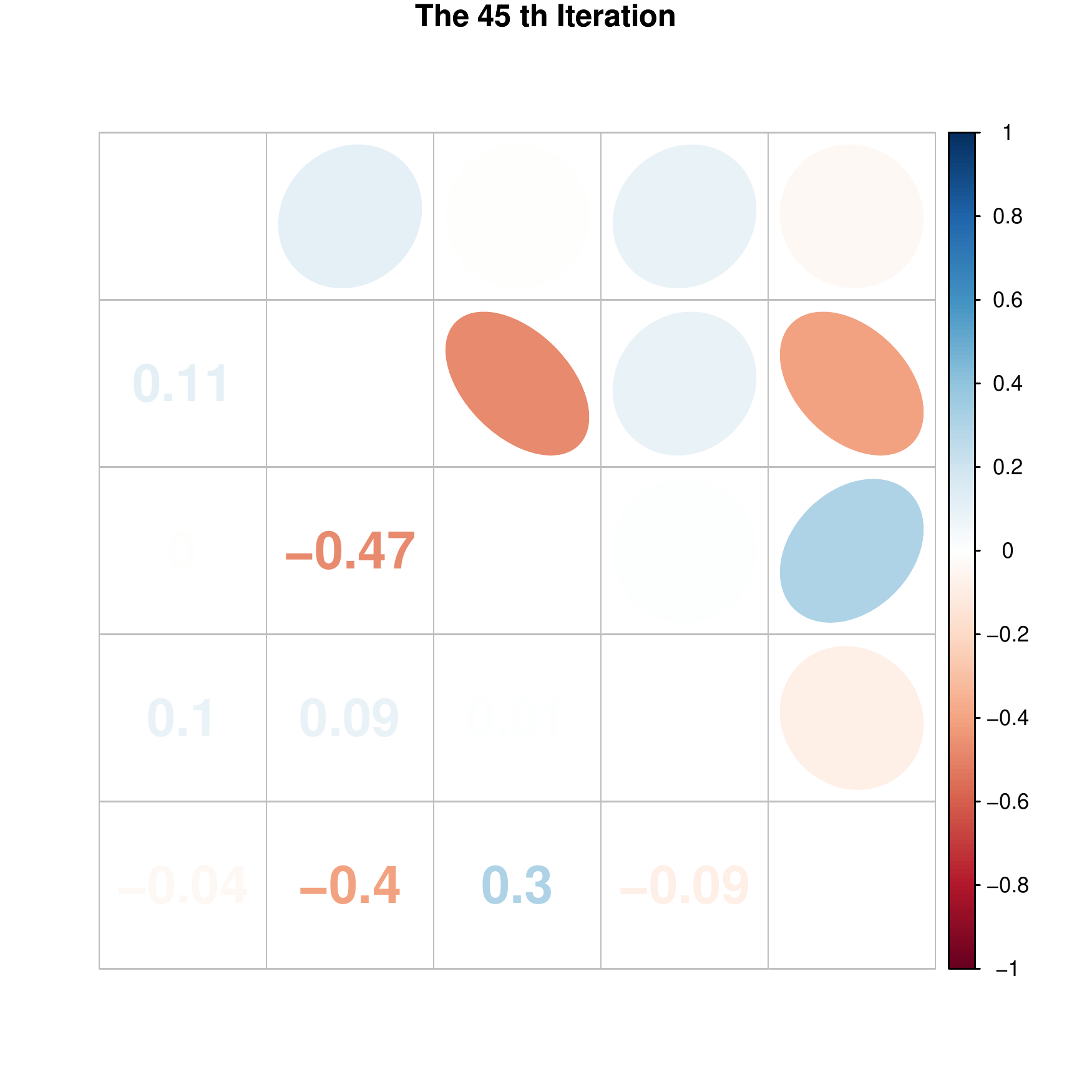}
\includegraphics[width=0.3\textwidth,height=0.2\textheight]{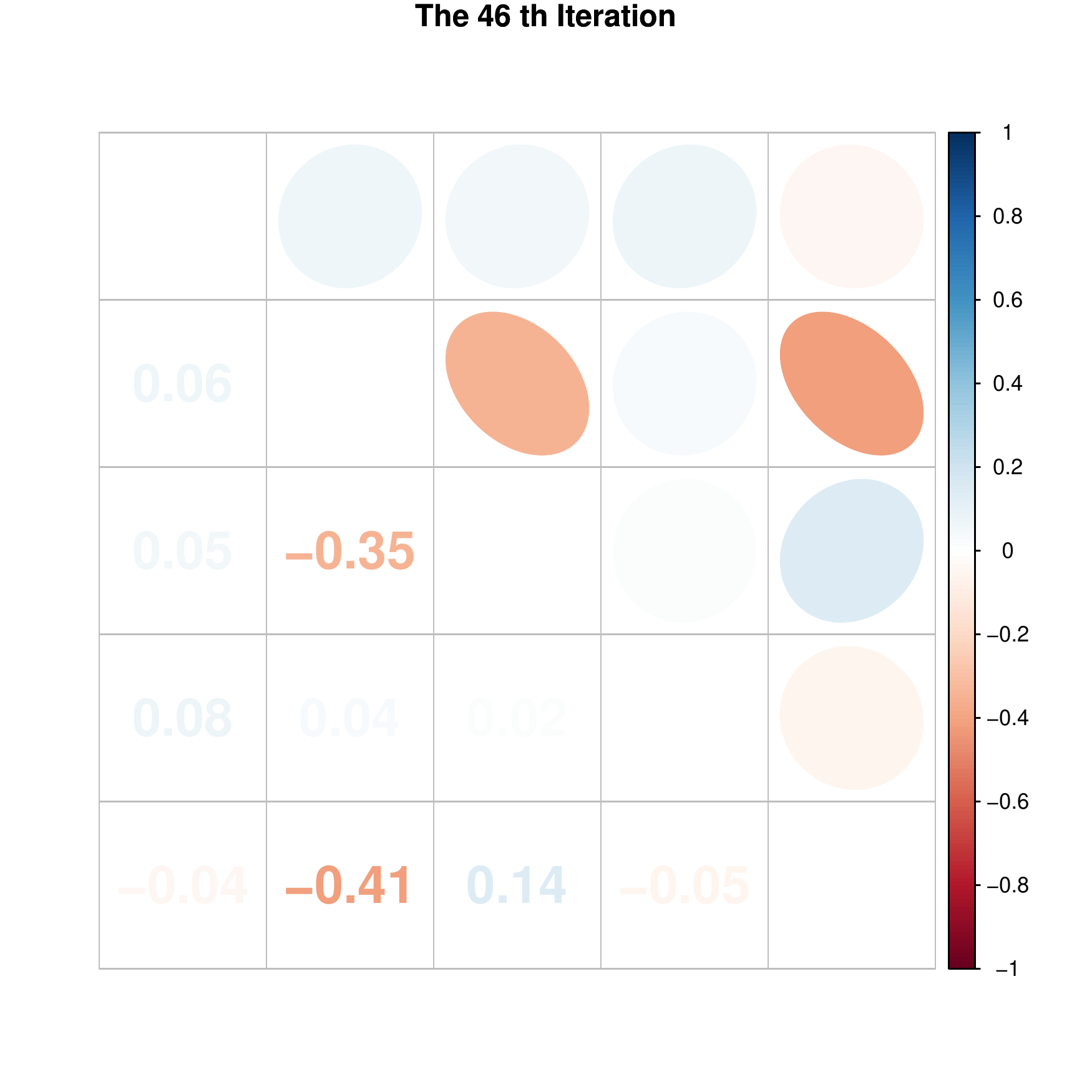}
\includegraphics[width=0.3\textwidth,height=0.2\textheight]{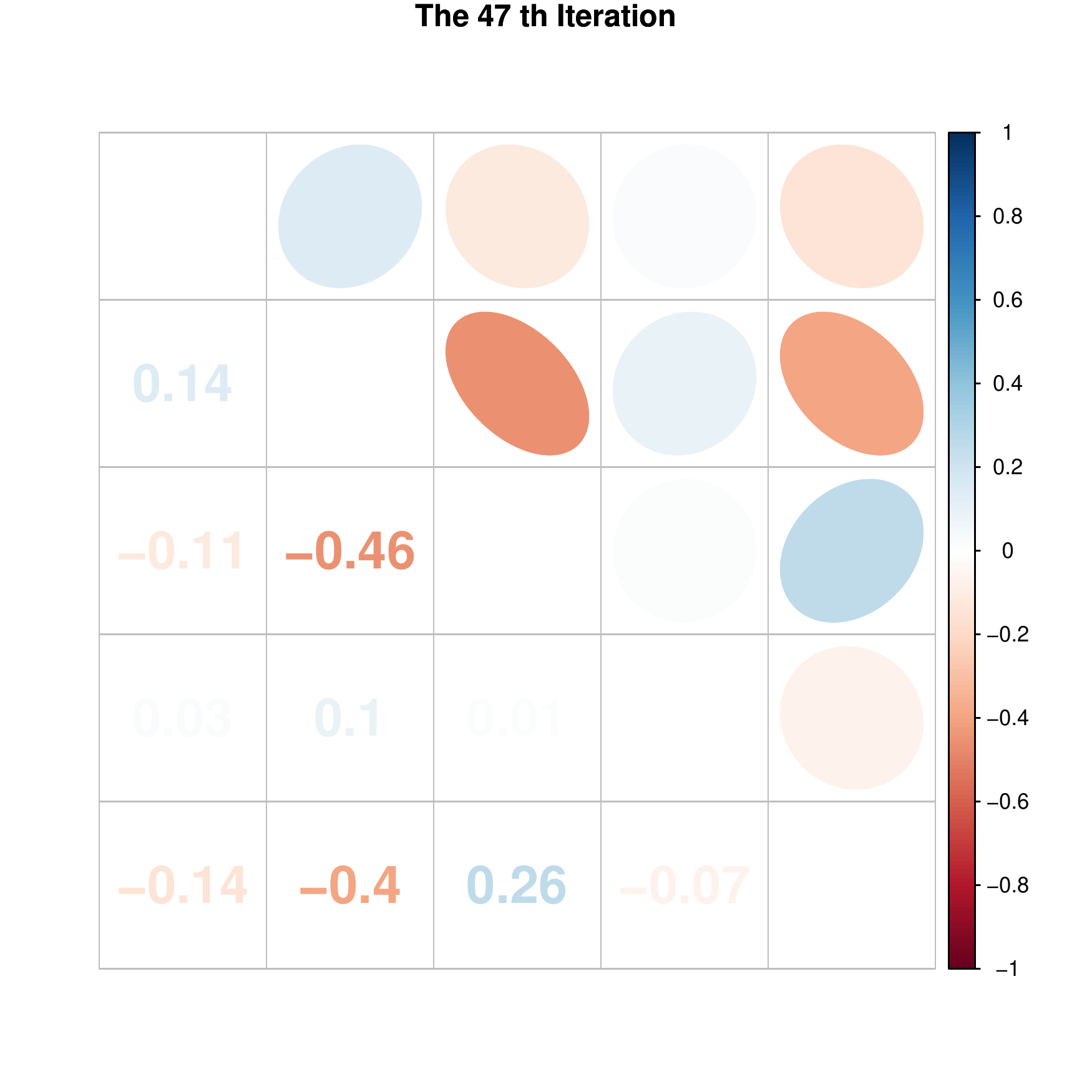}
\includegraphics[width=0.3\textwidth,height=0.2\textheight]{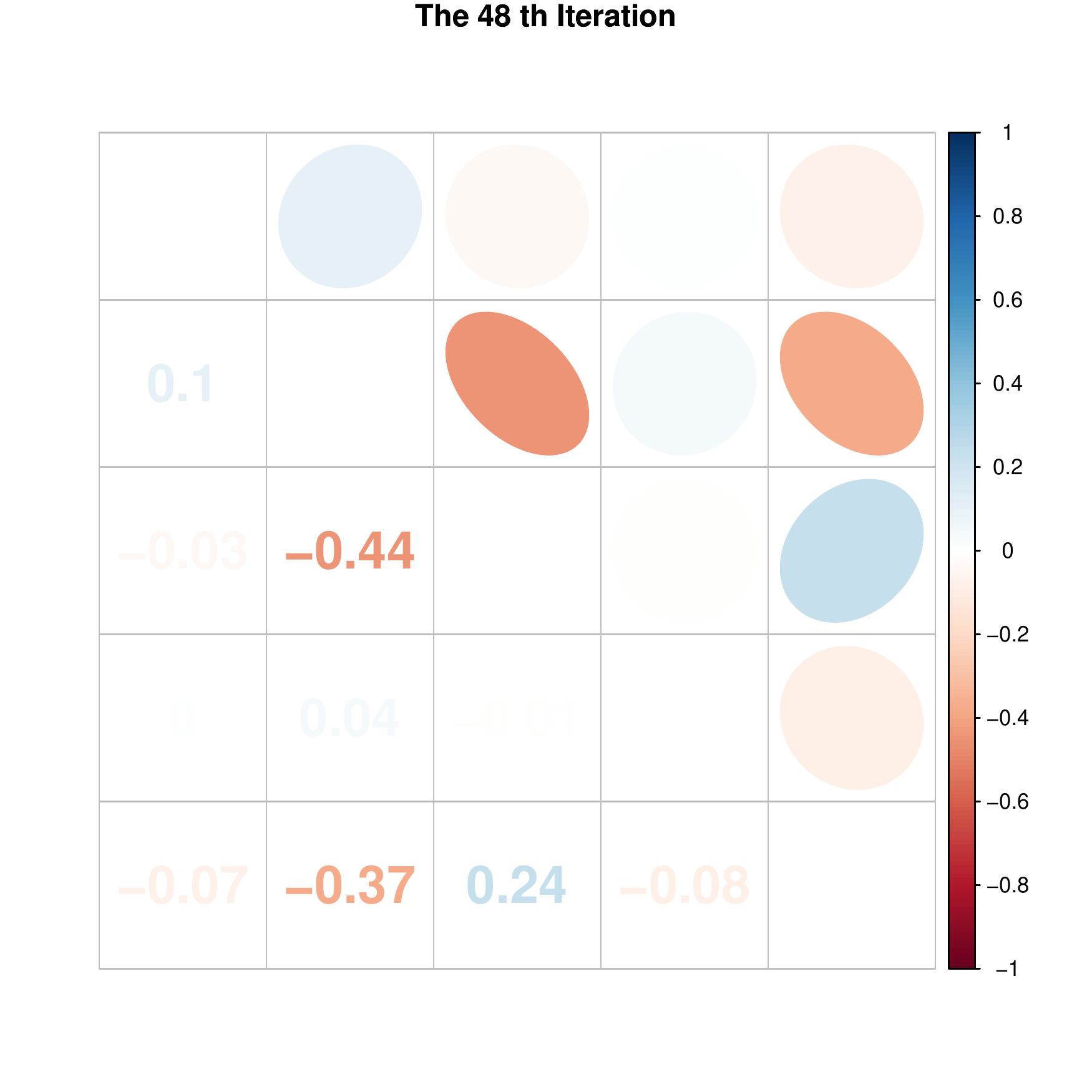}
\caption{Parameter Evolution Visualization. }
\end{figure}

\bibliographystyle{plain}

\begin{thebibliography}{10}

\bibitem{andrieu2008tutorial}
Christophe Andrieu and Johannes Thoms.
\newblock A tutorial on adaptive mcmc.
\newblock {\em Statistics and computing}, 18(4):343--373, 2008.

\bibitem{bartlett1951inverse}
Maurice~S Bartlett.
\newblock An inverse matrix adjustment arising in discriminant analysis.
\newblock {\em The Annals of Mathematical Statistics}, 22(1):107--111, 1951.

\bibitem{bedard2007weak}
Mylene B{\'e}dard.
\newblock Weak convergence of metropolis algorithms for non-iid target
  distributions.
\newblock {\em The Annals of Applied Probability}, pages 1222--1244, 2007.

\bibitem{beskos2009optimal}
Alexandros Beskos, Gareth Roberts, and Andrew Stuart.
\newblock Optimal scalings for local metropolis-hastings chains on nonproduct
  targets in high dimensions.
\newblock {\em The Annals of Applied Probability}, pages 863--898, 2009.

\bibitem{bodewig1959matrix}
E~Bodewig.
\newblock Matrix calculus, north, 1959.

\bibitem{box2011bayesian}
George~EP Box and George~C Tiao.
\newblock {\em Bayesian inference in statistical analysis}, volume~40.
\newblock John Wiley \& Sons, 2011.

\bibitem{carvalho2010particle}
Carlos~M Carvalho, Michael~S Johannes, Hedibert~F Lopes, Nicholas~G Polson,
  et~al.
\newblock Particle learning and smoothing.
\newblock {\em Statistical Science}, 25(1):88--106, 2010.

\bibitem{casella2004generalized}
George Casella, Christian~P Robert, and Martin~T Wells.
\newblock Generalized accept-reject sampling schemes.
\newblock {\em Lecture Notes-Monograph Series}, pages 342--347, 2004.

\bibitem{christen2005markov}
J~Andr{\'e}s Christen and Colin Fox.
\newblock Markov chain monte carlo using an approximation.
\newblock {\em Journal of Computational and Graphical statistics},
  14(4):795--810, 2005.

\bibitem{christen2010general}
J~Andr{\'e}s Christen, Colin Fox, et~al.
\newblock A general purpose sampling algorithm for continuous distributions
  (the t-walk).
\newblock {\em Bayesian Analysis}, 5(2):263--281, 2010.

\bibitem{deng2011generalization}
Chun~Yuan Deng.
\newblock A generalization of the sherman--morrison--woodbury formula.
\newblock {\em Applied Mathematics Letters}, 24(9):1561--1564, 2011.

\bibitem{dongarra2000guest}
Jack Dongarra and Francis Sullivan.
\newblock Guest editors’ introduction: The top 10 algorithms.
\newblock {\em Computing in Science \& Engineering}, 2(1):22--23, 2000.

\bibitem{einstein1956investigations}
A~Einstein.
\newblock Investigations on the theory of the brownian movement edited with
  notes by, r. f̈urth, translated by ad cowper dover publications, reprinted
  from (1905).
\newblock {\em Ann. Phys}, 17:549--560, 1956.

\bibitem{gelman2006prior}
Andrew Gelman et~al.
\newblock Prior distributions for variance parameters in hierarchical models
  (comment on article by browne and draper).
\newblock {\em Bayesian analysis}, 1(3):515--534, 2006.

\bibitem{gelman2008weakly}
Andrew Gelman, Aleks Jakulin, Maria~Grazia Pittau, and Yu-Sung Su.
\newblock A weakly informative default prior distribution for logistic and
  other regression models.
\newblock {\em The Annals of Applied Statistics}, pages 1360--1383, 2008.

\bibitem{gelman1996efficient}
Andrew Gelman, Gareth~O Roberts, Walter~R Gilks, et~al.
\newblock Efficient metropolis jumping rules.
\newblock {\em Bayesian statistics}, 5(599-608):42, 1996.

\bibitem{geman1984stochastic}
Stuart Geman and Donald Geman.
\newblock Stochastic relaxation, gibbs distributions, and the bayesian
  restoration of images.
\newblock {\em IEEE Transactions on pattern analysis and machine intelligence},
  (6):721--741, 1984.

\bibitem{geweke1989bayesian}
John Geweke.
\newblock Bayesian inference in econometric models using monte carlo
  integration.
\newblock {\em Econometrica: Journal of the Econometric Society}, pages
  1317--1339, 1989.

\bibitem{geyer1992practical}
Charles~J Geyer.
\newblock Practical markov chain monte carlo.
\newblock {\em Statistical science}, pages 473--483, 1992.

\bibitem{gilks1995markov}
Walter~R Gilks, Sylvia Richardson, and David Spiegelhalter.
\newblock {\em Markov chain Monte Carlo in practice}.
\newblock CRC press, 1995.

\bibitem{gong2016practical}
Lei Gong and James~M Flegal.
\newblock A practical sequential stopping rule for high-dimensional markov
  chain monte carlo.
\newblock {\em Journal of Computational and Graphical Statistics},
  25(3):684--700, 2016.

\bibitem{graves2011automatic}
Todd~L Graves.
\newblock Automatic step size selection in random walk metropolis algorithms.
\newblock {\em arXiv preprint arXiv:1103.5986}, 2011.

\bibitem{haario1999adaptive}
Heikki Haario, Eero Saksman, and Johanna Tamminen.
\newblock Adaptive proposal distribution for random walk metropolis algorithm.
\newblock {\em Computational Statistics}, 14(3):375--396, 1999.

\bibitem{hammersley1964percolation}
John~M Hammersley and DC~Handscomb.
\newblock Percolation processes.
\newblock In {\em Monte Carlo Methods}, pages 134--141. Springer, 1964.

\bibitem{hastings1970monte}
W~Keith Hastings.
\newblock Monte carlo sampling methods using markov chains and their
  applications.
\newblock {\em Biometrika}, 57(1):97--109, 1970.

\bibitem{jaynes1983papers}
Edwin~T Jaynes.
\newblock Papers on probability.
\newblock {\em Statistics and Statistical Physics}, 1983.

\bibitem{jeffries1961theory}
Harold Jeffries.
\newblock Theory of probability, 1961.

\bibitem{kass1998markov}
Robert~E Kass, Bradley~P Carlin, Andrew Gelman, and Radford~M Neal.
\newblock Markov chain monte carlo in practice: a roundtable discussion.
\newblock {\em The American Statistician}, 52(2):93--100, 1998.

\bibitem{kijima1997markov}
Masaaki Kijima.
\newblock {\em Markov processes for stochastic modeling}, volume~6.
\newblock CRC Press, 1997.

\bibitem{kitagawa1998self}
Genshiro Kitagawa.
\newblock A self-organizing state-space model.
\newblock {\em Journal of the American Statistical Association}, pages
  1203--1215, 1998.

\bibitem{lindley1972bayes}
Dennis~V Lindley and Adrian~FM Smith.
\newblock Bayes estimates for the linear model.
\newblock {\em Journal of the Royal Statistical Society. Series B
  (Methodological)}, pages 1--41, 1972.

\bibitem{liu2001combined}
Jane Liu and Mike West.
\newblock Combined parameter and state estimation in simulation-based
  filtering.
\newblock In {\em Sequential Monte Carlo methods in practice}, pages 197--223.
  Springer, 2001.

\bibitem{lopes2011particle}
Hedibert~F Lopes and Ruey~S Tsay.
\newblock Particle filters and bayesian inference in financial econometrics.
\newblock {\em Journal of Forecasting}, 30(1):168--209, 2011.

\bibitem{martino2010generalized}
Luca Martino and Joaqu{\'\i}n M{\'\i}guez.
\newblock Generalized rejection sampling schemes and applications in signal
  processing.
\newblock {\em Signal Processing}, 90(11):2981--2995, 2010.

\bibitem{mathew2012bayesian}
Boby Mathew, AM~Bauer, P~Koistinen, TC~Reetz, J~L{\'e}on, and
  MJ~Sillanp{\"a}{\"a}.
\newblock Bayesian adaptive markov chain monte carlo estimation of genetic
  parameters.
\newblock {\em Heredity}, 109(4):235, 2012.

\bibitem{medova2008bayesian}
Elena Medova.
\newblock {\em Bayesian Analysis and Markov Chain Monte Carlo Simulation}.
\newblock Wiley Online Library, 2008.

\bibitem{metropolis1953equation}
Nicholas Metropolis, Arianna~W Rosenbluth, Marshall~N Rosenbluth, Augusta~H
  Teller, and Edward Teller.
\newblock Equation of state calculations by fast computing machines.
\newblock {\em The journal of chemical physics}, 21(6):1087--1092, 1953.

\bibitem{muller1991generic}
Peter M{\"u}ller.
\newblock {\em A generic approach to posterior integration and Gibbs sampling}.
\newblock Purdue University, Department of Statistics, 1991.

\bibitem{polson2008practical}
Nicholas~G Polson, Jonathan~R Stroud, and Peter M{\"u}ller.
\newblock Practical filtering with sequential parameter learning.
\newblock {\em Journal of the Royal Statistical Society: Series B (Statistical
  Methodology)}, 70(2):413--428, 2008.

\bibitem{robert2004monte}
Christian~P Robert.
\newblock {\em Monte carlo methods}.
\newblock Wiley Online Library, 2004.

\bibitem{roberts1997weak}
Gareth~O Roberts, Andrew Gelman, Walter~R Gilks, et~al.
\newblock Weak convergence and optimal scaling of random walk metropolis
  algorithms.
\newblock {\em The annals of applied probability}, 7(1):110--120, 1997.

\bibitem{roberts2001optimal}
Gareth~O Roberts, Jeffrey~S Rosenthal, et~al.
\newblock Optimal scaling for various metropolis-hastings algorithms.
\newblock {\em Statistical science}, 16(4):351--367, 2001.

\bibitem{sherlock2013optimal}
Chris Sherlock.
\newblock Optimal scaling of the random walk metropolis: general criteria for
  the 0.234 acceptance rule.
\newblock {\em Journal of Applied Probability}, 50(1):1--15, 2013.

\bibitem{sherlock2010random}
Chris Sherlock, Paul Fearnhead, and Gareth~O Roberts.
\newblock The random walk metropolis: linking theory and practice through a
  case study.
\newblock {\em Statistical Science}, pages 172--190, 2010.

\bibitem{sherlock2016adaptive}
Chris Sherlock, Andrew Golightly, and Daniel~A Henderson.
\newblock Adaptive, delayed-acceptance mcmc for targets with expensive
  likelihoods.
\newblock {\em Journal of Computational and Graphical Statistics},
  (just-accepted), 2016.

\bibitem{sherlock2009optimal}
Chris Sherlock, Gareth Roberts, et~al.
\newblock Optimal scaling of the random walk metropolis on elliptically
  symmetric unimodal targets.
\newblock {\em Bernoulli}, 15(3):774--798, 2009.

\bibitem{sherlock2015efficiency}
Chris Sherlock, Alexandre Thiery, and Andrew Golightly.
\newblock Efficiency of delayed-acceptance random walk metropolis algorithms.
\newblock {\em arXiv preprint arXiv:1506.08155}, 2015.

\bibitem{sherman1950adjustment}
Jack Sherman and Winifred~J Morrison.
\newblock Adjustment of an inverse matrix corresponding to a change in one
  element of a given matrix.
\newblock {\em The Annals of Mathematical Statistics}, 21(1):124--127, 1950.

\bibitem{smith1973general}
Adrian~FM Smith.
\newblock A general bayesian linear model.
\newblock {\em Journal of the Royal Statistical Society. Series B
  (Methodological)}, pages 67--75, 1973.

\bibitem{smith1993bayesian}
Adrian~FM Smith and Gareth~O Roberts.
\newblock Bayesian computation via the gibbs sampler and related markov chain
  monte carlo methods.
\newblock {\em Journal of the Royal Statistical Society. Series B
  (Methodological)}, pages 3--23, 1993.

\bibitem{sokal1997monte}
A~Sokal.
\newblock Monte carlo methods in statistical mechanics: foundations and new
  algorithms.
\newblock In {\em Functional integration}, pages 131--192. Springer, 1997.

\bibitem{storvik2002particle}
Geir Storvik.
\newblock Particle filters for state-space models with the presence of unknown
  static parameters.
\newblock {\em IEEE Transactions on signal Processing}, 50(2):281--289, 2002.

\bibitem{stroud2007sequential}
Jonathan~R Stroud and Thomas Bengtsson.
\newblock Sequential state and variance estimation within the ensemble kalman
  filter.
\newblock {\em Monthly Weather Review}, 135(9):3194--3208, 2007.

\bibitem{stroud2016bayesian}
Jonathan~R Stroud, Matthias Katzfuss, and Christopher~K Wikle.
\newblock A bayesian adaptive ensemble kalman filter for sequential state and
  parameter estimation.
\newblock {\em arXiv preprint arXiv:1611.03835}, 2016.

\bibitem{tandeo2011linear}
Pierre Tandeo, Pierre Ailliot, and Emmanuelle Autret.
\newblock Linear gaussian state-space model with irregular sampling:
  application to sea surface temperature.
\newblock {\em Stochastic Environmental Research and Risk Assessment},
  25(6):793--804, 2011.

\bibitem{tierney1994markov}
Luke Tierney.
\newblock Markov chains for exploring posterior distributions.
\newblock {\em the Annals of Statistics}, pages 1701--1728, 1994.

\bibitem{vaughan2015goodness}
Audrey Vaughan.
\newblock Goodness of fit test: Ornstein-uhlenbeck process.
\newblock 2015.

\bibitem{vieira2016online}
Rui Vieira and Darren~J Wilkinson.
\newblock Online state and parameter estimation in dynamic generalised linear
  models.
\newblock {\em arXiv preprint arXiv:1608.08666}, 2016.

\bibitem{woodbury1950inverting}
Max~A Woodbury.
\newblock Inverting modified matrices.
\newblock {\em Memorandum report}, 42(106):336, 1950.

\end{thebibliography}

\end{document}